\documentclass[structabstract]{aa}  %A&A version
\usepackage{amstext,amsmath,mathtools}
\usepackage{graphicx}
\usepackage{subfigure}
\usepackage{textcomp}
\usepackage{txfonts}
\usepackage{natbib}
\usepackage{time}
\usepackage{subfig}
\usepackage{color}
\usepackage{url}
\usepackage{natbib}
\bibpunct{(}{)}{;}{a}{}{,}
\usepackage{overpic}
\usepackage{fancyvrb}
\usepackage{multirow}
\bibpunct{(}{)}{;}{a}{}{,}
\def\Blos{\ensuremath{B_\text{LOS}}}
\def\vlos{\ensuremath{v_\text{LOS}}}
\def\Ic{\ensuremath{I_\text{c}}}
\def\Icont{\ensuremath{I_\text{c}/\langle I_\text{c} \rangle}}
\def\Imin{\ensuremath{I_\text{min}}}
\def\mps{m\,s$^{-1}$}

\begin{document}

\title{Small-scale convection signatures associated with strong plage solar magnetic field}

\author{G. Narayan\inst{1,2} \and G.B. Scharmer\inst{1,2}}

\institute{Institute for Solar Physics, Royal Swedish Academy of
  Sciences, AlbaNova University Center, 106\,91 Stockholm, Sweden
  \\
  \email{[gautam;scharmer]@astro.su.se} \and Stockholm Observatory,
  Dept. of Astronomy, Stockholm University,
  AlbaNova University Center, 106\,91 Stockholm, Sweden\\
} \date{Draft: \now \today}

\abstract
{}
  {Solar convection in strong plage, having a magnetic field that is vertical and strong over \emph{extended} regions, but much weaker than in umbrae of large sunspots, has so far not been well studied. This is mostly due to lack of spectropolarimetric data at adequate spatial resolution. The combination of a large solar telescope, such as the Swedish 1-m Solar Telescope, adaptive optics, powerful image reconstruction techniques and a high-fidelity imaging spectropolarimeter is capable of producing such data. In this work, we study and quantify properties of strong-field small-scale convection and compare observed properties with those predicted by numerical simulations.}
  {We analyze spectropolarimetric 630.25~nm data from a unipolar ephemeral region near sun center. We use line-of-sight velocities and magnetic field measurements obtained with Milne-Eddington inversion techniques along with measured continuum intensities and Stokes $V$ amplitude asymmetry at a spatial resolution of 0\farcs15 to establish statistical relations between the measured quantities. We also study these properties for different types of distinct magnetic features, such as micropores, bright points, ribbons, flowers and strings.}
  {We present the first direct observations of a small-scale granular magneto-convection pattern within extended regions of strong (more than 600~G average) magnetic field. Along the \emph{boundaries} of the flux concentrations we see mostly downflows and asymmetric Stokes $V$ profiles, consistent with synthetic line profiles calculated from MHD simulations. We note the frequent occurrence of \emph{bright downflows} along these boundaries. In the \emph{interior} of the flux concentrations, we observe an up/down flow pattern that we identify as small-scale magnetoconvection, appearing similar to that of field-free granulation but with scales 4 times smaller. Measured RMS velocities are 70\% of those of nearby field-free granulation, even though the average radiative flux is not reduced. The interiors of these flux concentrations are dominated by upflows.}
{}


   \maketitle

\section{Introduction}
\label{sec:introduction}

The interaction of convection with an existing initially weak magnetic field is the primary step in the production of the strong small-scale flux concentration features observed in the solar photosphere. When the average field strength is small enough, the horizontal flow component of convection sweeps the magnetic field to the intergranular lanes, leading to expulsion of field from the center of granules and flux concentrations at the intergranular lanes. These flux concentrations reduce convective heat transfer, leading to super adiabatic cooling and collapse into a strong-field state \citep{1978ApJ...221..368P,1979SoPh...61..363S}, referred to traditionally as a flux tube or, when more appropriate, as a flux sheet. Once formed, a flux tube remains in a strong-field state for some time, even though it evolves continuously as a result of surrounding granular convection.  If the filling factor of these flux tubes is sufficiently small, the result is a granular pattern that is relatively unaffected by the flux tubes but with adjacent downflows driven by enhanced radiative cooling through the optically thin flux tubes. The appearance of flux tubes, in particular when observed away from disk center or in spectral lines, are as ``bright points'' or faculae when the flux concentrations are larger. The formation and evolution of such isolated flux concentrations are now well understood from observations using the Swedish 1-m Solar Telescope \cite[SST;][]{scharmer03new} combined with 3D MHD simulations \citep{keller04origin,carlsson04observational,2005A&A...430..691S}.

In contrast to bright small-scale magnetic features, the largest magnetic flux concentrations, sunspots, inhibit most of the convective energy flux from below and lead to intermittent nearly field-free convection in the form of narrow plumes, observed as umbral dots, when the field is strong and nearly vertical \citep{2006ApJ...641L..73S}.

In the present paper, we discuss observations of magnetic structures with scales corresponding to bright points and larger but with field strengths lower than those of sunspots. These small scale structures
include abnormal granulation \citep{1973SoPh...33..281D,1992ApJ...393..782T}, micropores and other
distinct small-scale magnetic features that have been dubbed ribbons, flowers and strings \citep{berger04solar}. The impact of the magnetic field on the convection within and adjacent to these magnetic field concentrations has so far not been well studied, mostly due to lack of adequate data at sufficient spatial resolution. By means of spectropolarimetric data and inversion techniques, we describe
convective signatures separately for abnormal granulation and the various types of small-scale magnetic structures described in \citet{berger04solar} and compare these observations to existing 3D MHD simulations.

\section{Observations and data reduction}
\label{sec:observ-data-reduct}

\subsection{Instrumentation}
\label{sec:instrumentation}

The observational data were recorded with the SST and the CRisp
Imaging SPectropolarimeter \citep[CRISP; ][]{scharmer08crisp}. CRISP
is a telecentric dual Fabry--Perot etalon system used in combination
with two nematic liquid crystal variable retarders (LCVRs) and a
polarizing beam splitter. CRISP uses a unique combination of a
high-resolution, high-reflectivity etalon with a low-resolution
\emph{low-reflectivity} etalon. This produces a high-fidelity
transmission profile with high transmission and small variations of
its \emph{shape} over the field-of-view (FOV)
\citep{scharmer06comments}. The polarizing beam splitter separates the
beam from CRISP into two orthogonally polarized states, allowing
strong suppression of seeing-induced cross-talk from Stokes $I$ to
$Q$, $U$ and $V$. The two linearly polarized images are recorded with
separate CCD's and cover about 70\arcsec$\times$70\arcsec\ at an image
scale of 0\farcs071/pixel (the image scale was changed to
0\farcs059/pixel in 2009). A third CCD is used to record broad-band
images through the prefilter of CRISP, providing needed support for
reconstruction and alignment of the narrowband images obtained at
different wavelengths and polarization states (see below). Images
recorded through the three CCD's are exposed simultaneously by means
of a rotating chopper. To improve image quality and image restoration,
short exposure times are used. The frame rate was set by the chopper
at 36 Hz, corresponding to an exposure time of about 17 ms and a
``dark'' (CCD readout) time of 11 ms. To compensate for the relatively
poor signal-to-noise of each short exposure frame, a large number of
frames are needed at each wavelength and polarization state.

\subsection{Observations}
\label{sec:observations}

We observed a unipolar (within the FOV) ephemeral region, located
close to disk center (heliocentric coordinates S9 E11) on 26 May 2008.
CRISP was used to record images at 9 wavelengths between $-19.2$ and
$+19.2$~pm in steps of 4.8~pm from the line center of the \ion{Fe}{i}
630.25~nm line and also at an adjacent continuum wavelength. At each
wavelength position images were obtained with 4 different LCVR tunings
to allow measurements of the complete Stokes vector. For each
wavelength and LCVR state, 16 images were recorded per camera. Each
sequence processed thus consists of up to 640 images per CCD (1920
images in total) but in practice missing or corrupted images reduced
the number of useful frames per camera to about 630. Each data set
corresponds to a wall clock time of approximately 21 s, including time
for tuning CRISP. A time series of about 40 minutes was recorded in
reasonably good seeing conditions during which the SST adaptive optics
system \citep{scharmer03adaptive} indicated a lock rate between 50\%
and 90\% most of the time. In this paper, we discuss only one of the
best snapshots recorded.

\subsection{Data reduction}
\label{sec:data-reduction}

The images from each 21~sec scan were processed as a single MOMFBD
data set using the image reconstruction method developed by
\citet{lofdahl02multi-frame} and implemented by \citet{noort05solar}.
In addition to compensating residual seeing partially compensated by
the AO system, this processing produces restored images at different
wavelengths and polarization states that are accurately co-aligned
with respect to each other. The processed images were demodulated and
corrected for telescope polarization, using the telescope polarization
model of \citet{selbing05sst}. This model typically gives residual cross-talks, or errors in the M\"uller matrix elements, that are on the order of 1\%, the dominant effect of which is cross-talk from Stokes $I$ to $Q$, $U$, and $V$. This cross-talk was compensated for with the aid of the Stokes images recorded in the continuum. There is most likely also cross-talk from $V$, contaminating the $Q$ and $U$ profiles (and vice-versa). However, due to the proximity of this region to sun center, the $Q$ and $U$ profiles are in any case very weak and neither provide meaningful information about field inclinations nor field azimuths. We therefore did not attempt to compensate for cross-talk from $V$ to $Q$ and $U$ and in the following discuss only the Stokes $V$ profiles and the associated LOS component of the magnetic field.

To visualize (by enhanced contrast) the various small-scale magnetic structures discussed in Sect.~\ref{sec:small-scale-velocity}, we also calculated maps of the 630.25~nm line minimum intensity. This was done by fitting the center part of the Stokes $I$ line profiles to a second-order polynomial and from the coefficients of these fits calculating the minimum intensity at each pixel.

\subsection{Inversions}
\label{sec:inversions}

Inversions of the data were carried out using the Milne--Eddington
inversion code Helix \citep{2004A&A...414.1109L}. The inversions
involve calculation of the synthetic Stokes profiles based on
Unno--Rachkovsky solutions of the polarized radiative transfer
equations with the fitting and optimization controlled by the genetic algorithm Pikaia \citep{1995ApJS..101..309C}. To account for the limited spectral resolution of CRISP, about 6.4~pm at 630~nm, the synthetic profiles were convolved with the theoretical CRISP spectral transmission profile at each iteration. The following free parameters were solved for with Helix:
The field strength $B$, the azimuth angle, the inclination angle
$\gamma$, the line-of-sight (LOS) velocity $\vlos$, the Doppler width and the gradient of the source function. The fixed parameters were the line-strength (ratio of line to continuum opacity), set to 16, and the damping parameter, set to 1. This choice was made based on numerous tests with Helix using the data and inversions discussed by \citet{scharmer08crisp}. Fits were made with different weights for the 4 components of the Stokes vector. Following the suggestion of A.~Lagg (private communication), we used weights for $I$ and $V$ that were twice as large as for $Q$ and $U$. Extensive tests with Helix and the CRISP data discussed by \citet{scharmer08crisp} failed to produce stable estimates of the magnetic filling factor $f$, but demonstrated that $f$ multiplied by the field strength was consistently close to what was obtained by setting $f$ to unity. Also for the present inversions, we therefore set $f$ to unity, although already the limited spatial resolution implies the existence of spatial straylight for unresolved magnetic structures. In addition, a diffuse component of straylight must exist to explain the measured granulation contrast (see Sect.~\ref{sec:data-quality-fits}). The measured LOS component of the magnetic field \Blos, discussed in the following sections of the paper, therefore corresponds to a magnetic flux density or a (poorly defined) spatial average. This measured average most likely is an underestimate in magnitude near the centers of magnetic flux concentrations and probably an overestimate at or outside their magnetic boundaries because of straylight. Due to the proximity of the observed region to sun center and the absence of systematic differences between the disk center and limb side of the observed quantities discussed, we did not transform the measured magnetic field vectors to the solar frame.

\begin{figure*}[!htb]
  \begin{center}
    \includegraphics[bb=80 30 870 830, clip,height=0.45\linewidth]{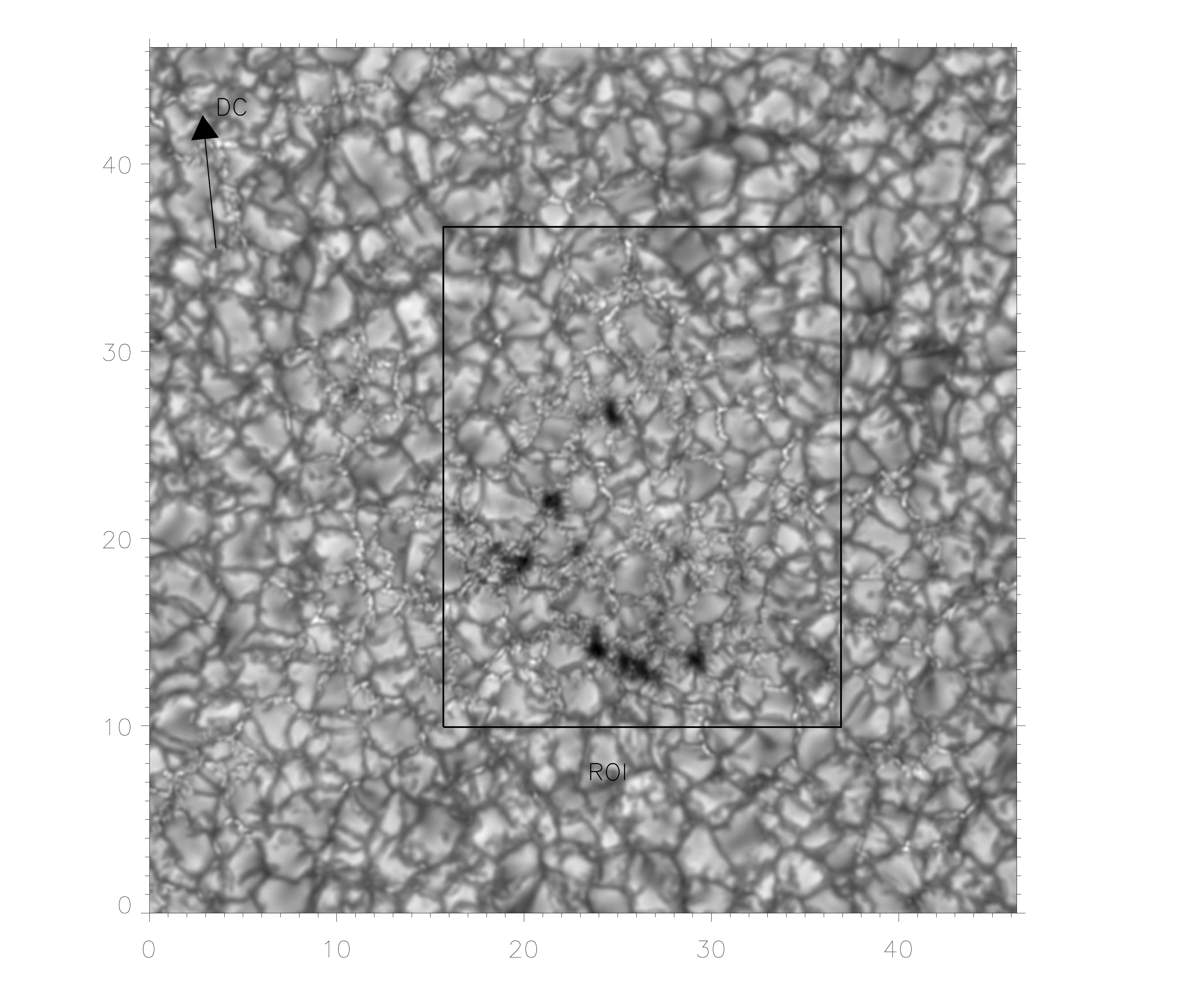}
    \includegraphics[bb=80 30 950 760, clip,height=0.45\linewidth]{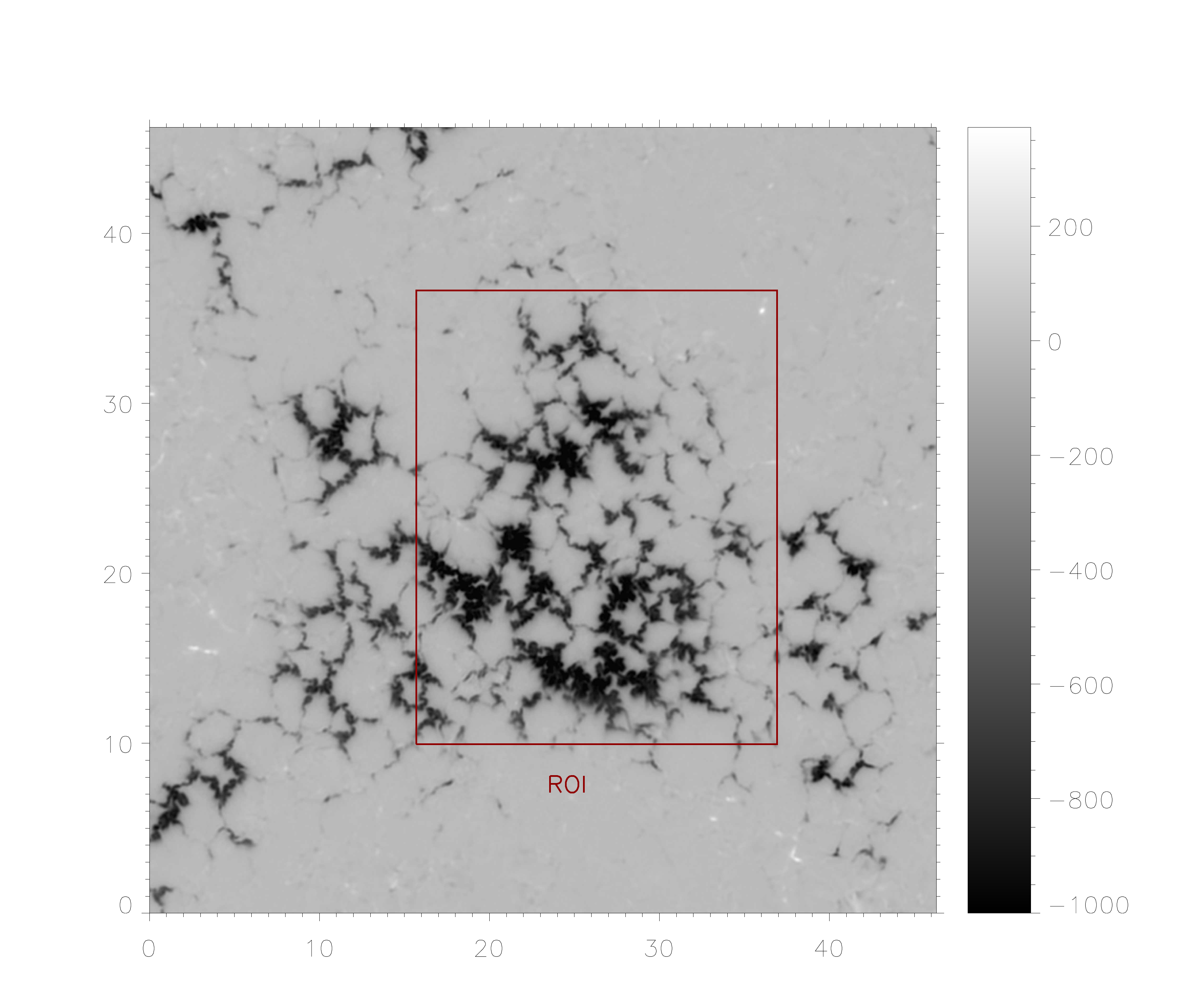}
    \includegraphics[bb=80 30 870 830, clip,height=0.45\linewidth]{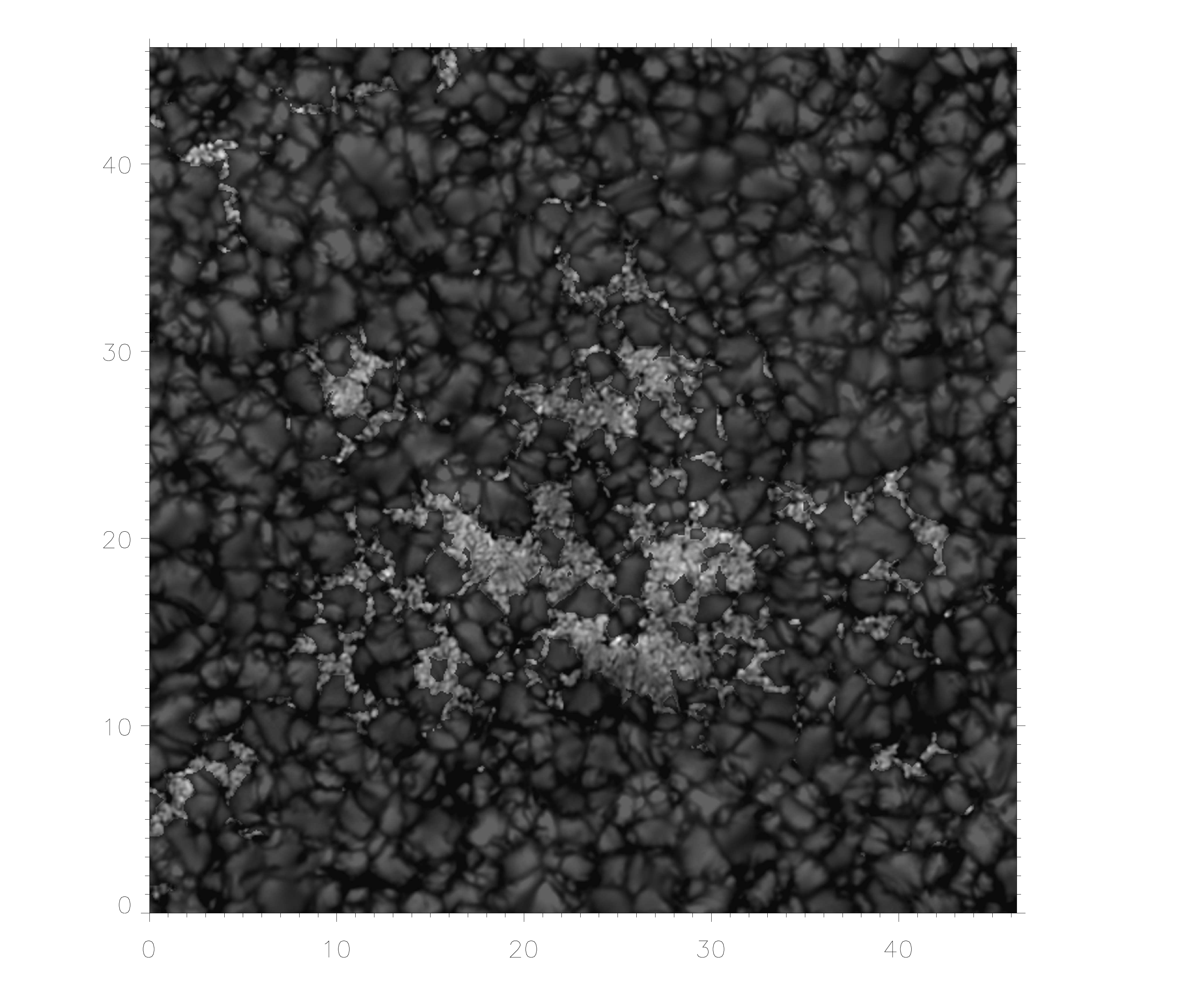}
    \includegraphics[bb=80 30 950 760, clip,height=0.45\linewidth]{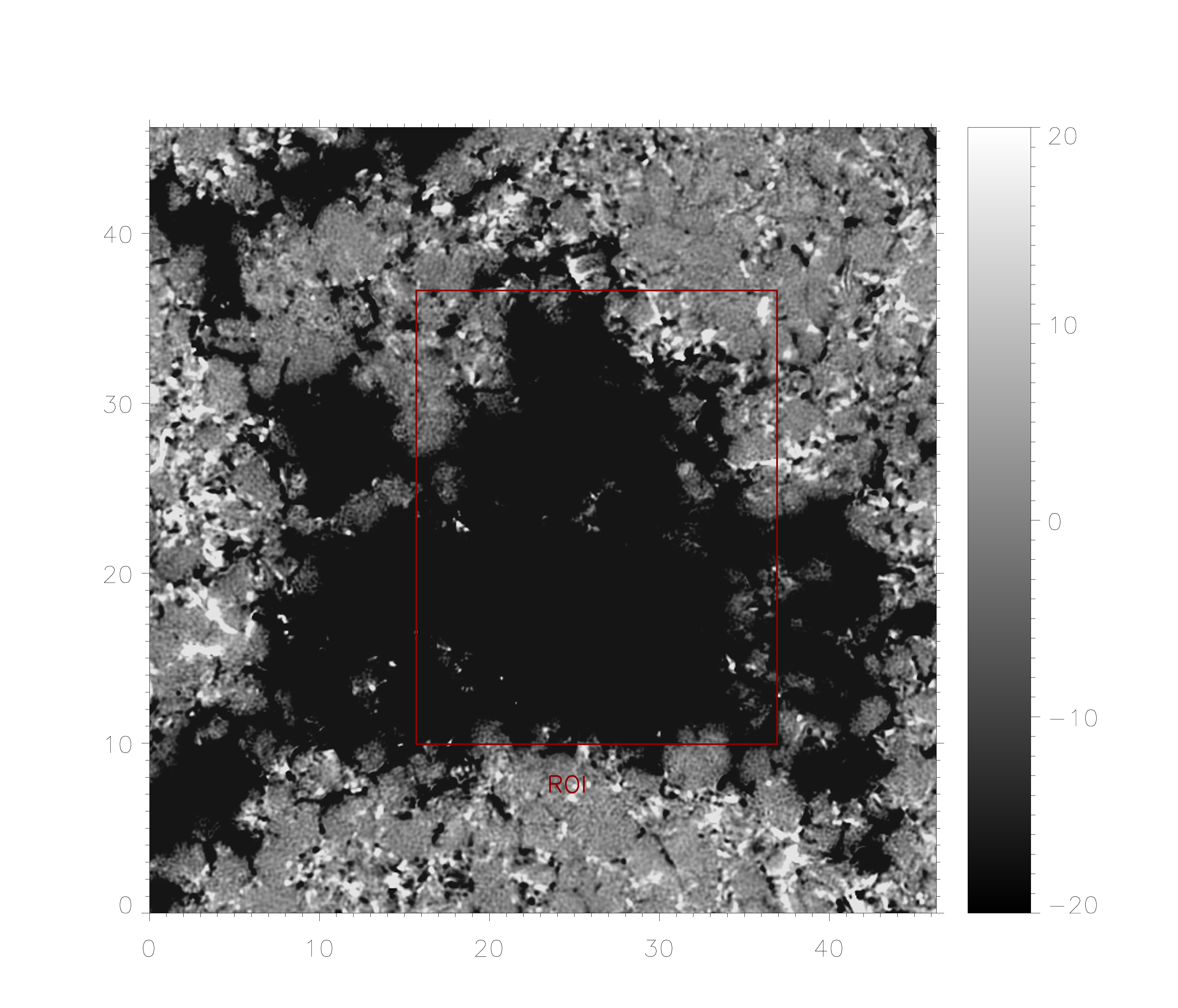}
    \caption{Full Field of view showing the unipolar ephemeral region with small pores with the 21\arcsec$\times$27\arcsec region of interest (ROI) marked as a red or black rectangle. Tick marks are in units of arcsec. The upper row shows the restored CRISP continuum image.The other three panels show quantities returned by the inversion code: the LOS velocity (lower-left) and LOS magnetic field to enhance the strong (upper-right) and weak (lower-left) field. The LOS velocities are shown enhanced by adding an offset to the velocities within the 200~G contour. The grey level bars on the right hand side of the figure indicate the signed LOS magnetic field in units of Gauss.}
    \label{fig:fullFOV}
  \end{center}
\end{figure*}

\subsection{Velocity calibration}
\label{sec:velocity-calibration}

To compensate for the wavelength shifts due to cavity errors dominated
by the high-resolution etalon, we used the flat-field images, obtained
1 hour after recording the science data. At each pixel, we made a fit
of a second-order polynomial to the center part of the 630.2~nm line
profile and measured the wavelength shifts. These wavelength shifts
(converted to equivalent Doppler shifts) were subtracted from the LOS
velocities returned by Helix. The zero point of the corrected LOS
velocities was fixed by assuming the dark parts of the pores to be at
rest and using a mask based on the restored continuum image to
identify the pores. Adopting this zero point and calculating the
average LOS velocity outside the center 400$\times$400 pixels, we
obtained a convective blueshift of $-270$~\mps. The obtained blueshift
is close to the convective blueshift of $-210$~\mps\ obtained by
\citet{scharmer08crisp}, based on a similar set of CRISP data also
recorded close to sun center and also using the Helix inversion code.
We also compared to convective blueshifts obtained from 3D convection
simulations (de la Cruz Rodriguez, private communication). An average
630.2~nm line profile calculated from these simulations produces a
bisector that midways between the continuum and the deepest part is
shifted by about $300$~\mps, which is close to the value we obtained.
This suggests that our velocity calibration is accurate to within
about $100$~\mps\ but the obtained convective blueshift appears on the
high side and we cannot exclude an error of $200$~\mps. A more
accurate assessment would require inversions with Helix using
synthetic line profiles calculated from the convection simulations and
degraded to the spectral resolution and sampling of the CRISP data.

\begin{figure*}[!htb]
  % \subfigure[]{
  \begin{center}
    \includegraphics[bb=11 20 532 392,clip, height=0.185\linewidth]{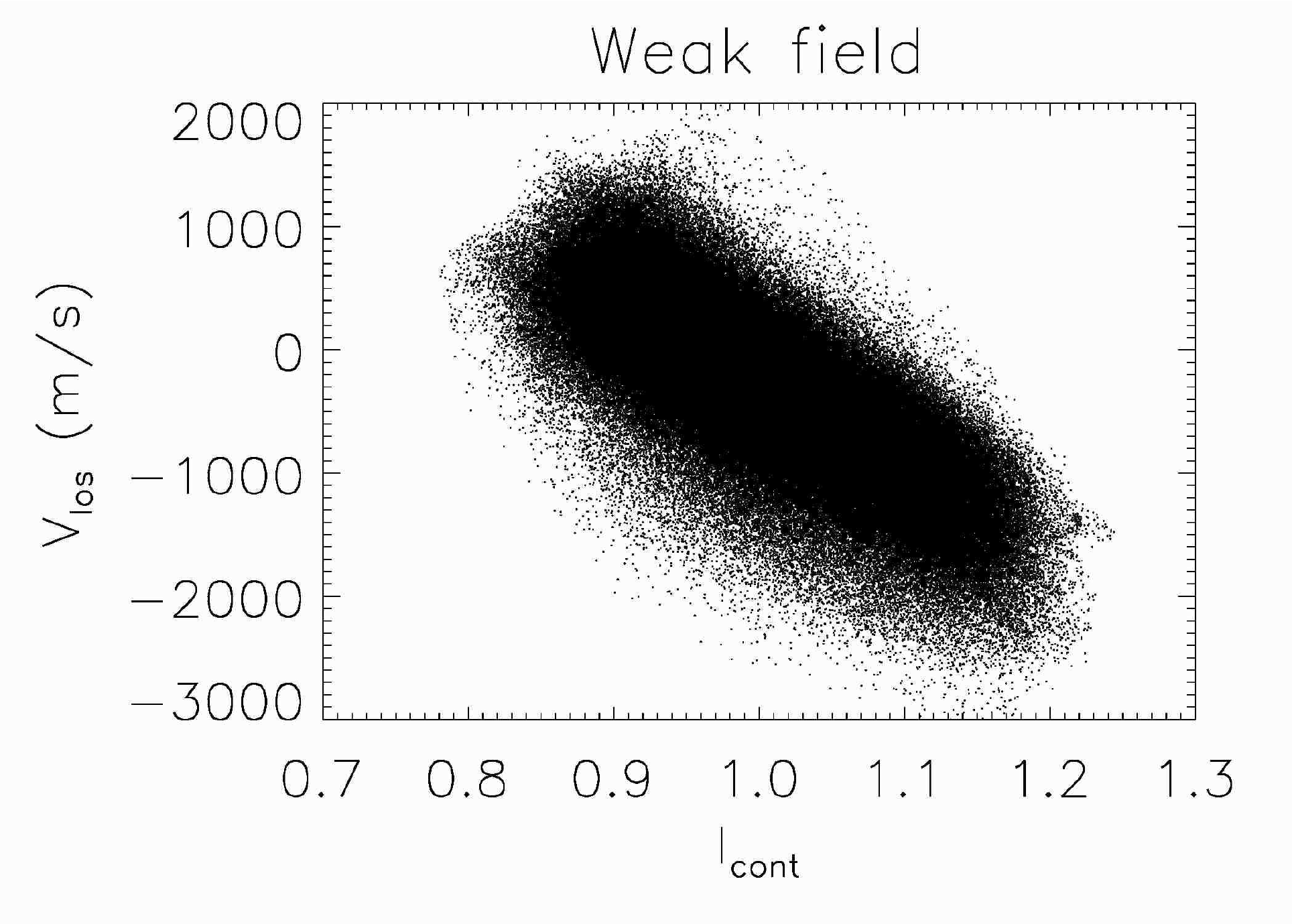}
    \includegraphics[bb=52 20 532 392,clip,height=0.185\linewidth]{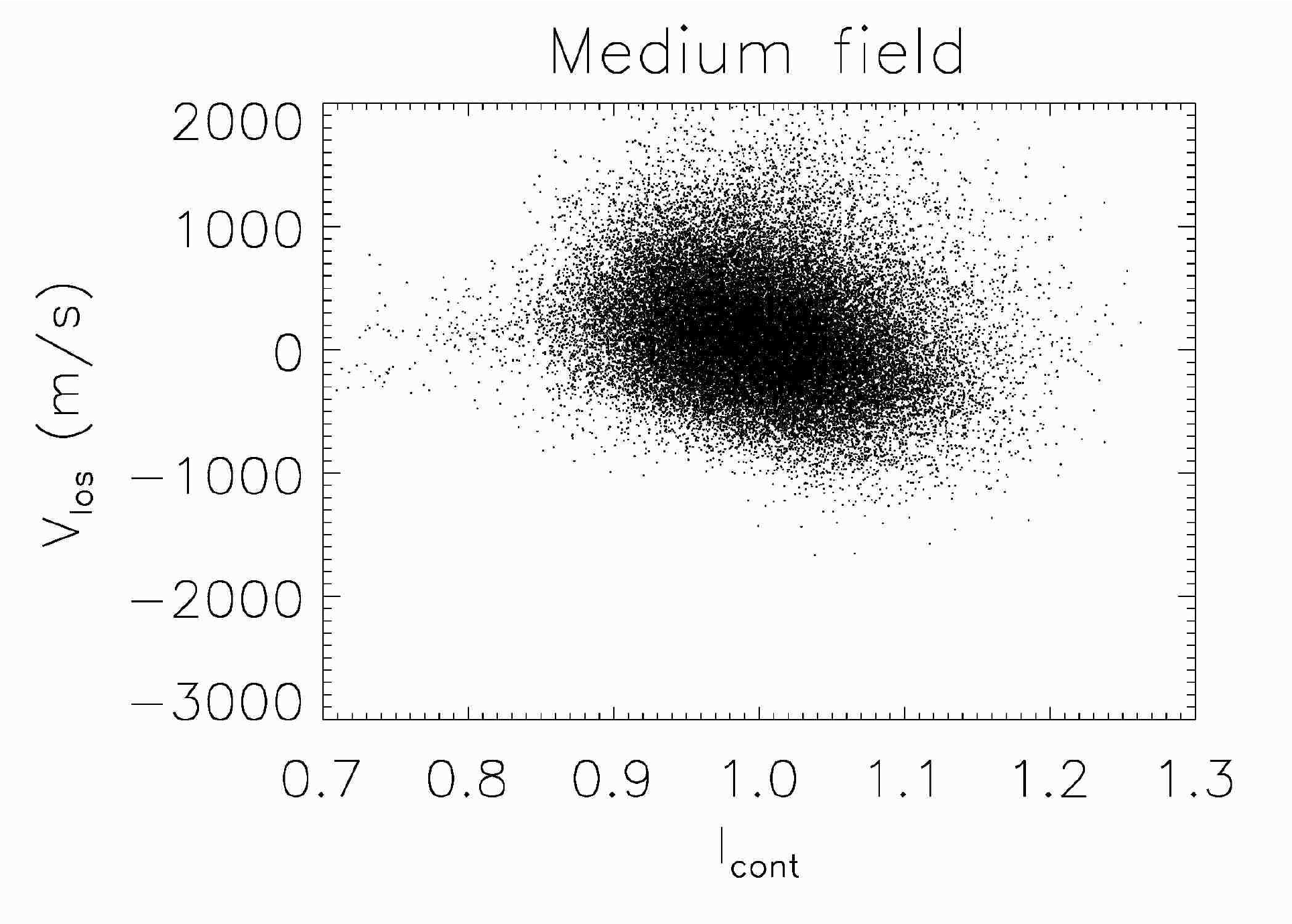}
    \includegraphics[bb=52 20 532 392,clip,height=0.185\linewidth]{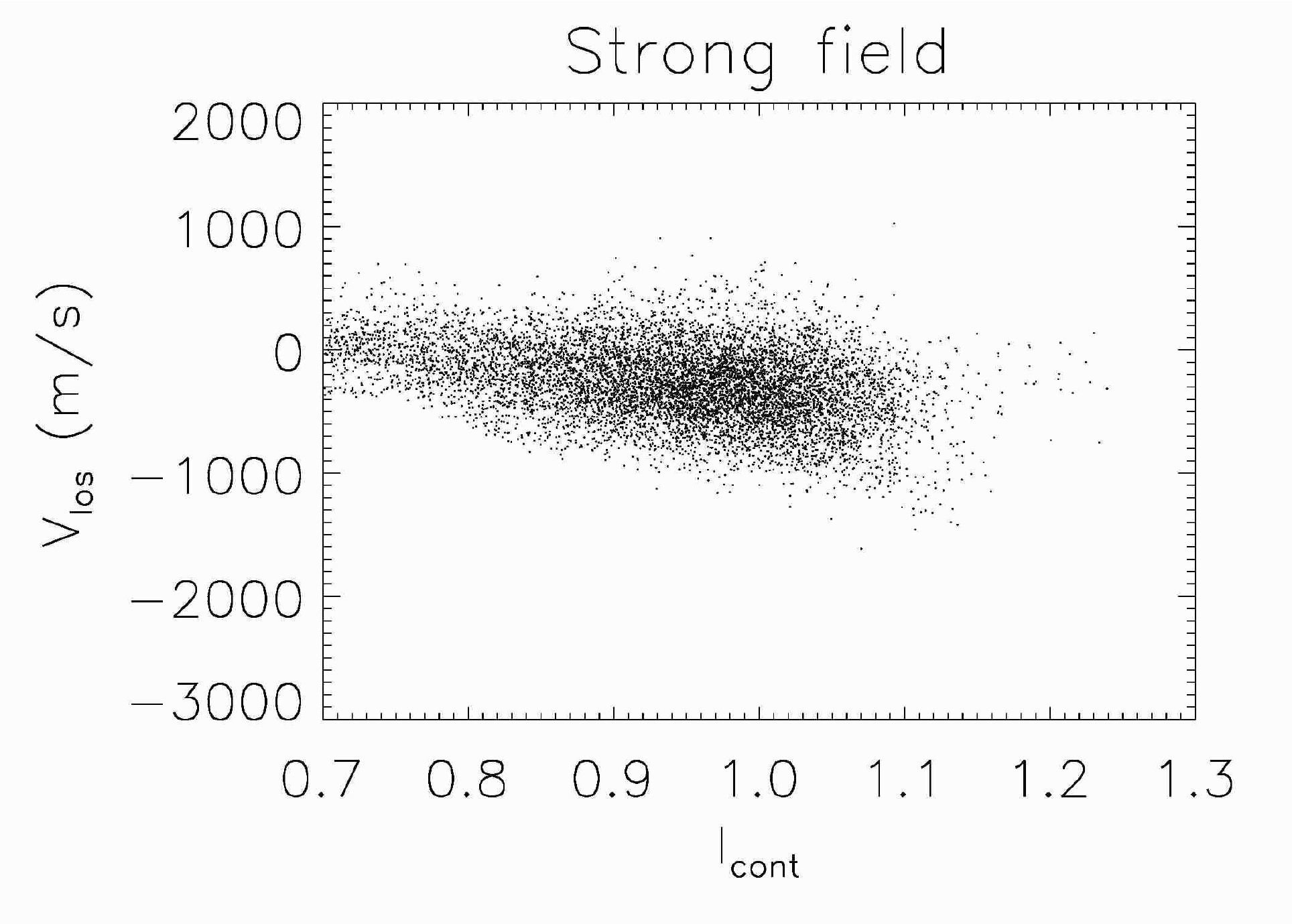}
    \includegraphics[bb=28 15 532 392,clip,height=0.185\linewidth]{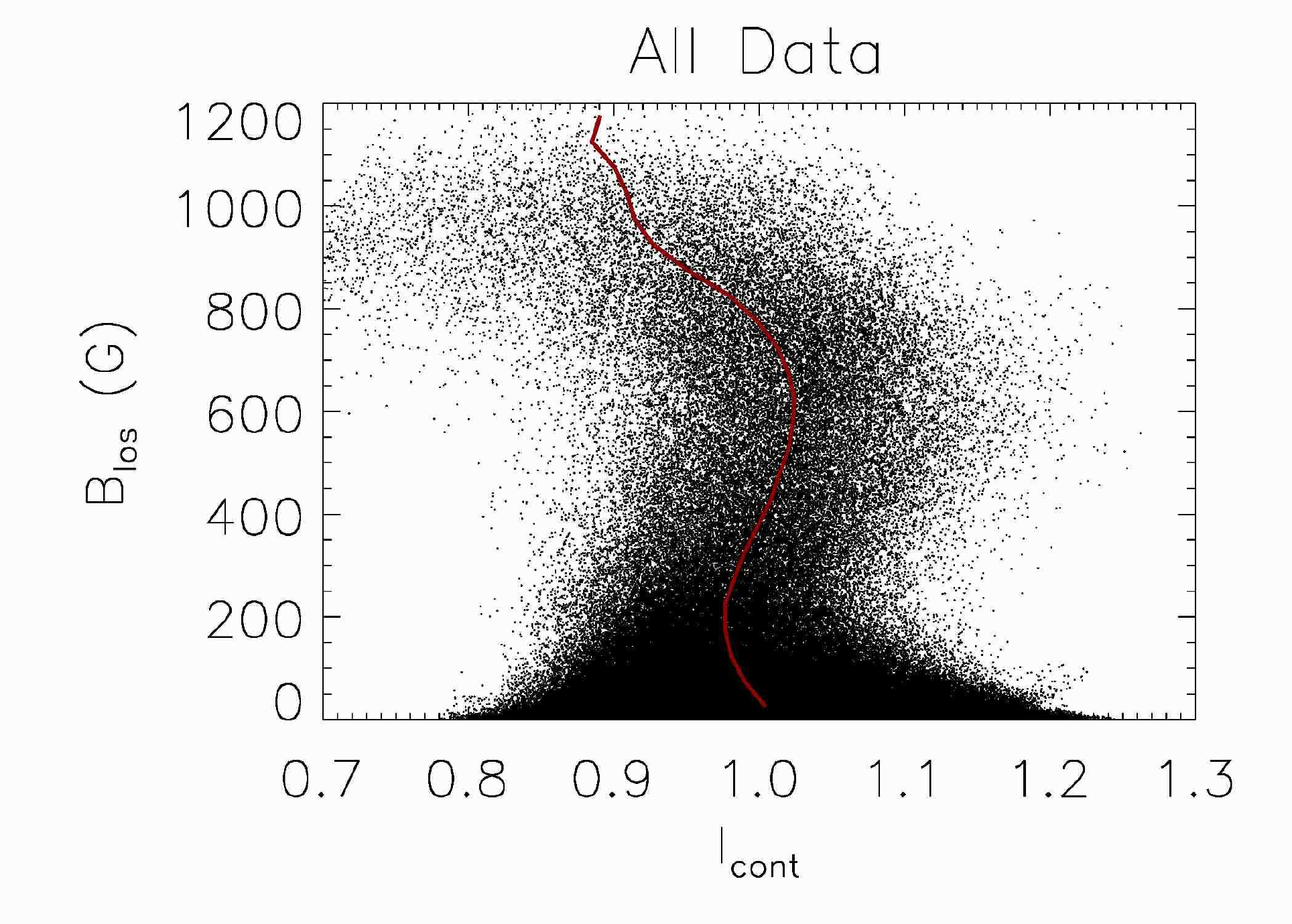}
    \includegraphics[bb=28 17 539 392,clip,height=0.19\linewidth]{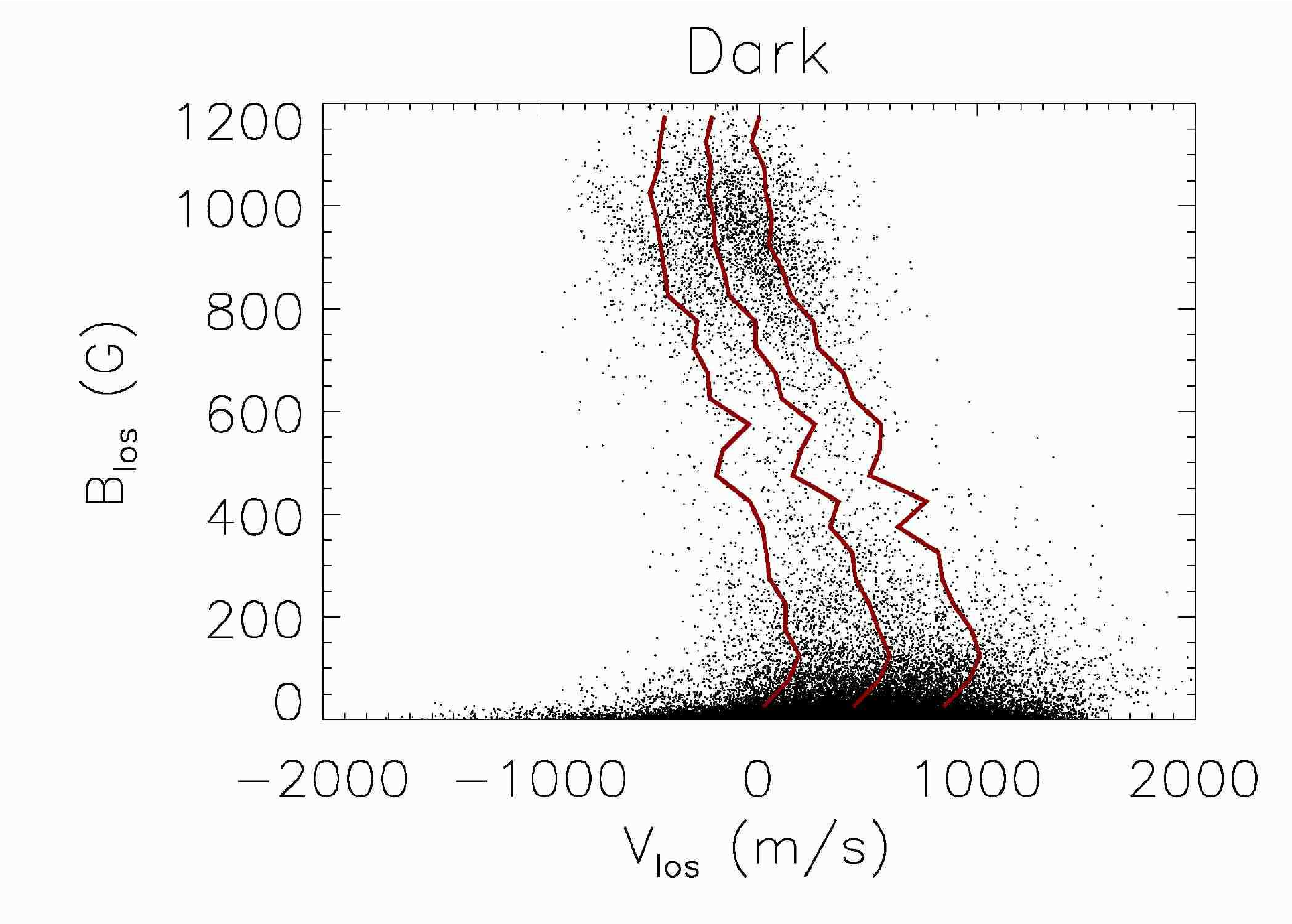}
    \includegraphics[bb=65 17 544 392,clip,height=0.19\linewidth]{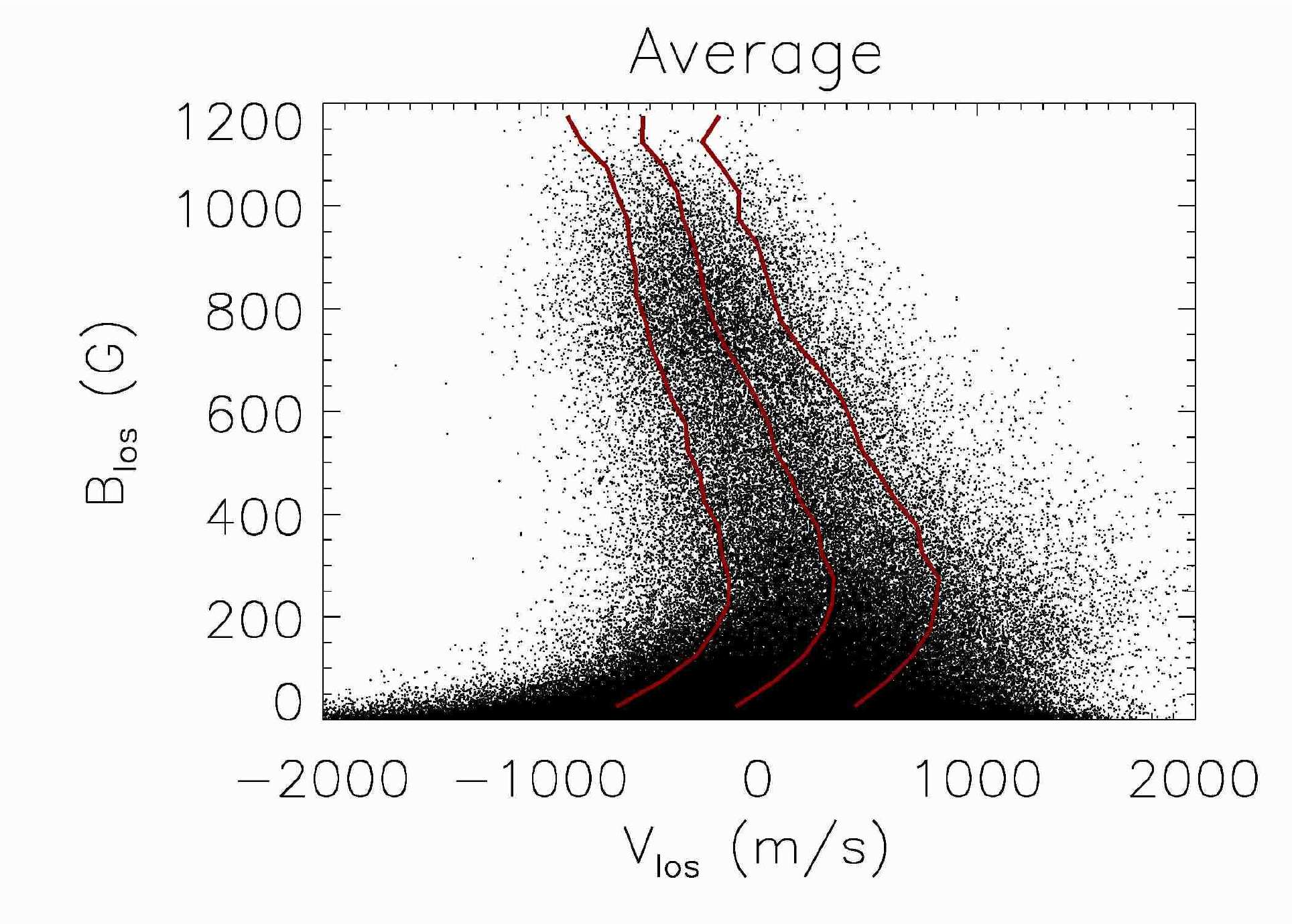}
    \includegraphics[bb=65 17 544 392,clip,height=0.19\linewidth]{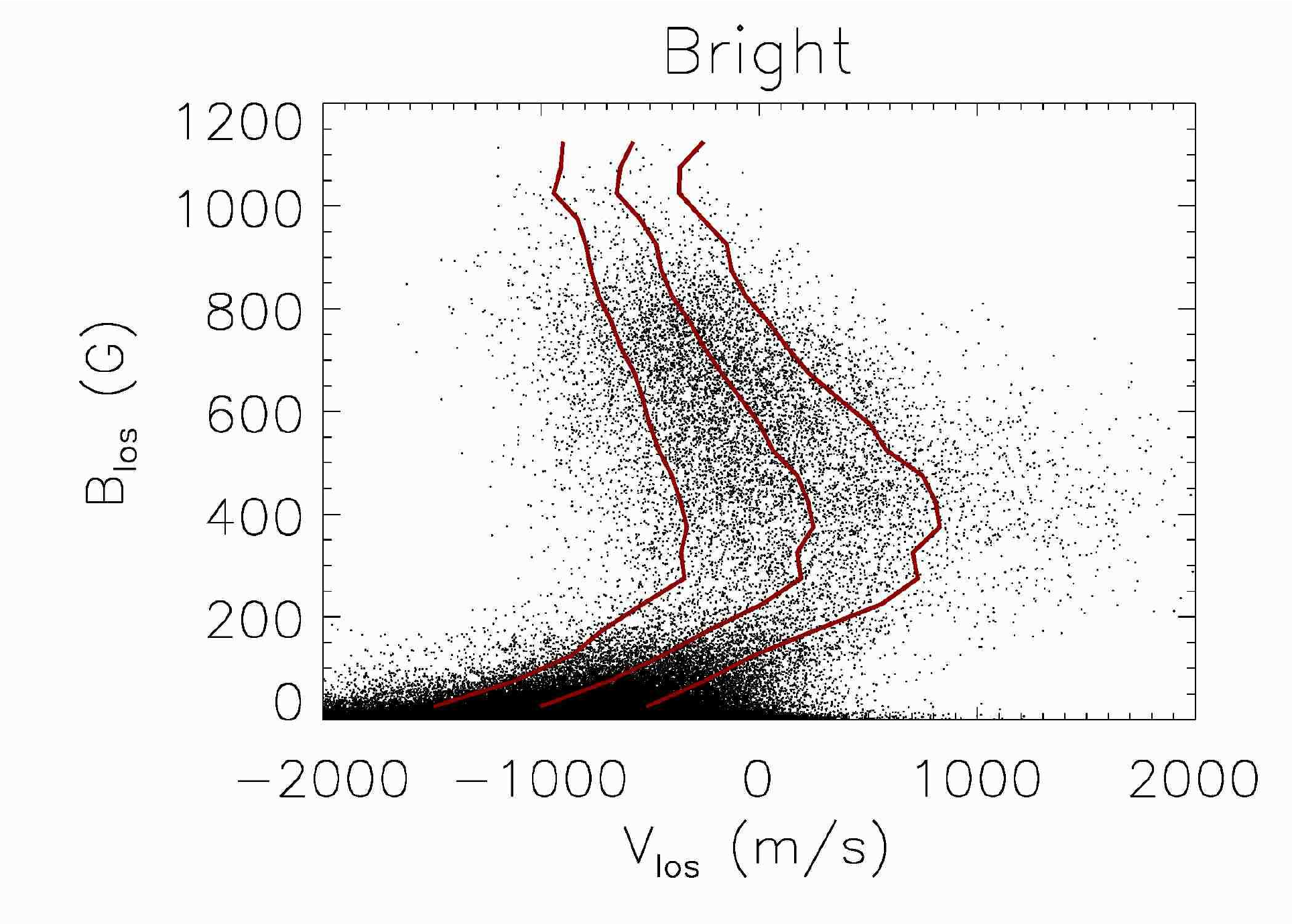}
    \includegraphics[bb=65 17 544 392,clip,height=0.19\linewidth]{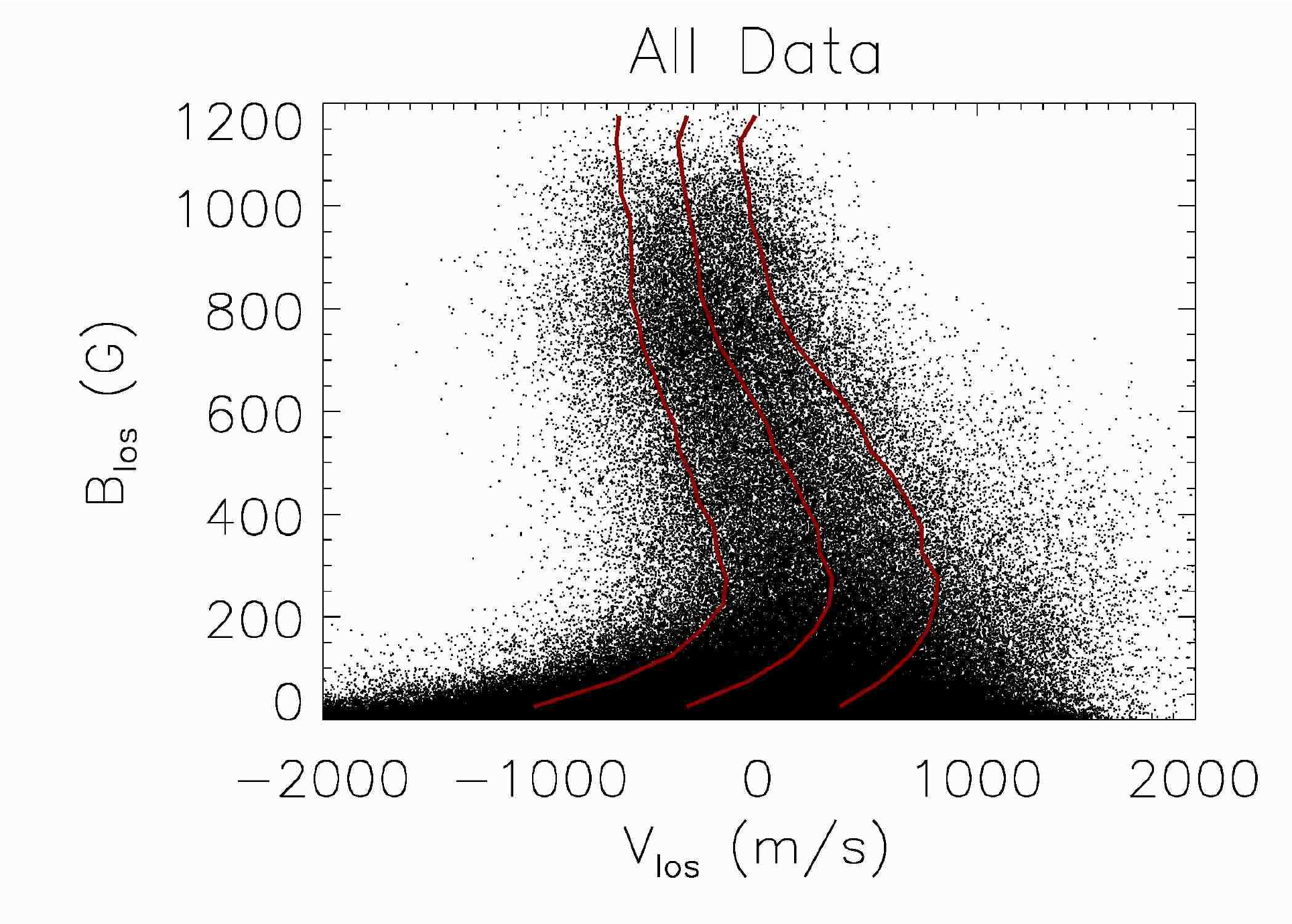}
%compressed files:
    %\includegraphics[bb=24 30 497 365,clip,height=0.18\linewidth]{fg2a_c.pdf}
    %\includegraphics[bb=61 30 497 365,clip,height=0.18\linewidth]{fg2b_c.pdf}
    %\includegraphics[bb=61 30 497 365,clip,height=0.18\linewidth]{fg2c_c.pdf}
    %\includegraphics[bb=15 30 497 365,clip,height=0.18\linewidth]{fg2d_c.pdf}
    %\includegraphics[bb=42 30 507 367,clip,height=0.19\linewidth]{fg2e_c.pdf}
    %\includegraphics[bb=80 30 507 367,clip,height=0.19\linewidth]{fg2f_c.pdf}
    %\includegraphics[bb=80 30 507 367,clip,height=0.19\linewidth]{fg2g_c.pdf}
    %\includegraphics[bb=80 30 507 367,clip,height=0.19\linewidth]{fg2h_c.pdf}
    \caption{The top row shows correlations between continuum
      intensity and LOS velocity for weak ($|\Blos|<$50~G), medium strong ($200\text{ G}<|\Blos|<800\text{ G}$) and strong ($800\text{ G}<|\Blos|$) magnetic field and the correlation between the continuum intensity and LOS magnetic field. The bottom row shows correlations between LOS velocity and LOS magnetic field for dark ($\Icont<0.9$), average ($0.9<\Icont<1.05$) and bright ($1.05<\Icont$) structures. Red lines show binned averages and in the lower row also $\pm1$ standard
      deviation of the variations.}
    \label{fig:vlosplots}
  \end{center}
\end{figure*}

\subsection{Data quality and fits}
\label{sec:data-quality-fits}

The polarimetric noise was estimated from a $500\times500$ pixel area
in the Stokes $Q, U$ and $V$ images recorded at the continuum
wavelength and corresponds to $1.2\times 10^{-3}$, $1.4\times
10^{-3}$, and $1.0\times 10^{-3}$ for Q, U and V respectively. The continuum
Stokes images appear completely clean from granulation features or
other artifacts, suggesting very low levels of seeing induced
cross-talk. We also measured the RMS noise in \Blos\ over several
20$\times$20 pixel boxes that appear free of polarization features and
obtained noise levels in the range 2.2--3.2~G, suggesting ``safe''
detection levels of about 10--20~G.

Figure~\ref{fig:fullFOV} shows a map of \Blos\ scaled from $-1000$~G
to $+370$~G (upper right panel) and from $-20$~G to $+20$~G (lower
panel). When scaled to show weak field, the center part of the FOV
appears to have magnetic field nearly everywhere. This may be in part straylight from uncompensated high-order aberrations in MOMFBD
processing or other sources of straylight in the telescope and/or the
Earth's atmosphere. The presence of straylight is clearly indicated by the measured RMS contrast from the CRISP continuum image. The high continuum RMS contrast measured, about 8.3\% in the region without significant magnetic field, is still far from the value obtained from 3D convection simulations, 14\%. This is not a spatial resolution issue: the main contributions to the granulation RMS contrast come from spatial frequencies that are well below the SST diffraction limit. We also note that the strongest LOS component of the magnetic field shown in Fig.~\ref{fig:fullFOV} is only about 1.2~kG which appears low. Including the transverse magnetic field components, the highest field strength is about 1.3~kG. Inversions based on Hinode spectropolarimetric data from a similar region with small pores \citep{2008A&A...481L..29M} gave peak mean field strengths (magnetic filling factor multiplied by the field strength) of 1.5--1.6~kG. 

Most observed $V$ profiles, some of which are discussed in
Sect.~\ref{sec:small-scale-velocity}, are rather symmetric, justifying the use of a Milne--Eddington inversion technique to analyze the data. Only Stokes $V$ profiles with very low amplitudes are poorly fitted. We also note a few cases where the shapes of the observed Stokes $I$ and $V$ profiles are well reproduced by the fits, but where the (very weak) observed Stokes $V$ profiles
are strongly redshifted with respect to the synthetic Stokes $V$
profiles. This implies that the obtained LOS velocities primarily
refer to the non-magnetic rather than the magnetic atmosphere or are
contaminated with spatial straylight, where a difference exists. The
bias toward estimates of LOS velocities from the Stokes $I$ rather
than the $V$ profiles is not surprising, given the equal weights on
Stokes $I$ and $V$ used with the inversions and the very weak Stokes
$V$ signatures for these features.

There is also a tendency for the synthetic Stokes $I$ profiles to be
somewhat redshifted with respect to the observed profiles. This is
almost certainly due to the telluric blend in the red wing of the
630.2~nm line, which is not compensated for in the inversions.
However, there is a similar systematic influence of the telluric line
also on the line profiles from the pores, used to calibrate the LOS
velocities, such that we expect the net effect of the telluric blend
on the measured LOS velocities to be small.

\section{Results}
\label{sec:results}

\subsection{Overview}
\label{sec:overview}

The restored CRISP continuum image is shown in the top panel of
Fig.~\ref{fig:fullFOV}. The disk center (DC) direction is indicated
with an arrow and the region of interest (ROI) discussed in
Sect.~\ref{sec:small-scale-velocity} is indicated with a rectangle.
Due to the proximity of this region to sun center, the transverse
components of the magnetic field obtained with Helix are mostly weak and very noisy.
In the following, we discuss the LOS component of the magnetic
field, calculated as $\Blos=B \cos \gamma$ and scaled to show both the
strong and weak fields in Fig.~\ref{fig:fullFOV}. In the MDI
magnetograms this region is indicated as having positive polarity, but
for greater clarity positive polarities are here indicated as dark.
The lower right panel demonstrates the high sensitivity of the data
and the absence of spurious magnetic features.

The observed region contains micropores in various stages of evolution
and numerous flux concentrations, outlining a network of strong field
located mostly in intergranular lanes. The region close to the pores
contain more \emph{extended contiguous regions of strong magnetic
field where the granulation pattern is small-scale and irregular},
so-called abnormal granulation \citep{1973SoPh...33..281D}.

We note that the regions shown as dark in the lower-right panel of Fig.~\ref{fig:fullFOV} contain no \emph{strong} field of opposite polarity to that of this region and only a few cases with weak such field. Possibly, this is an effect of polarized straylight from neighboring strong field canceling the $V$ signal from weak field of opposite polarity.

\subsection{Evidence for small-scale magneto-convection}
\label{sec:evidence-small-scale}

The lower left panel of Fig.~\ref{fig:fullFOV} shows the LOS velocity
obtained from the inversions, but with a mask used to enhance the
regions where \Blos\ is stronger than 200~G. The unsigned average of
\Blos\ within this mask is 530~G, the average field strength is
640~G. A mask with a threshold of 100~G (unsigned average of \Blos\
400~G, average field strength 510~G) or 300~G (unsigned average of
\Blos\ 620~G, average field strength 720~G) would outline nearly the
same regions but the 100~G mask would include also some of the larger
granules embedded within the stronger field. We remark that the
precise value of the threshold used to define the mask is of no
particular physical significance for an \emph{extended}
region. Because of the finite spatial resolution and the critical
sampling, any discontinuous change in the field will be observed as a
continuous transition. The chosen threshold simply defines the
boundary between the nearly field-free and strong-field regions as
that for which $|\Blos|$ is somewhat less than half the average of the
adjacent strong field.

There is a dramatic difference in the velocity structures inside and
outside the mask. Within the 200~G mask (average field strength
640~G), we see a small-scale velocity pattern that appears similar to
the granular velocity field outside the mask, but with characteristic
scales of about 0\farcs3, or 4 times smaller than field-free granules.
Also the continuum image shows distinctly small-scale, but more
irregular, structures within the mask. The average LOS velocity within
the mask is about $70$~\mps. The RMS velocity within the 200~G mask is
$490$~\mps\ compared to $700$~\mps\ outside the mask. There appears to
be some contributions to this RMS from large-scale velocity fields
(see also lower-right panel in Fig.~\ref{fig:zoommosaic}), possibly
from 300-sec oscillations, implying that the RMS estimated for
small-scale velocity field represents an upper limit. Using a
2\arcsec$\times$2\arcsec\ boxcar average to reduce the contributions
from large-scale velocity fields reduces the RMS velocity within the
200~G mask to $450$~\mps. We conclude that the small-scale velocity
pattern corresponds to upward/downward flows of relatively large
amplitude that averages approximately to zero. This RMS velocity
decreases with increasing LOS field and is only $220$~\mps, if the
2\arcsec$\times$2\arcsec\ boxcar average is first subtracted, for the
strongest fields measured.

The RMS continuum intensity within the 200~G mask but excluding the
pores (defined by $\Icont<0.8$, where $\langle I_\text{c} \rangle$ is the continuum intensity averaged over the entire FOV) is 6.5\% which is only 22\% lower than that of the field-free granulation (8.3\%). The top row of plots in
Fig.~\ref{fig:vlosplots} shows the correlation between continuum
intensity and LOS velocity. The red curves indicate binned averages
and $\pm$ one standard deviation of the variations. The left-most plot
shows the correlation for nearly field-free convection ($|\Blos| <
50$~G), the second panel the same correlation with $|\Blos|$ in the
range 200--800~G and the third plot with $|\Blos|$ stronger than
800~G. As shown in these plots, also the magnetic parts of the FOV
show a (weak) correlation between the continuum intensity and the LOS
velocity in the same sense as for field-free convection. This
correlation is stronger if the 2\arcsec$\times$2\arcsec\ boxcar
average is used to reduce the effects of large-scale flows and 300-sec
oscillations, but such averaging was not used to produce the plots
shown. The relatively weak correlation found for the magnetic part of
the FOV is obviously also related to the morphological differences
between the continuum intensity and LOS velocity maps; only the LOS
velocity map shows a clear ``granulation-like'' pattern within the
200~G mask. These morphological differences are very likely related to
the formation height of the 630.25~nm line above the photosphere, the
small horizontal scale of magnetic field variations inducing a
similarly small vertical scale of such variations and the increasing
dominance with height of the magnetic field, constraining the
convective flows above the photosphere.

The small-scale LOS velocity pattern is barely visible at a resolution
close to $\sim$0\farcs15. Quite obviously, the RMS velocities (as well as RMS intensity variations) measured within the magnetic mask must be strongly underestimated, in particular since this is true already for the much larger field-free granulation pattern. The small-scale granular velocity pattern would be difficult, if not impossible, to observe with a significantly smaller telescope. It is thus not surprising that \citet{2008A&A...481L..29M}, using the 50-cm SOT on Hinode, did not report direct evidence for such small-scale
convection, but interpreted their results as suppression of convection.

\citet{2004ApJ...604..906R} plotted the correlation between LOS
velocities measured from two wavelengths in the wings of the 630.25~nm
and the line wing intensity for selected bright points \cite[Fig.
7]{2004ApJ...604..906R}. This correlation plot appears quite similar
to the first and third panel in the upper row of
Fig.~\ref{fig:vlosplots}. Rimmele also plotted the correlation between
$\Blos$, estimated from single-wavelength magnetograms in the blue
wing of the 630.25~nm line and the line wing intensity as well as the
correlation between $v_{LOS}$ and $\Blos$ \cite[Fig.
6]{2004ApJ...604..906R}. Also these plots appear similar to the
corresponding plots in Fig.~\ref{fig:vlosplots}.

We believe that the measured properties of the small-scale velocity field and the correlations shown in Fig.~\ref{fig:vlosplots} justify the identification of the velocity pattern inside the mask as a \emph{small-scale magneto-convection pattern}. At the SST resolution
it appears similar to that of field-free convection, with upflows in
the centers and downflows at the edges. However, the correlation between continuum intensity and LOS velocity is much weaker for this small-scale velocity pattern than for field-free granulation. This may in part be due to the small horizontal scales of these structures combined with the height of formation of the 630.25~nm line, more easily leading to decorrelations between these measured properties than for large (field-free) granules. We speculate that the main reason for this decorrelation is that small-scale convection does not overshoot to the same heights as field-free convection and leaves only very weak traces at the line forming and perhaps even continuum forming layers. This would imply that the measured RMS velocity is not a good indicator of the convective flux for these structures and that possibly there are contributions to the velocity signatures from dynamics above the photosphere that is not of convective origin. In this context it is important to emphasize that despite the large RMS velocities measured for field-free granulation at the formation height of the 630.25~nm line, the convective flux in the quiet sun is negligibly small 100~km above the photosphere \cite[Fig. 21]{2009LRSP....6....2N}. The RMS velocity measured with the 630.25~nm line is therefore only a tell-tale of field-free convection actually peaking below the visible photosphere and there is no a priori reason for expecting the same signature for small-scale magneto-convection.

Given the uncertainty of an appropriate threshold to define the mask outlying this small-scale velocity pattern and the likely influence of straylight on our estimates of \Blos, we conclude that there is a \emph{transition to small-scale magneto-convection} when the field covers a sufficiently large area and reaches an average strength of 600--800~G.

3D magneto-convection simulations with similar average field strengths
over a large FOV (6$\times$6~Mm, corresponding to 115$\times$115 of
our pixels) were discussed by
\citet{2004RvMA...17...69V,2005MmSAI..76..842V}, while
\citet{2003ASPC..286..121S} presented 12$\times$12~Mm simulations made
with an initial average field strength of 400~G. \citet[Fig.
10]{2004RvMA...17...69V} discussed snapshots from simulations with
average field strengths of 400~G and 800~G. By comparison, \emph{the
largest average field strength within a 6$\times$6~Mm box of our
data is 600~G}. The 400~G simulation of
\citet{2004RvMA...17...69V,2005MmSAI..76..842V} shows granules with
somewhat reduced size but otherwise looking similar to field-free
granules. The intergranular lanes are wider than for field-free
convection, contain nearly all the magnetic field and show flows of
strongly reduced magnitude. The 800~G simulations show a much more
small-scale intensity pattern and the velocity map shows small-scale
strong upflows superimposed on a background with small velocities.
Strong downflows are found only in thin sheets at the boundaries of
the strong magnetic field, also constituting the boundaries of small
bright granules. The map of the vertical magnetic field made from the
simulations appears different from our \Blos\ map. The simulated map
shows very small nearly field-free granules, corresponding to the
locations of upflows, superimposed on a background of very strong
(nearly 2~kG) field. These differences may be primarily related to the
inadequate spatial resolution of the SST data and/or straylight but
also due to the differences in height for the synthetic map (made at a
continuum optical depth $\tau_{500}=1$) and the map from the Helix
inversions (representative of approximately
$\tau_{500}\approx0.01$--0.1). We are not aware of any publications
presenting similar synthetic maps at heights relevant for the
formation of the 630.25~nm line from simulations with strong average
field strengths, needed for further interpretation of our data.

\begin{figure}
  \centering
  \includegraphics[bb=80 35 870 825, clip, width=\linewidth]{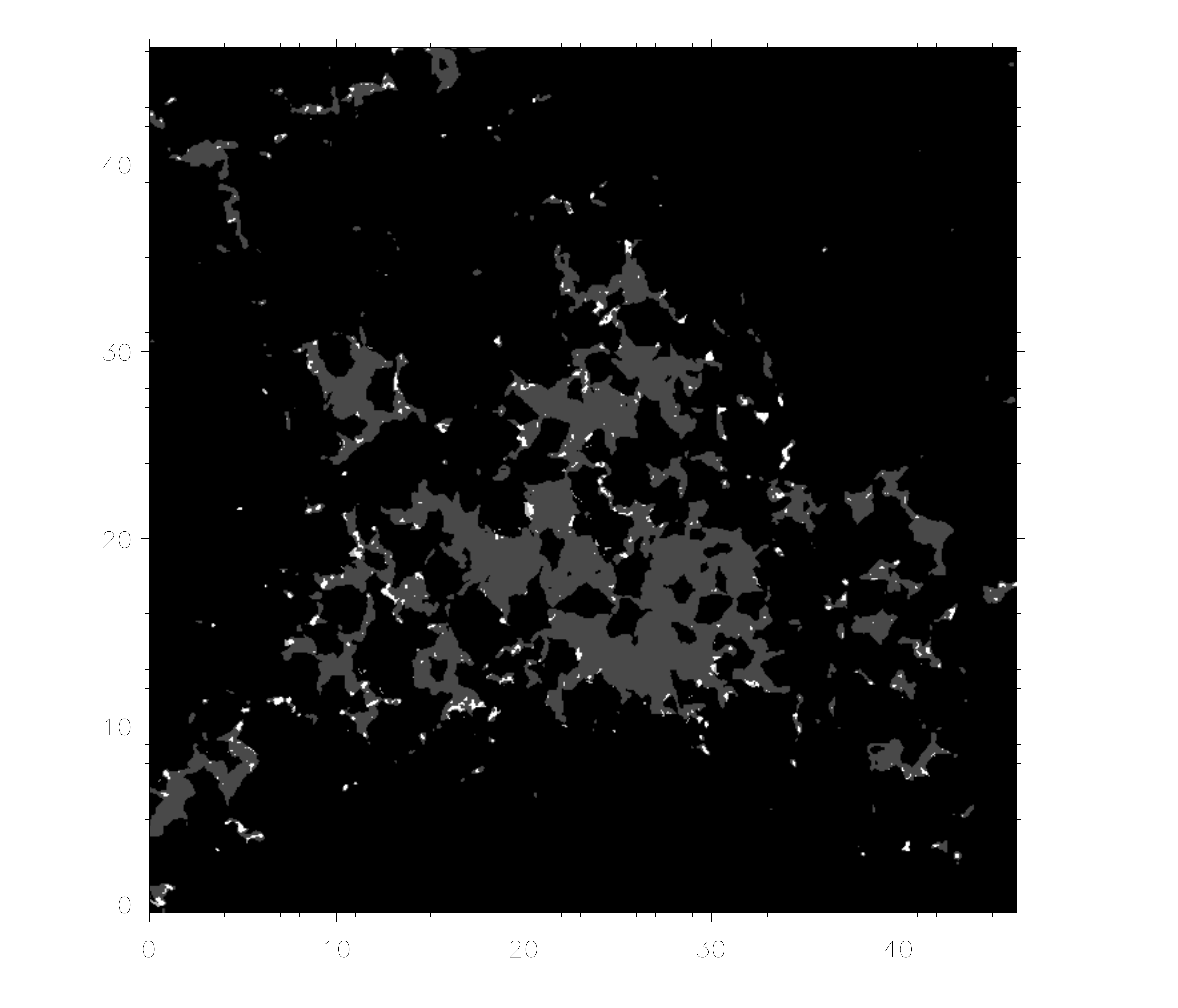} 
  \caption{Locations of magnetic bright downflows (coded as white)
    superimposed on a binary map outlining the region with
    $|\Blos|>200$~G, shown gray. Tick marks are in units of arcsec.}
  \label{fig:locations}
\end{figure}

\begin{figure}
  \centering
  \includegraphics[bb=31 15 702 839,width=0.99\linewidth]{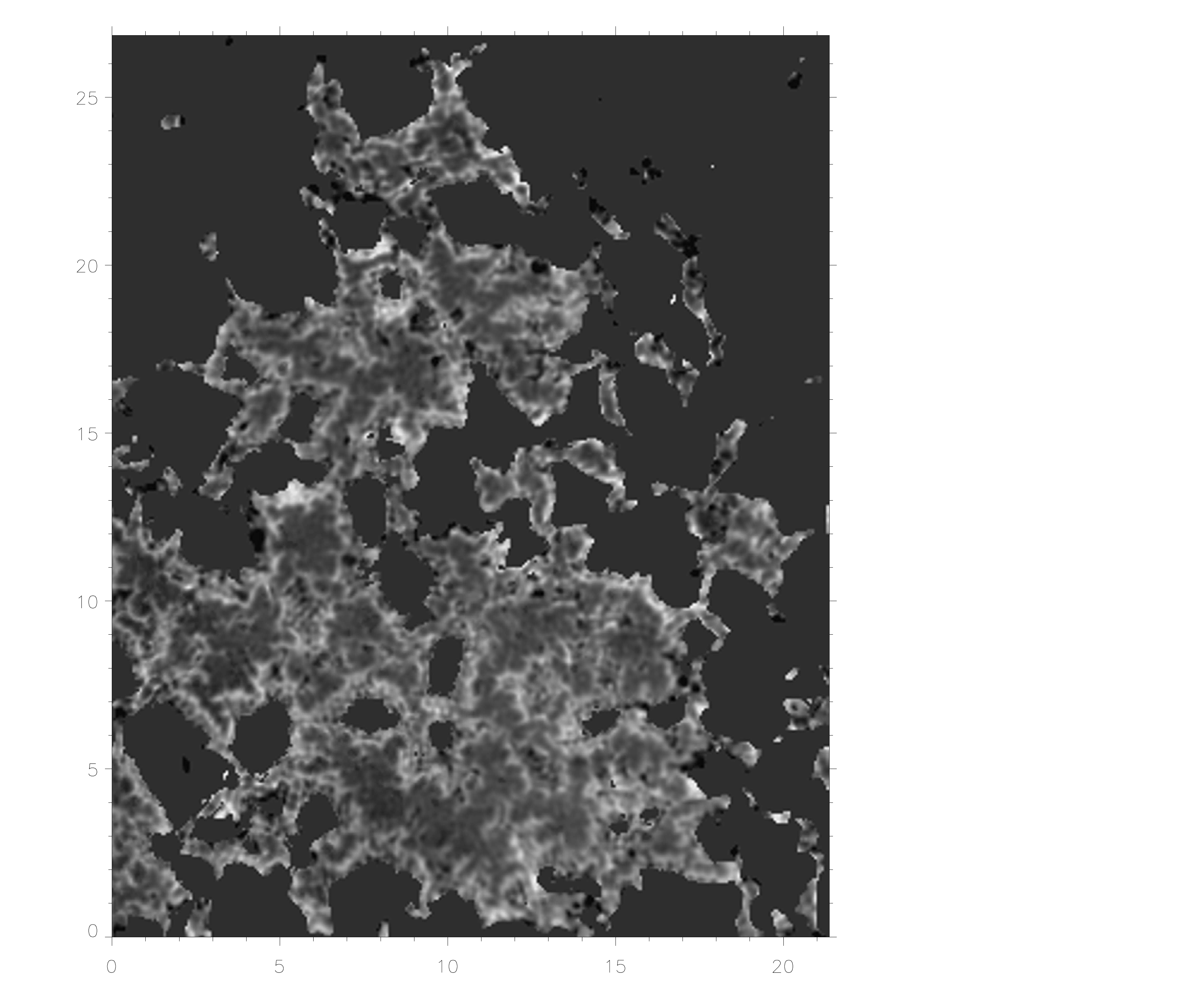}
  \caption{Map of the Stokes $V$ amplitude asymmetry, scaled from
    $-0.1$ to $+0.5$ within a mask defined $|\Blos|>75$~G. The FOV
    shown corresponds to the ROI marked in Fig~\ref{fig:fullFOV}. Tick marks are in units of arcsec.}
  \label{fig:sv_asymm1}
\end{figure}

\begin{figure*}[!htb]
  \begin{center}
    \includegraphics[bb=35 15 550 360,clip, height=0.172\linewidth]{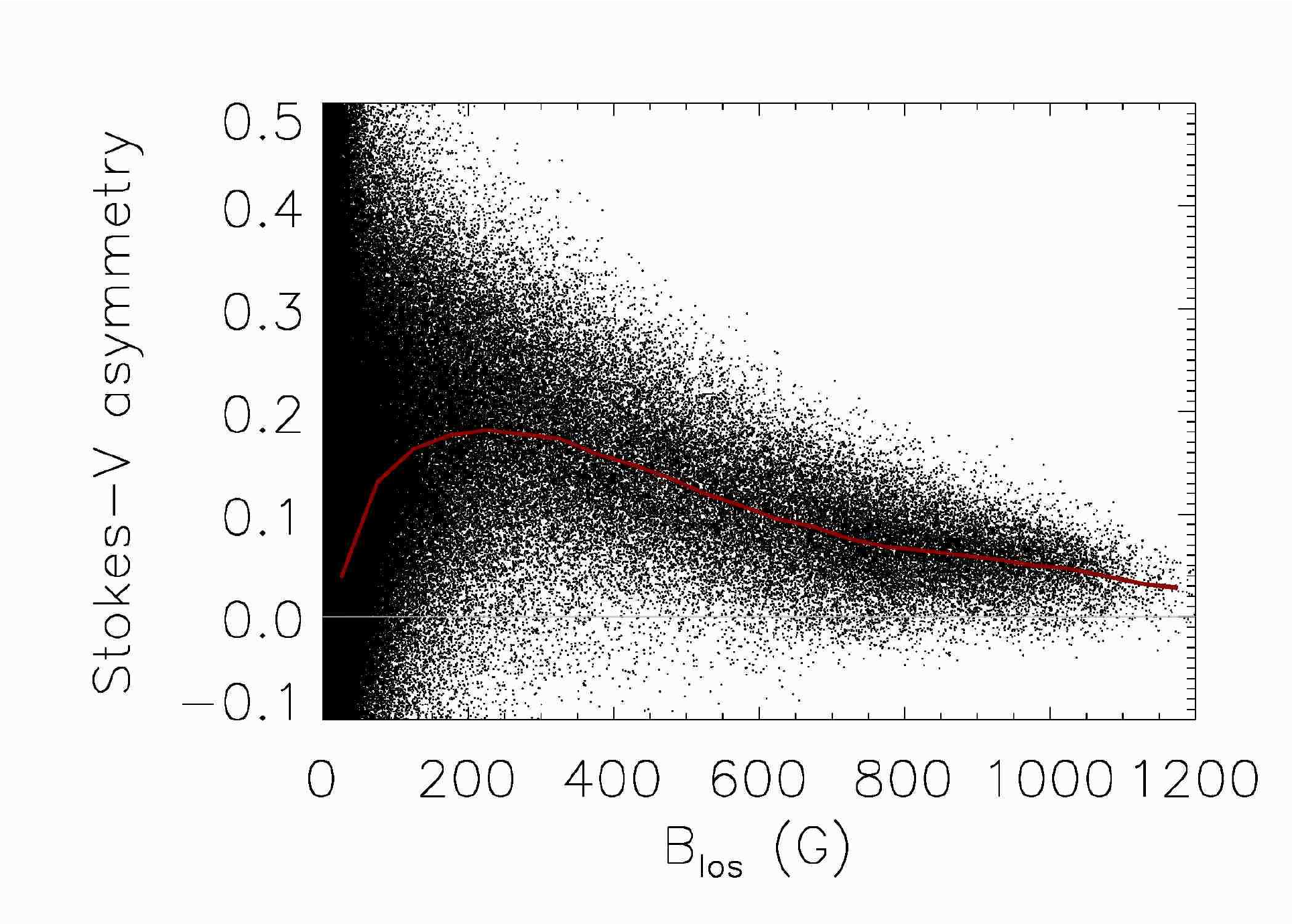}
    \includegraphics[bb=70 15 550 360,clip,height=0.172\linewidth]{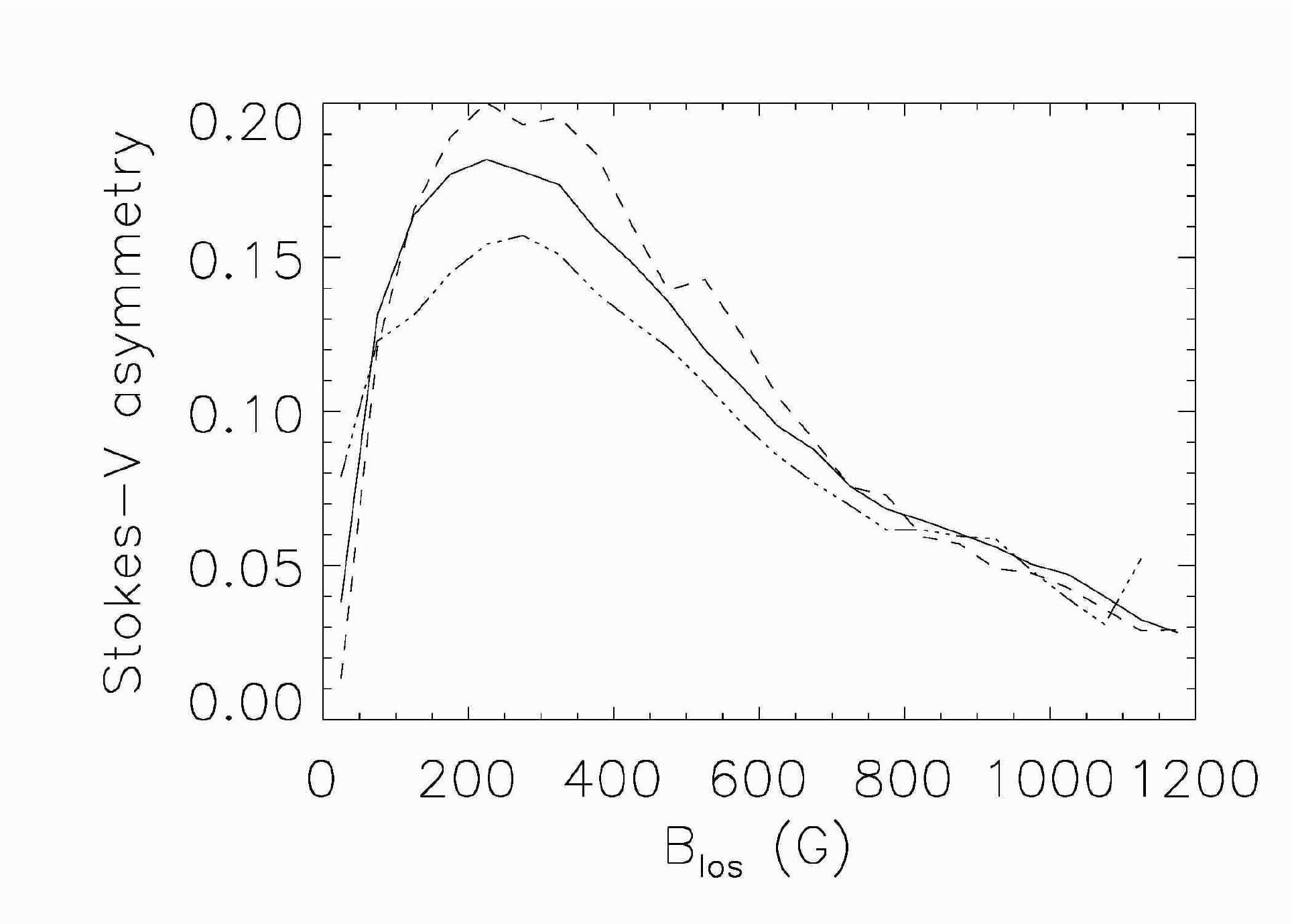}
    \includegraphics[bb=80 15 550 360,clip, height=0.172\linewidth]{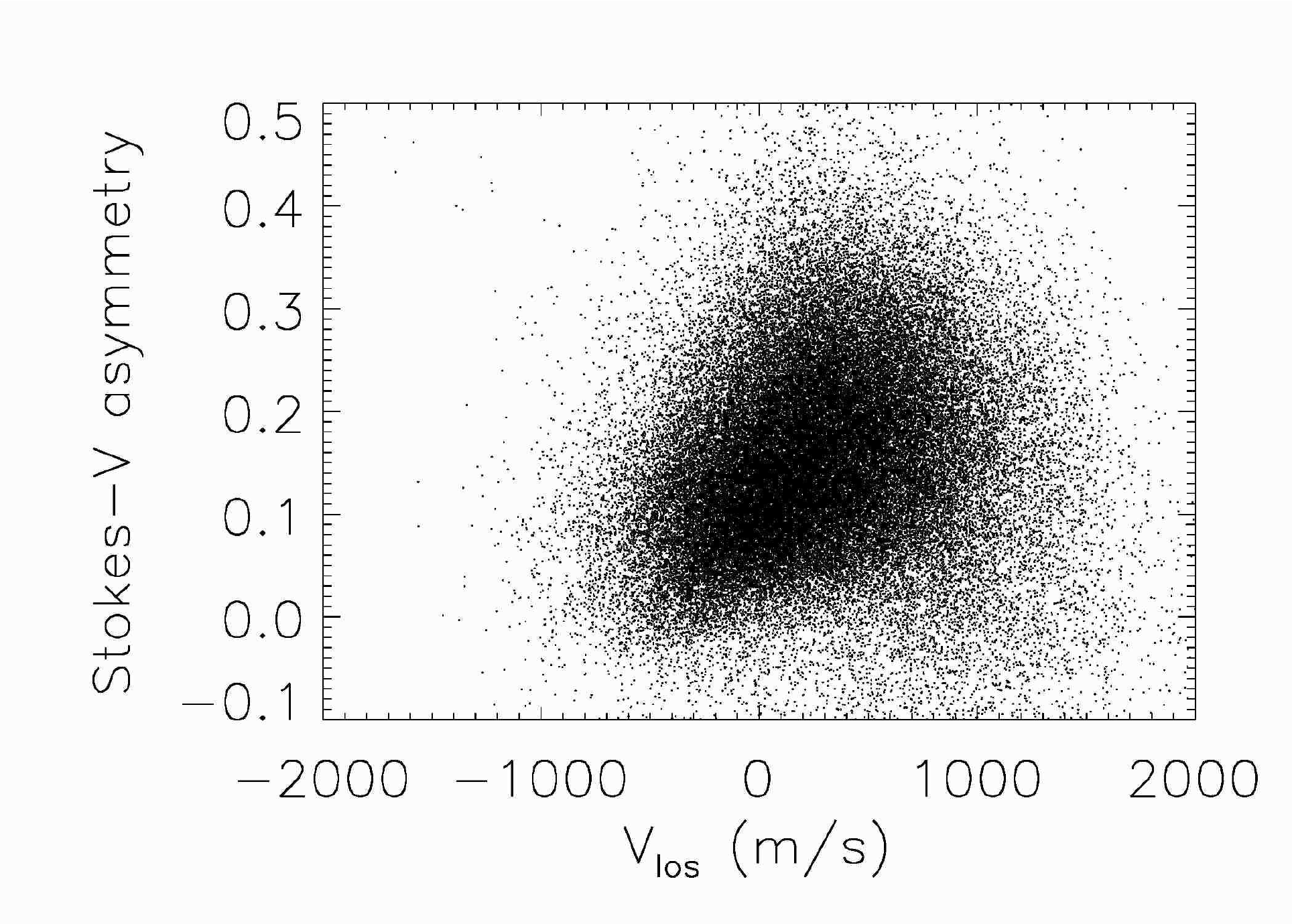}
    \includegraphics[bb=80 15 550 360,clip, height=0.172\linewidth]{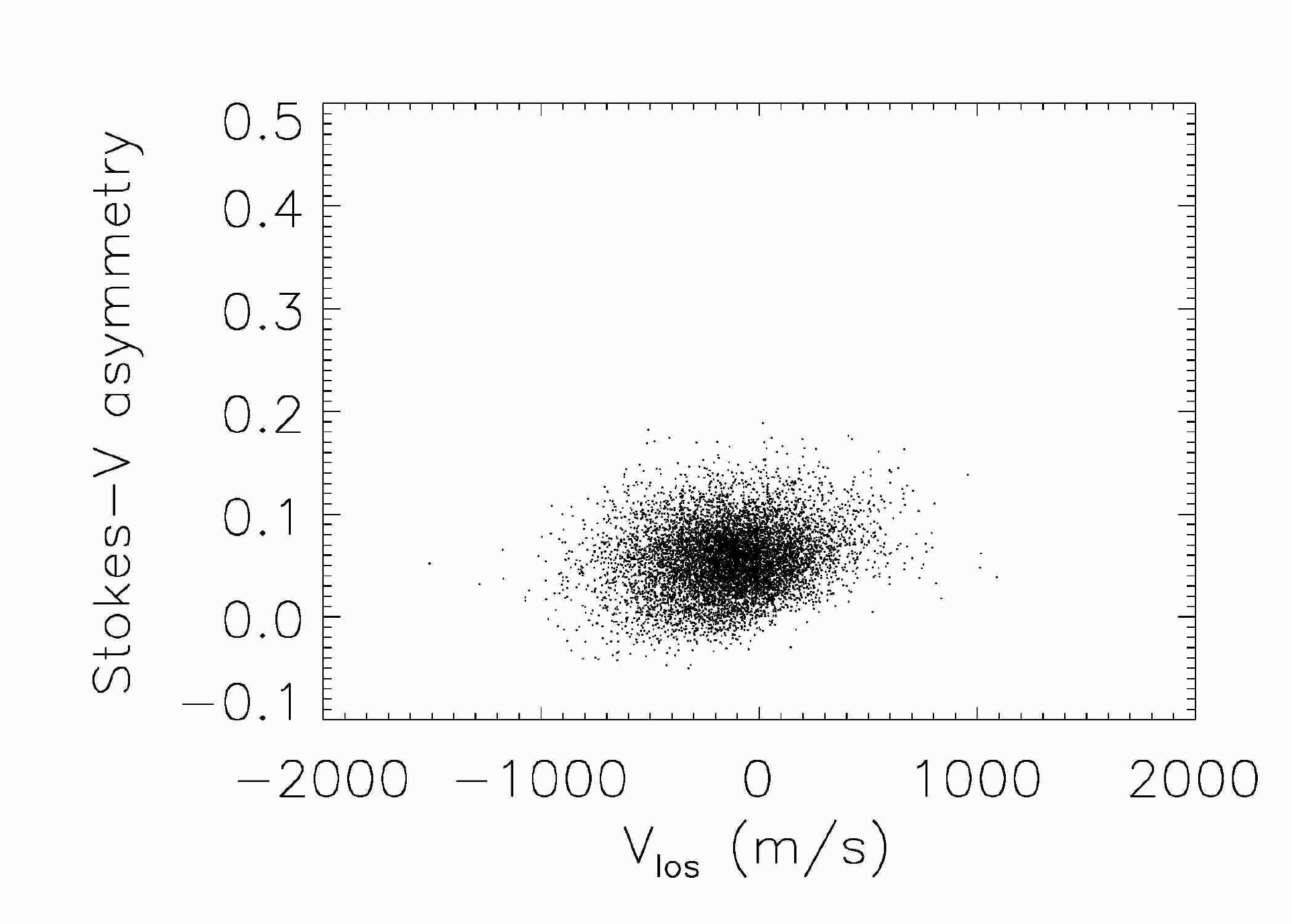}
%compressed figures:
    %\includegraphics[bb=45 30 510 340,clip, height=0.172\linewidth]{\figdir/sv_asym_B1_c.pdf}
    %\includegraphics[bb=80 30 510 340,clip,height=0.172\linewidth]{\figdir/sv_asym_B2_c.pdf}
    %\includegraphics[bb=80 30 510 340,clip, height=0.172\linewidth]{\figdir/sv_asym_V_c.pdf}
    %\includegraphics[bb=80 30 510 340,clip, height=0.172\linewidth]{\figdir/sv_asym_V2_c.pdf}
    \caption{Statistical properties of Stokes $V$ amplitude asymmetry,
      $\delta a$. From left to right the correlation of $\delta a$
      with the strength of the LOS magnetic field, $|\Blos|$, for
      all data (first plot) and the averaged $\delta a$ for dark
      (upper curve), all (middle curve) and bright (lower curve)
      structures (second plot). The third plot shows the correlation
      between LOS velocity and $\delta a$ for $200\text{
        G}<|\Blos|<800$~G, the fourth plot for $|\Blos|>800$~G.}
    \label{fig:sv_asymm2}
  \end{center}
\end{figure*}

\citet{2005A&A...429..335V} calculated the averaged bolometric disk
center intensity for 6$\times$6~Mm simulation ``boxes'' with different
average field strengths. He found that average field strengths up to
about 250~G leads to a small enhancement of the bolometric intensity
by approximately 1\%, but for stronger average fields the intensity is
reduced. At an average field strength of 400~G, the reduction in
intensity is only 2\% but for 800~G average field, the disk center
intensity is reduced by 10\%. The average intensity within our mask,
including the pores, is 98.6\% of that for the field-free region.
Excluding the pores, this average intensity is 99.6\%. Our magnetic
region thus appears much brighter than expected from simulations. This
may be a coincidental disagreement or indicate a problem of relaxation
of the simulation model, its deeper structure or the lower boundary.
The fourth panel in the top row of Fig.~\ref{fig:vlosplots} shows a
complicated correlation between the measured continuum intensity and
the strength of the LOS magnetic field. For $|\Blos| < 200$~G, the
intensity decreases with increasing field strength. For $|\Blos|$
larger than 200~G but smaller than 600~G, the intensity increases with
$|\Blos|$. For stronger fields the intensity again decreases with
$|\Blos|$. \citet{2005A&A...429..335V} investigated similar relations
between the disk center intensity and the vertical component of the
magnetic field, using simulations with average field strengths of 0,
50, 200, 400, and 800~G. He found that the intensity increases
monotonously with field strength for all simulations, i.e., the
strongest fields always appear to be the brightest. This is explained
as an opacity effect, shifting the $\tau=1$ level downward into hotter
regions, where the field strength is higher. This opacity effect is
large when the \emph{average} field strength is low and amounts to
only a few percent when the \emph{average} field is strong. A similar
variation of the field strength with temperature (at $\tau_{500}=0.1$)
was found by \citet{2007A&A...469..731S} from a 200~G simulation. In
our observations, regions with $|\Blos|$ stronger than 900~G
correspond mostly to micropores, the formation of which is missing in
the simulations of V\"ogler and Shelyag et al. Such pores did however
form spontaneously in the 400~G simulations of
\citet{2003ASPC..286..121S} and were also found in the 200~G
simulations of \citet{2005A&A...429..335V}.

\subsubsection{Magnetic downflows}
\label{sec:magnetic-downflows}

The lower three panels in Fig.~\ref{fig:vlosplots} show the
correlation between the LOS velocity \vlos\ and $|\Blos|$ for dark
($\Icont<0.9$), average ($0.9<\Icont<1.05$ and bright $\Icont>1.05$)
features. The dark structures mostly correspond to weak
($|\Blos|<100$~G) or strong $|\Blos|>800$~G fields. At locations of
very weak fields, the dark, average and bright structures show the
expected signatures of convection: dark downflows and bright upflows.
However, as $|\Blos|$ increases, the average LOS velocity increases
such that \emph{all structures show a strong tendency for downflows
when $|\Blos|$ is in the range 100--400~G}. The most striking plot
is that for the bright structures (third panel in the lower row of
Fig.~\ref{fig:vlosplots}). At low field strengths, bright structures
show average upflows of 1~k\mps\ but \emph{300--500~G bright
structures show average downflows} of 300~\mps.
Figure~\ref{fig:locations} shows the locations of all downflows (coded
as white) stronger than 300~\mps, having continuum intensities
stronger than 1.05 and $|\Blos|>50$~G. Such bright magnetic downflows
are over 30 times more frequent per unit area than for $|\Blos|<50$~G.
Also shown as gray is the region defined by the $|\Blos|>200$~G mask.
It is evident that \emph{virtually all bright magnetic downflows occur
near the boundary to the strong field}. At these locations
``leakage'' of polarized straylight from the limited spatial
resolution and straylight is likely to have a significant influence,
such that the actual field strengths for these features may be
significantly lower than measured. Our result is qualitatively
consistent with the simulations of \citet{2004RvMA...17...69V},
showing strong downflows only in thin sheets at the interface between
field-free granular structures protruding into the stronger field. The
brightness of these downflows is explained by the transparency of the
magnetic gas, allowing us to see deeper into hotter layers of the
downflowing (nearly field-free) gas
\citep{keller04origin,carlsson05high}. The fourth plot in the lower
row of Fig.~\ref{fig:vlosplots} shows the relation between \vlos\ and
$|\Blos|$ based on all data points. This is similar to Figs. 5 and 6
of \citet{2008A&A...481L..29M}, also showing the tendency for more
downward flows as $|\Blos|$ increases from zero to about 400~G.
However, the amplitude of the \vlos\ variation is much larger for the
SST data than for the Hinode data. Apart from the higher spatial
resolution of the SST data, a difference between the analysis of
Morinaga et al. and ours is that their LOS velocities were obtained by
fitting the line core of the 630.15 nm line, thus corresponding to
velocities at a higher layer in the atmosphere than for our data.
Also, Morinaga et al. calibrated their LOS velocities to the average
value for the surrounding quiet sun (showing convective blueshift),
whereas we use the pores within the FOV as reference. Compensating for
the differences in zero point, we obtain qualitatively similar
results: an increased LOS velocity for field strengths in the range
0--300~G. We also note the presence of a weak population with
\emph{strong} downflows for bright structures with $|\Blos|$ in the
range 200--600~G. 

Finally, we note (third panel in upper row of
Fig.~\ref{fig:vlosplots}) that both dark and bright features with the
LOS magnetic field stronger than 600~G on the average show upflows,
reaching up to $-500$~\mps\ for the strongest fields.

\subsubsection{Stokes $V$ asymmetries}
To exploit observed properties of the Stokes $V$ profiles ignored in
Milne--Eddington inversions, we discuss also their asymmetries.
Several such measures of asymmetries are used in the literature: the
net circular polarization, the area asymmetry and the amplitude
asymmetry. The limited wavelength range scanned with CRISP for these
data (c.f., Fig.~\ref{fig:fitobs}) prevents accurate estimates of the
first two of these measures. Here, we discuss only the Stokes $V$
amplitude asymmetry $\delta a$, defined as
\begin{equation}
  \delta a = (a_\text{b}-a_\text{r})/(a_\text{b}+a_\text{r})
\end{equation}
where $a_b$ and $a_r$ are the absolute values of the maximum and
minimum values of the Stokes $V$ profile in the blue and red wings
respectively. These were estimated by fitting second order polynomials to $V$
at the three wavelengths closest to the extrema and from the fitted
coefficients determine max/min values. For a small percentage of
pixels, the extrema are at the first or last wavelength scanned by
CRISP; for these we simply used the maximum or minimum values found to
estimate $\delta a$. Since we use only three wavelengths for the fits
to each of the extrema, the measured amplitude asymmetry is noisy for
field strengths below about 100~G. In Fig.~\ref{fig:sv_asymm1} we show
$\delta a$ scaled from $-0.1$ to $+0.5$ with a mask to eliminate
pixels where the strength of the LOS magnetic field is less than 75~G. Above this threshold, we find no multi-lobed $V$ profiles among the hundreds of profiles inspected.
For nearly all pixels shown, the Stokes $V$ amplitude asymmetry is
positive. Strongly enhanced $\delta a$ forms irregular connected
``networks'', shown bright in the figure. Blinking the $\delta a$ map
with the continuum image, we note that \emph{strong Stokes V amplitude
asymmetry is found almost exclusively at the boundary between
regions with small-scale abnormal granulation and much larger
``normal-looking'' granules}. The left-most plot in
Fig.~\ref{fig:sv_asymm2} shows the correlation between $|\Blos|$ and
$\delta a$ and includes all pixels in the FOV, the red curve
corresponds to binned averages. Quite clearly, $\delta a$ peaks for
the weaker fields and Stokes amplitude asymmetries are very small for
the strongest fields. The second plot shows averaged values for dark
($\Icont<0.9$, upper curve), and bright structures ($\Icont>1.05$,
lower curve). This demonstrates that darker structures have somewhat
more asymmetric $V$ profiles than brighter structures at small
$|\Blos|$. The third plot shows that downflows on the average have
much more asymmetric $V$ profiles than upflows when $|\Blos|$ is in
the range 200--800~G, although the scatter is large. The fourth plot
shows only a weak correlation for $|\Blos|$ stronger than 800~G, but
with a clear tendency for some upflows to have negative $V$
asymmetries. We note that the scatter in $\delta a$ is much smaller
for large $|B_{LOS}|$ than for weak fields, this is not an effect of
noise.

The first plot in Fig.~\ref{fig:sv_asymm2} corresponds to Fig. 15 of
\citet{2007A&A...469..731S}, showing the variation of the 630.25~nm
Stokes $V$ amplitude asymmetry with field strength at $\tau=0.1$ from
250~G simulation data. This plot looks strikingly similar to ours,
except that the simulated data shows an amplitude asymmetry that is
50\% larger for small field strengths than for our data. It appears
plausible that this difference is related to the lower spatial and/or
spectral resolution of the CRISP/SST data.

We note that the Stokes $V$ asymmetry peaks at roughly the same low
field strengths for which dark as well as bright gas on the average
shows downflows c.f., Fig.~\ref{fig:vlosplots}. This strongly supports
the explanation of the $V$ asymmetries by \citet{2007A&A...469..731S}
as due to the peripheral parts of magnetic flux concentrations
expanding with height and being surrounded with (strong) downflows. At
these locations, the LOS cuts through the magnetopause such that the
magnetic field strength \emph{decreases} with depth while the downflow
velocity increases in magnitude with depth, consistent with $\delta a$
being positive. Interpretations along these lines were first
proposed by \citet{1988A&A...201L..37S} and supported by, e.g., \citet{1988A&A...206L..37G,1989A&A...224..225S, 1997ApJ...474..810M}.
%and were supported, e.g., by 1\arcsec\ resolution ASP data \citep{1997ApJ...474..810M}. 
Our present high-resolution data confirms
this nicely by locating the peaking of $\delta a$ precisely at the
boundary between abnormal and normal-looking granules, where we expect
magnetic canopies to expand over the more field-free gas. The similar
correlations for \emph{weak} plage simulation data and the present
observed data from strong plage suggests that the dynamics close to
the \emph{boundary} of flux concentrations is similar for small and
more extended flux concentrations.

The fourth plot in Fig.~\ref{fig:sv_asymm2} shows the correlation
between $V$ asymmetry and LOS velocity for field strengths stronger
than 800~G, corresponding to the \emph{interior} of the flux
concentrations. Here, the gas shows mostly upflows (see lower four
plots of Fig.~\ref{fig:vlosplots}) and we expect the field to expand
with height and its strength thus to decrease with height. In this
region $\delta a$ remains mostly positive, as expected with upflow
velocities decreasing in magnitude with height. The correlation
between $\vlos$ and $\delta a$ is weak, but in the sense of being
stronger for downflows than for upflows, for pixels with $|\Blos|$
stronger than 800~G. There is also a small fraction of pixels with
\emph{negative} $V$ asymmetry where upflows are stronger than
500~\mps. Whether this fits within a magneto-convection scenario with
strong field remains to be investigated.

\begin{figure*}[!htb]
  \begin{center}
    \includegraphics[bb=31 15 702 839,width=0.45\linewidth]{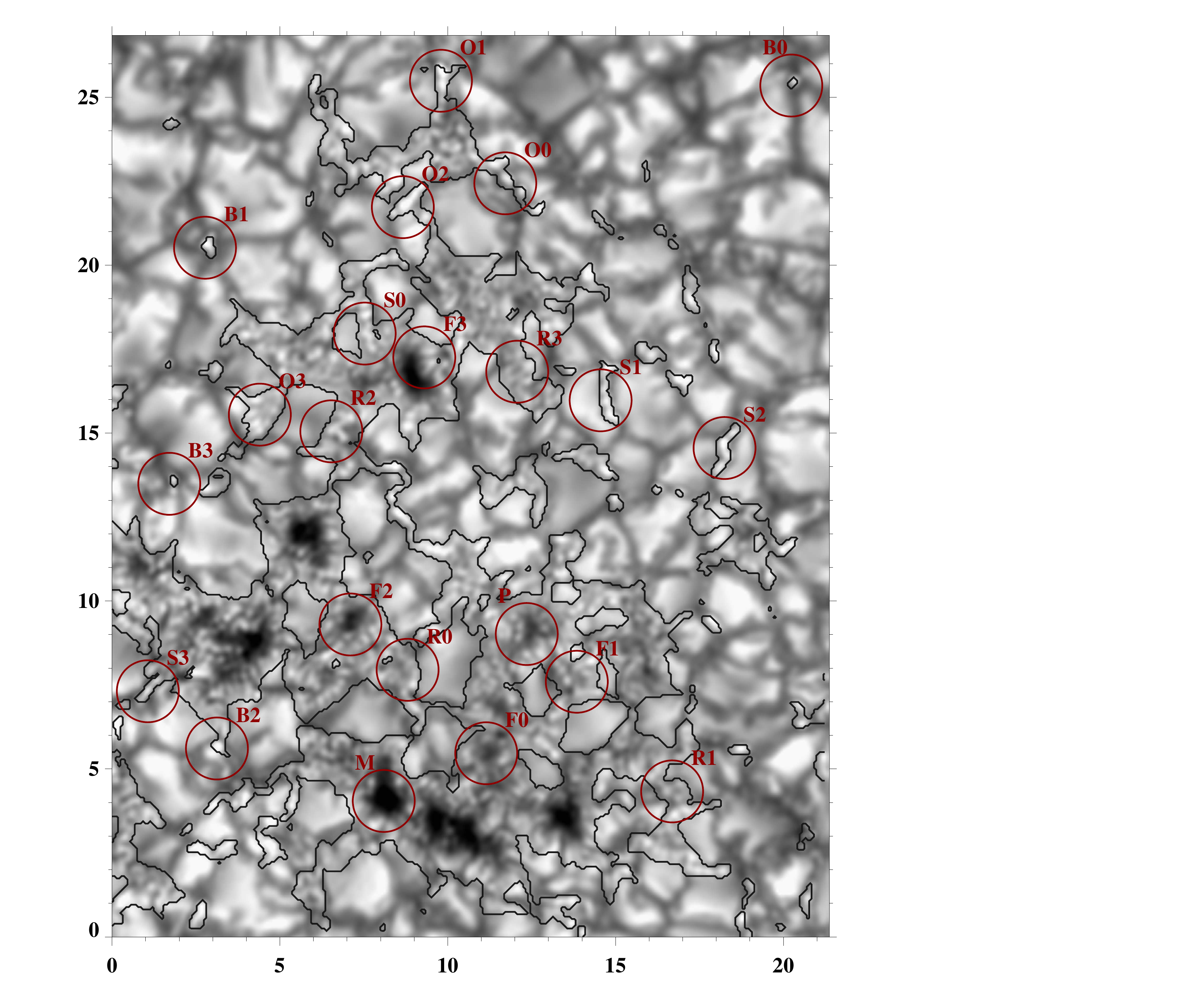}
    \includegraphics[bb=31 15 702 839,width=0.45\linewidth]{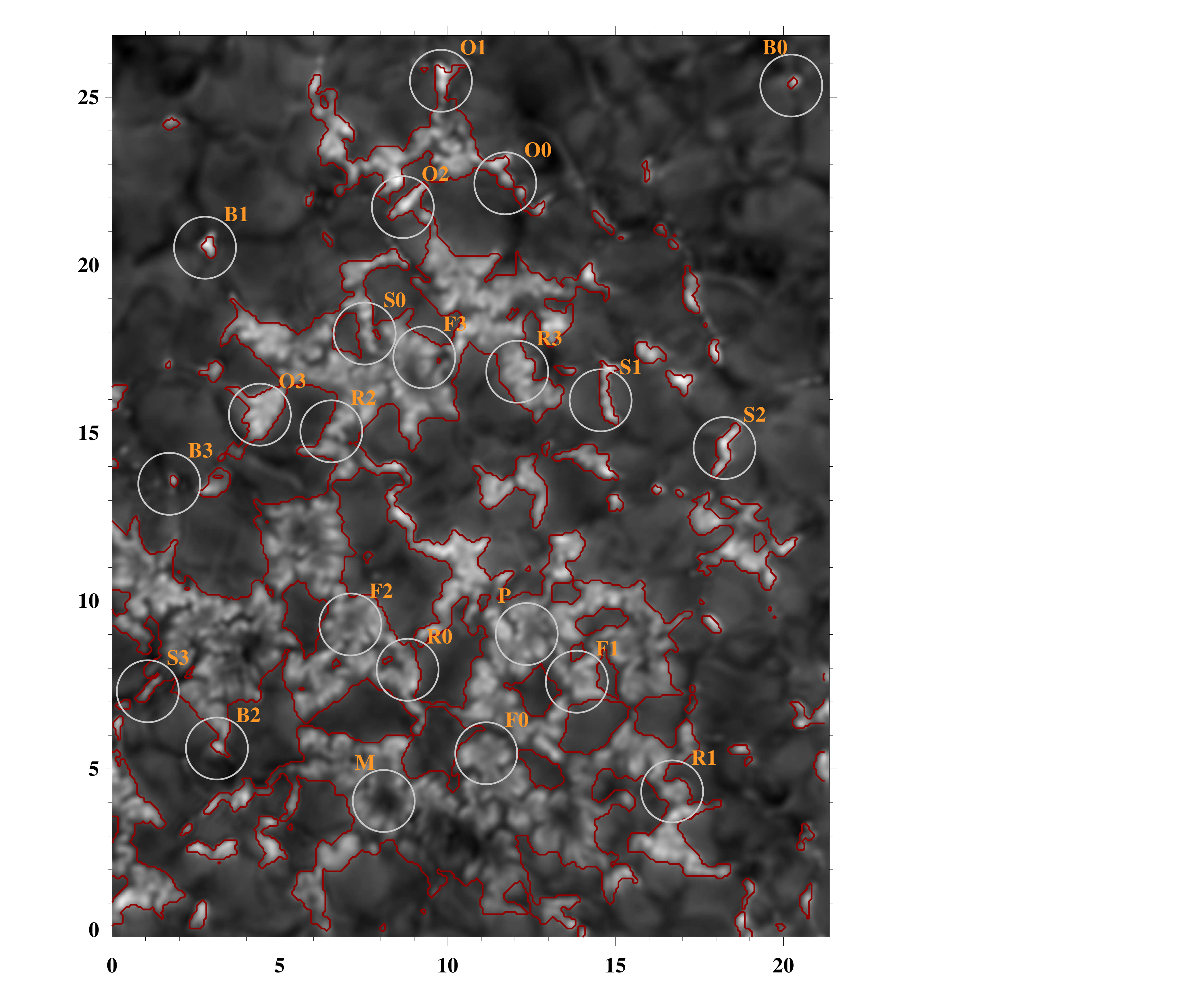}
    \includegraphics[bb=31 15 702 839,width=0.45\linewidth]{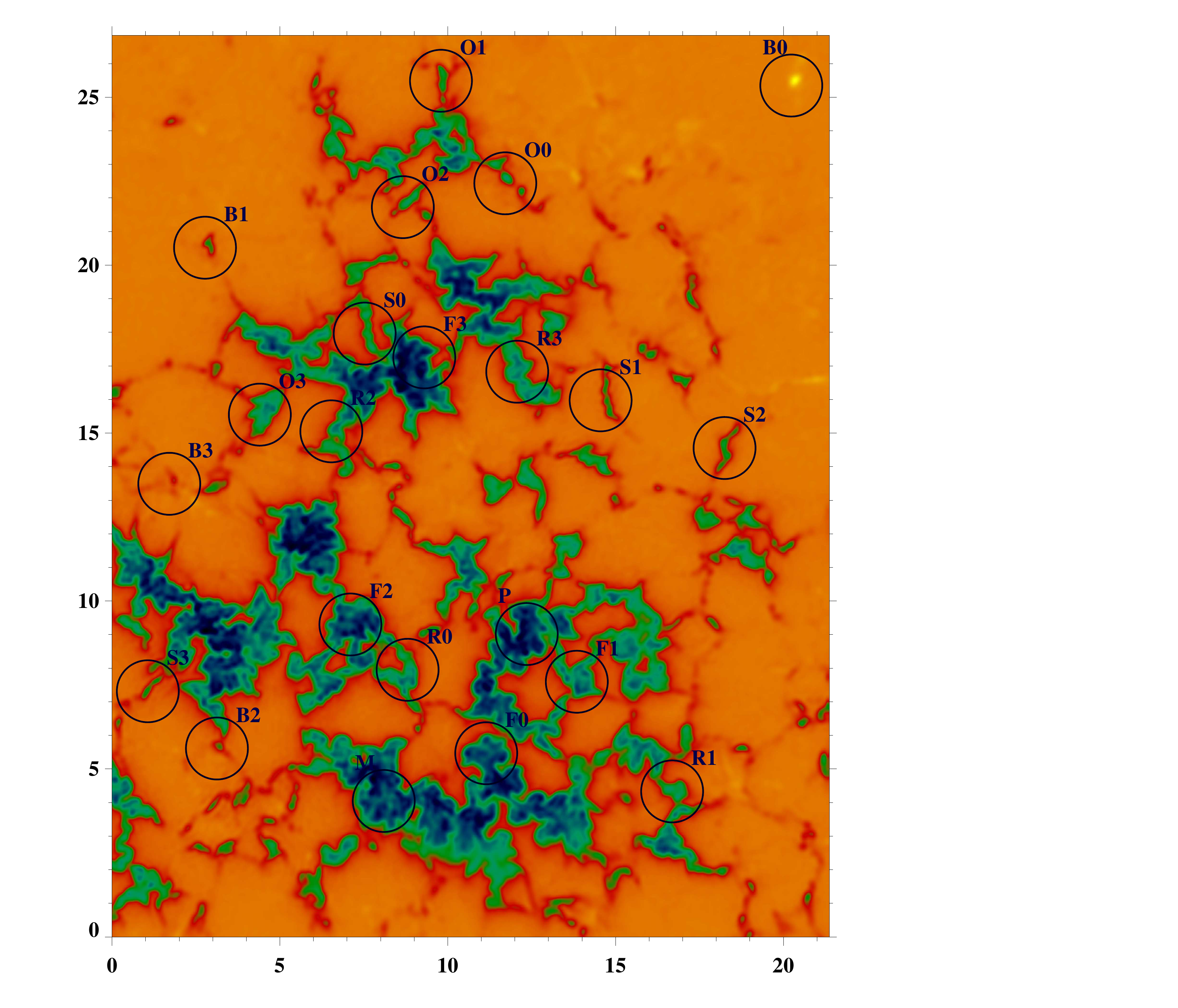}
    \includegraphics[bb=31 15 702 839,width=0.45\linewidth]{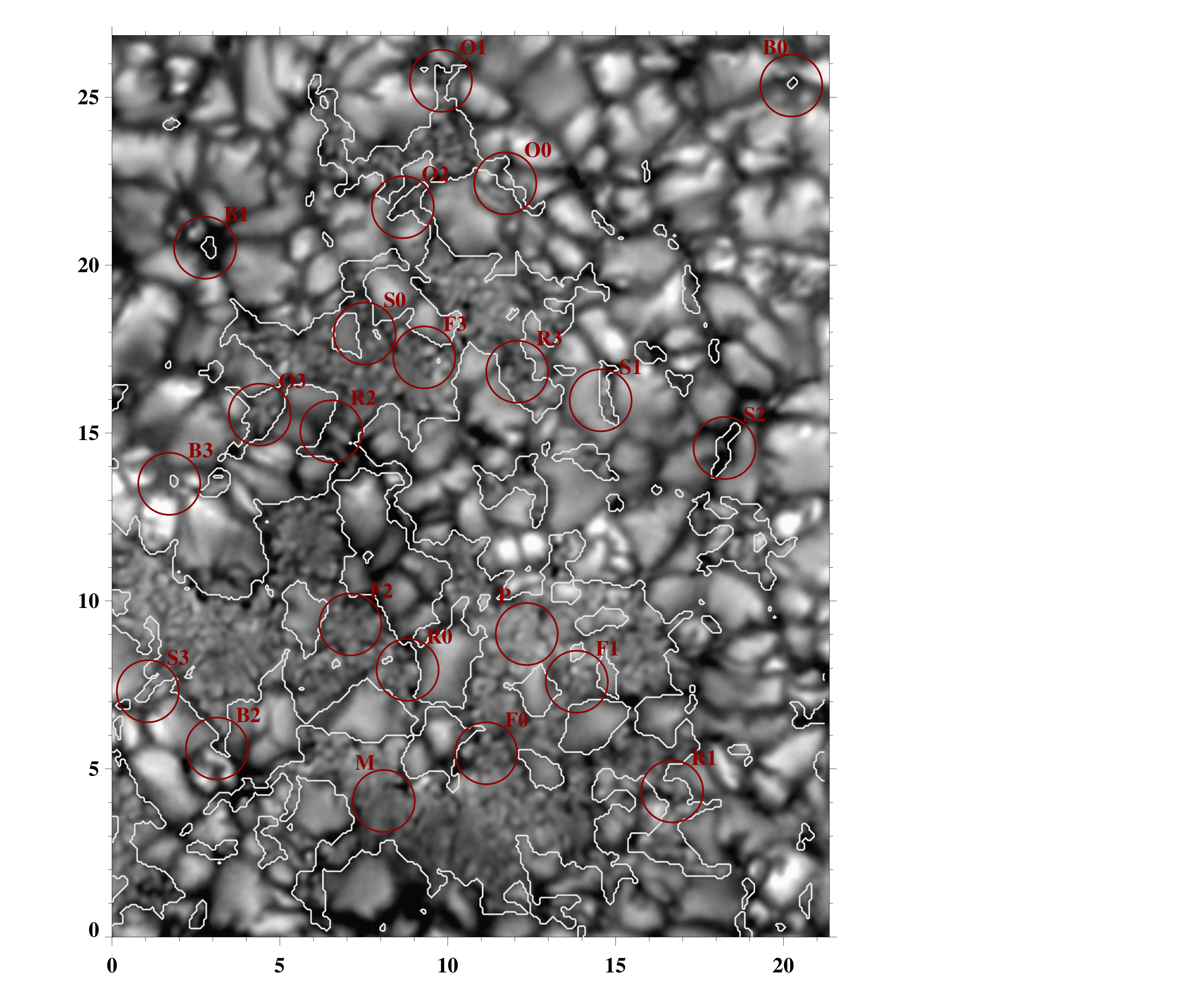}
    \includegraphics[bb=2 744 976 807,clip,width=0.45\linewidth]{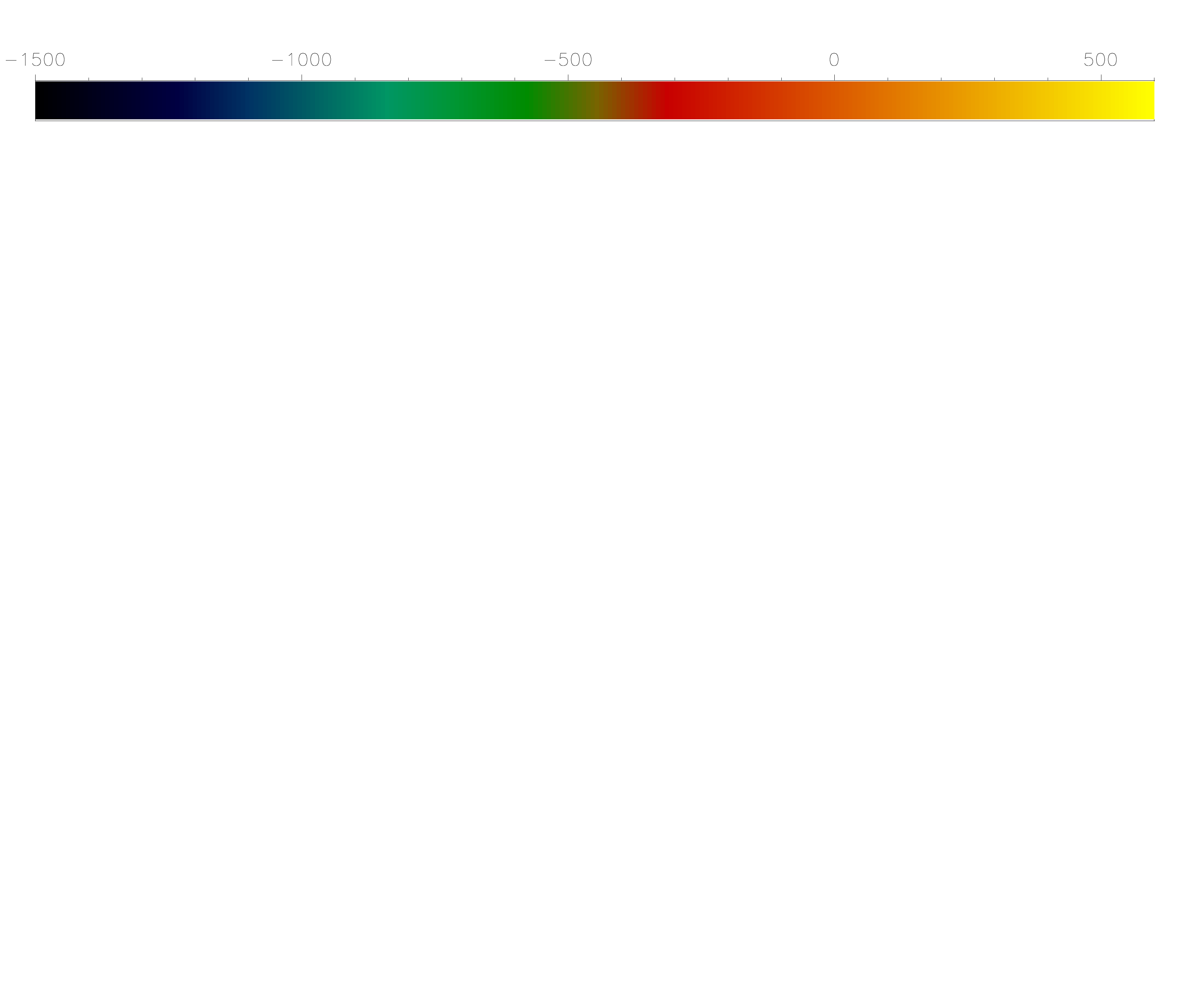}
    \hspace{0.4cm}
    \includegraphics[bb=2 744 992 807,width=0.45\linewidth]{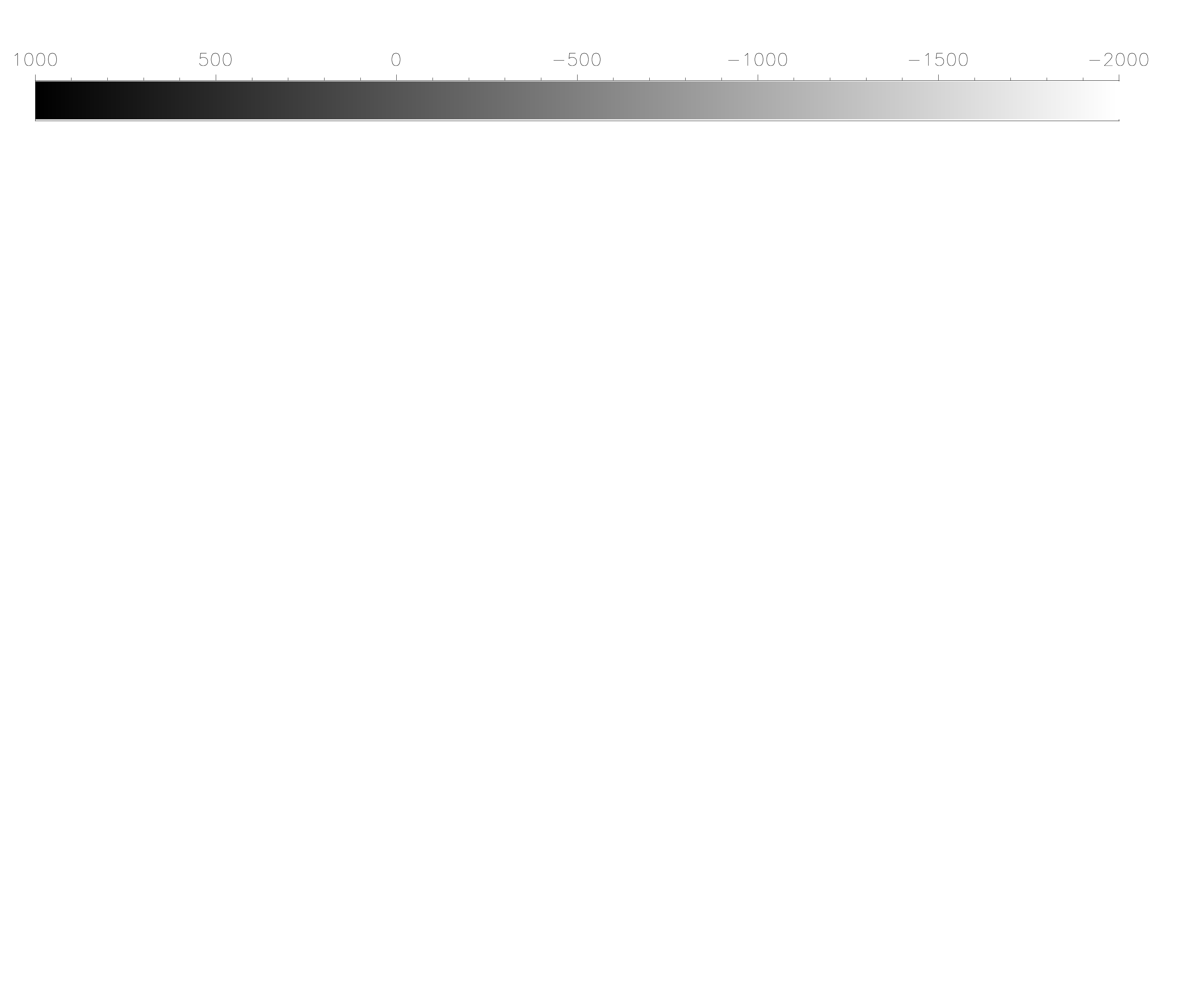}
    \caption{The ROI shown in Fig.~\ref{fig:fullFOV}, used to identify
      isolated bright points (B0--B3), ribbons (R0--R3), flowers
      (F0--F3), strings (S0--S3) and other features (O0--O3). The
      upper row shows the continuum image and the \Imin\ map, the
      lower row \Blos\ and \vlos\ maps obtained from inversions. The
      white/black/red contours correspond to $|\Blos|=200$~G. Tick marks are in units of arcsec. The color bar indicates the signed LOS magnetic field in Gauss, the grey scale bar the LOS velocity in \mps.}
    \label{fig:zoommosaic}
  \end{center}
\end{figure*}

\subsection{Small-scale velocity features}
\label{sec:small-scale-velocity}

\renewcommand{\thesubfigure}{{}}
\begin{figure*}[htb!]
  \subfigure[ B0]{
    \includegraphics[bb=45 95 1133 640,clip,width=0.240\linewidth]{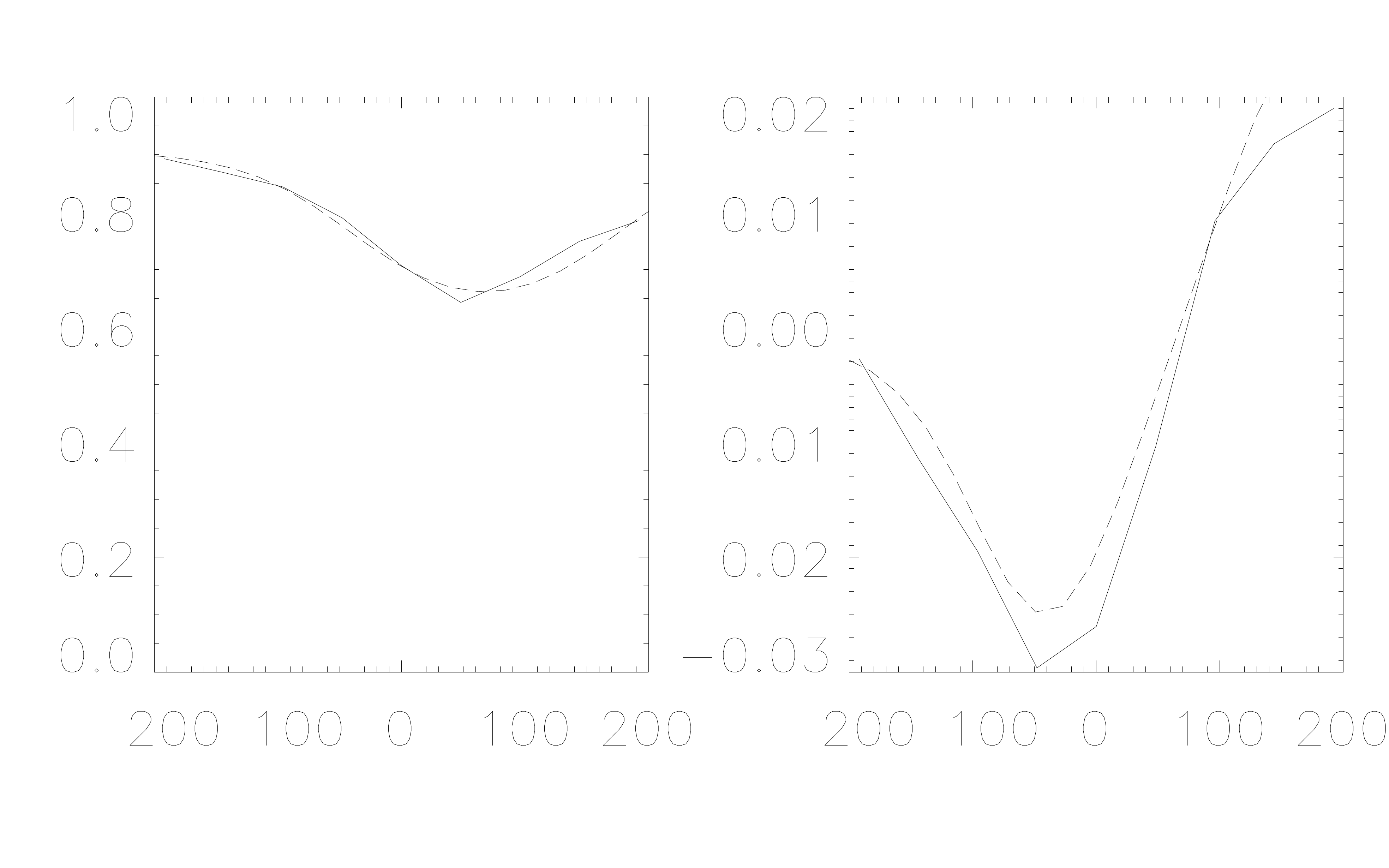}}
  \subfigure[ B1]{
    \includegraphics[bb=45 95 1133 640,clip,width=0.240\linewidth]{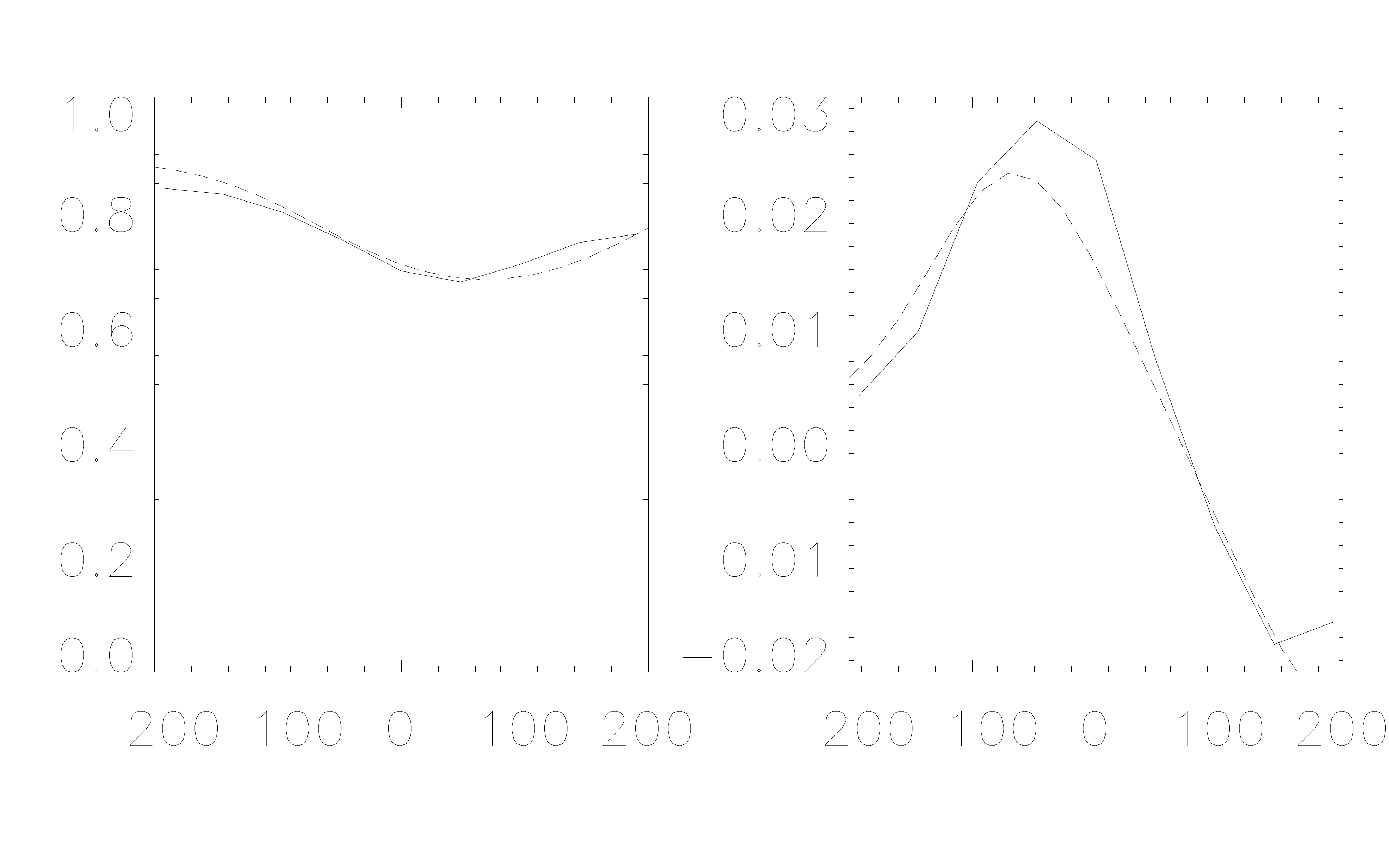}}
  \subfigure[ B2]{
    \includegraphics[bb=45 95 1133 640,clip,width=0.240\linewidth]{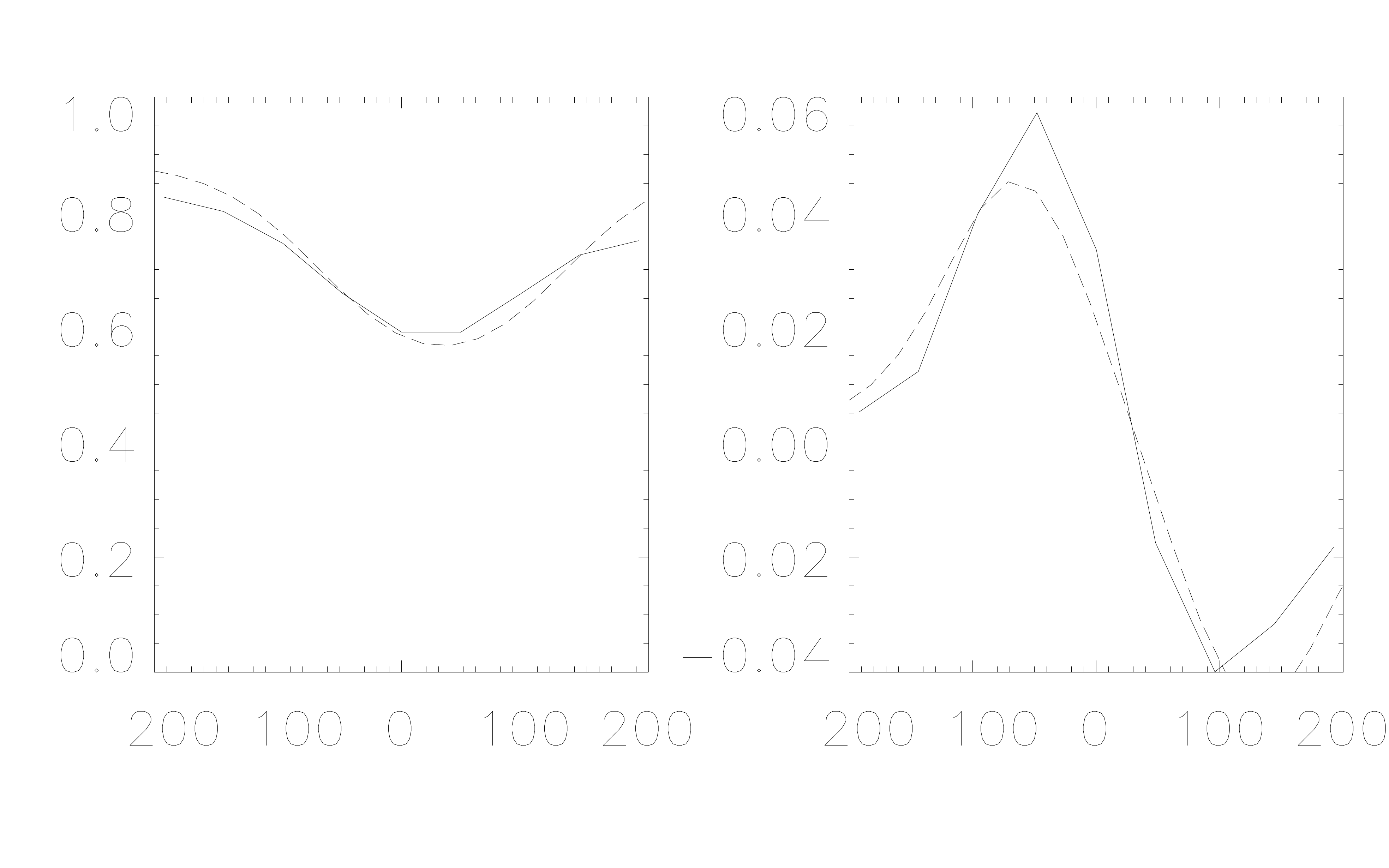}}
  \subfigure[ B3]{
    \includegraphics[bb=45 95 1133 640,clip,width=0.240\linewidth]{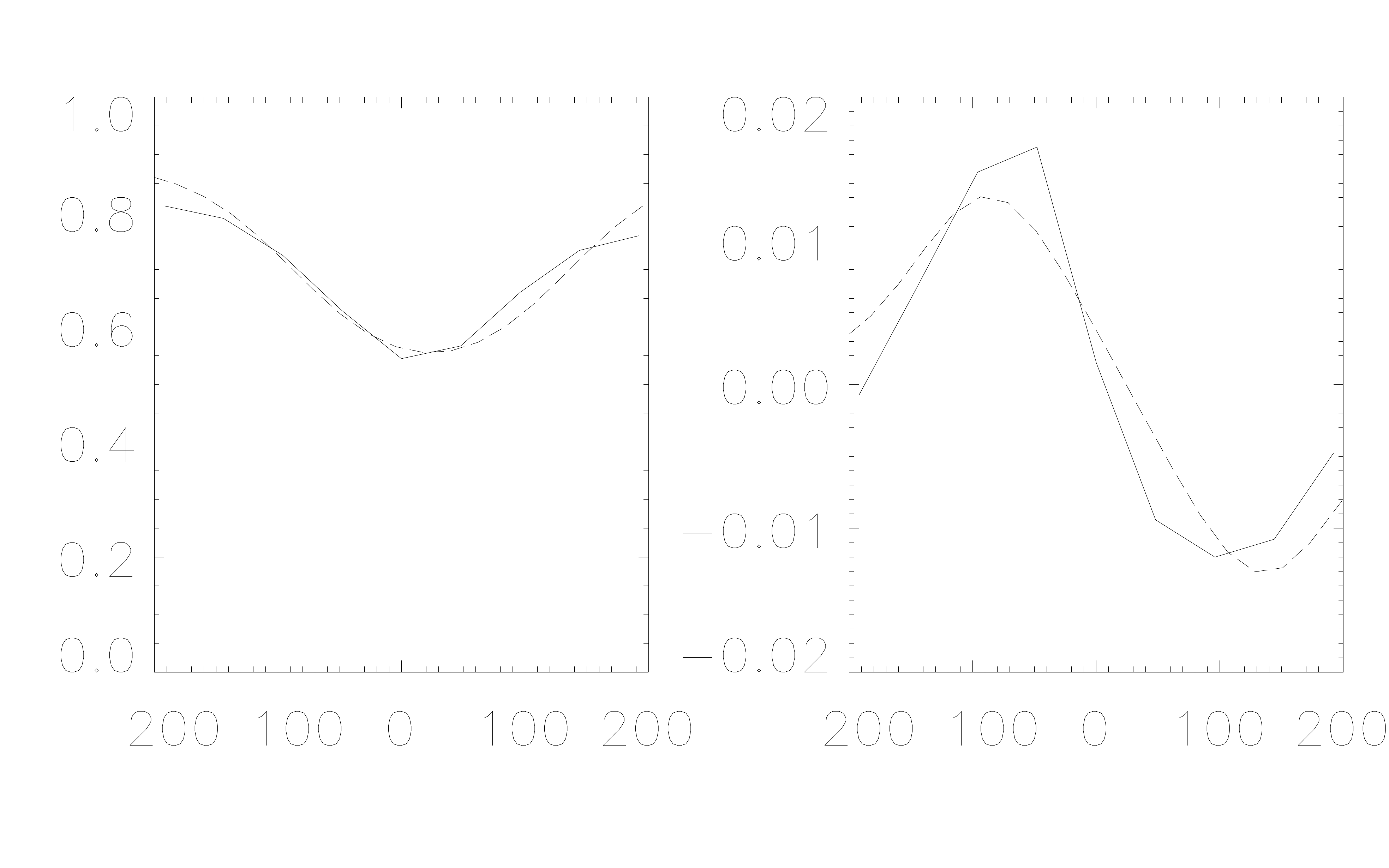}}
  \subfigure[ R0]{
    \includegraphics[bb=45 95 1133 640,clip,width=0.240\linewidth]{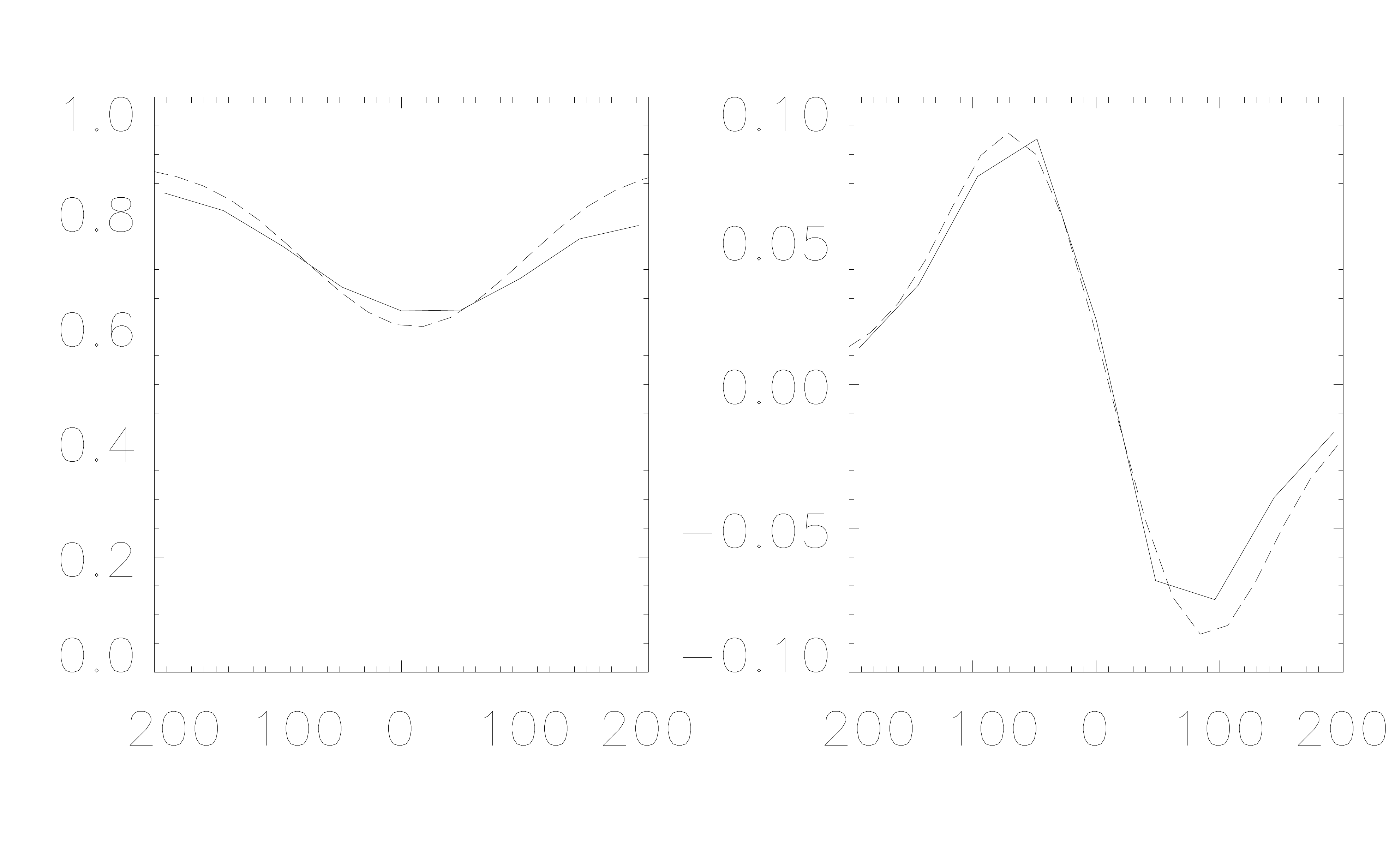}}
  \subfigure[ R1]{
    \includegraphics[bb=45 95 1133 640,clip,width=0.240\linewidth]{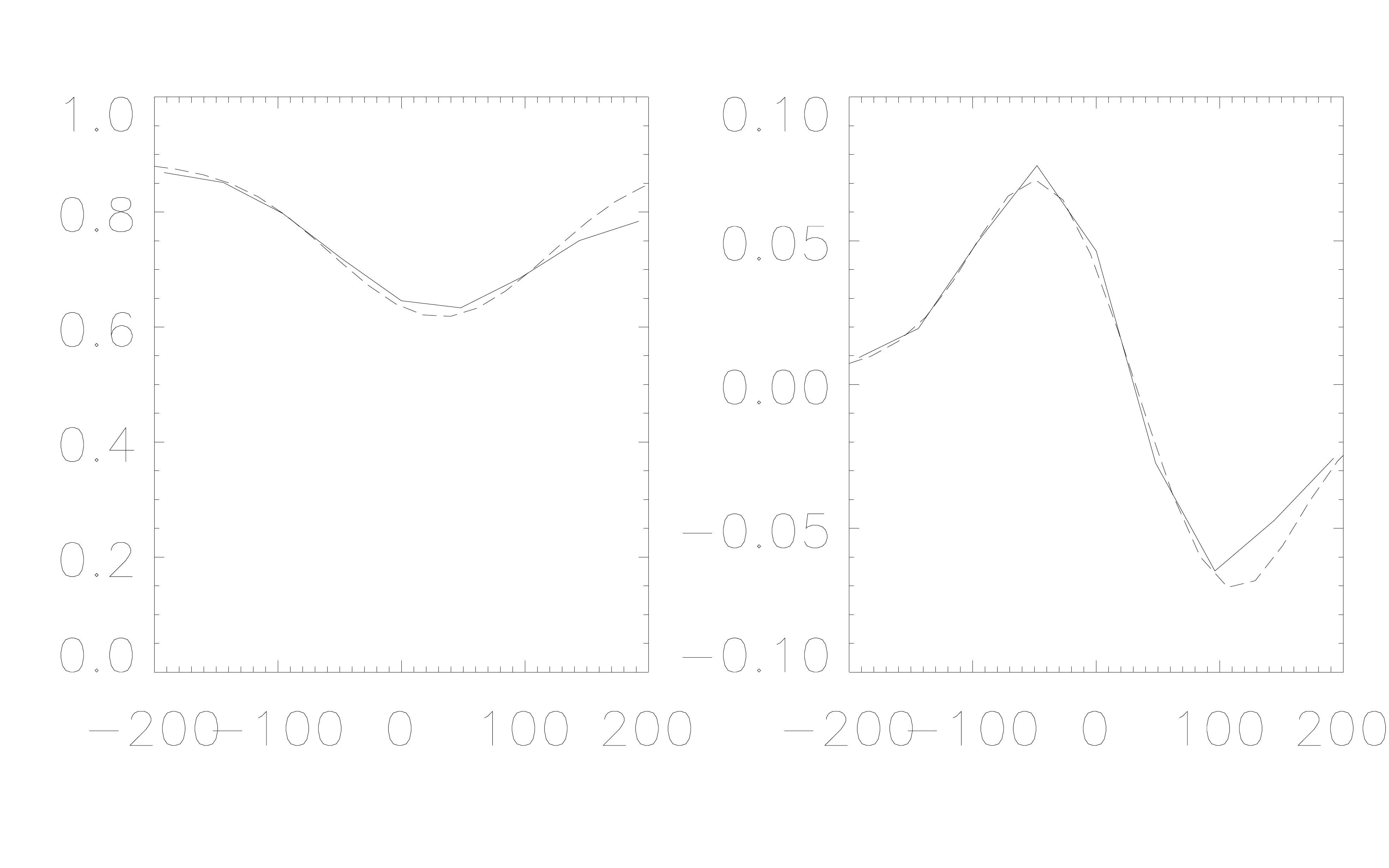}}
  \subfigure[ R2]{
    \includegraphics[bb=45 95 1133 640,clip,width=0.240\linewidth]{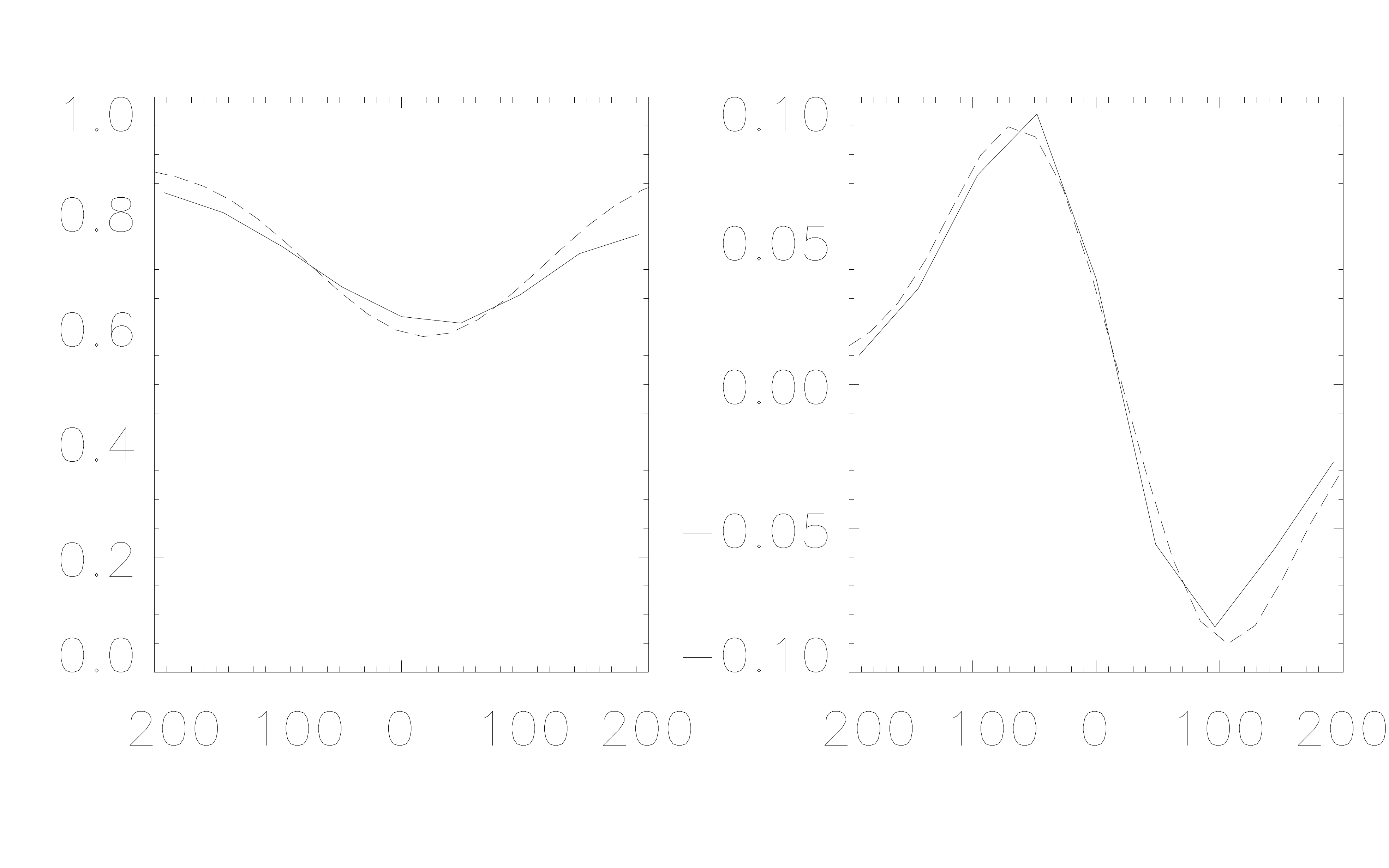}}
  \subfigure[ R3]{
    \includegraphics[bb=45 95 1133 640,clip,width=0.240\linewidth]{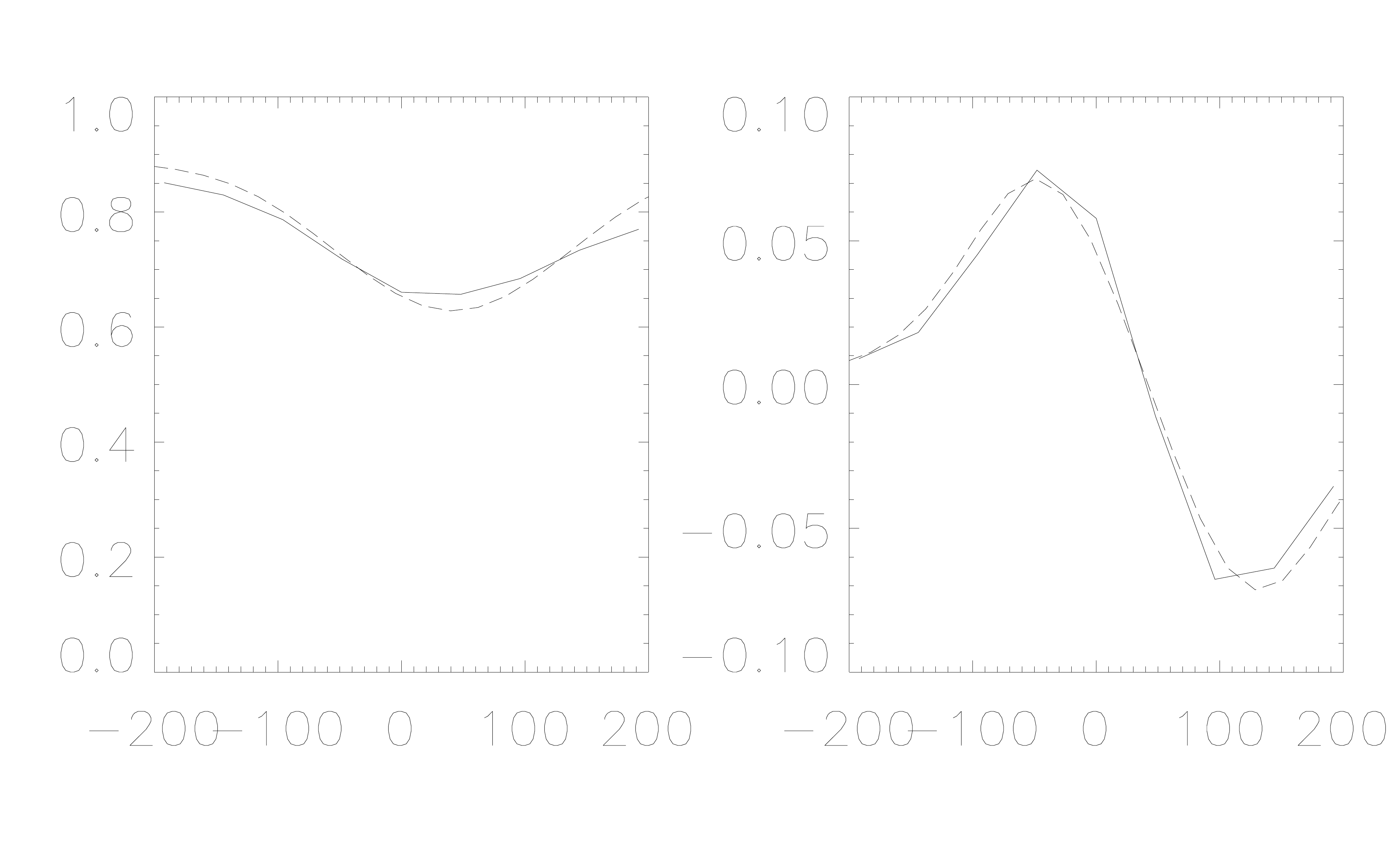}}
  \subfigure[ F0]{
    \includegraphics[bb=45 95 1133 640,clip,width=0.240\linewidth]{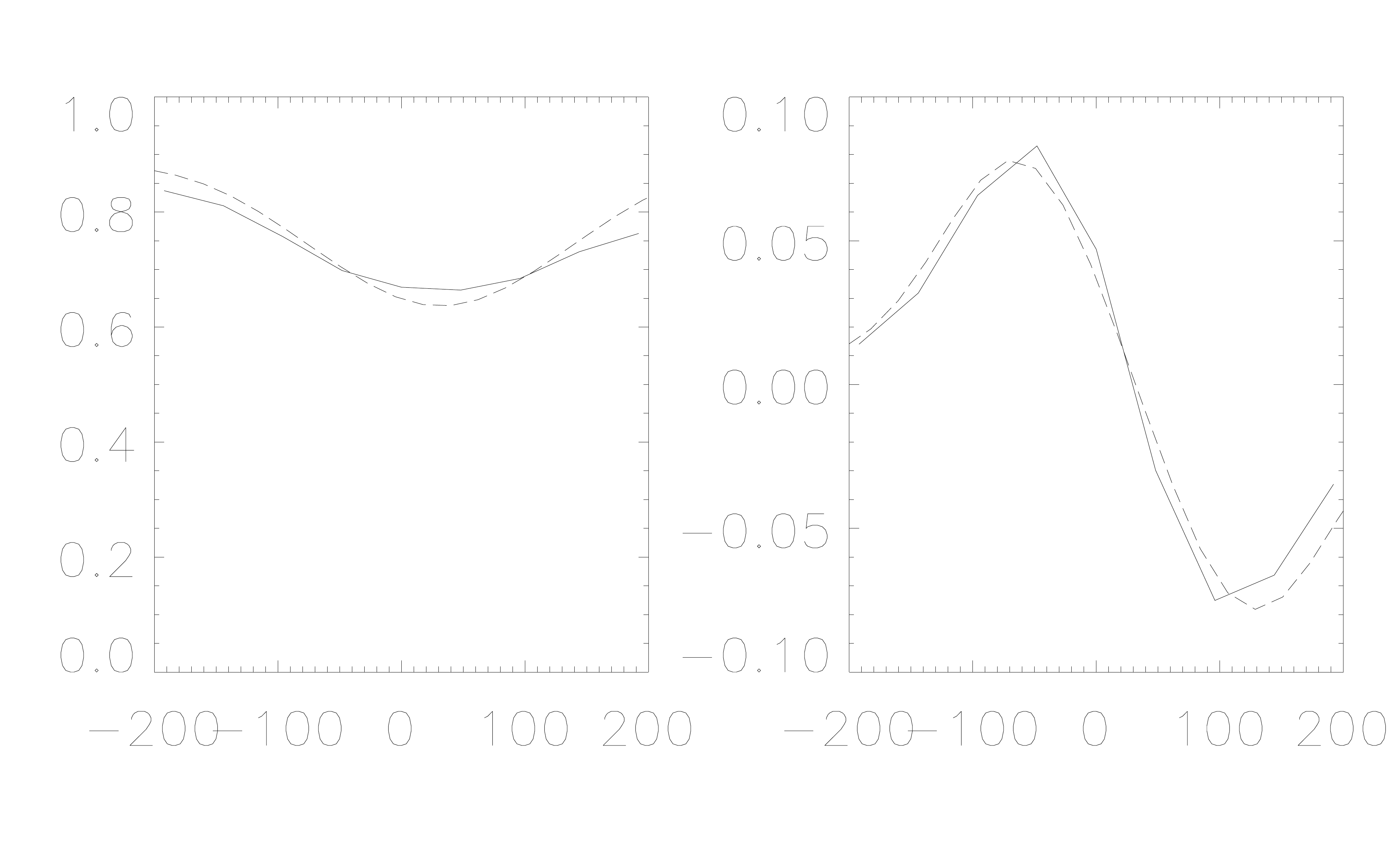}}
  \subfigure[ F1]{
    \includegraphics[bb=45 95 1133 640,clip,width=0.240\linewidth]{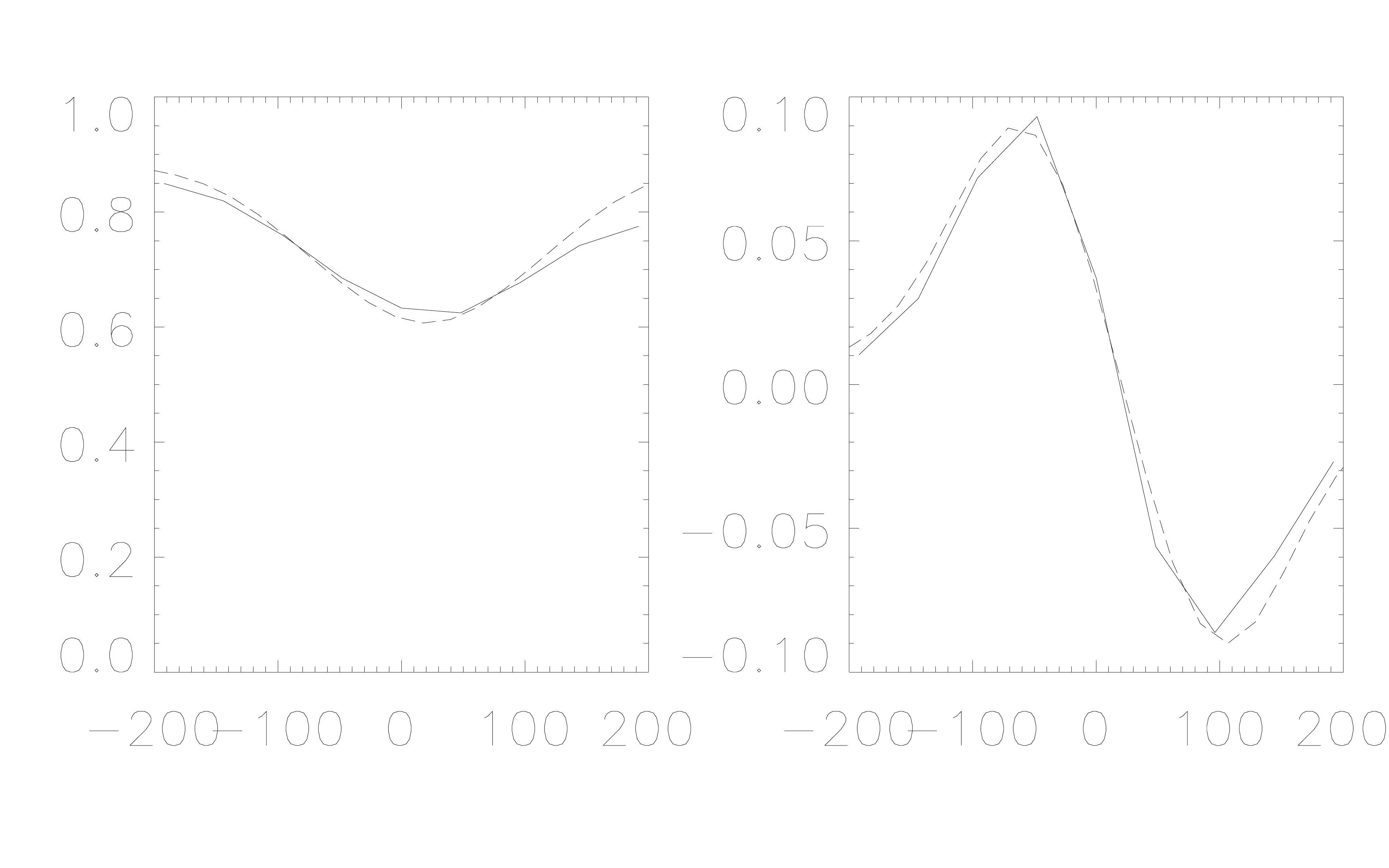}}
  \subfigure[ F2]{
    \includegraphics[bb=45 95 1133 640,clip,width=0.240\linewidth]{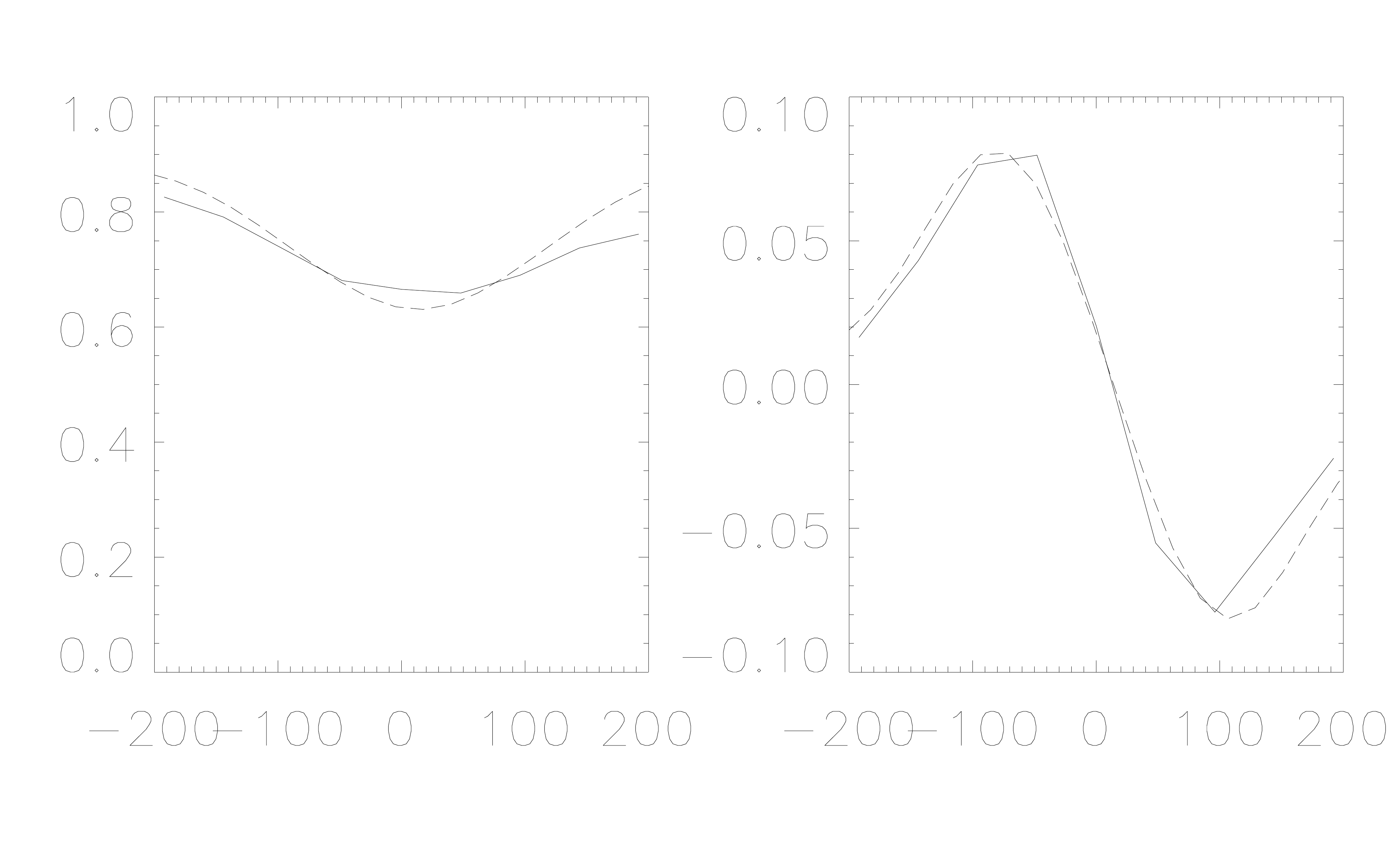}}
  \subfigure[F3]{
    \includegraphics[bb=45 95 1133 640,clip,width=0.240\linewidth]{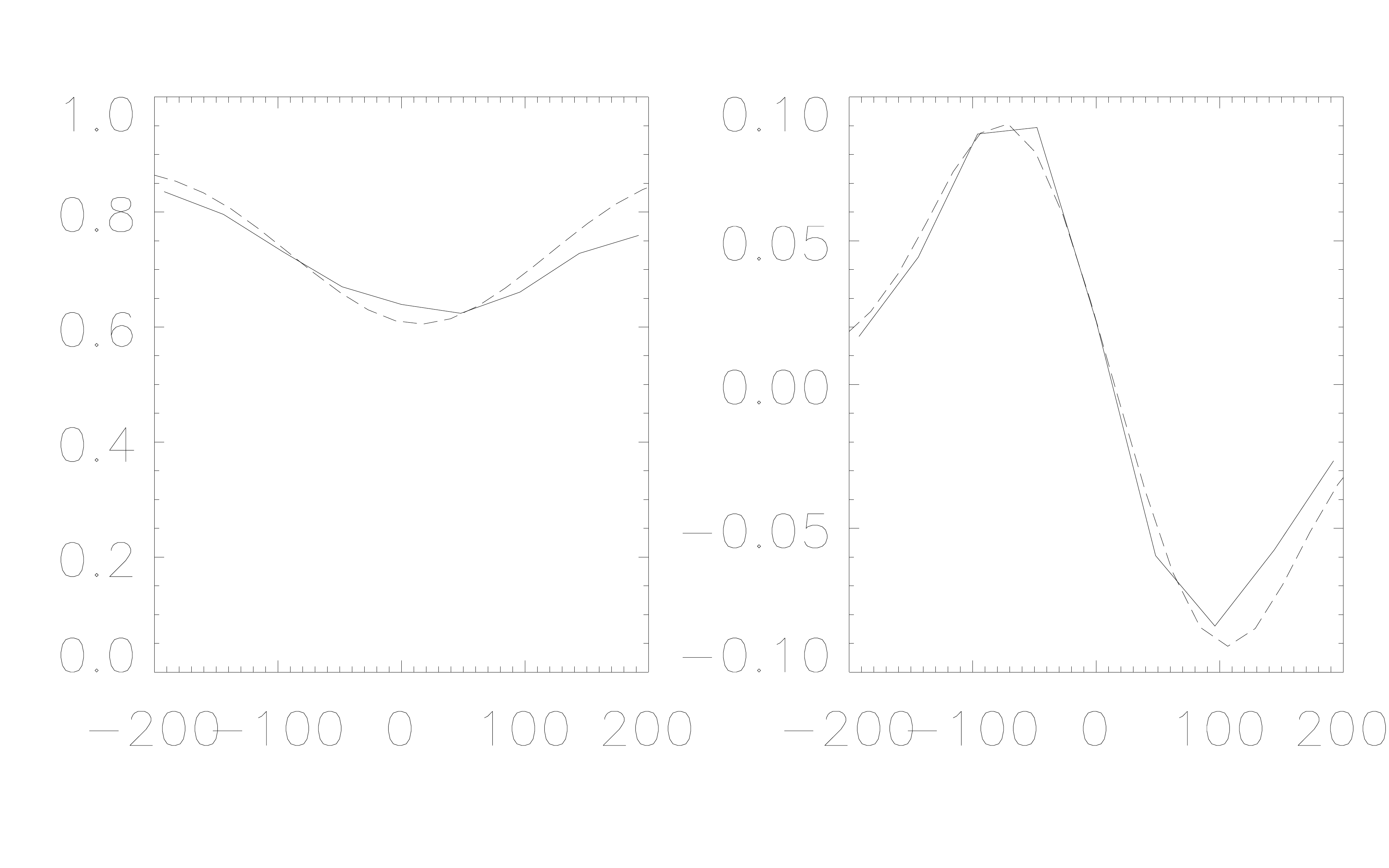}}
  \subfigure[ S0]{
    \includegraphics[bb=45 95 1133 640,clip,width=0.240\linewidth]{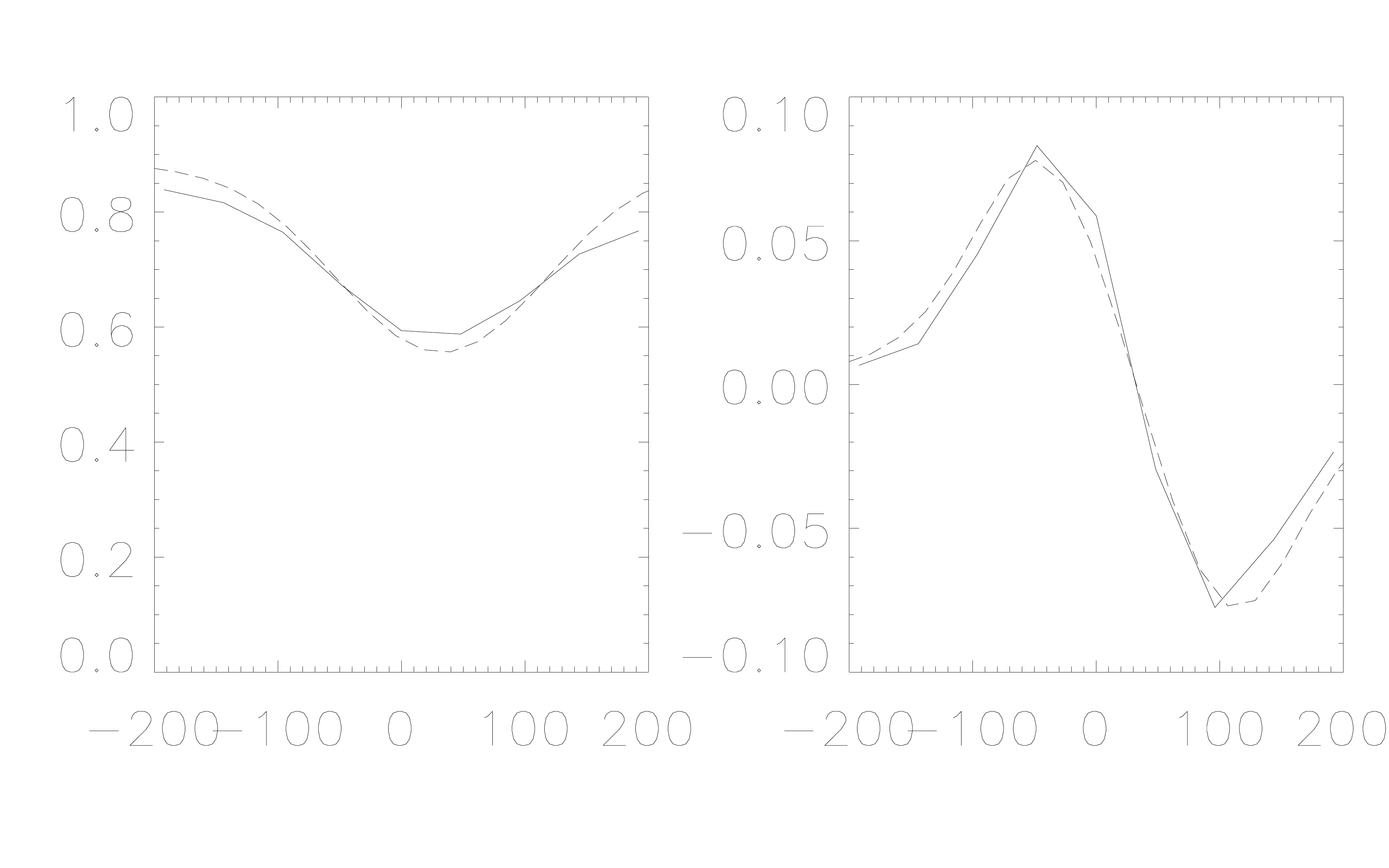}}
  \subfigure[ S1]{
    \includegraphics[bb=45 95 1133 640,clip,width=0.240\linewidth]{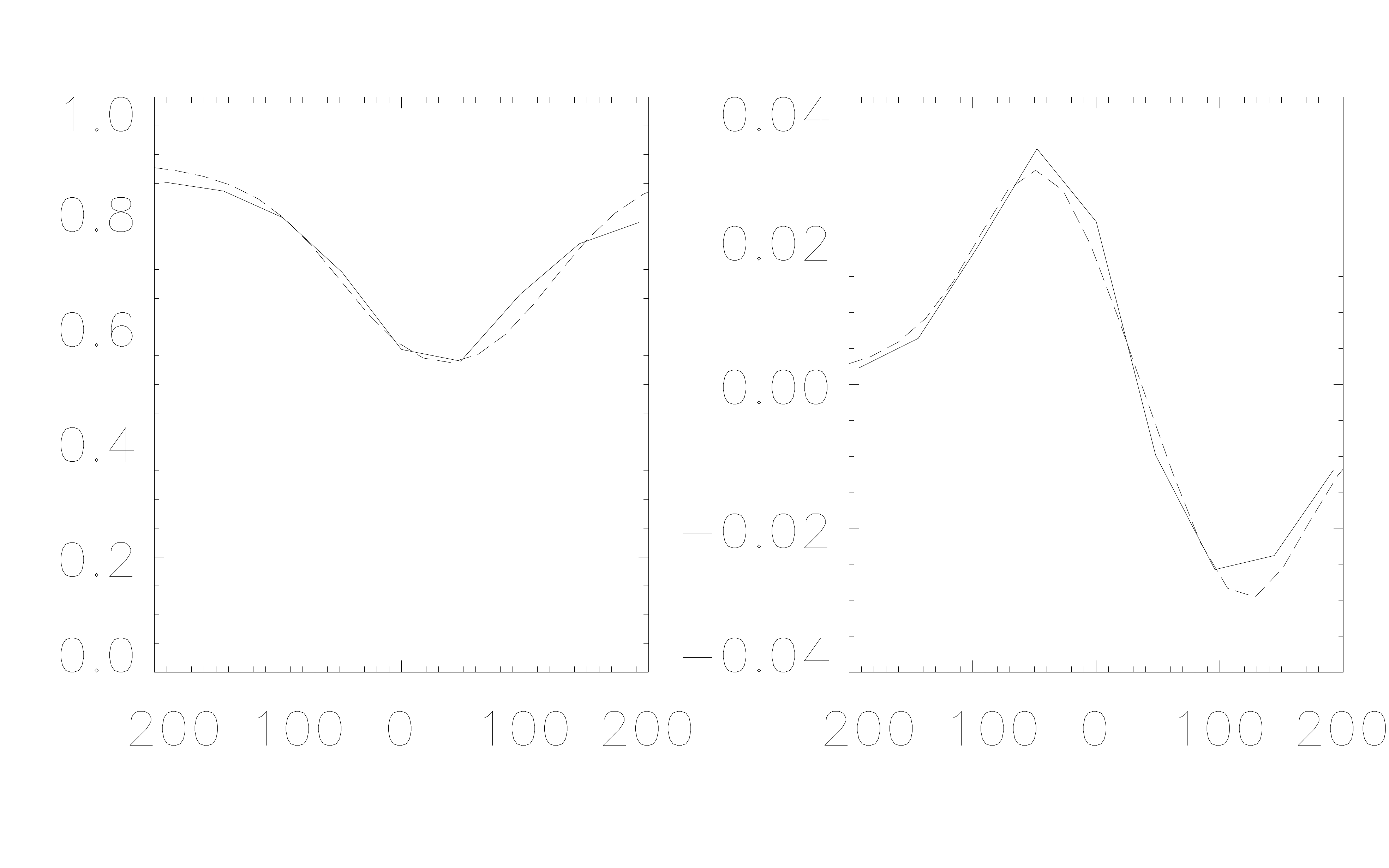}}
  \subfigure[ S2]{
    \includegraphics[bb=45 95 1133 640,clip,width=0.240\linewidth]{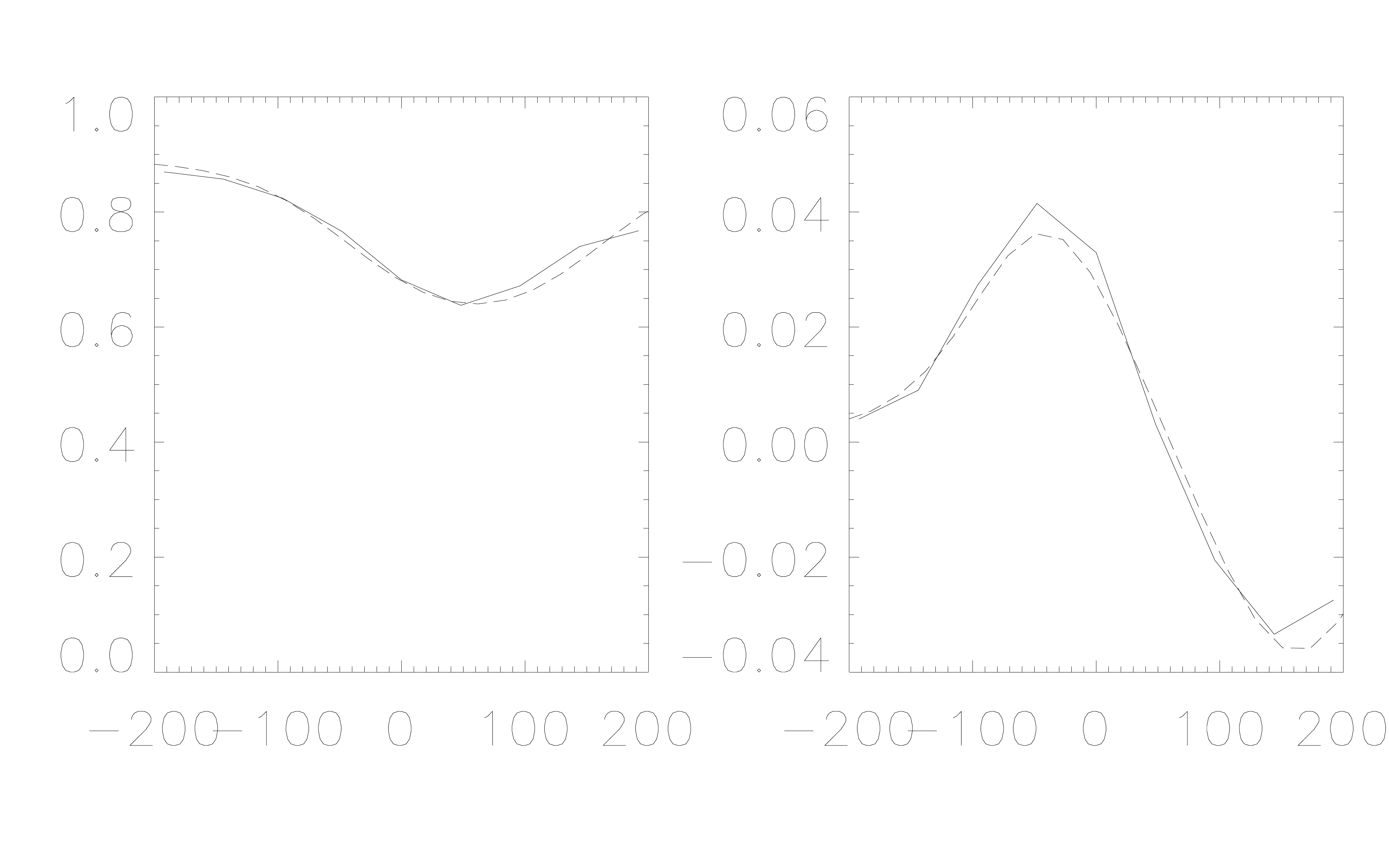}}
  \subfigure[ S3]{
    \includegraphics[bb=45 95 1133 640,clip,width=0.240\linewidth]{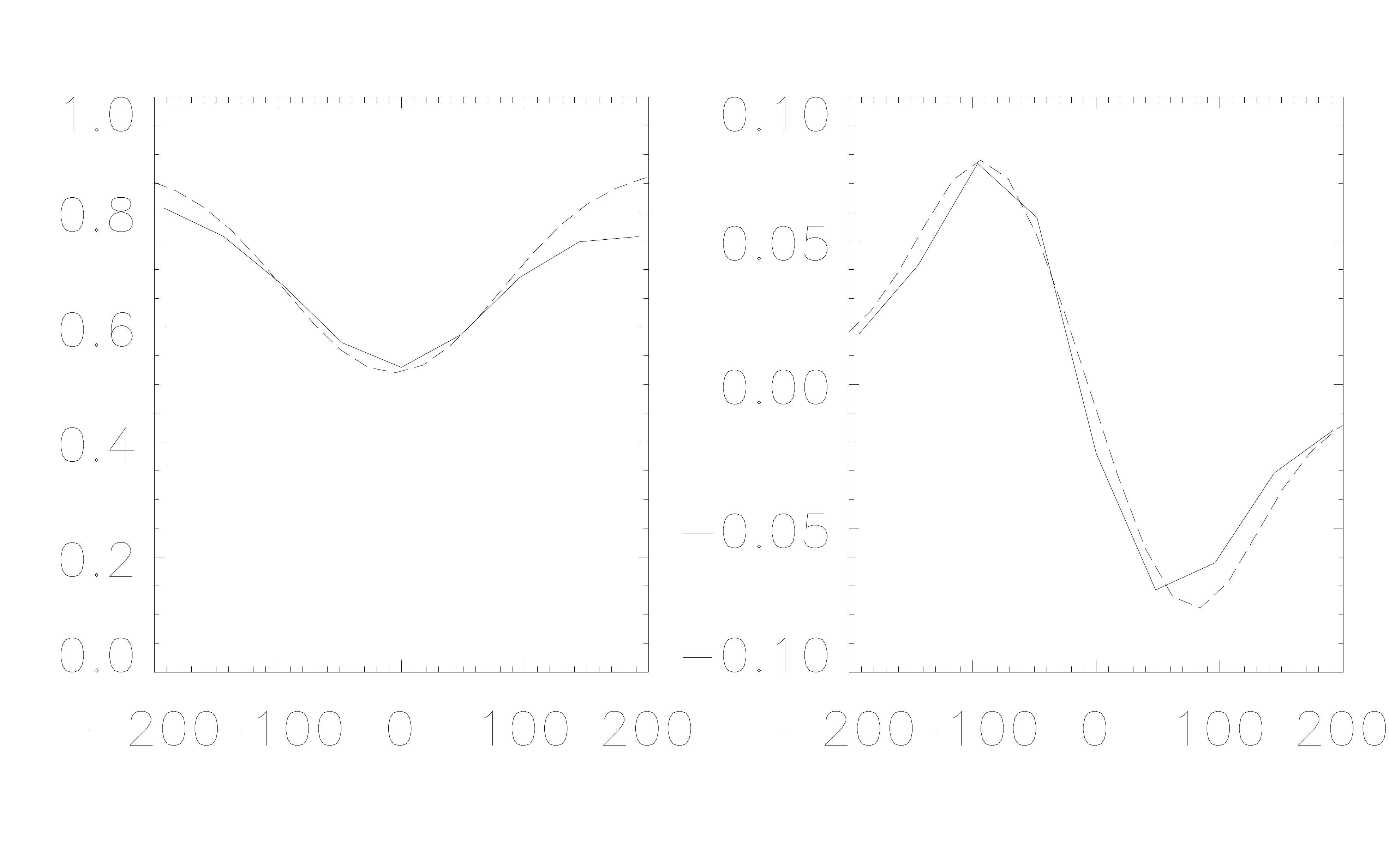}}
  \subfigure[ O0]{
    \includegraphics[bb=45 95 1133 640,clip,width=0.240\linewidth]{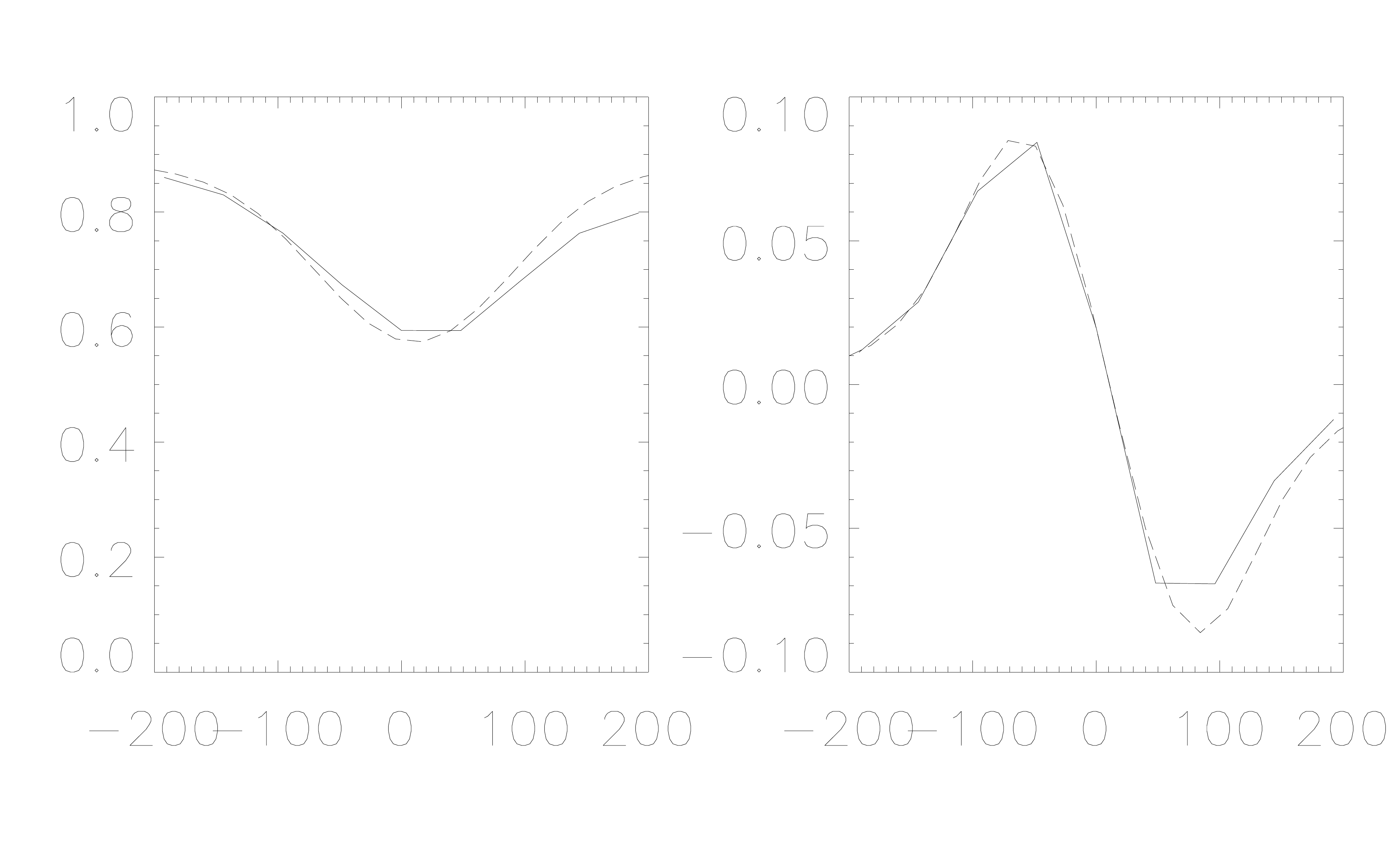}}
  \subfigure[ O1]{
    \includegraphics[bb=45 95 1133 640,clip,width=0.240\linewidth]{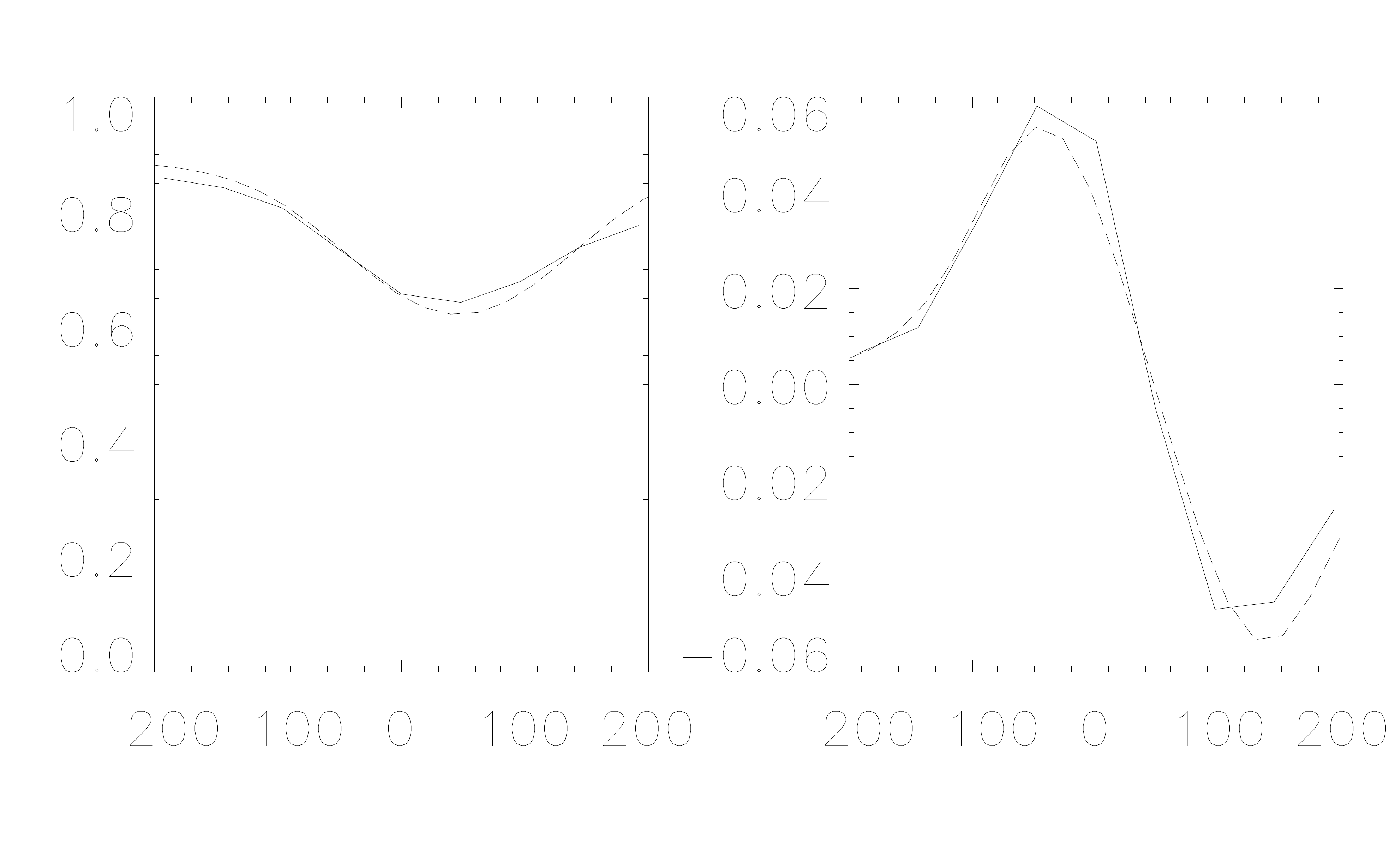}}
  \subfigure[ O2]{
    \includegraphics[bb=45 95 1133 640,clip,width=0.240\linewidth]{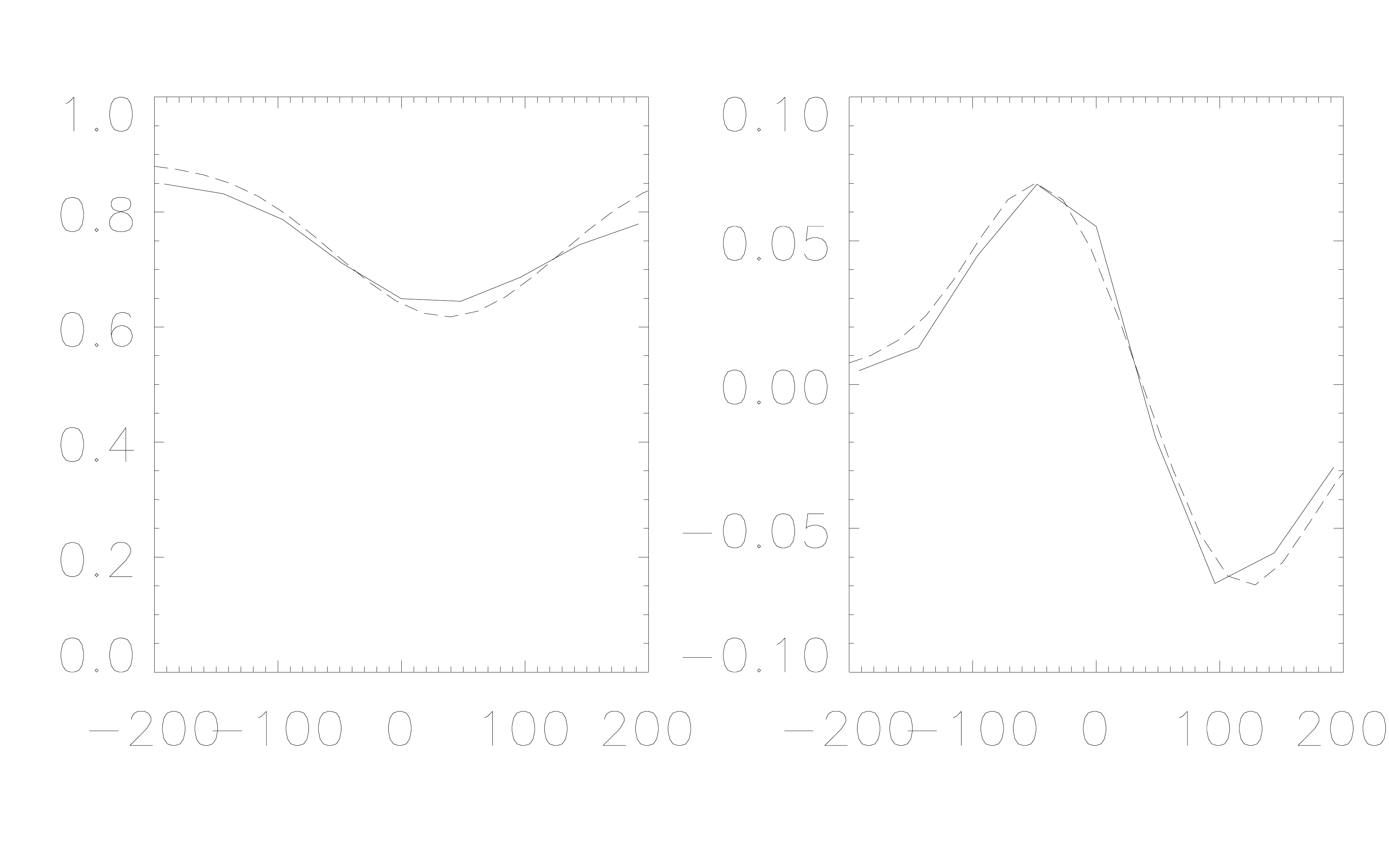}}
  \subfigure[ O3]{
    \includegraphics[bb=45 95 1133 640,clip,width=0.240\linewidth]{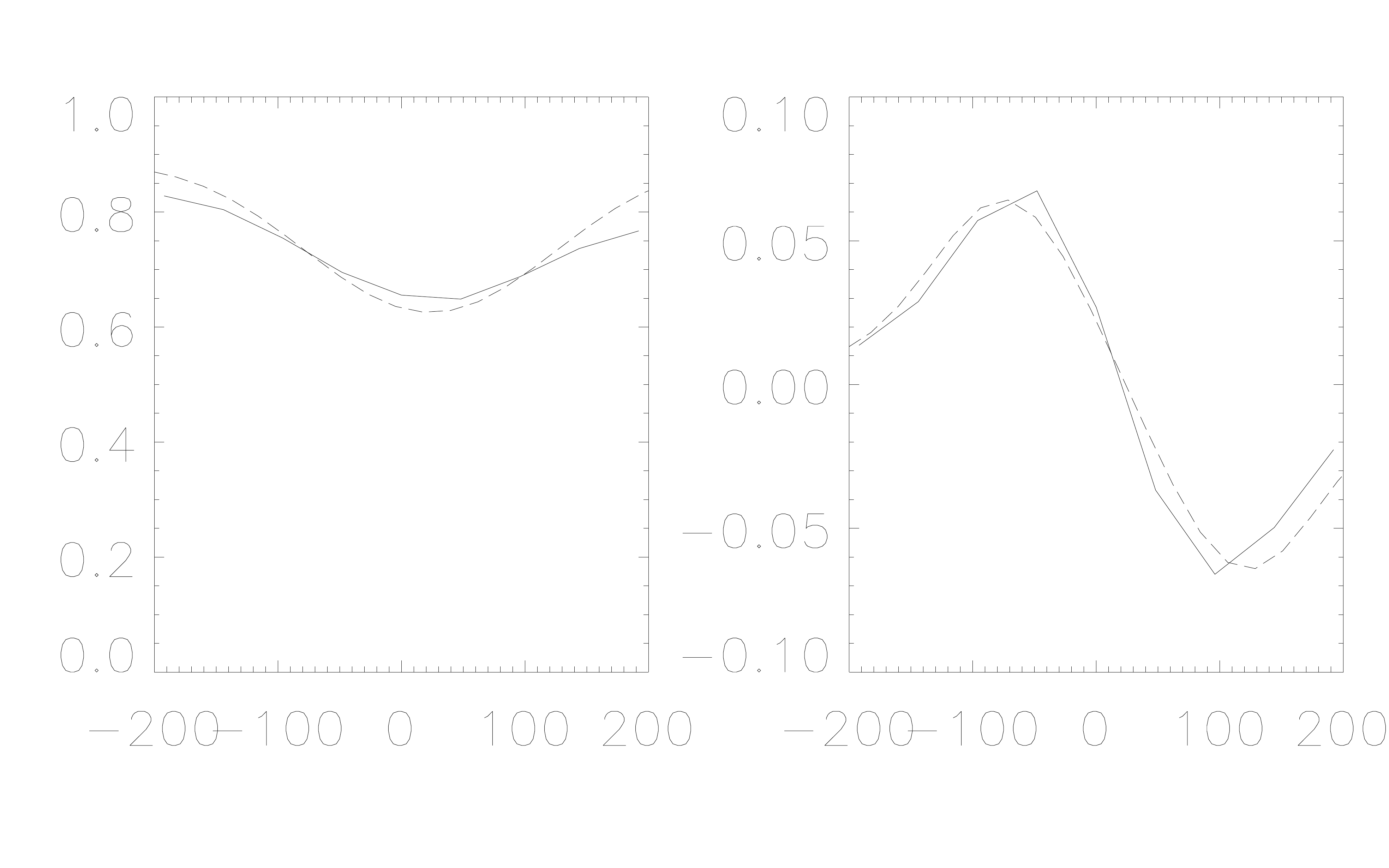}}
  \caption{Observed (solid) and fitted (dashed) Stokes $I$ and $V$
    profiles for isolated bright points (B0--B3), ribbons (R0--R3),
    flowers (F0--F3), strings (S0--S3) and other features (O0--O3)
    indicated in Fig.~\ref{fig:zoommosaic}. Both Stokes $I$ and $V$ are normalized to the continuum intensity, the wavelength of which is outside the range plotted. Wavelengths along the x-axis are in units of m{\AA}.}
  \label{fig:fitobs}
\end{figure*}
\renewcommand{\thesubfigure}{{}}

\begin{figure*}[!htbp]
\subfigure[ B0]{
\includegraphics[bb=11 47 453 323,clip,width=0.24\linewidth]{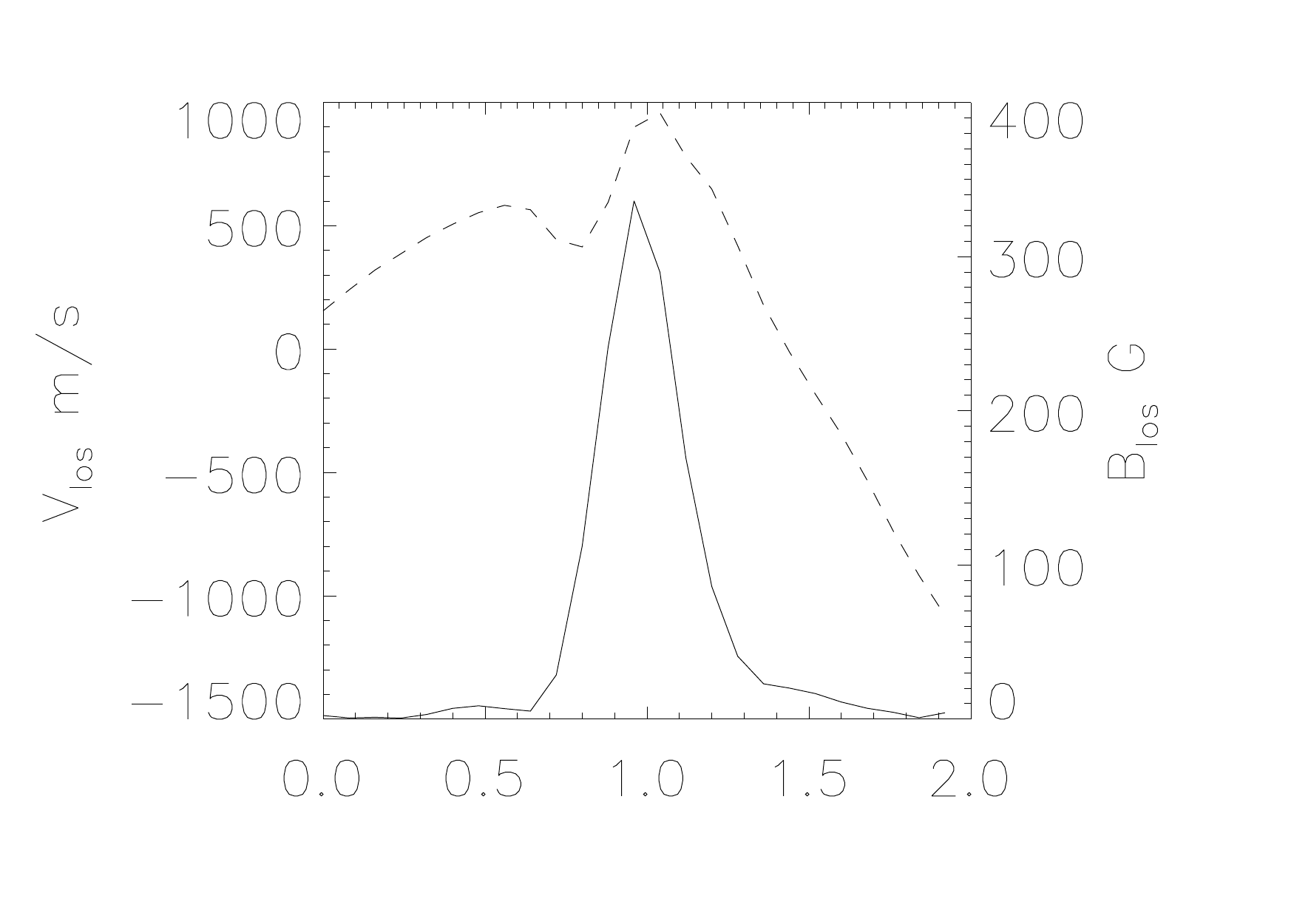}}
\subfigure[ B1]{
\includegraphics[bb=11 47 453 323,clip,width=0.24\linewidth]{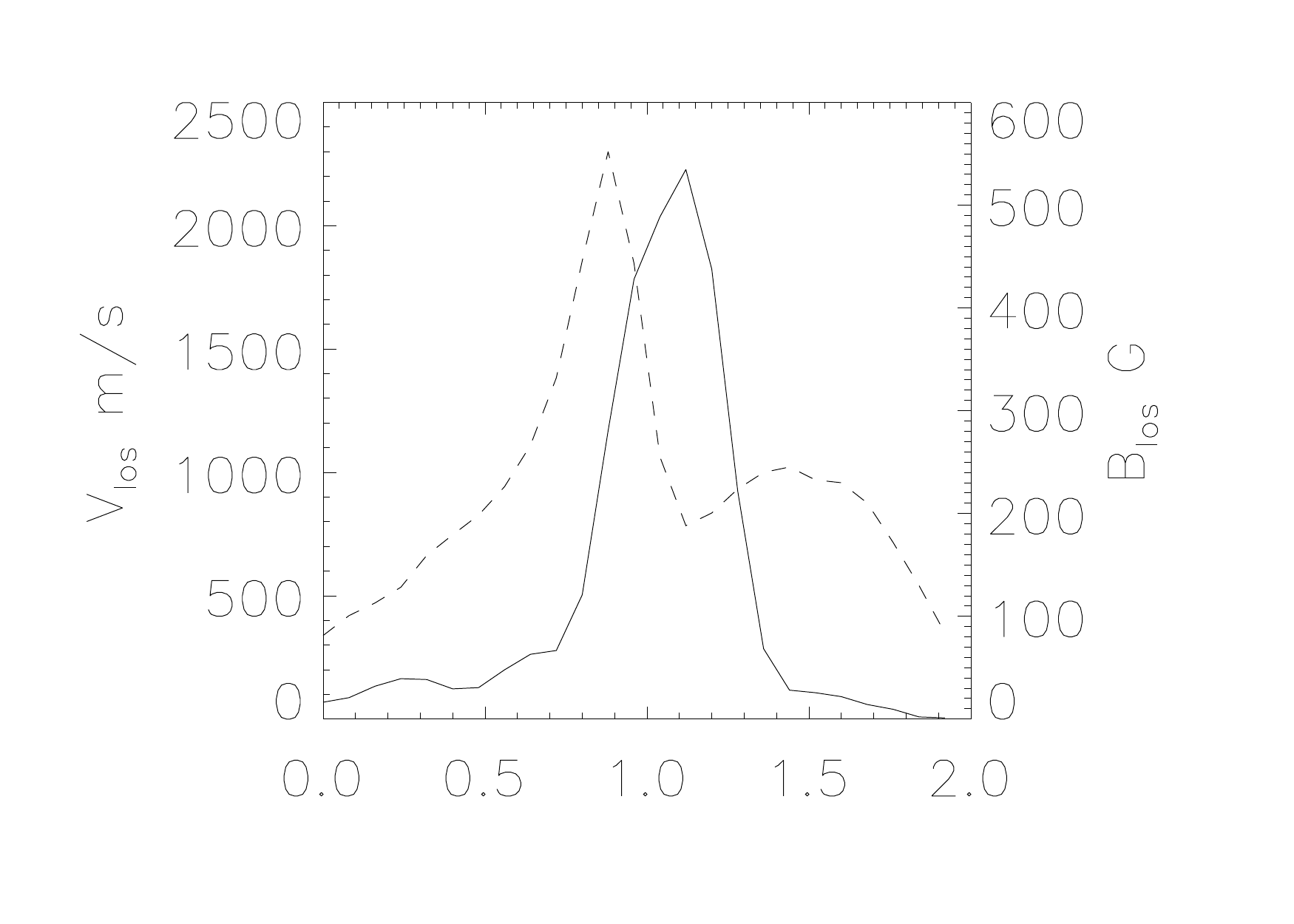}}
\subfigure[ B2]{
\includegraphics[bb=11 47 453 323,clip,width=0.24\linewidth]{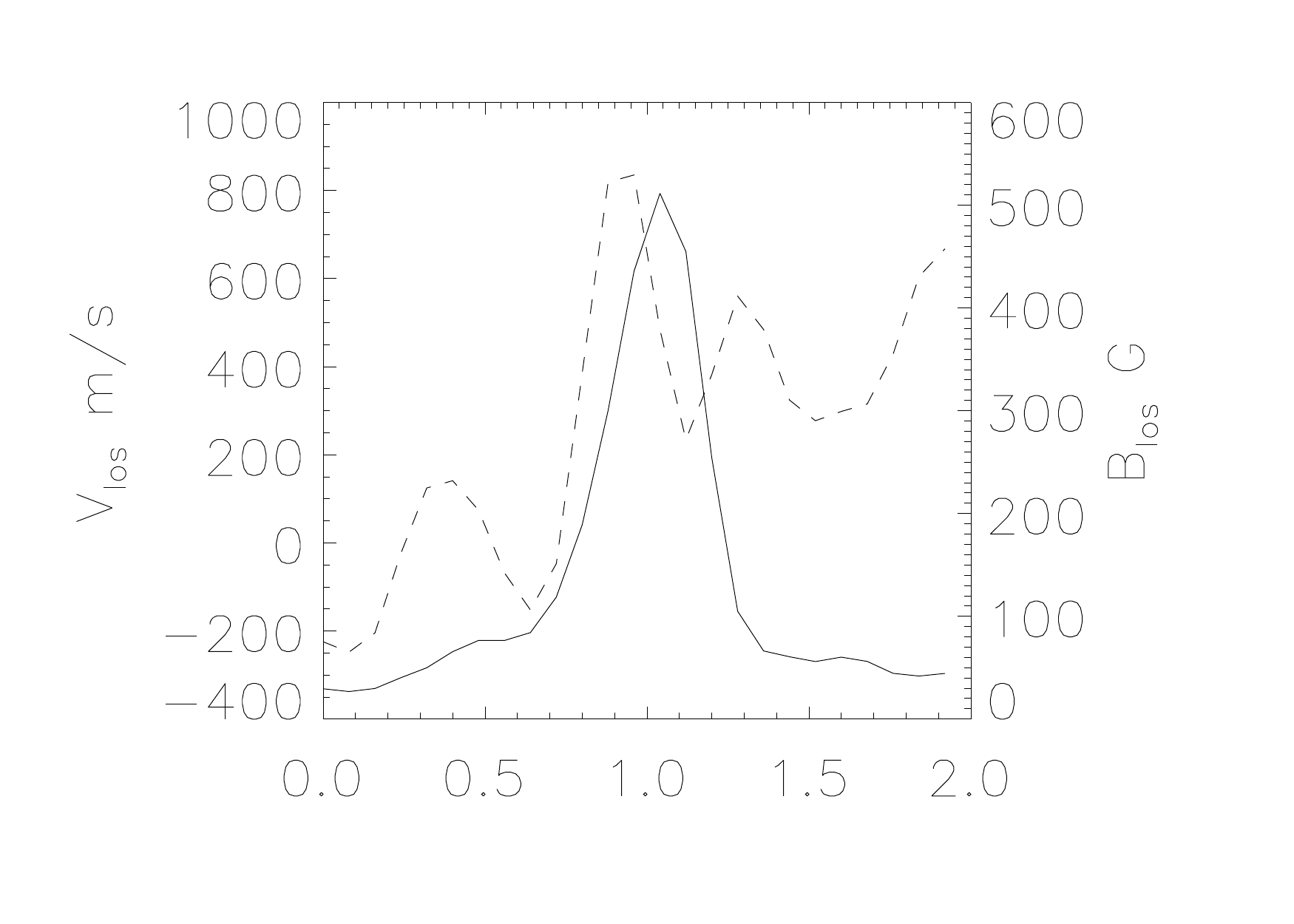}}
\subfigure[ B3]{
\includegraphics[bb=11 47 453 323,clip,width=0.24\linewidth]{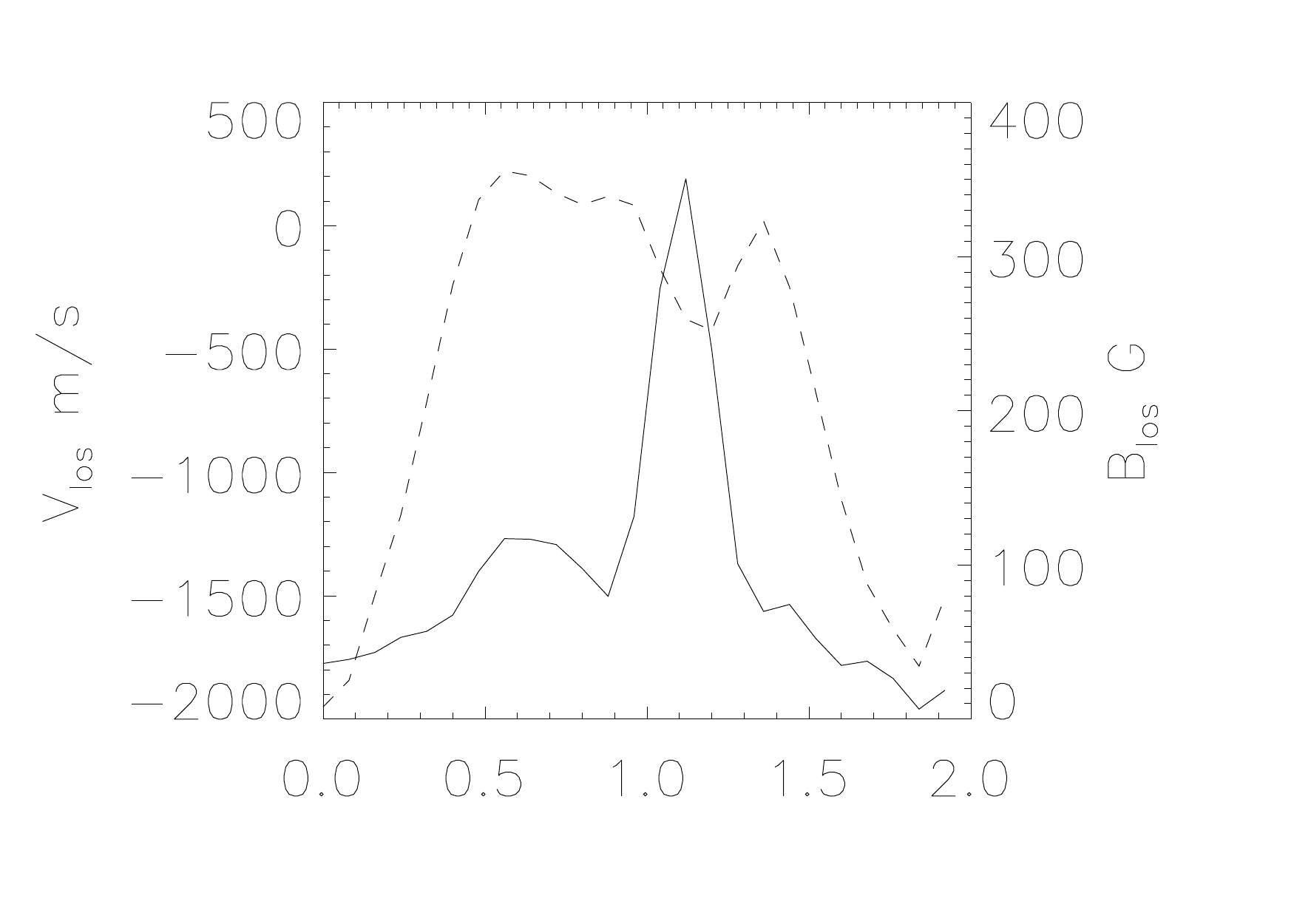}}
\subfigure[ R0]{
\includegraphics[bb=11 47 453 323,clip,width=0.24\linewidth]{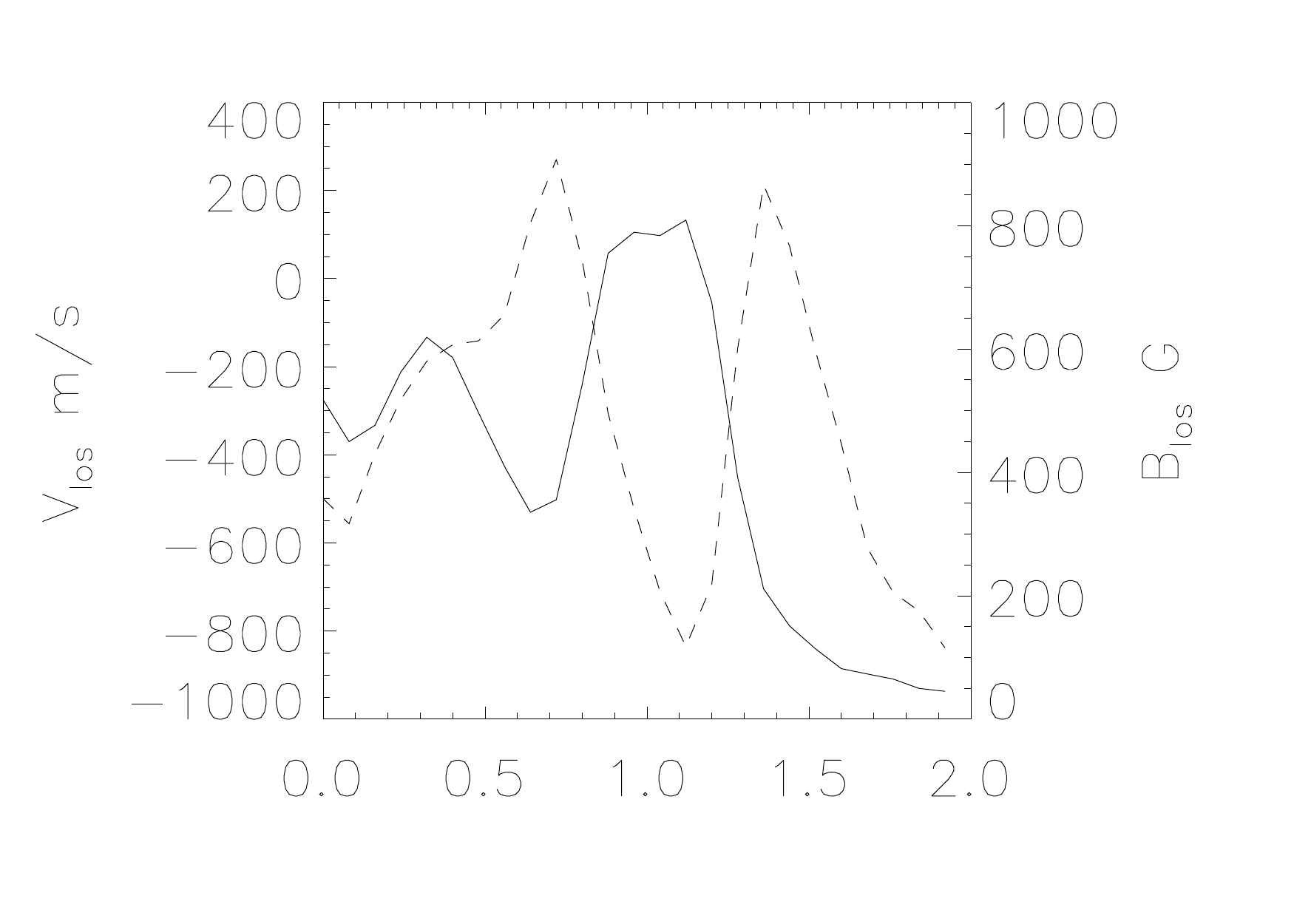}}
\subfigure[ R1]{
\includegraphics[bb=11 47 453 323,clip,width=0.24\linewidth]{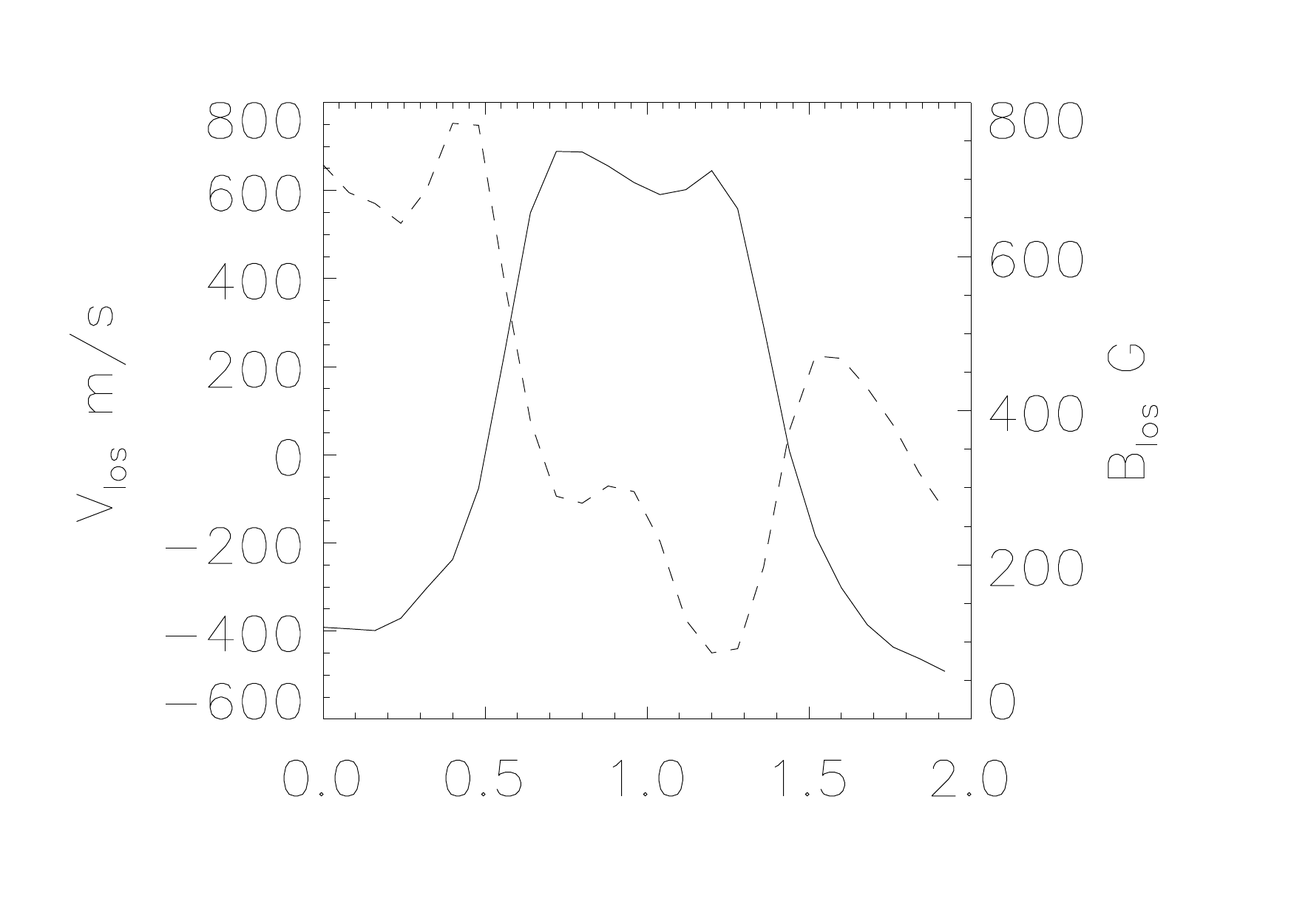}}
\subfigure[ R2]{
\includegraphics[bb=11 47 453 323,clip,width=0.24\linewidth]{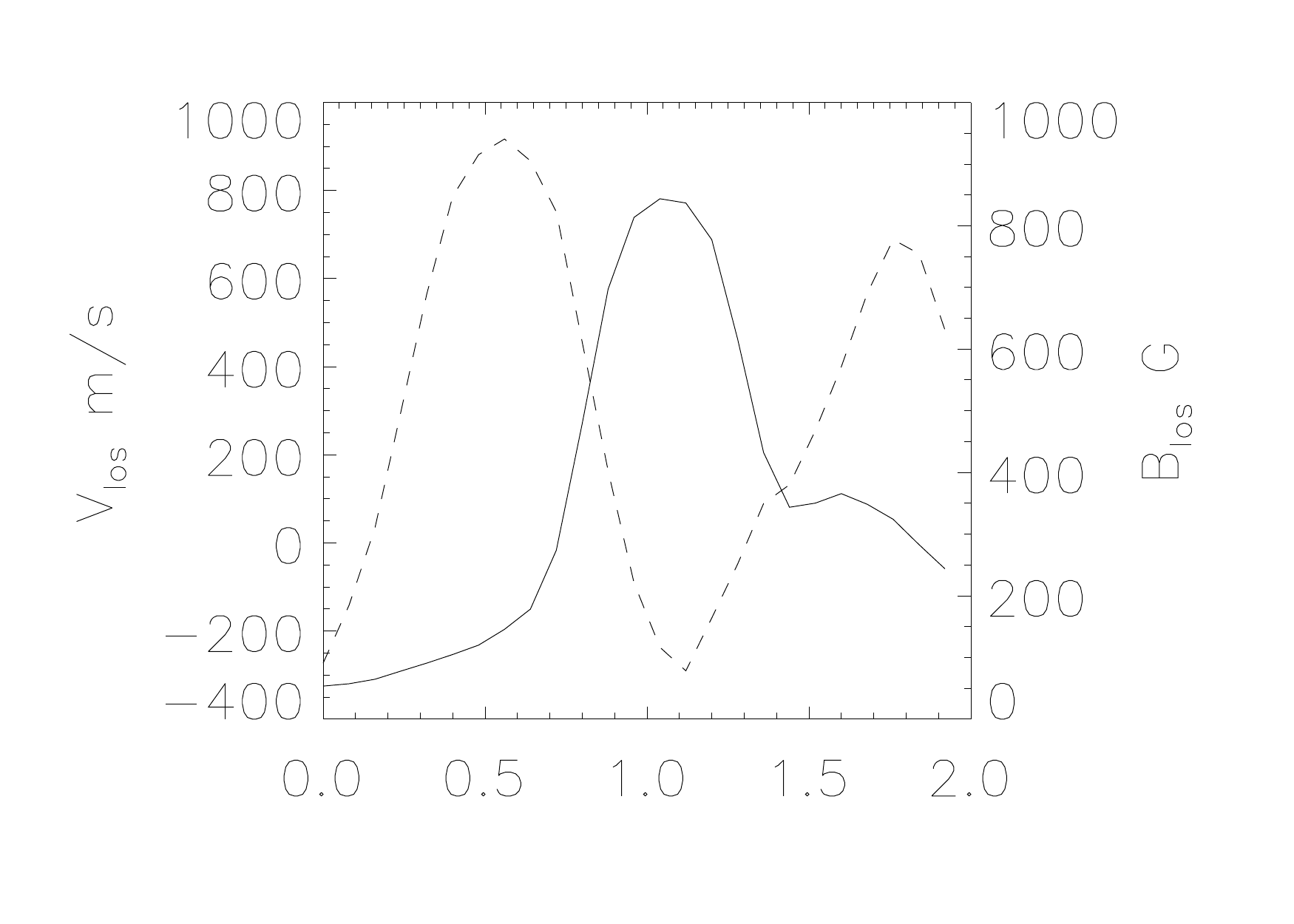}}
\subfigure[ R3]{
\includegraphics[bb=11 47 453 323,clip,width=0.24\linewidth]{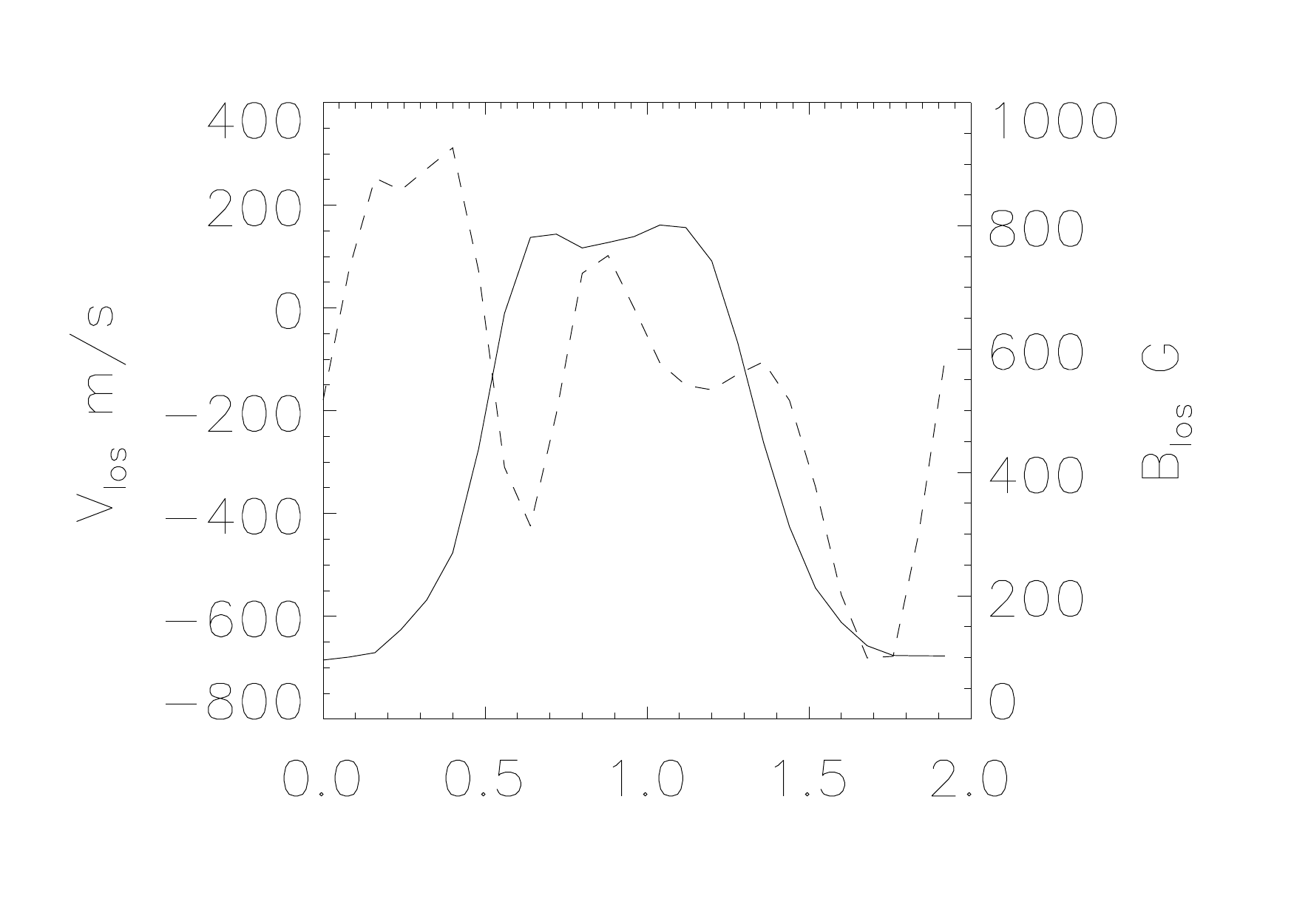}}
\subfigure[ F0]{
\includegraphics[bb=11 47 453 323,clip,width=0.24\linewidth]{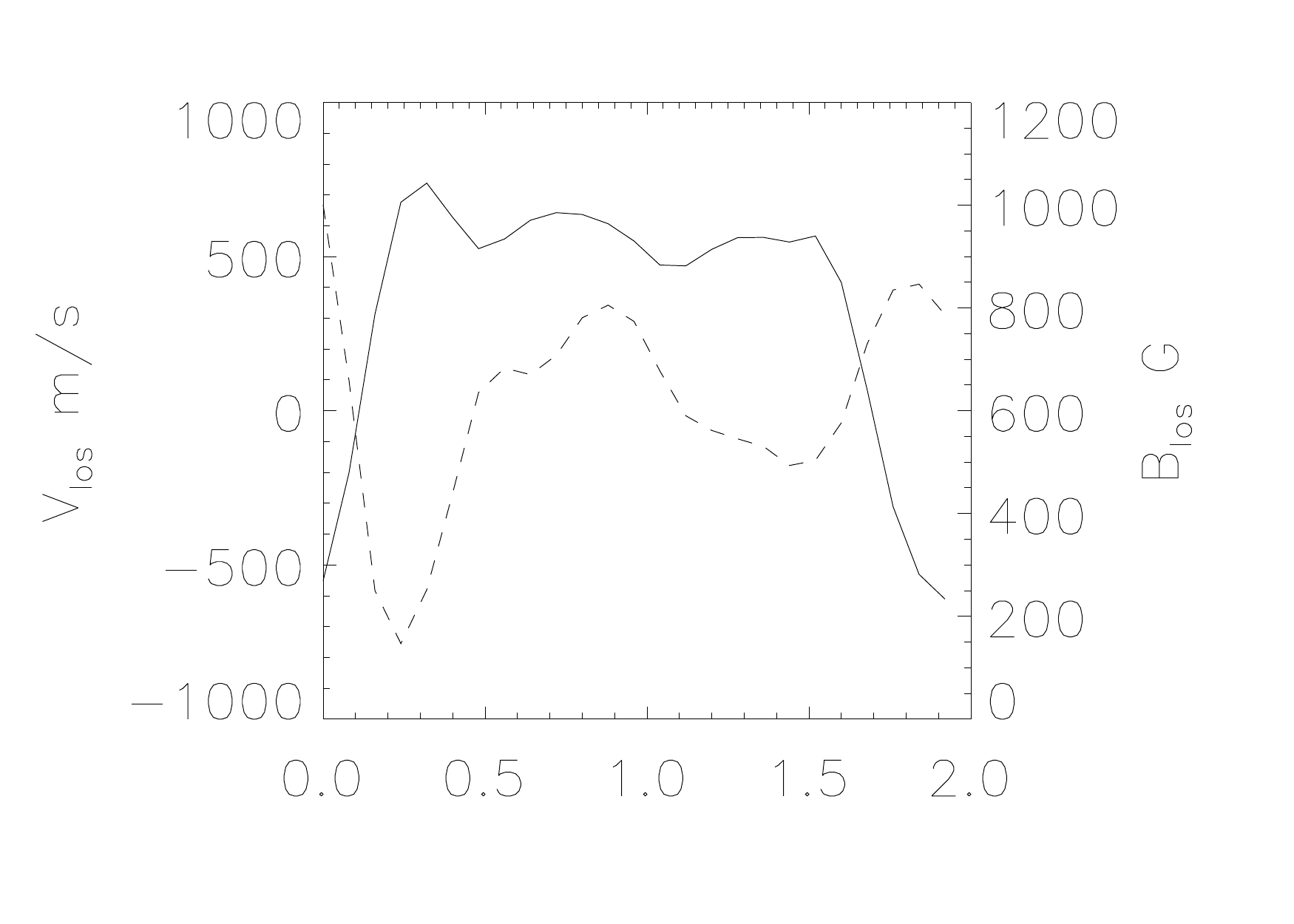}}
\subfigure[ F1]{
\includegraphics[bb=11 47 453 323,clip,width=0.24\linewidth]{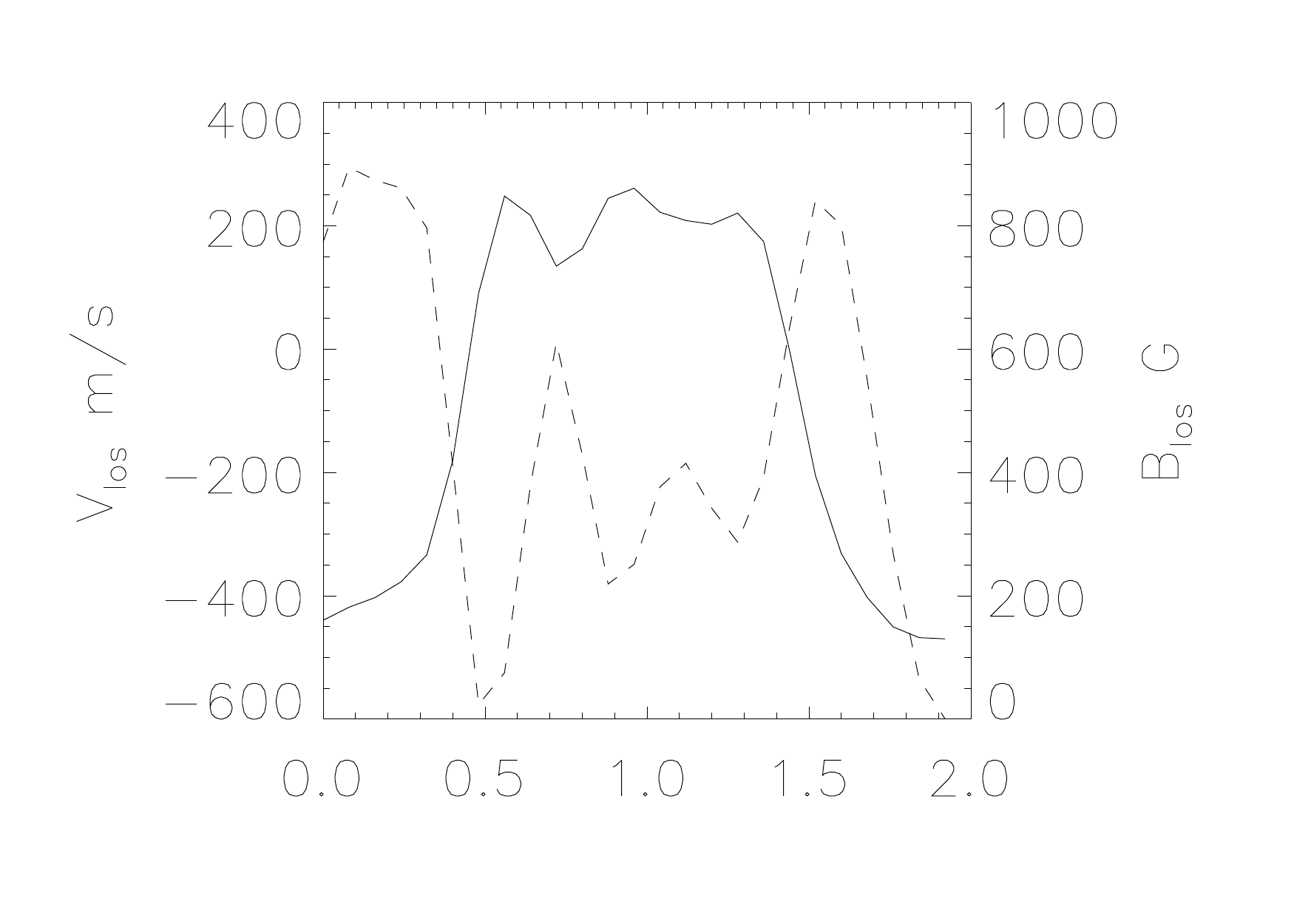}}
\subfigure[ F2]{
\includegraphics[bb=11 47 453 323,clip,width=0.24\linewidth]{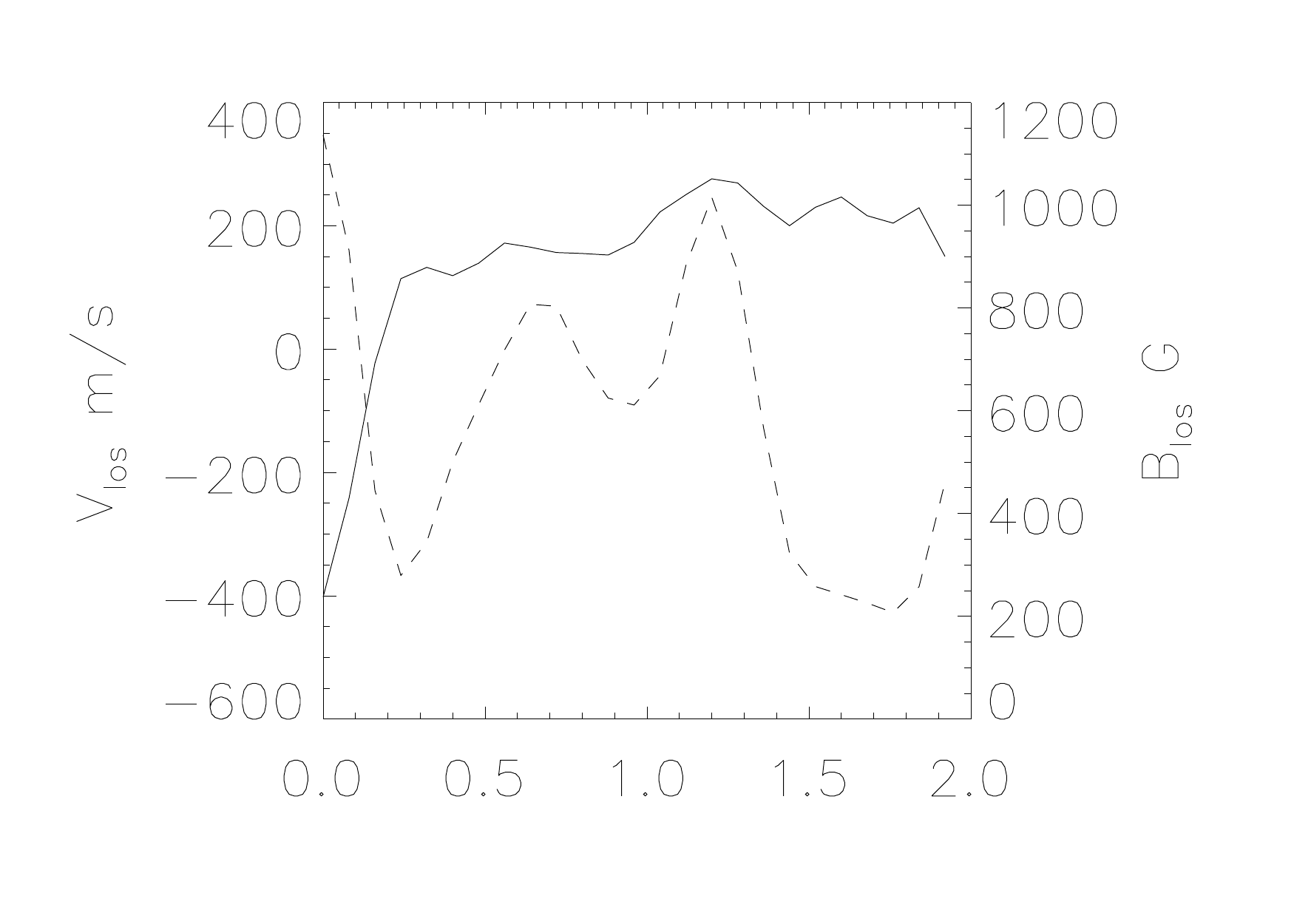}}
\subfigure[F3]{
\includegraphics[bb=11 47 453 323,clip,width=0.24\linewidth]{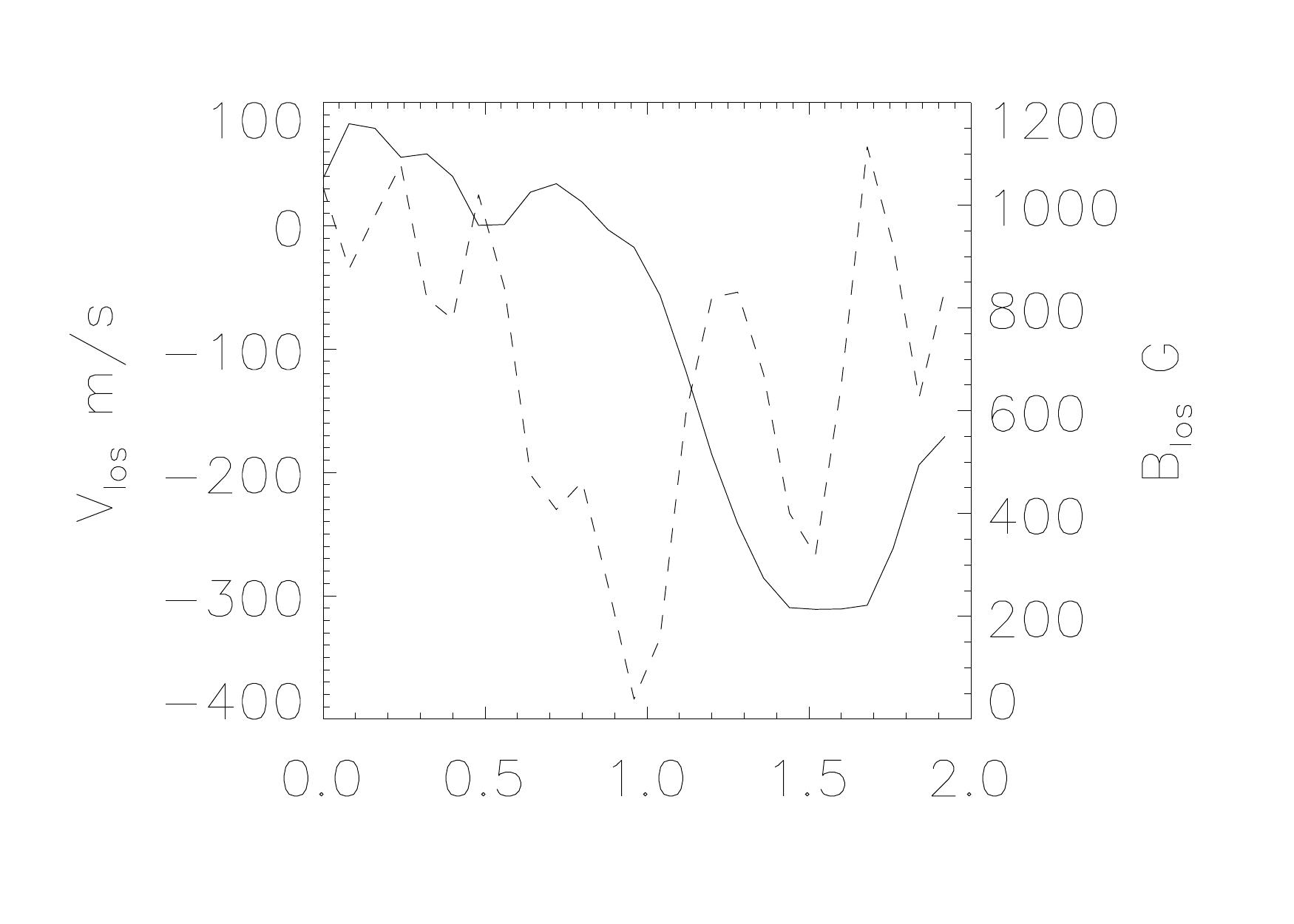}}
\subfigure[ S0]{
\includegraphics[bb=11 47 453 323,clip,width=0.24\linewidth]{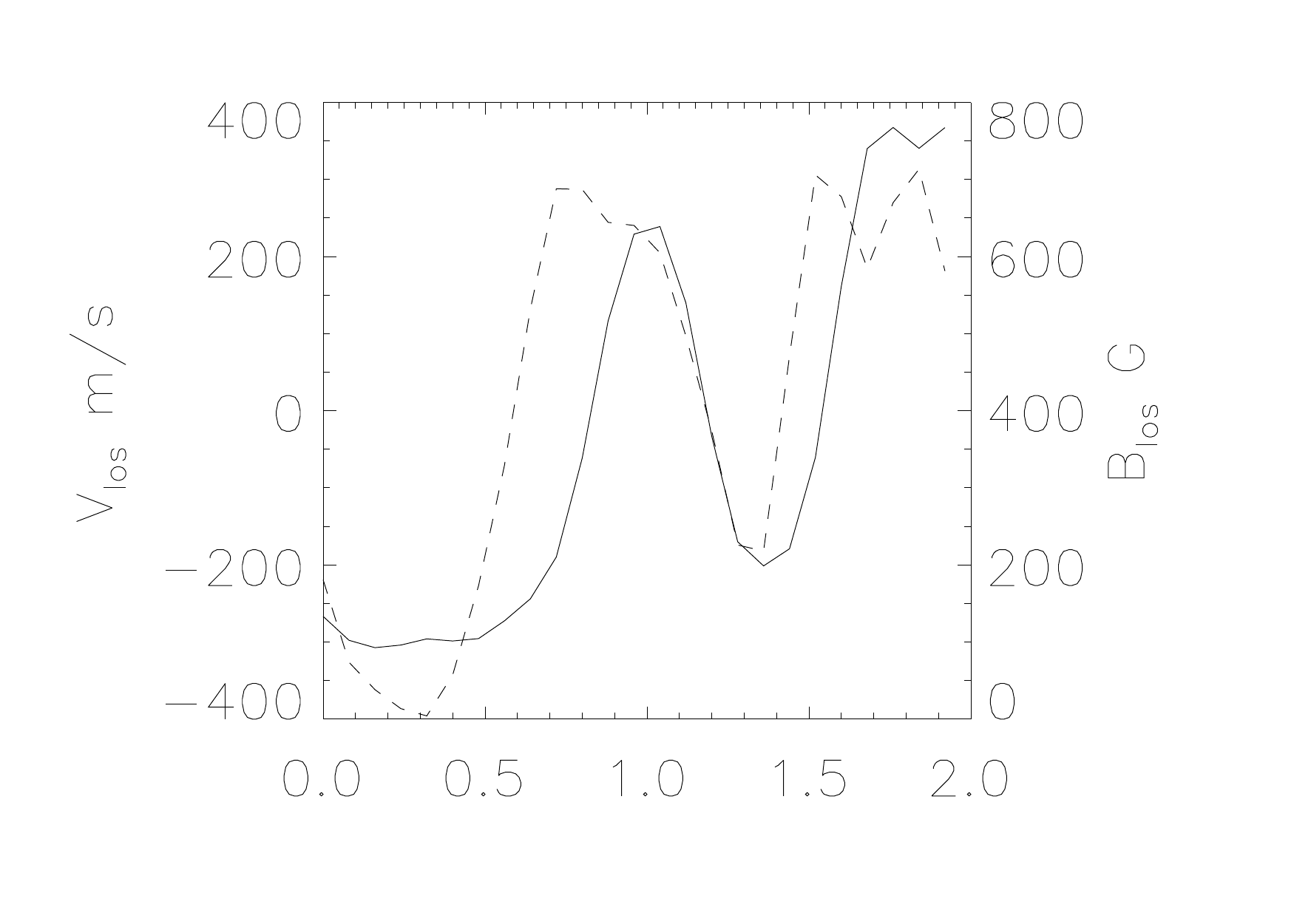}}
\subfigure[ S1]{
\includegraphics[bb=11 47 453 323,clip,width=0.24\linewidth]{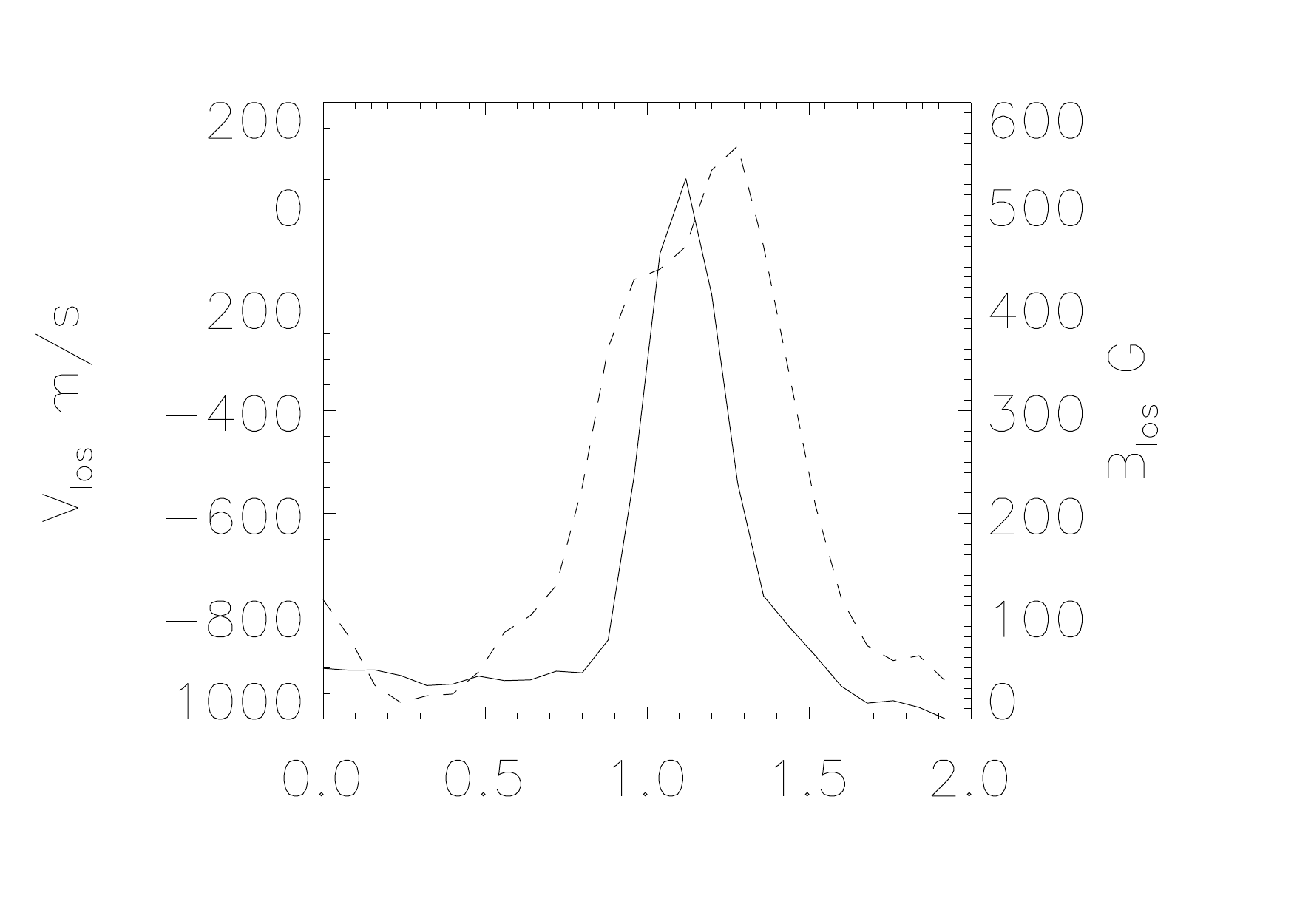}}
\subfigure[ S2]{
\includegraphics[bb=11 47 453 323,clip,width=0.24\linewidth]{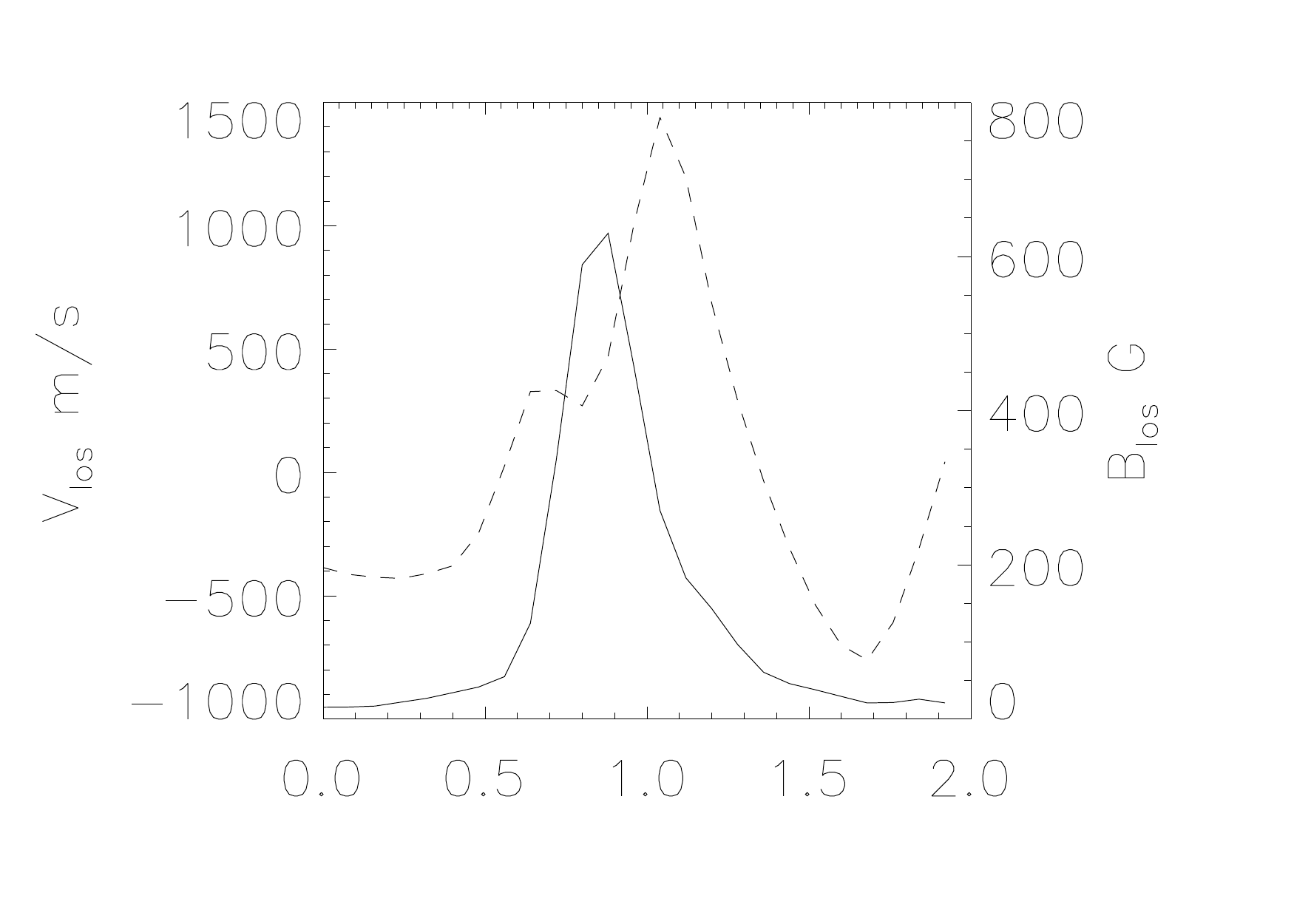}}
\subfigure[ S3]{
\includegraphics[bb=11 47 453 323,clip,width=0.24\linewidth]{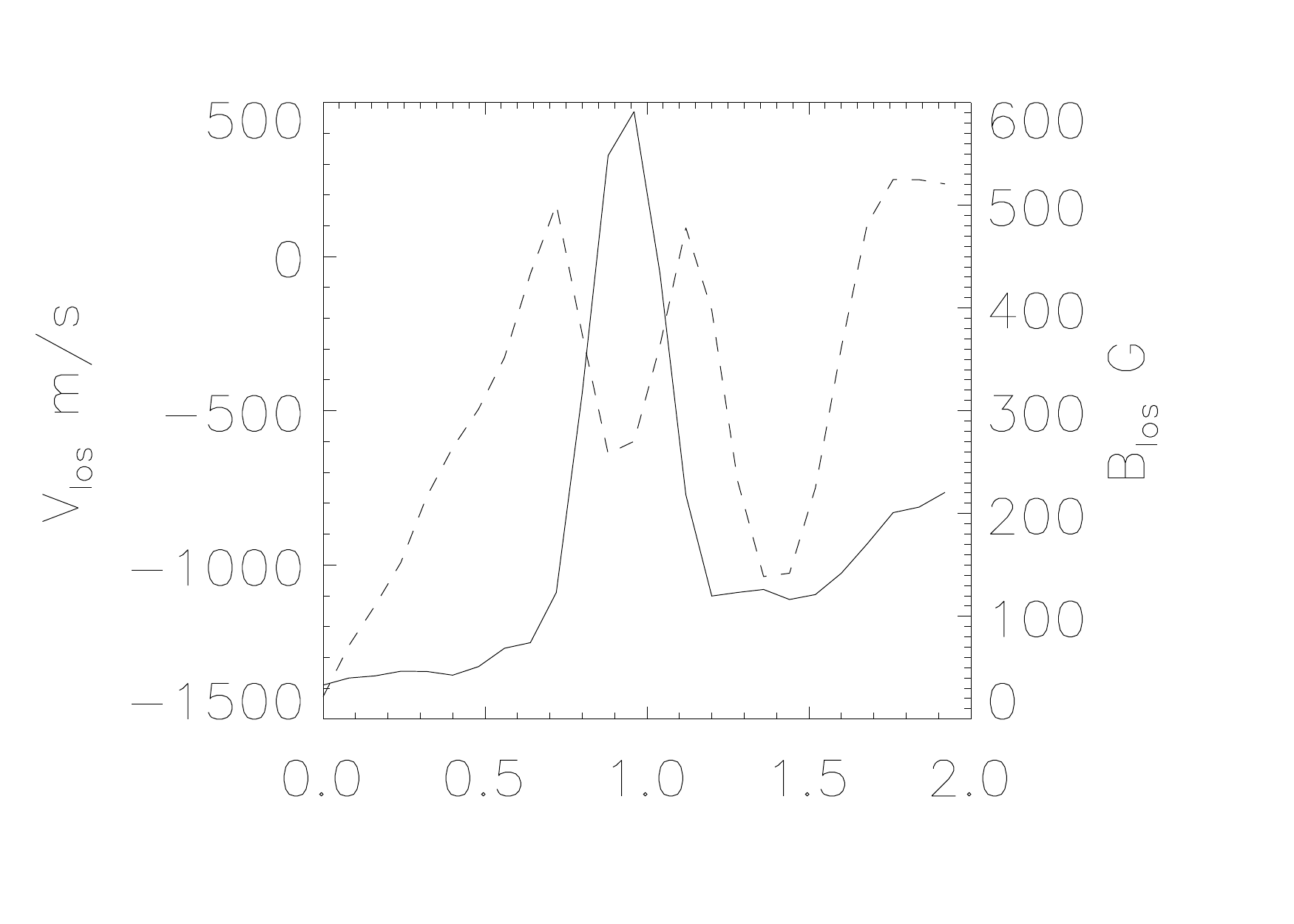}}
\subfigure[ O0]{
\includegraphics[bb=11 47 453 323,clip,width=0.24\linewidth]{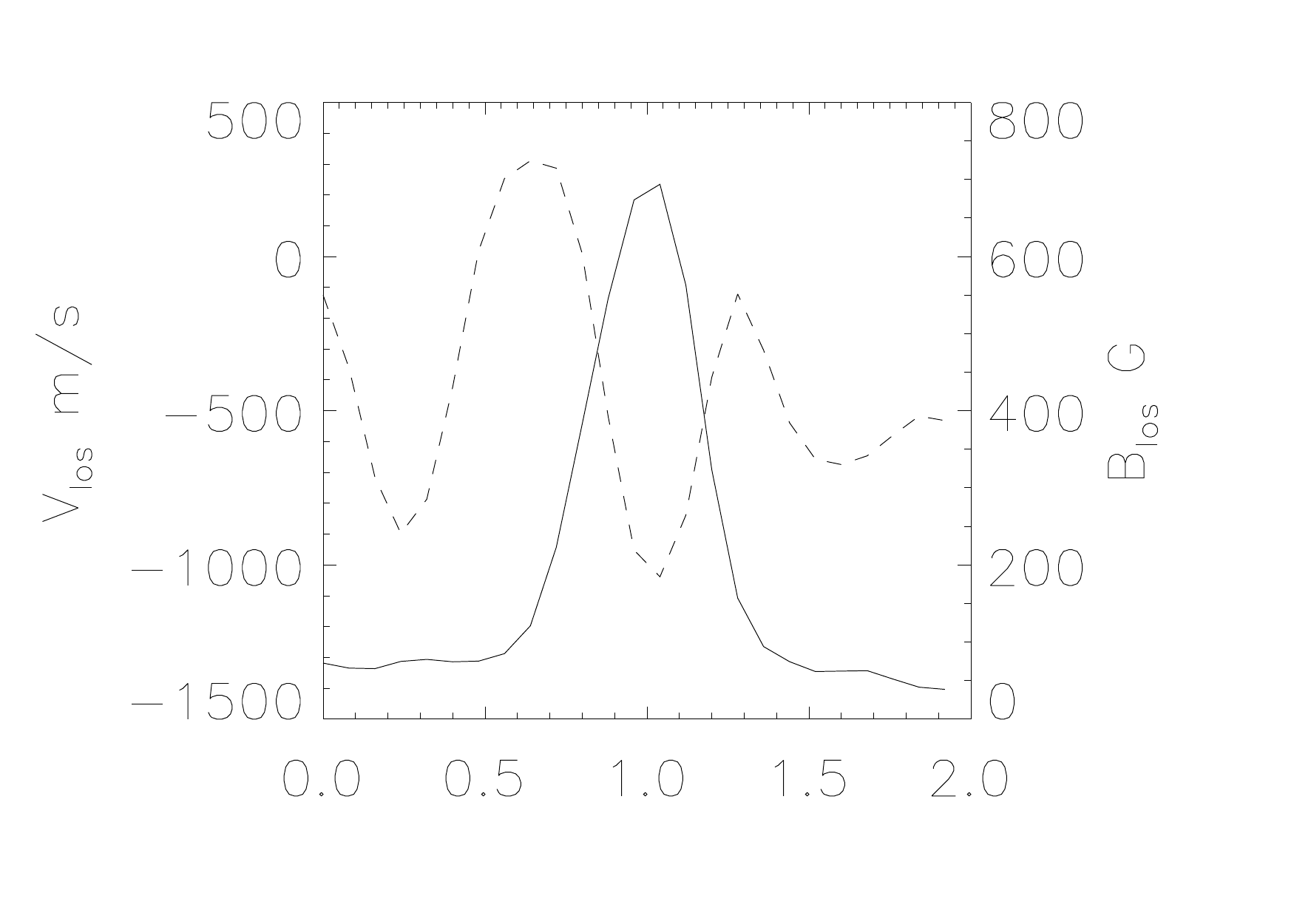}}
\subfigure[ O1]{
\includegraphics[bb=11 47 453 323,clip,width=0.24\linewidth]{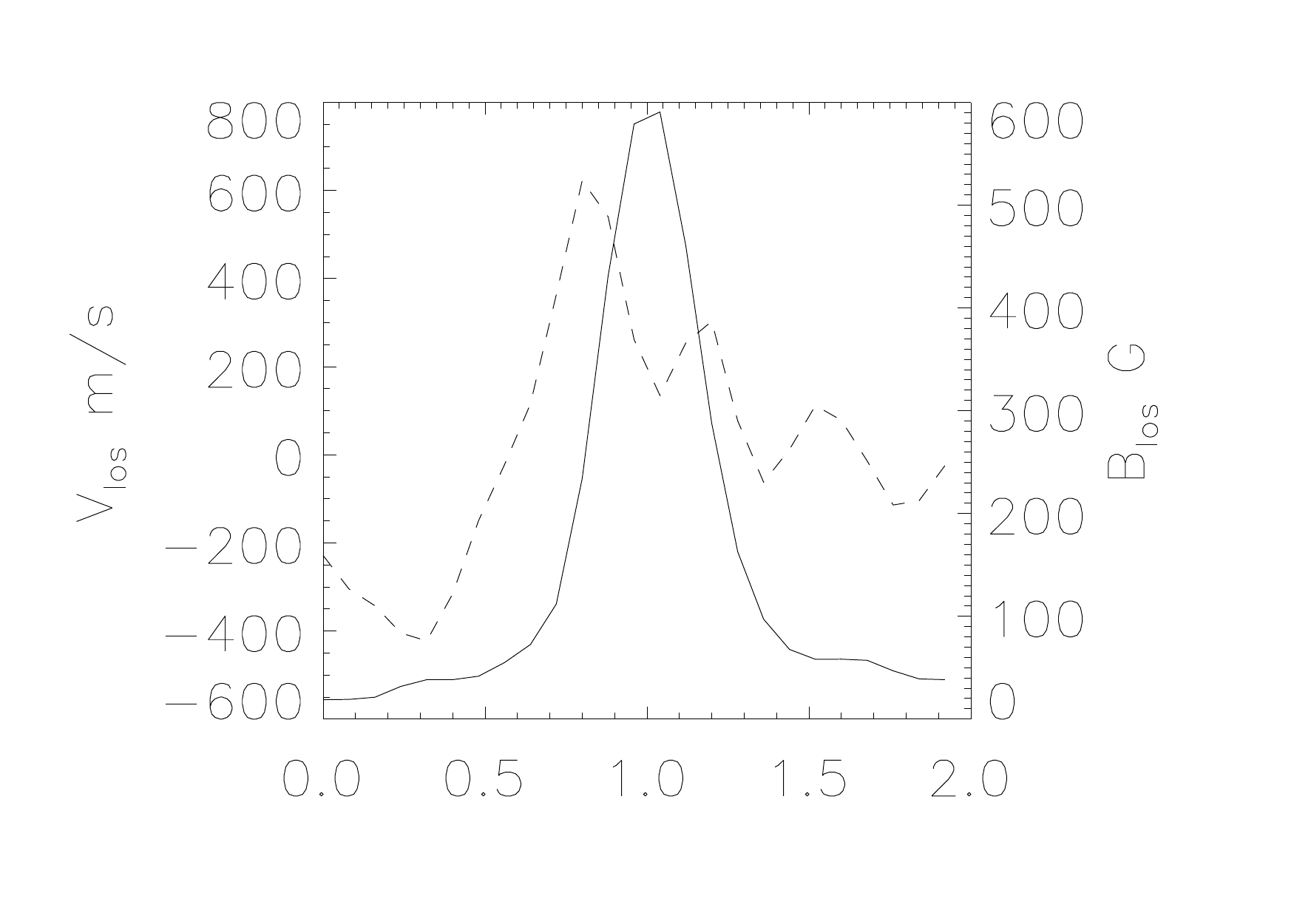}}
\subfigure[ O2]{
\includegraphics[bb=11 47 453 323,clip,width=0.24\linewidth]{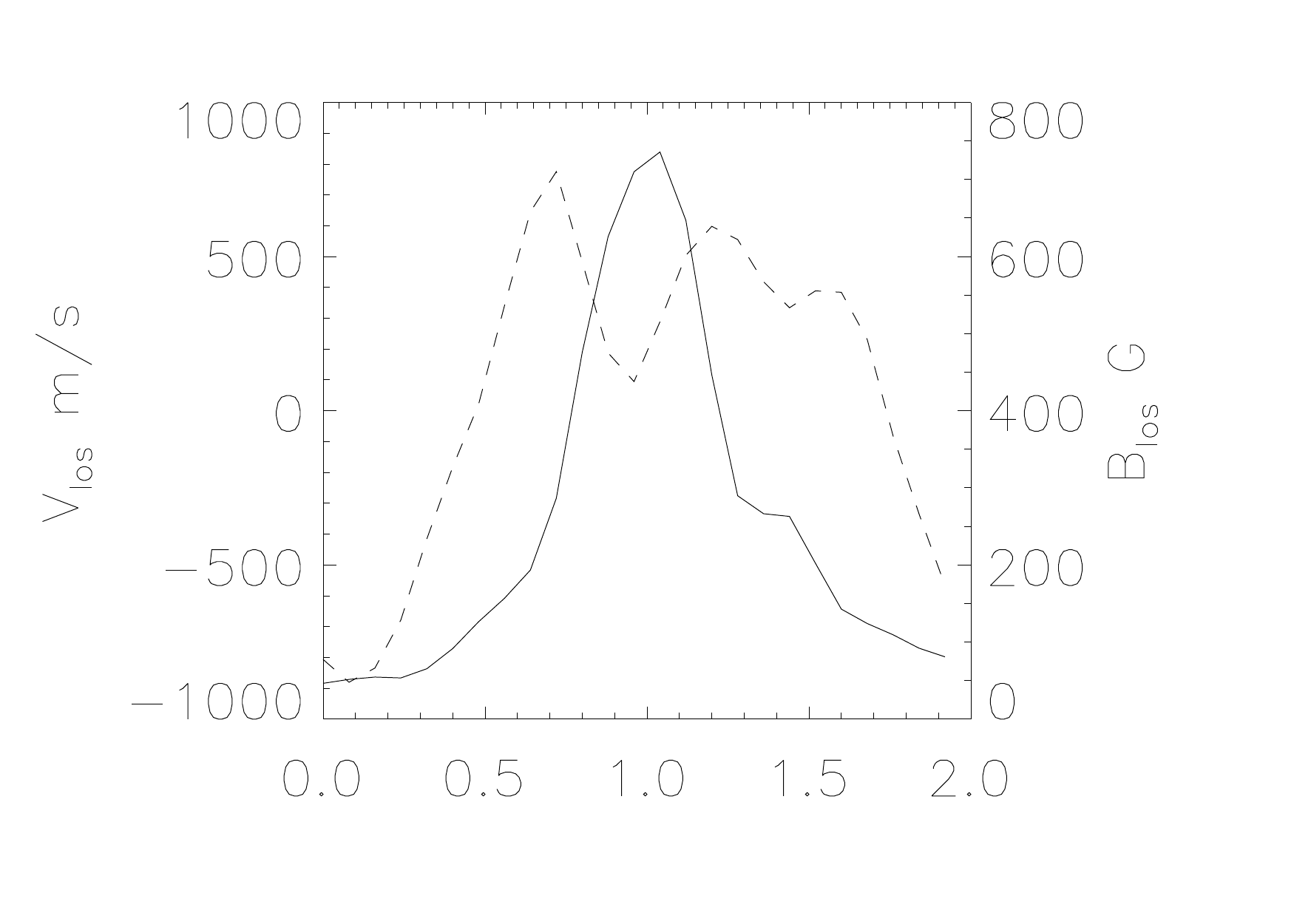}}
\subfigure[ O3]{
\includegraphics[bb=11 47 453 323,clip,width=0.24\linewidth]{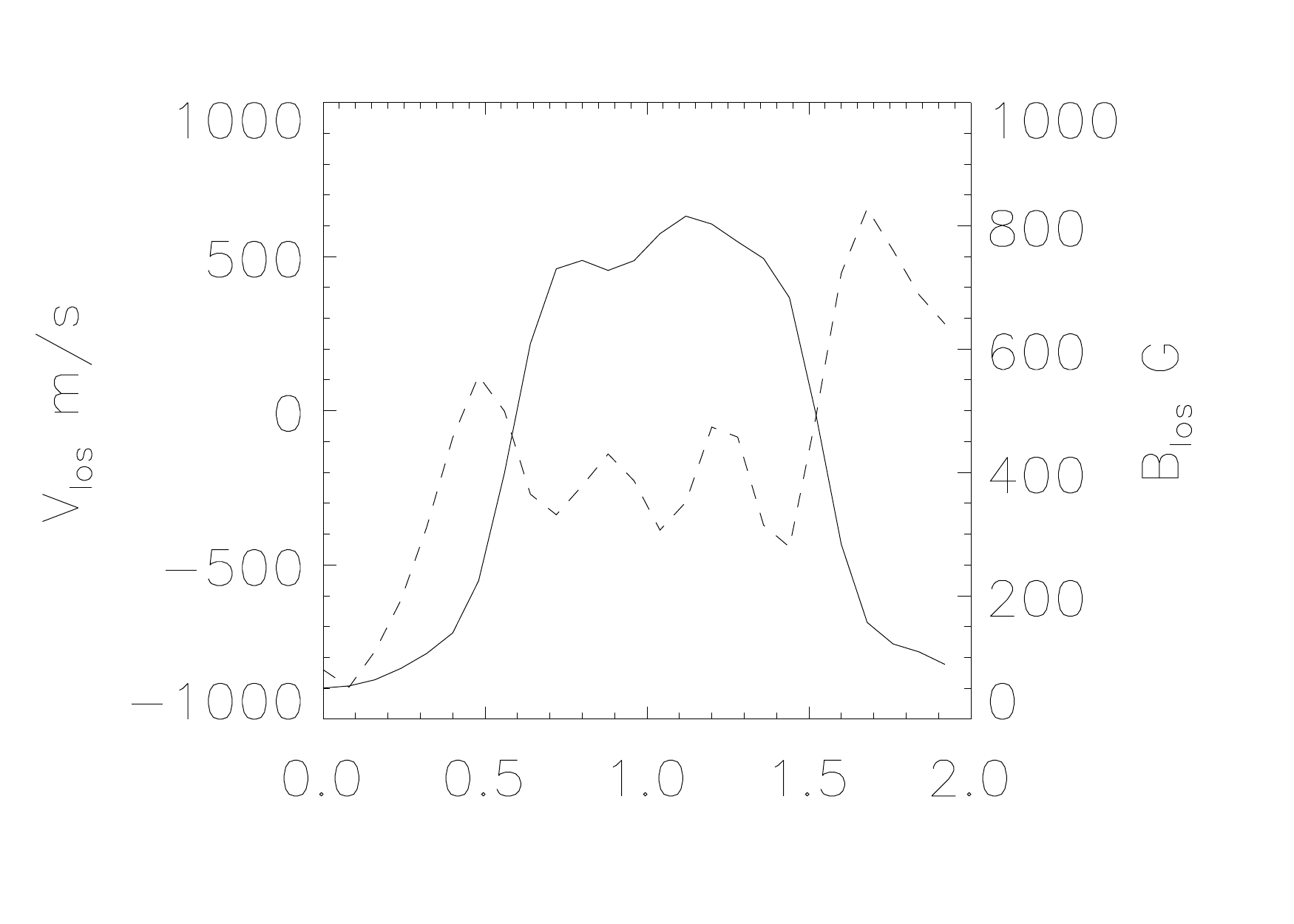}}
\caption{Horizontal cuts of LOS velocity (dashed) and strength of LOS
  magnetic field $|\Blos|$ (solid) for isolated bright points
  (B0--B3), ribbons (R0--R3), flowers (F0--F3), strings (S0--S3) and
  other features (O0--O3) indicated in Fig.~\ref{fig:zoommosaic}. Distances along the x-axis are in units of arcsec.} 
\label{fig:vlos-blos}
\end{figure*}

Figures~\ref{fig:zoommosaic} and \ref{fig:zoommosaic2} shows the ROI
chosen for identification and discussion of specific velocity
features. The two upper panels show the continuum intensity and line
minimum intensity \Imin\ in the 630.2~nm line. Comparing to
Fig.~\ref{fig:fullFOV} and the \Blos\ map in the upper-right panel of
Fig.~\ref{fig:zoommosaic} indicates that the line minimum intensity is
a good proxy for detecting strong-field magnetic features. This is in
part because of line weakening due to the larger temperature at equal
optical depth in the magnetic flux concentrations
\citep{1968SoPh....5..442C,2007A&A...469..731S}. The \vlos\ map, made
from data returned by Helix is shown in the lower-right panel of
Fig.~\ref{fig:zoommosaic}. The white/black/red contours in these
panels correspond to $|\Blos|=200$~G and outline the same regions as
high-lighted in the \vlos\ map of Fig.~\ref{fig:fullFOV}.

The line of sight velocity (\vlos) map in Fig.~\ref{fig:zoommosaic}
shows the signatures of field-free and magneto-convection discussed
earlier. There are a few areas with strong downflows reaching up to
3.5~k\mps; the strongest upflows are approximately $-1.3$~k\mps.

The bright features in the line minimum images appear sufficiently
similar to those in images recorded in the G-band and the
\ion{Ca}{ii}~H line to allow us to describe properties of the observed
features following the nomenclature of \citet{berger04solar} and
\citet{rouppe05solar}.  In the following, we discuss the results of
the inversions for a few selected features of each type, indicated
with circles in Fig.~\ref{fig:zoommosaic}. For each of these selected
features, we show the observed and fitted Stokes $I$ and $V$ profiles
at the center of the corresponding circle (Figs.~\ref{fig:fitobs}
and~\ref{fig:fitobs2}) and the inferred variation of \vlos\ and \Blos\
along a horizontal cut through the center of the same circle
(Figs.~\ref{fig:vlos-blos} and \ref{fig:vlos-blos2}).

In Figs.~\ref{fig:fitobs} and ~\ref{fig:fitobs2} we show observed and
fitted Stokes $I$ and $V$ profiles at center of the features discussed
in the following subsections. Fits for which the
Stokes $V$ signal is strong appear good and justify the use of
Milne--Eddington inversions neglecting LOS gradients of the magnetic
field and LOS velocity. However, for the profiles with the weakest $V$ signal shown in Fig.~\ref{fig:fitobs2}, the fits are quite poor.

\subsubsection{Pores}
\label{sec:pores}

The strongest flux concentrations appear as dark pores in the $\Ic$ image but are only marginally darker than their surroundings in the
\Imin\ map. In the following, we refer to tiny pores which appear as
distinctly dark in the continuum intensity ($\Ic$) maps as
micropores and those that have only hints of darkening in the
continuum intensity, but still show strong flux concentrations in the
\Blos\ map, as protopores. With this definition, the feature labeled M
in Fig.~\ref{fig:zoommosaic} is a micropore and the one labeled P is a
protopore. Most of the pores show at least one strong downflow channel
adjacent to their boundaries. This is in agreement with earlier
observational studies, e.g. by \citet{2001ApJ...552..354L},
\citet{2003ApJ...598..689S}, \citet{2003A&A...405..331H} and
\citet{2005A&A...432..319S}. Thin downflows lanes around pores were
also seen in the simulations of \citet{2007A&A...474..261C}.

\subsubsection{Isolated bright points}
\label{sec:isol-bright-points}

Circles labeled B0, B1 etc in Fig.~\ref{fig:zoommosaic} contain
examples of isolated bright points. Figure \ref{fig:vlos-blos} shows
the plots of \vlos\ and \Blos\ across the various features. The top
row of Fig.~\ref{fig:vlos-blos} show plots across the bright points.
For the feature marked B0 there is a strong downflow of about
960~\mps\ close to where the field is strongest. In B1 we can also see that there is a strong downflow of about 2.3~k\mps\ close to where the LOS field peaks. In B2 we see a similar though weaker downflow of 840~\mps. We note that the downflow occurs adjacent to but not precisely where the LOS field peaks. This is consistent with previous studies of G-band bright points \citep{berger04solar,2004ApJ...604..906R}. B3 seems to be an exception with no strong downflow but instead an upflow of about $-430$~\mps\ close to the location of the strongest LOS field.

\subsubsection{Ribbons and flowers}
\label{sec:ribbons-flowers}

The micropores are surrounded by several small scale magnetic features that form almost circular interfaces between the micropores and (the disturbed) surrounding granules. These small scale features are clearly visible in the line minimum images as bright structures having ribbon- and flower-like shapes. Similar features were first observed and described by \citet{berger04solar} and further studied by \citet{rouppe05solar}. Ribbons and flowers are similar structures, the difference being that flowers are somewhat circular and ribbons are rather elongated. When ribbons take a circular shape, they are here referred to as flowers. The features marked F0--F3 etc are examples of flowers and those marked R0--R3 are examples of ribbons. The \Imin\ maps shows that the ribbons and flowers have somewhat dark cores and brighter edges. The \Blos\ map confirms that the flowers and ribbons are magnetic. An aspect worth noting is that in our observations flowers are found only in the close vicinity of micropores and protopores. Ribbons are found mostly in the vicinity of micropores and protopores and in nearby network plage regions.

The second row of Fig.~\ref{fig:vlos-blos} shows plots across ribbons. R0 shows a strong upflow of about $-840$~\mps\ very close to the peaking of the field. This is also seen in R1 where the upflow peaks at $-450$~\mps. The region of strong \Blos\ seems slightly more extended in R1. Although we see similar variations in R2, the upflow is much weaker reaching only $-290$~\mps. The region of strong \Blos\ seems to be slightly more extended for R3 than for R2. We see a primary upflow of $-420$~\mps\ within the feature and a secondary upflow of $-680$~\mps, in contrast to the single upflows found for other ribbons.

The third row of Fig.~\ref{fig:vlos-blos} shows plots across flowers. F0 shows a strong upflow of about $-760$~\mps\ close to the peaking of the field. F1 is similar but with an upflow of $-580$~\mps. F2 shows similar variations but with a smaller upflow of around $-400$~\mps. F3 although shows a upflow close to $-400$~\mps\ seems to have different properties compared to the other flower features. This could be because it is at the boundary of a fully developed micropore and hence we do not see the peaking of the field at the flower as clearly. The extended nature of flowers is clear from these plots. Flowers show only a small variation \Blos\ but strong variation in \vlos. Another common feature is that at the boundary of the flowers there is often a
transition from a downflow to an upflow.

\begin{figure*}[!htb]
  \begin{center}
    \includegraphics[bb=31 15 702 839,width=0.45\linewidth]{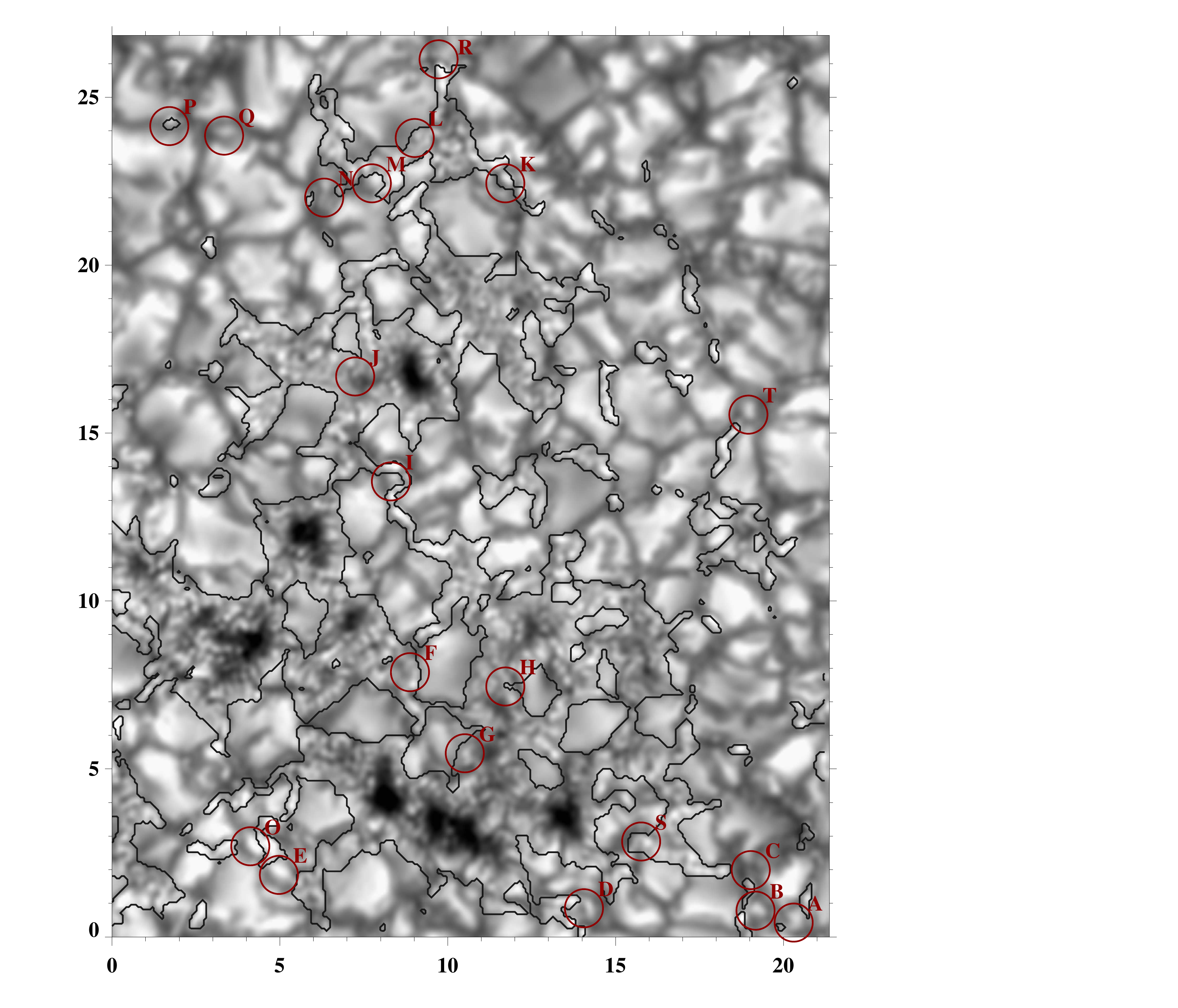}
    \includegraphics[bb=31 15 702 839,width=0.45\linewidth]{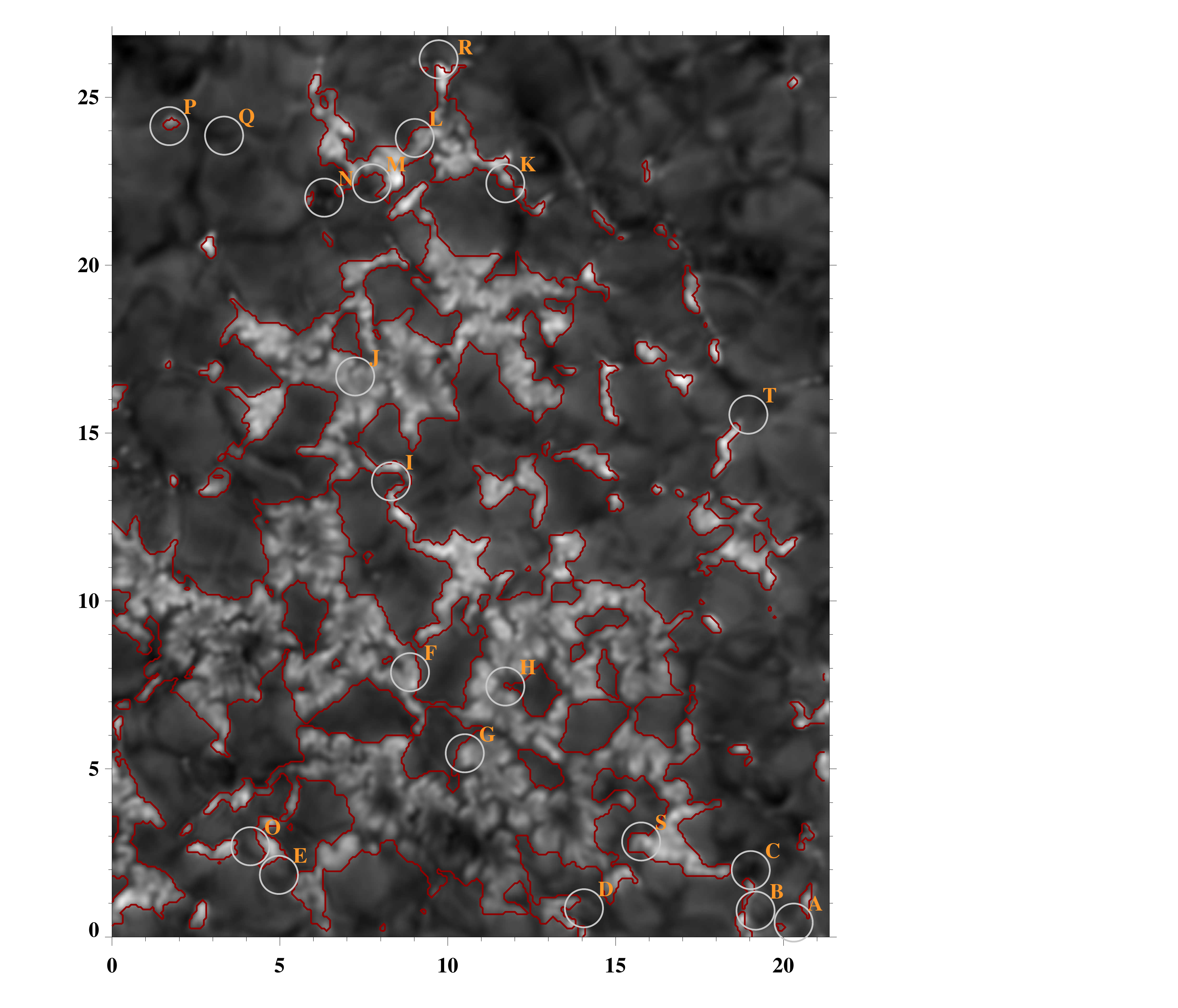}
    \includegraphics[bb=31 15 702 839,width=0.45\linewidth]{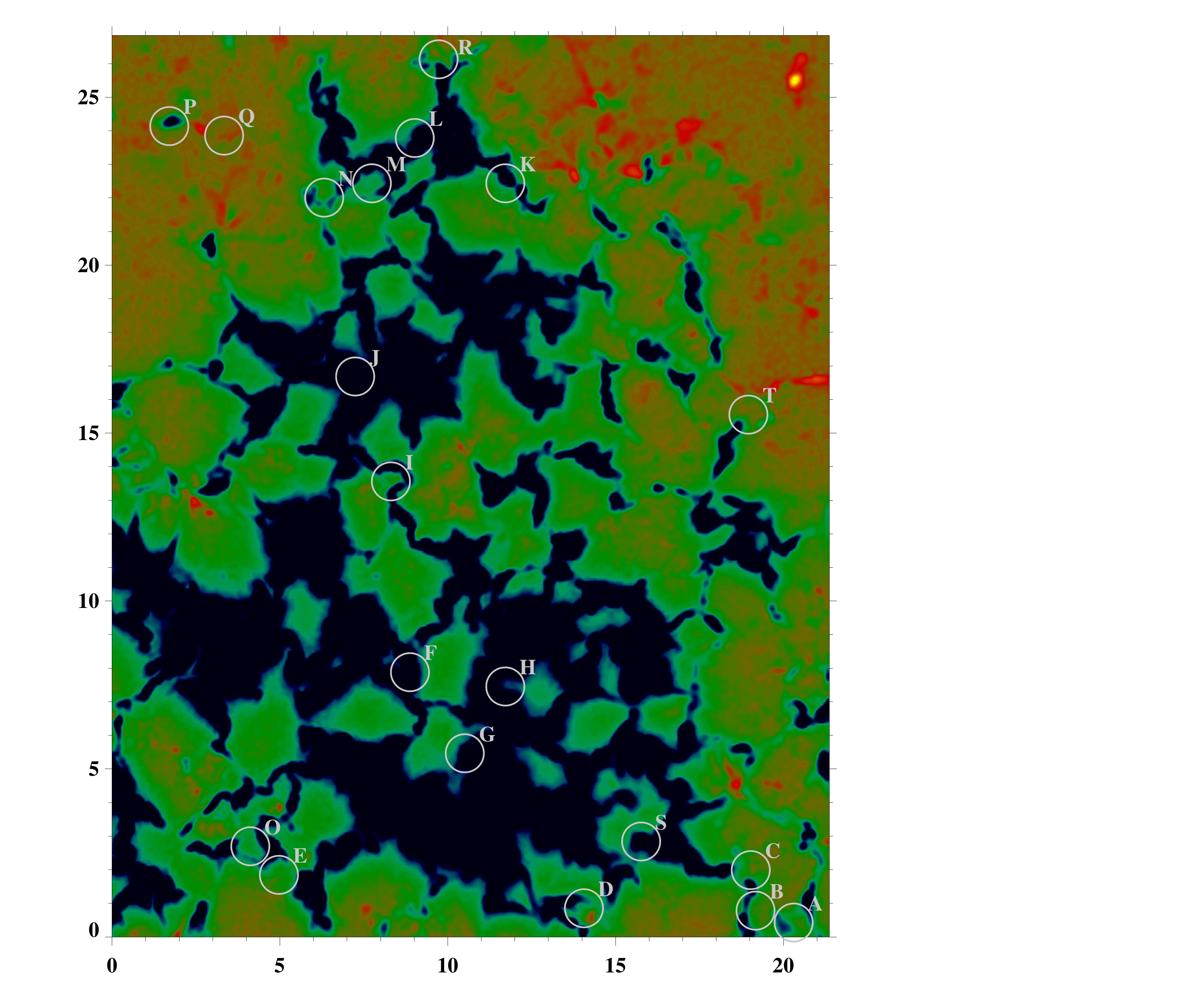}
    \includegraphics[bb=31 15 702 839,width=0.45\linewidth]{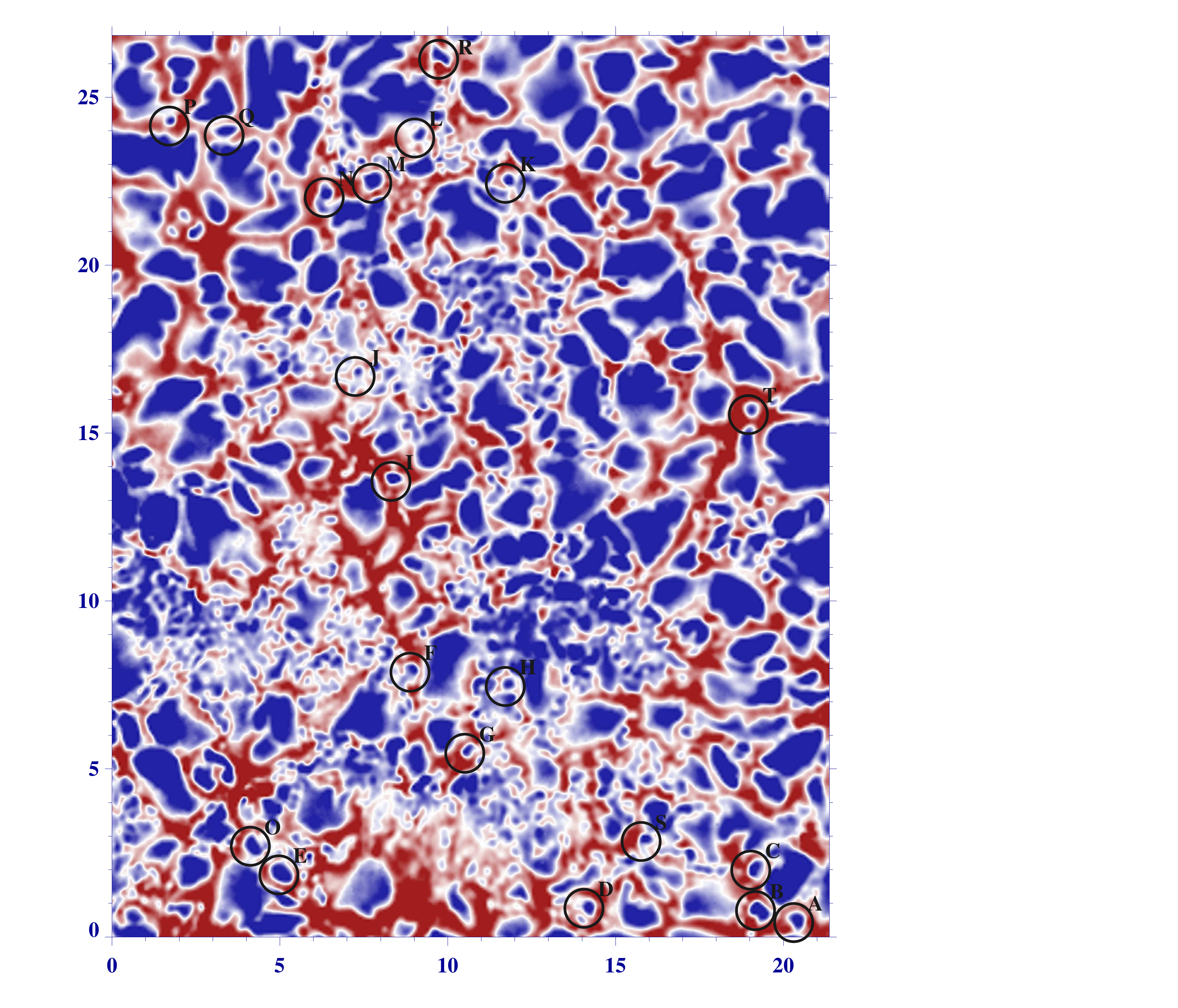}
    \hspace{0.6cm}
    \includegraphics[bb=2 744 976 807,clip,width=0.42\linewidth]{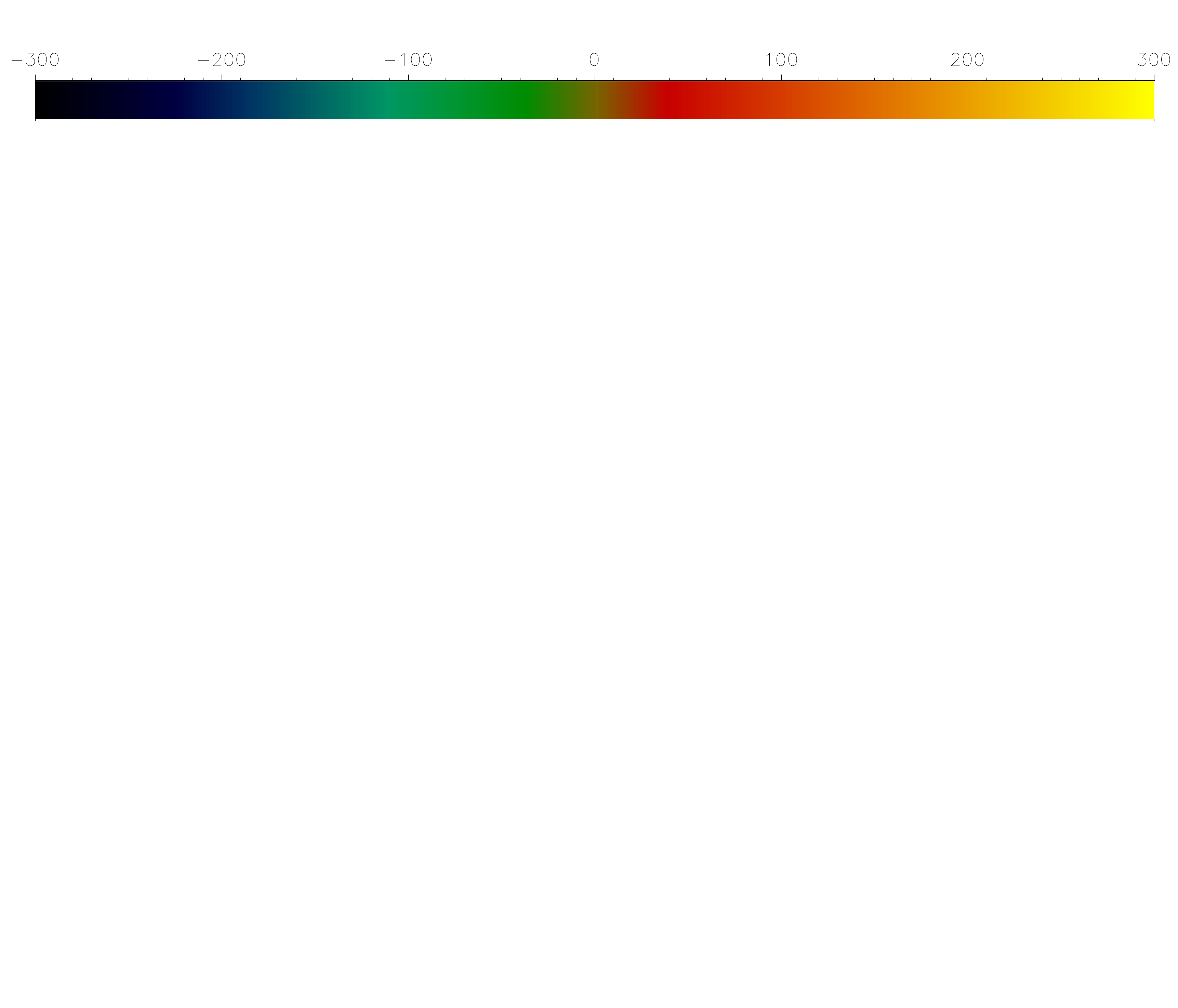}
    \hspace{0.4cm}
    \includegraphics[bb=2 744 992 807,width=0.42\linewidth]{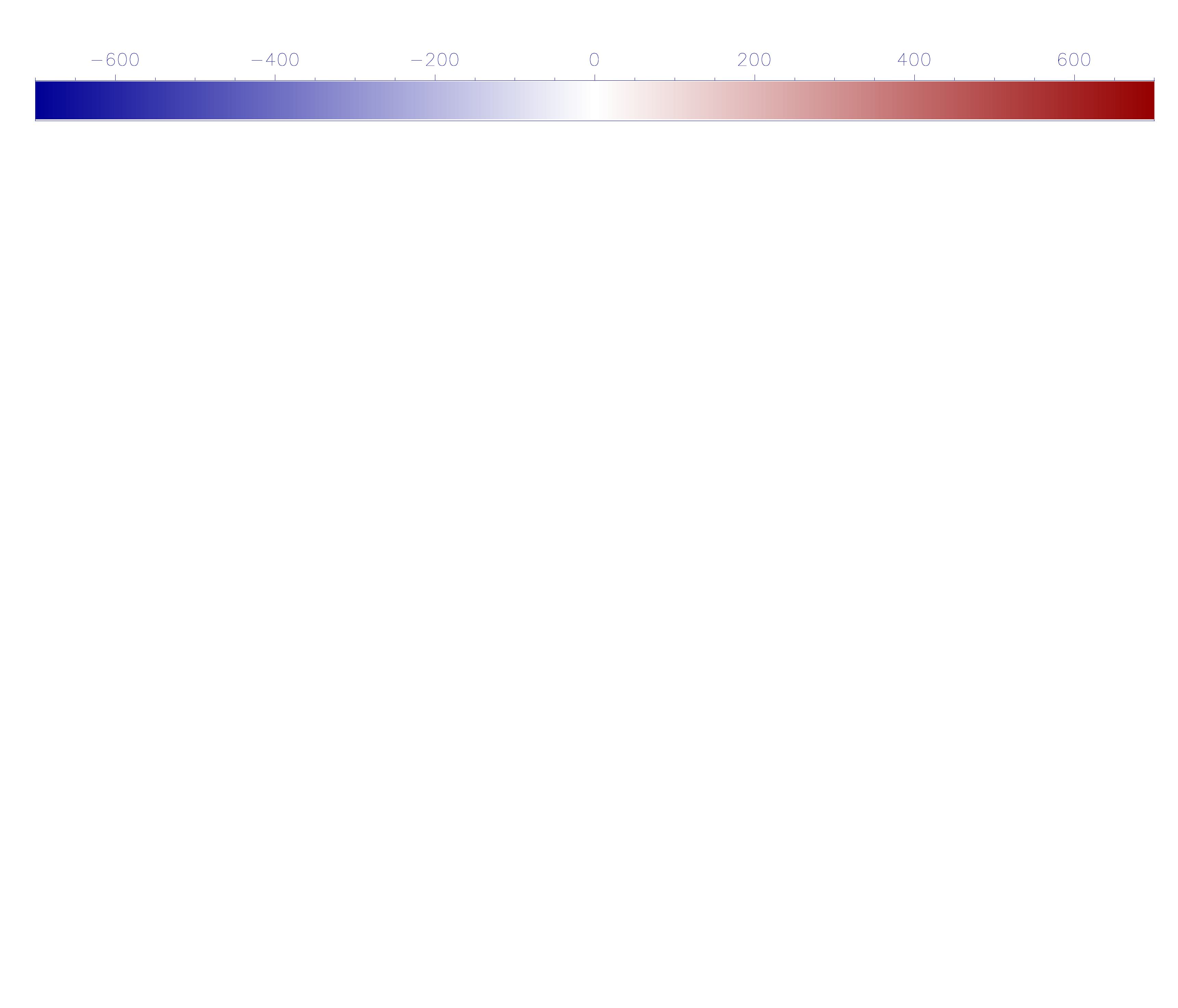}
    \caption{The ROI shown in Fig.~\ref{fig:fullFOV}, used to identify
      various upflow features (marked A--T). The upper row shows the
      continuum image and the \Imin\ map, the lower row \Blos\ and
      \vlos\ maps obtained from inversions. The black/red contours
      correspond to $|\Blos|=200$~G. Tick marks are in units of arcsec. The color bars indicates the signed LOS magnetic field in Gauss (left) and LOS velocity in \mps (right). Except for the color and grey scales used, this figure is identical to Fig.~\ref{fig:zoommosaic}.}
    \label{fig:zoommosaic2}
  \end{center}
\end{figure*}
\subsubsection{Strings}
\label{sec:strings}

The network also consists of other bright features such as thread-like
features classified as strings by \citet{rouppe05solar}. The features
labeled S0, S1 etc are example of strings.

The fourth row of Fig.~\ref{fig:vlos-blos} shows plots across strings.
In S0 we see a downflow of about 300~\mps\ close to the peak field. In
S1 we see a downflow of about 120~\mps. In S2 we see a strong downflow
of 1.4~k\mps. S3 is different from the other strings studied and shows
an upflow of about $-640$~\mps\ close to the position of the peak LOS
field. There is also a secondary upflow of about $-1.0$~k\mps\ nearby.
This secondary upflow appears to be related to a tiny bright structure
visible in the continuum map. It is not clear if this structure is a
tiny fragment from the neighboring granule. \Blos\ shows that it is a
weak magnetic structure.

\subsubsection{Other small-scale magnetic features}
\label{sec:other-small-scale}

The fifth row of Fig.~\ref{fig:vlos-blos} shows plots of a few
miscellaneous small-scale magnetic features which we refer to as
``other'' features. These are bright features which do not fit in any
of the above categories. Some examples are shown in circles labeled as
O1, O2 etc. O0 looks like a fragmented string and shows a strong
upflow of about $-1.05$~k\mps. O1 shows a variable downflow pattern
ranging between 100~\mps\ and 600~\mps, O2 a similar flow pattern in
the range 200--800~\mps. O3 is an extended structure with an upflow
varying between $-300$~\mps\ and $-450$~\mps.

\subsubsection{Small-scale upflow features}
\label{sec:small-scale-upflow}

\renewcommand{\thesubfigure}{{}}
\begin{figure*}[htb!]
\subfigure[A]{
\includegraphics[bb=45 95 1133 640,clip,width=0.240\linewidth]{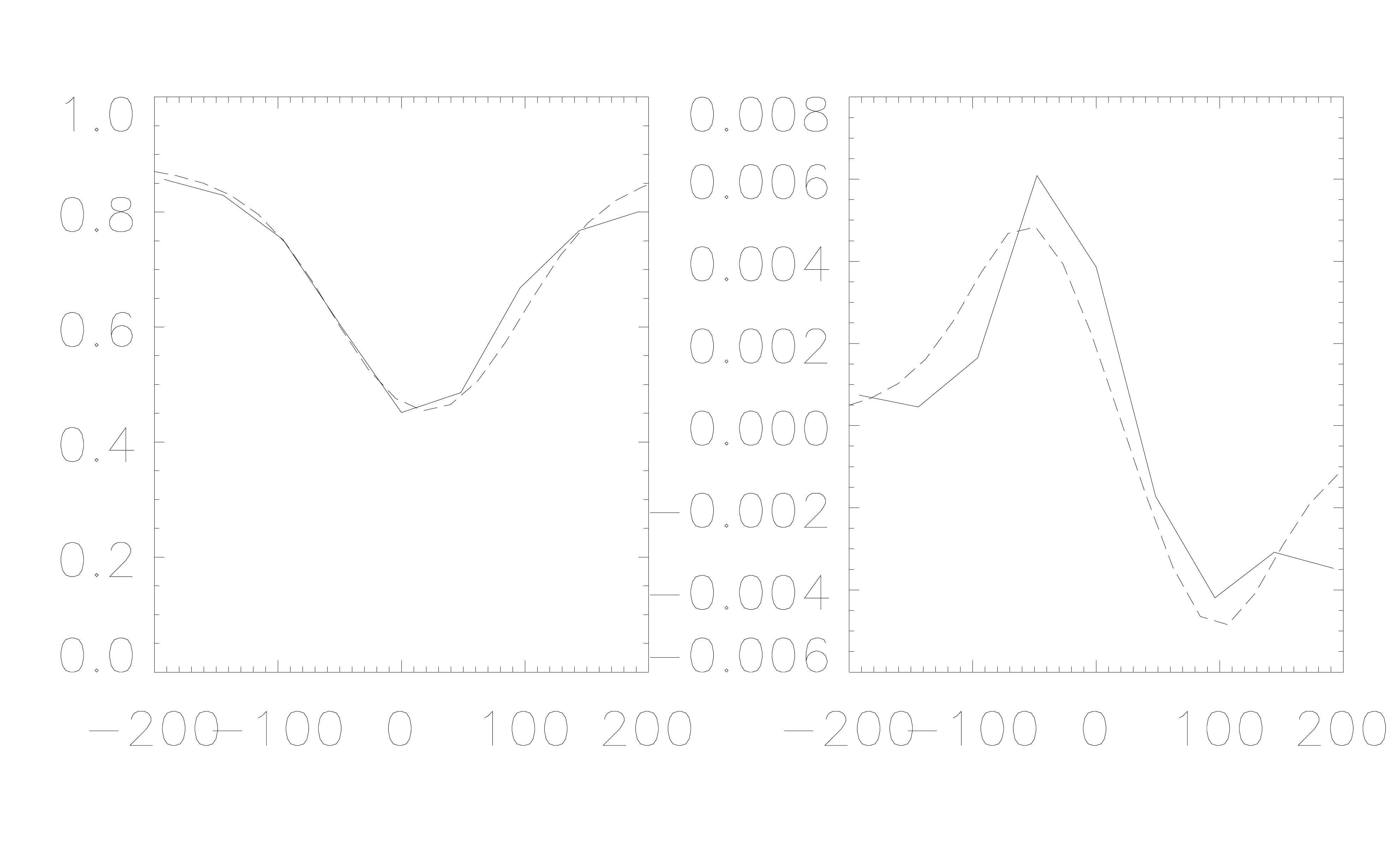}}
\subfigure[ B]{
\includegraphics[bb=45 95 1133 640,clip,width=0.240\linewidth]{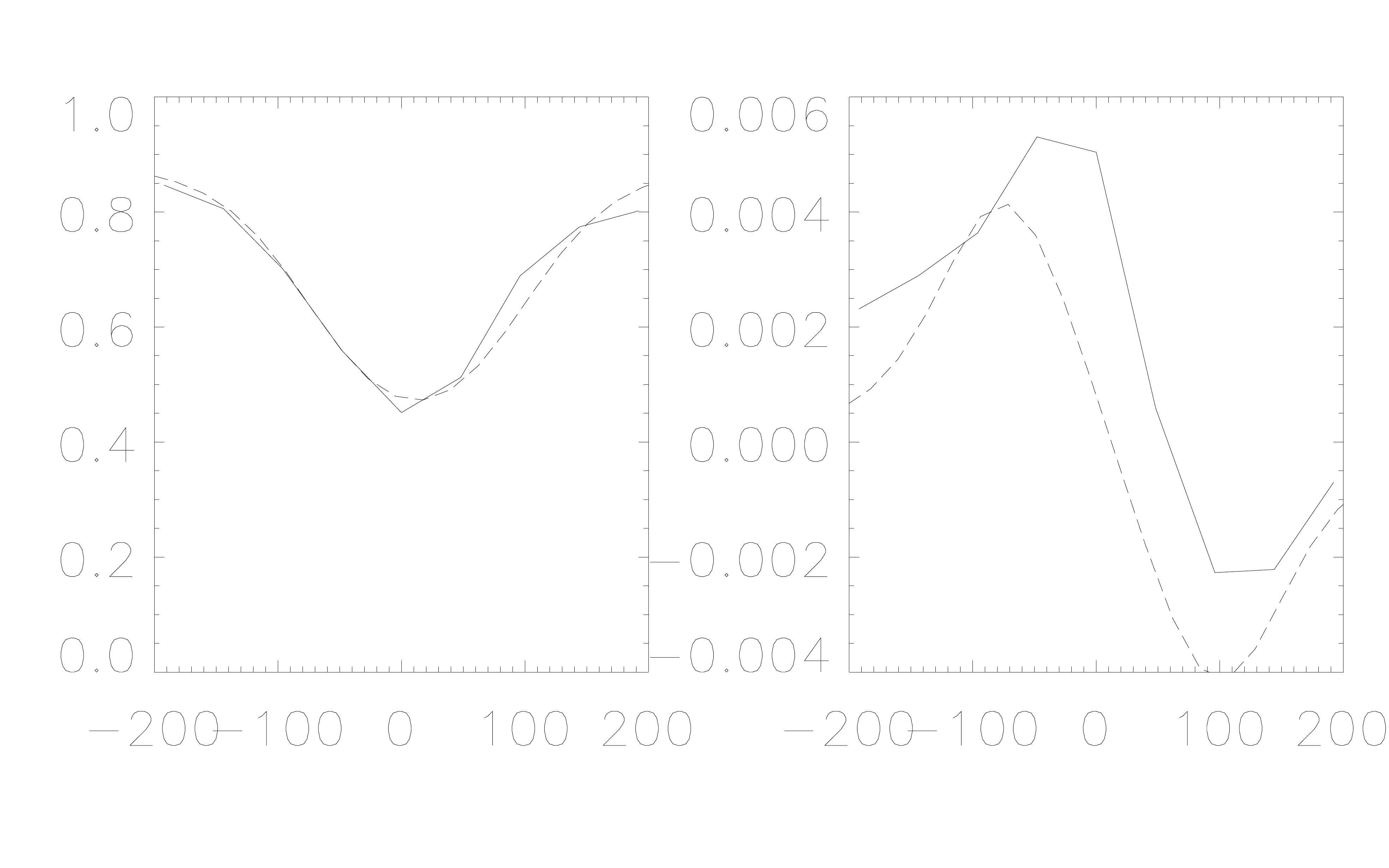}}
\subfigure[ C]{
\includegraphics[bb=45 95 1133 640,clip,width=0.240\linewidth]{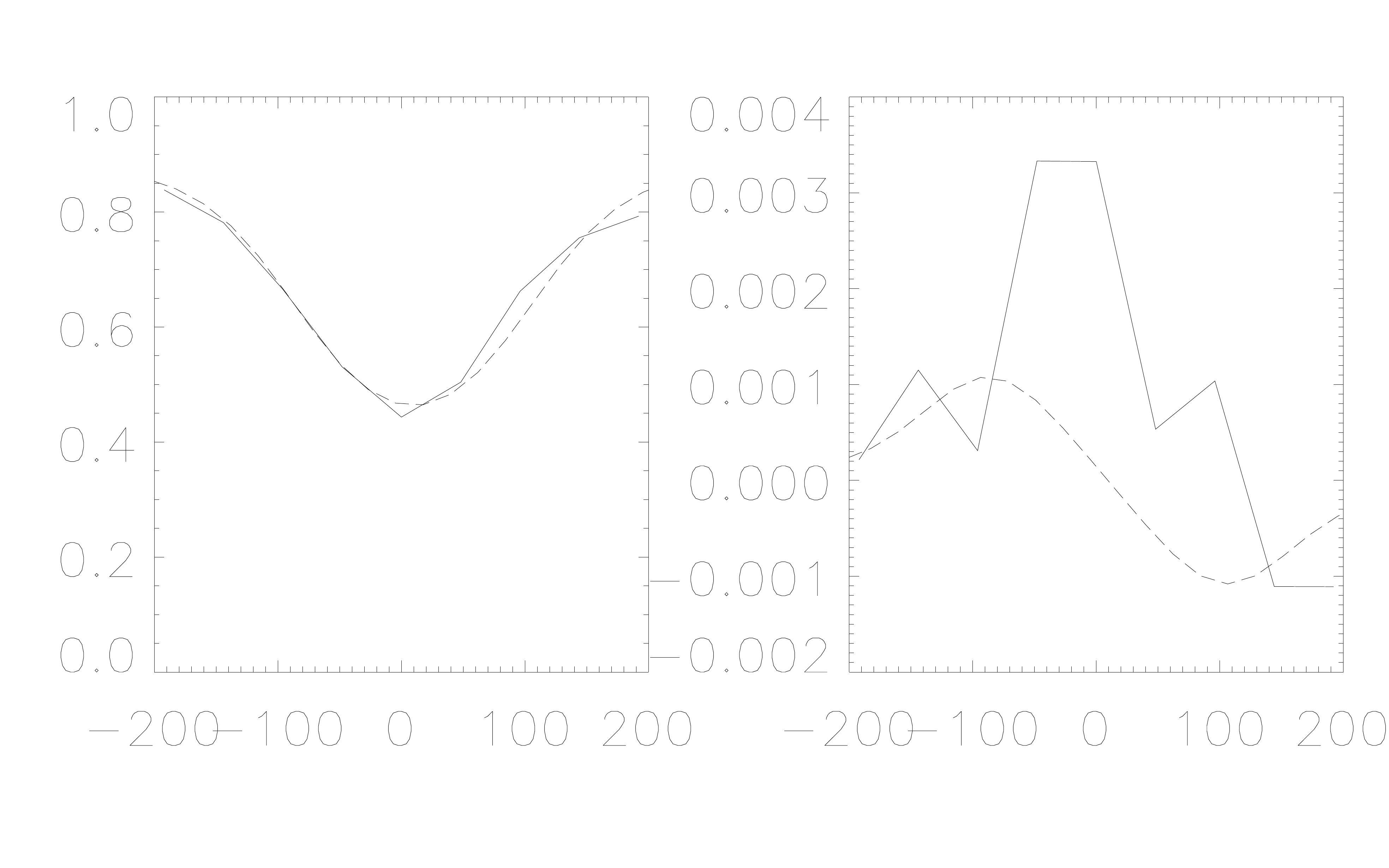}}
\subfigure[ D]{
\includegraphics[bb=45 95 1133 640,clip,width=0.240\linewidth]{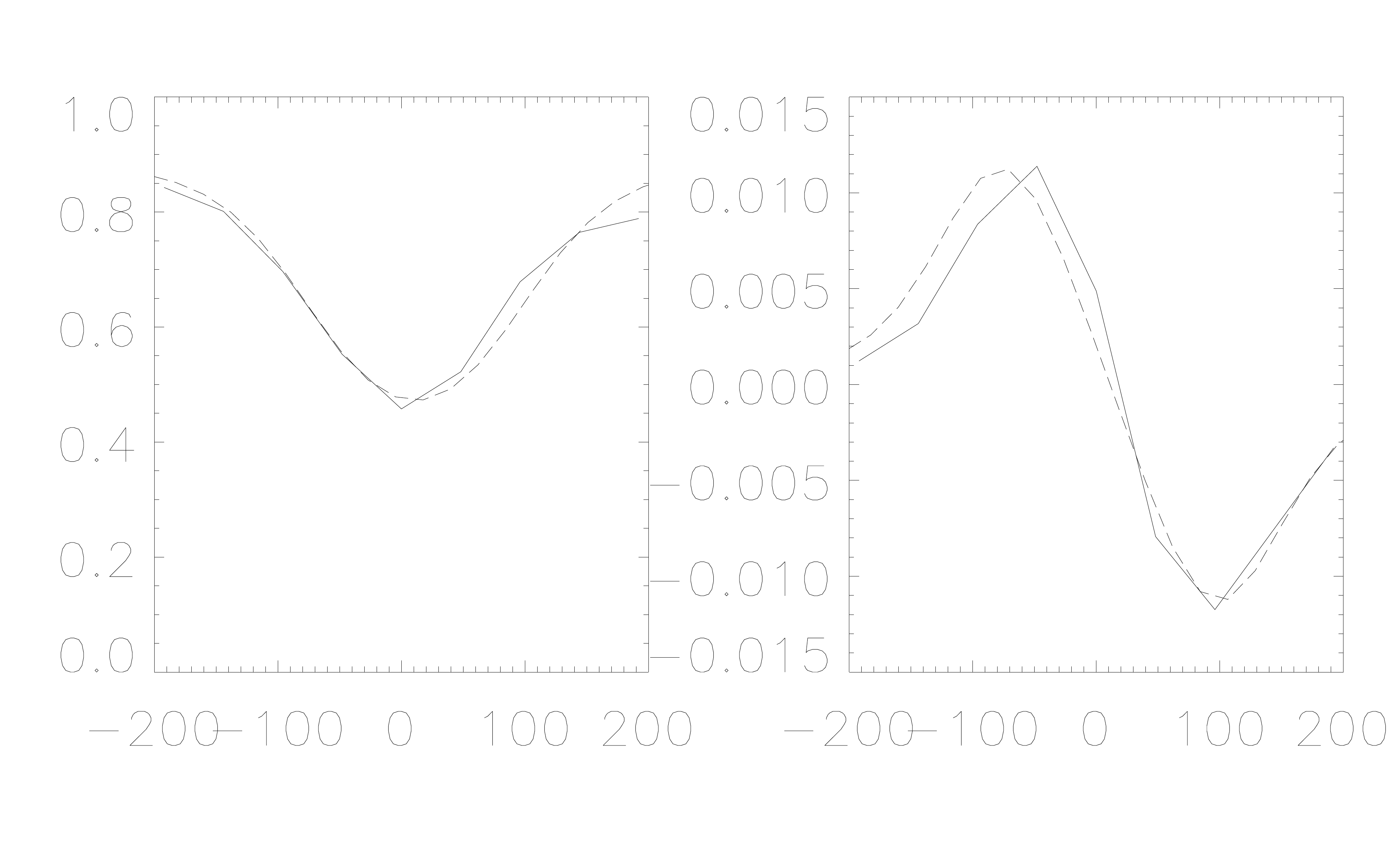}}
\subfigure[ E]{
\includegraphics[bb=45 95 1133 640,clip,width=0.240\linewidth]{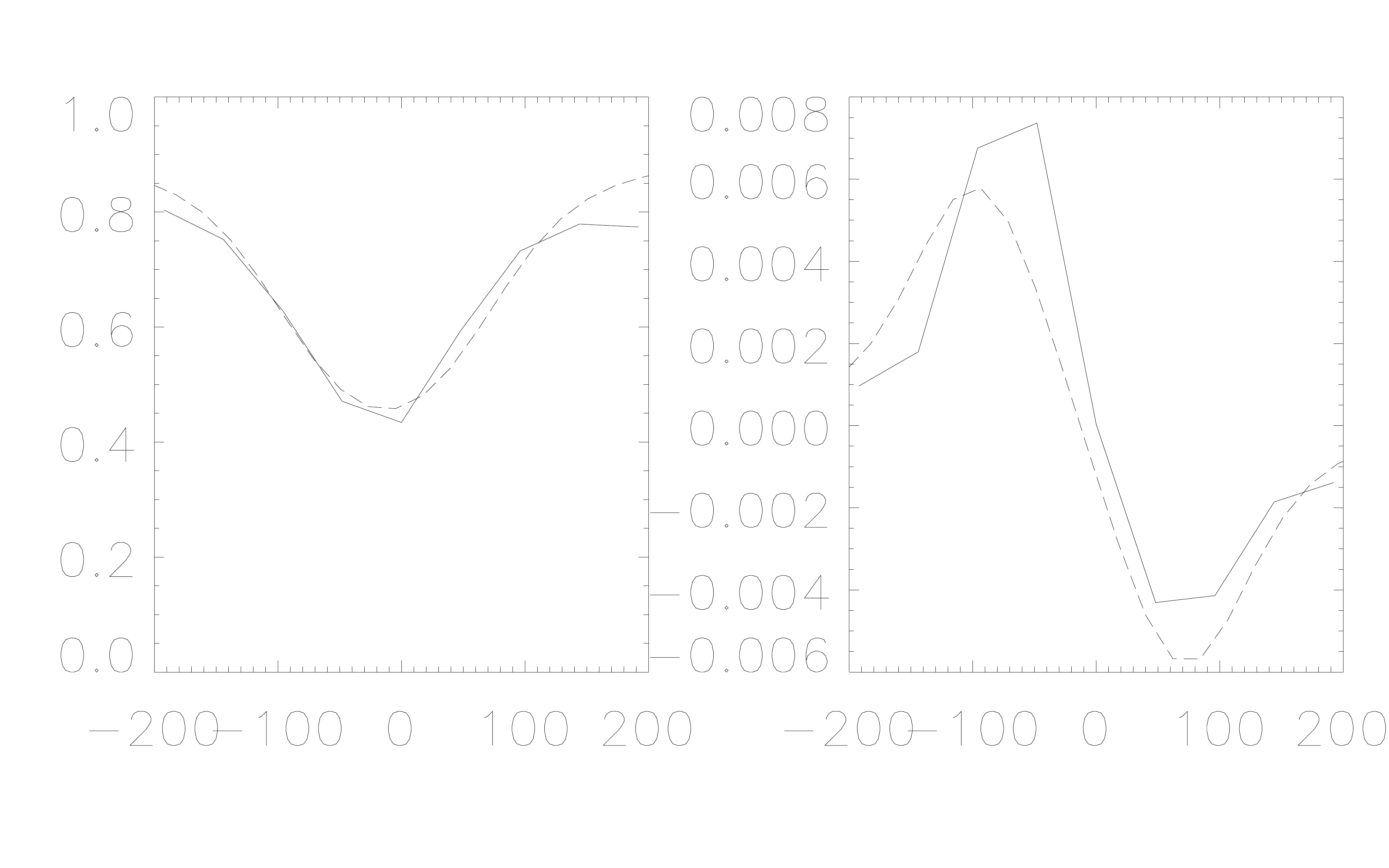}}
\subfigure[ F]{
\includegraphics[bb=45 95 1133 640,clip,width=0.240\linewidth]{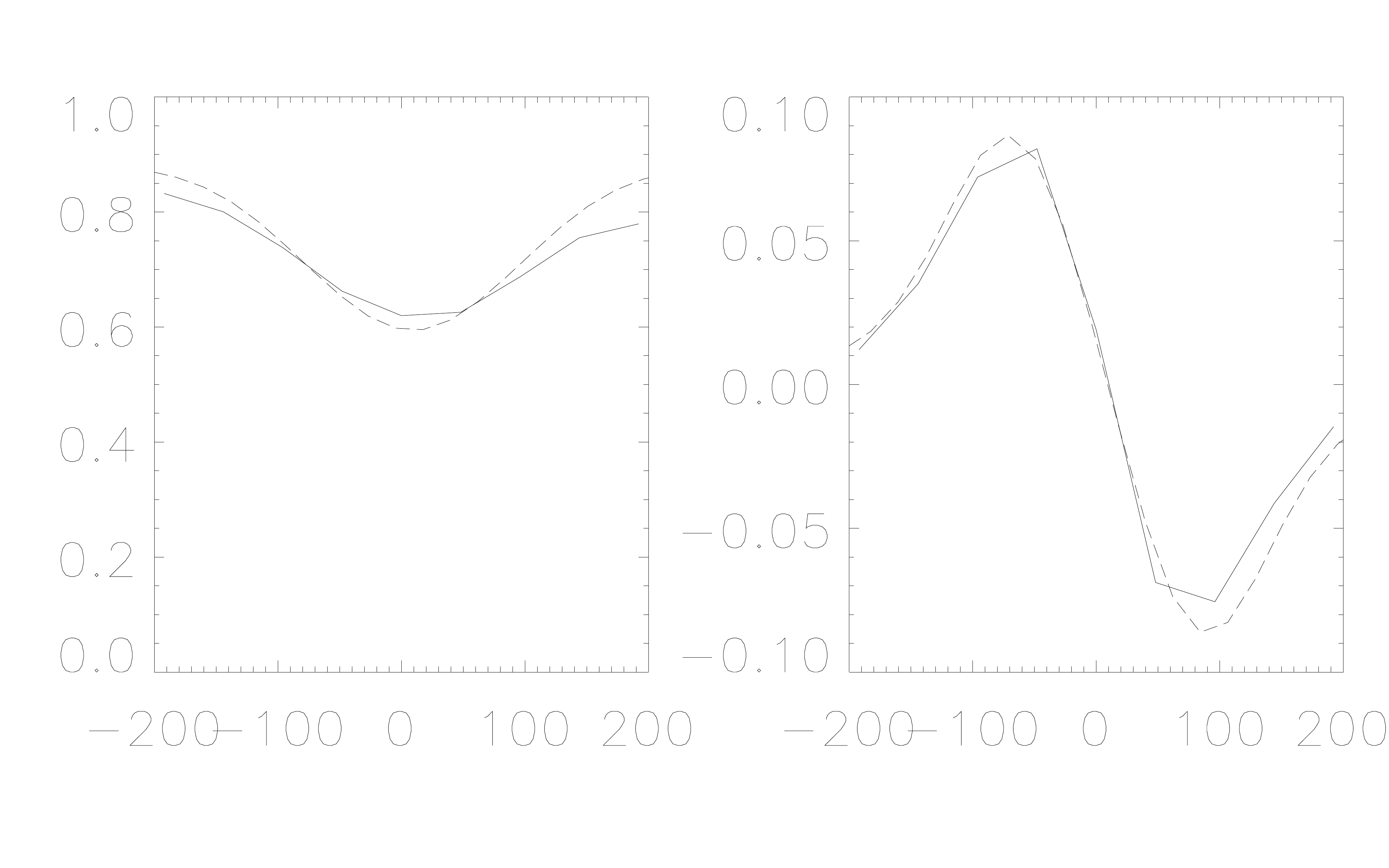}}
\subfigure[ G]{
\includegraphics[bb=45 95 1133 640,clip,width=0.240\linewidth]{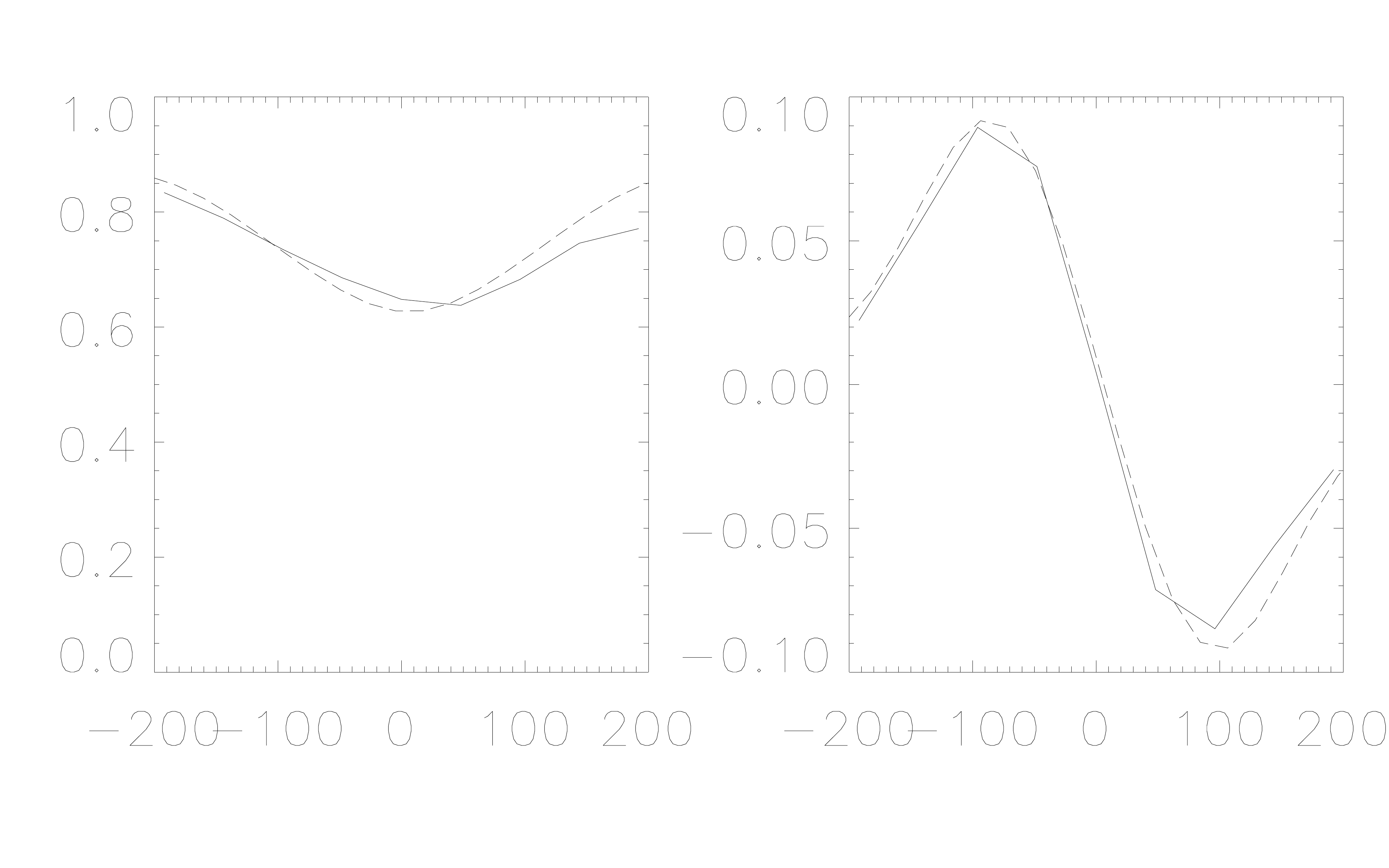}}
\subfigure[ H]{
\includegraphics[bb=45 95 1133 640,clip,width=0.240\linewidth]{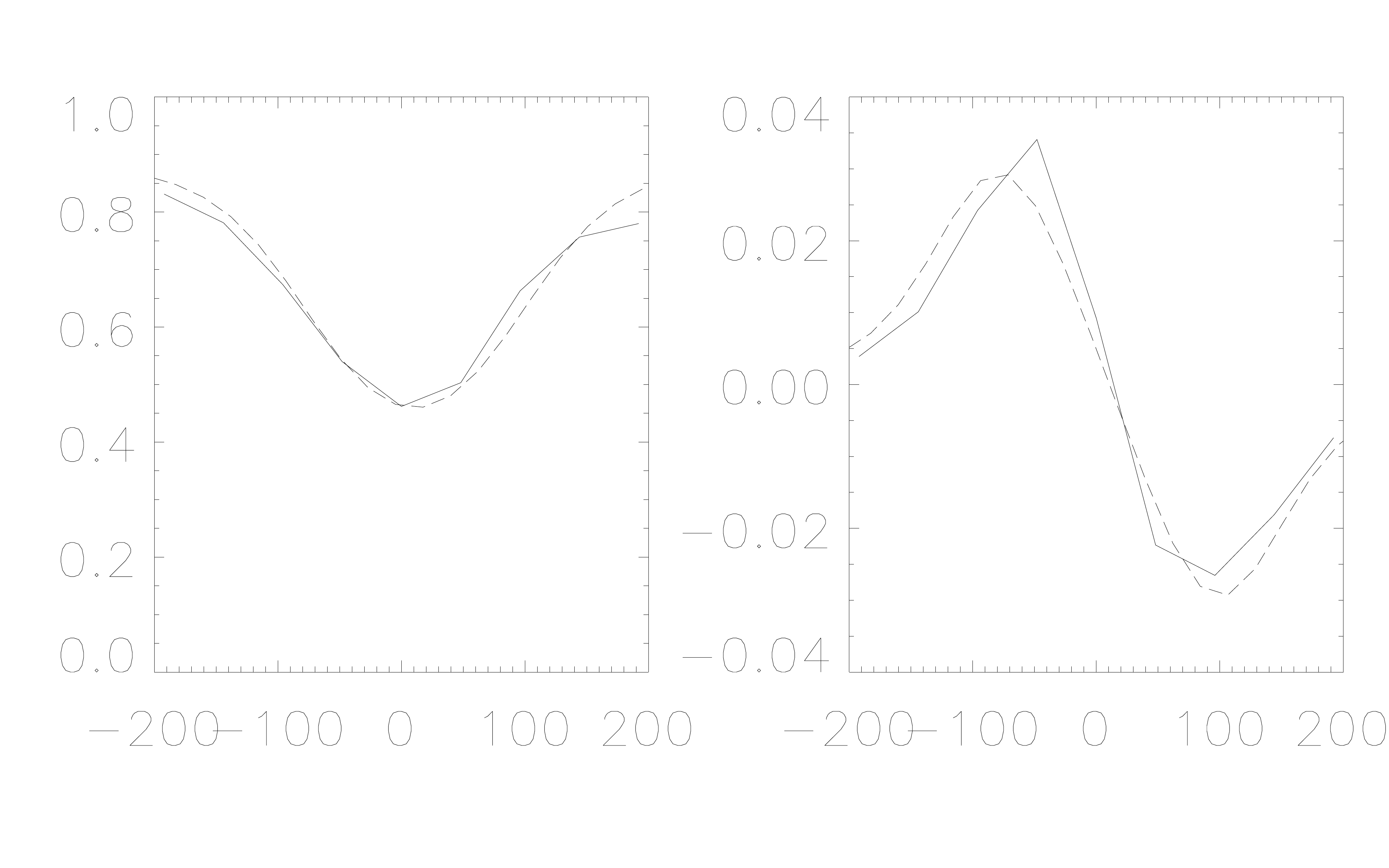}}
\subfigure[ I]{
\includegraphics[bb=45 95 1133 640,clip,width=0.240\linewidth]{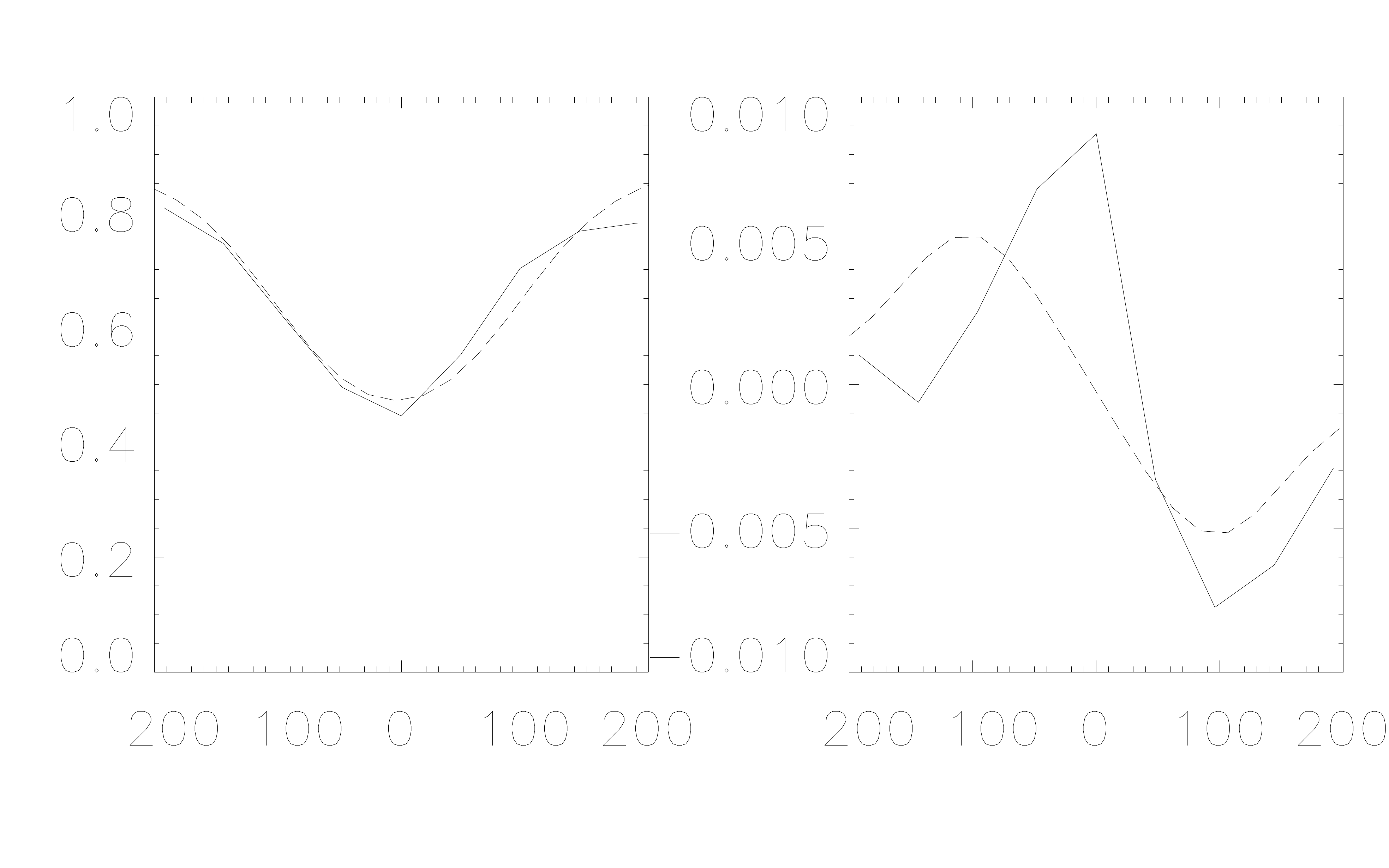}}
\subfigure[ J]{
\includegraphics[bb=45 95 1133 640,clip,width=0.240\linewidth]{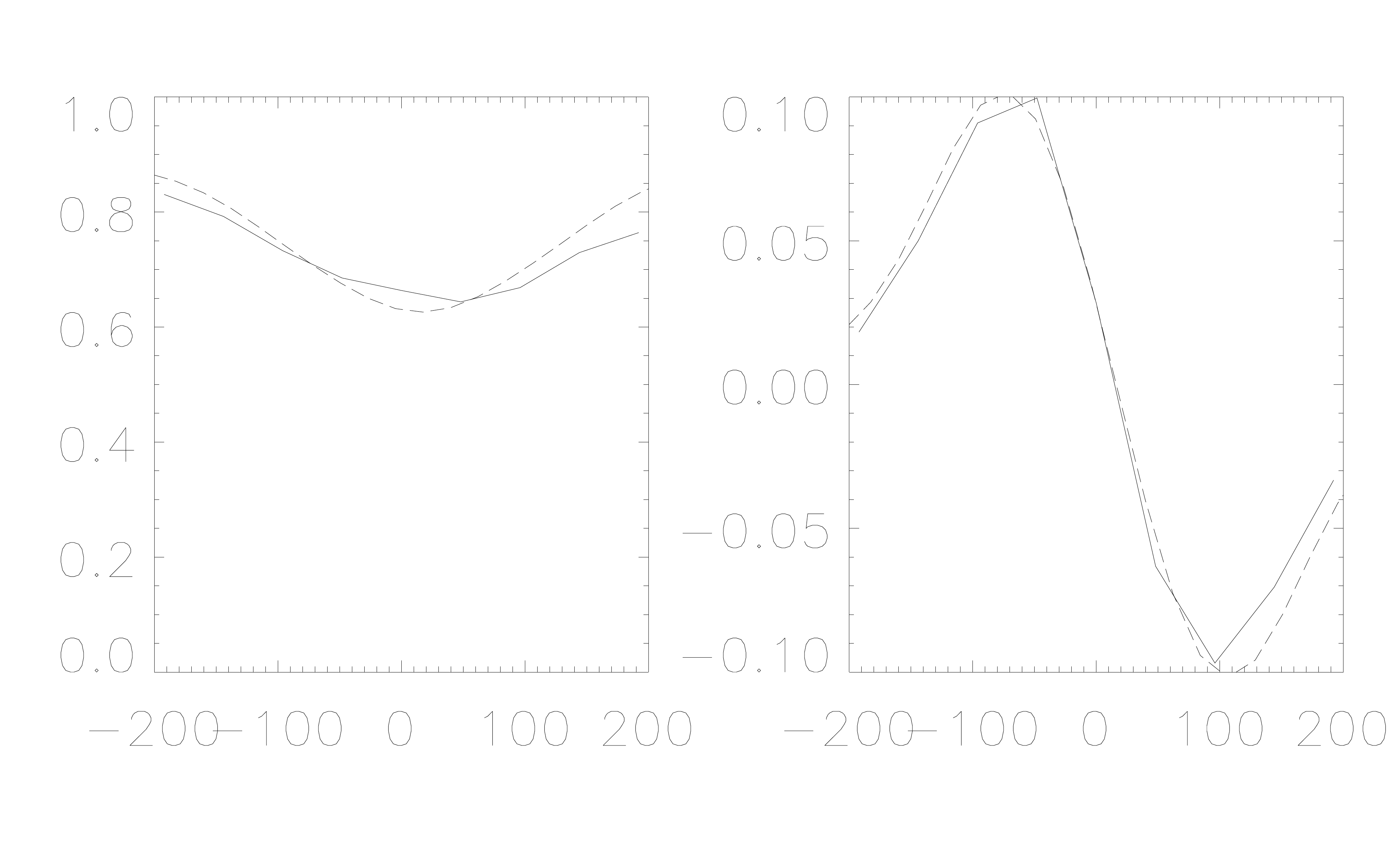}}
\subfigure[ K]{
\includegraphics[bb=45 95 1133 640,clip,width=0.240\linewidth]{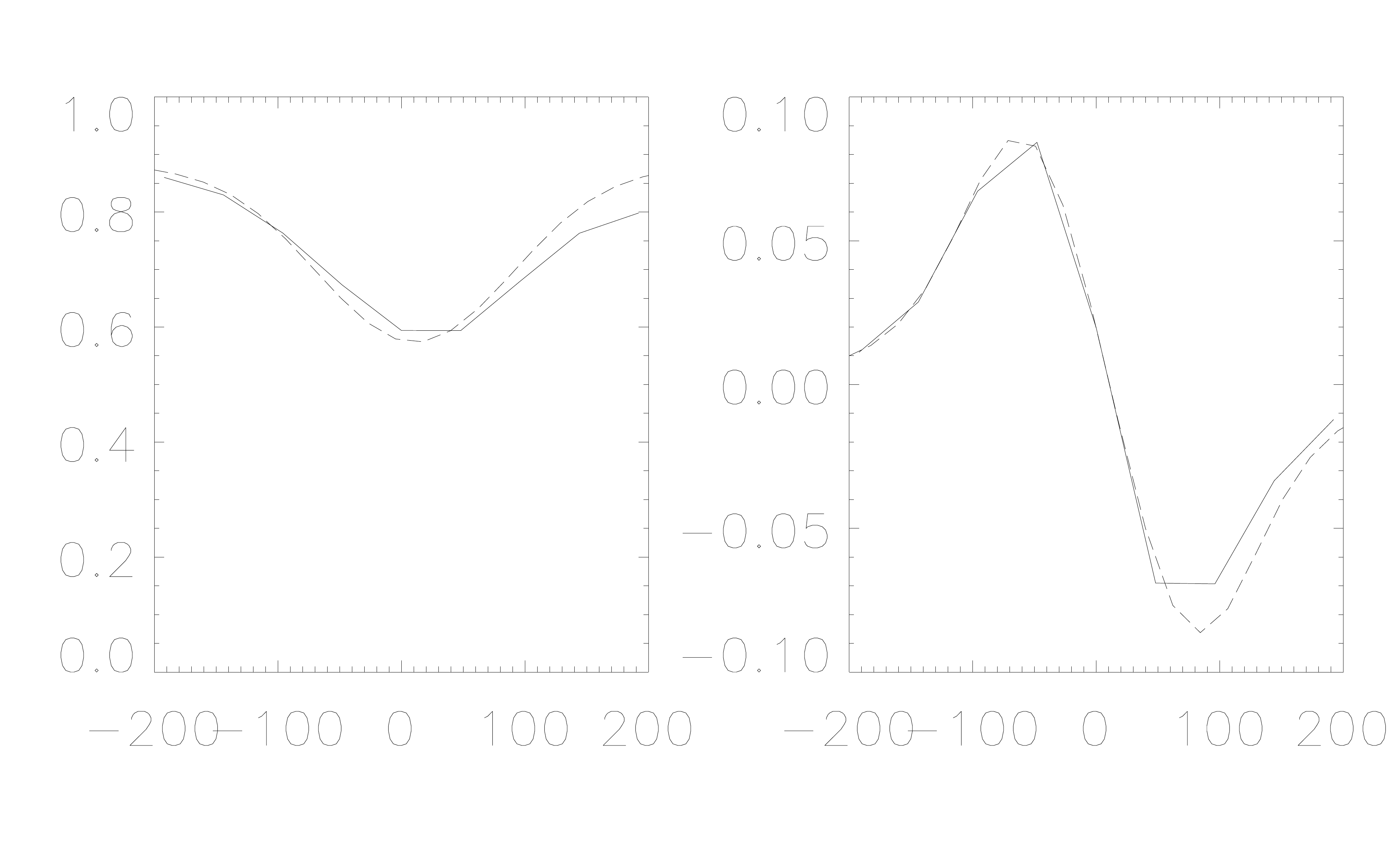}}
\subfigure[L]{
\includegraphics[bb=45 95 1133 640,clip,width=0.240\linewidth]{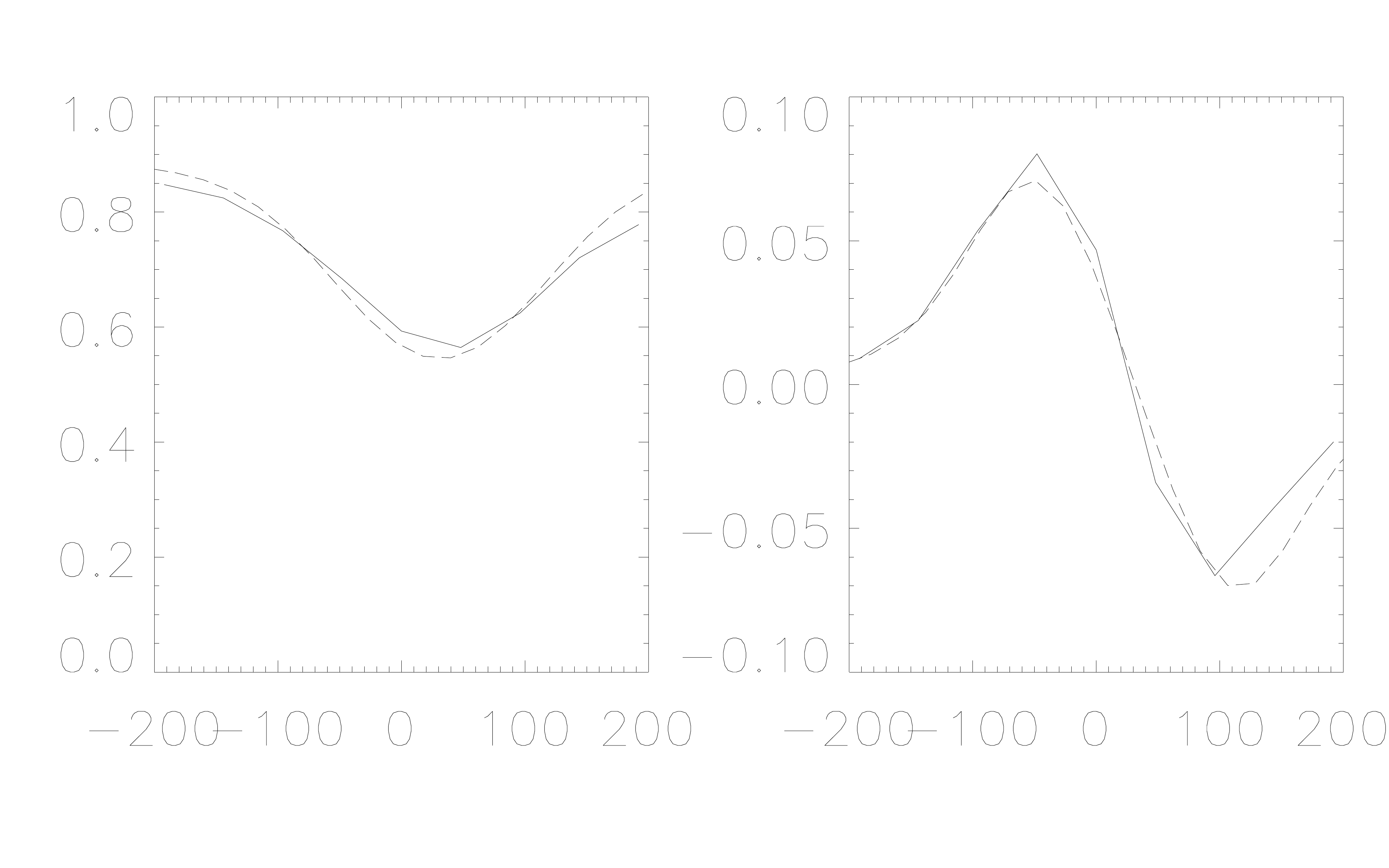}}
\subfigure[ M]{
\includegraphics[bb=45 95 1133 640,clip,width=0.240\linewidth]{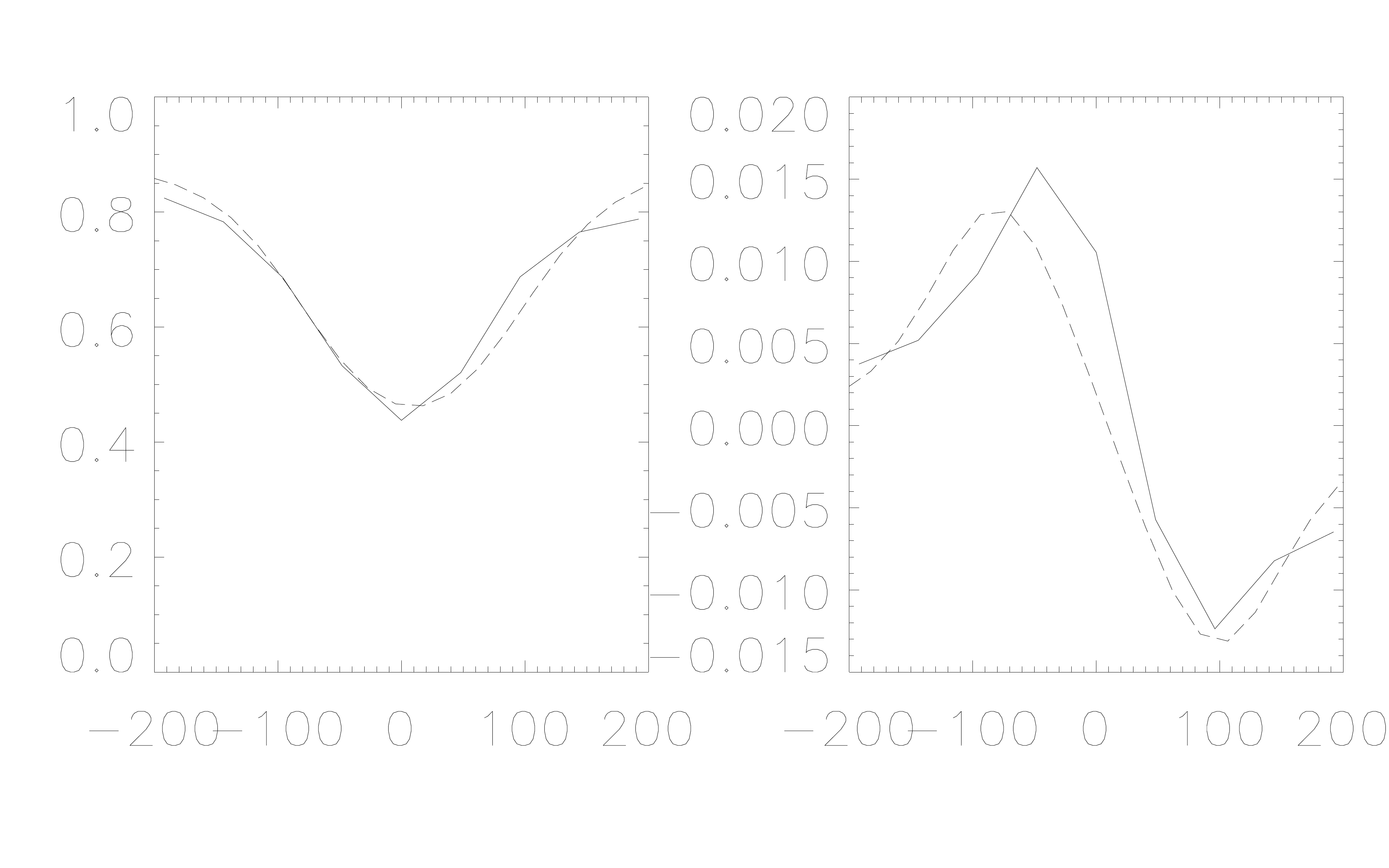}}
\subfigure[ N]{
\includegraphics[bb=45 95 1133 640,clip,width=0.240\linewidth]{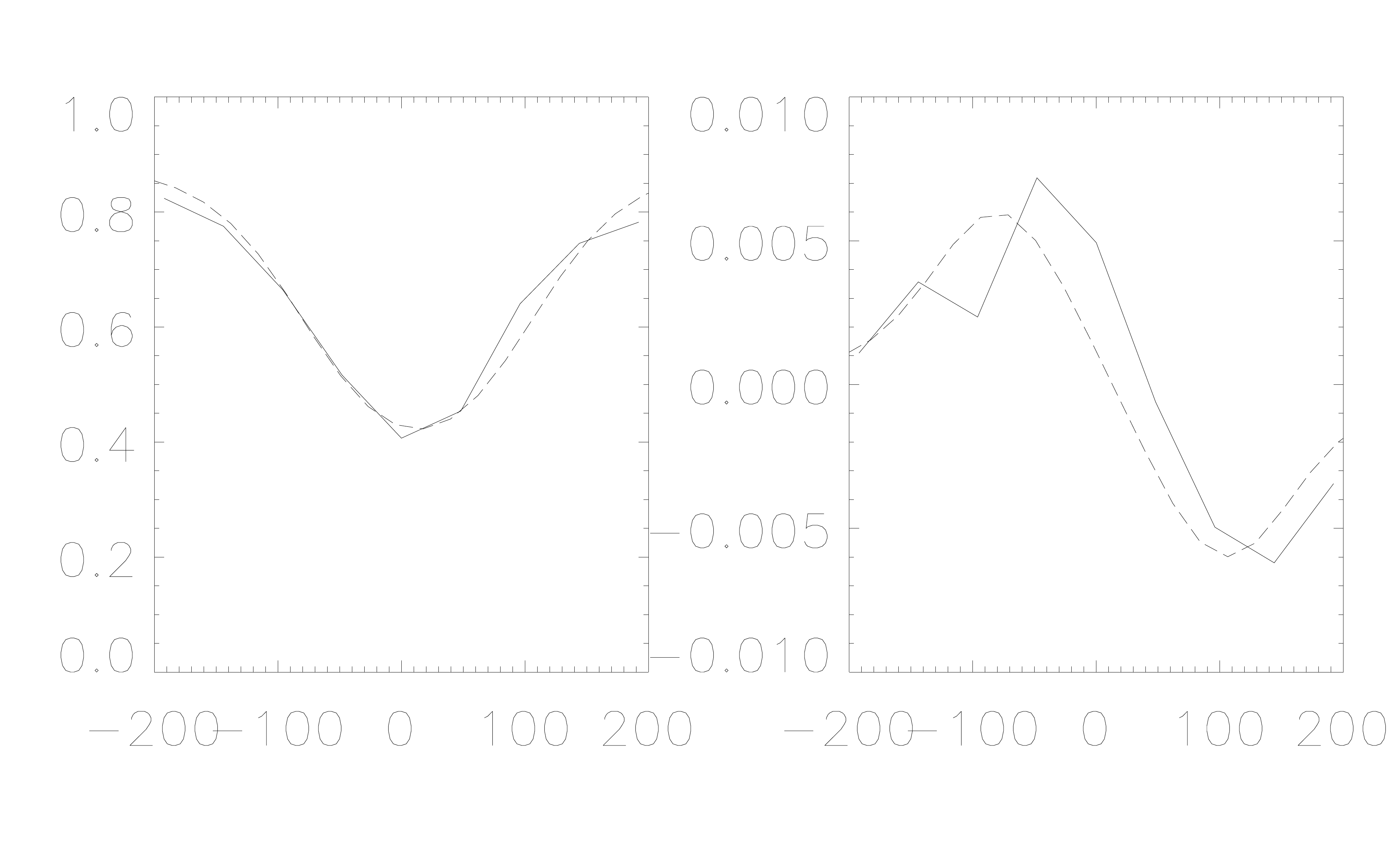}}
\subfigure[ O]{
\includegraphics[bb=45 95 1133 640,clip,width=0.240\linewidth]{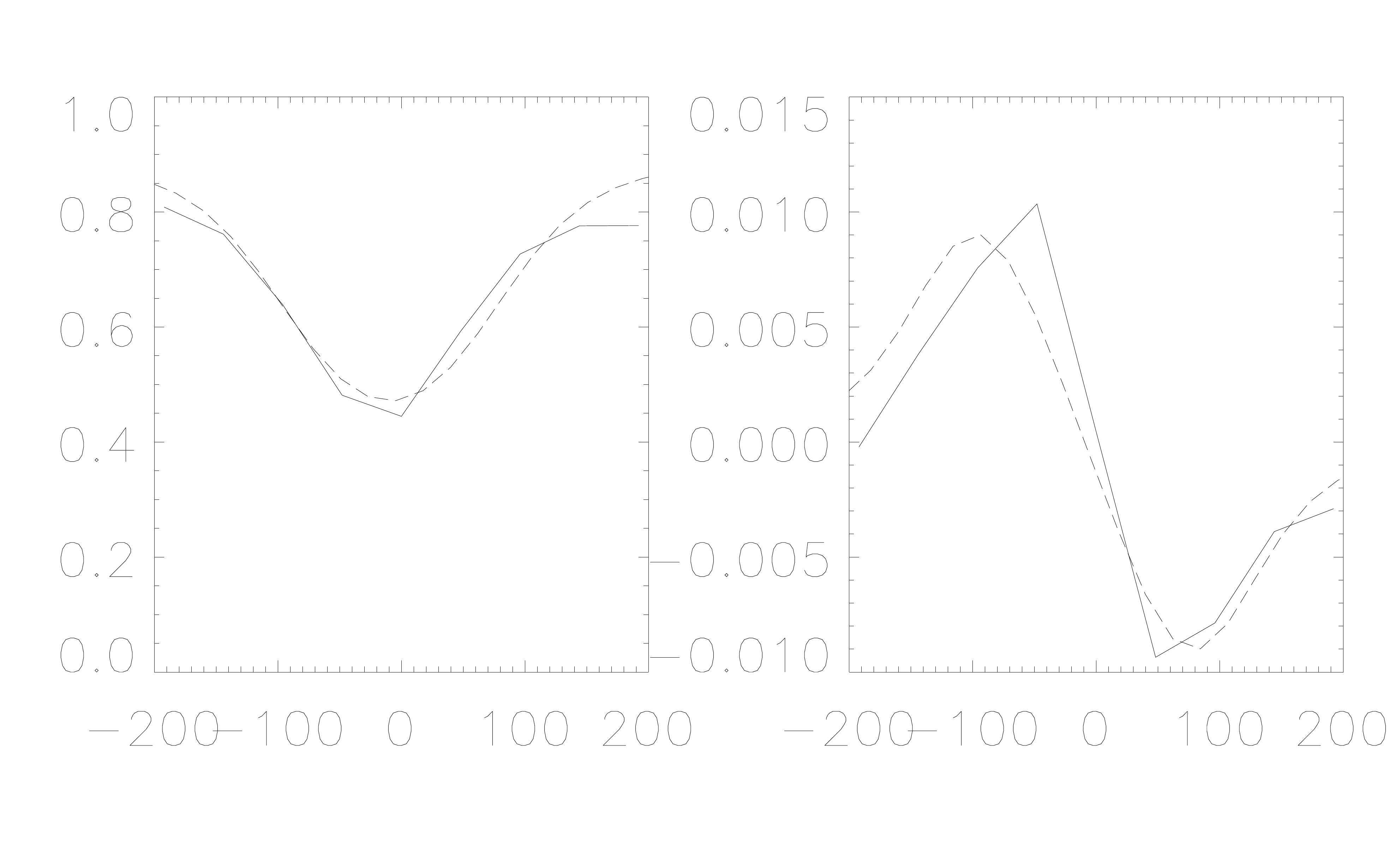}}
\subfigure[ P]{
\includegraphics[bb=45 95 1133 640,clip,width=0.240\linewidth]{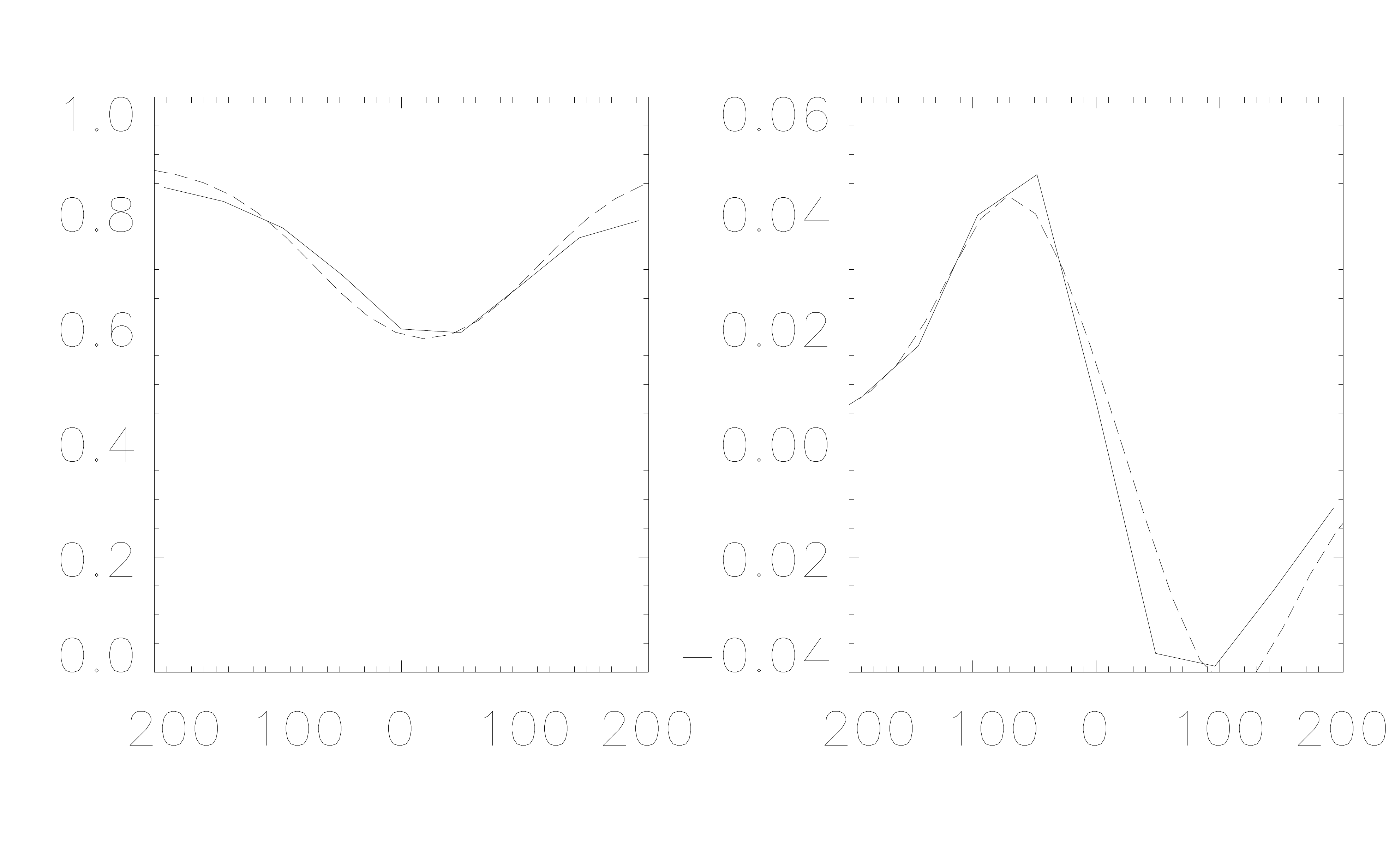}}
\subfigure[ Q]{
\includegraphics[bb=45 95 1133 640,clip,width=0.240\linewidth]{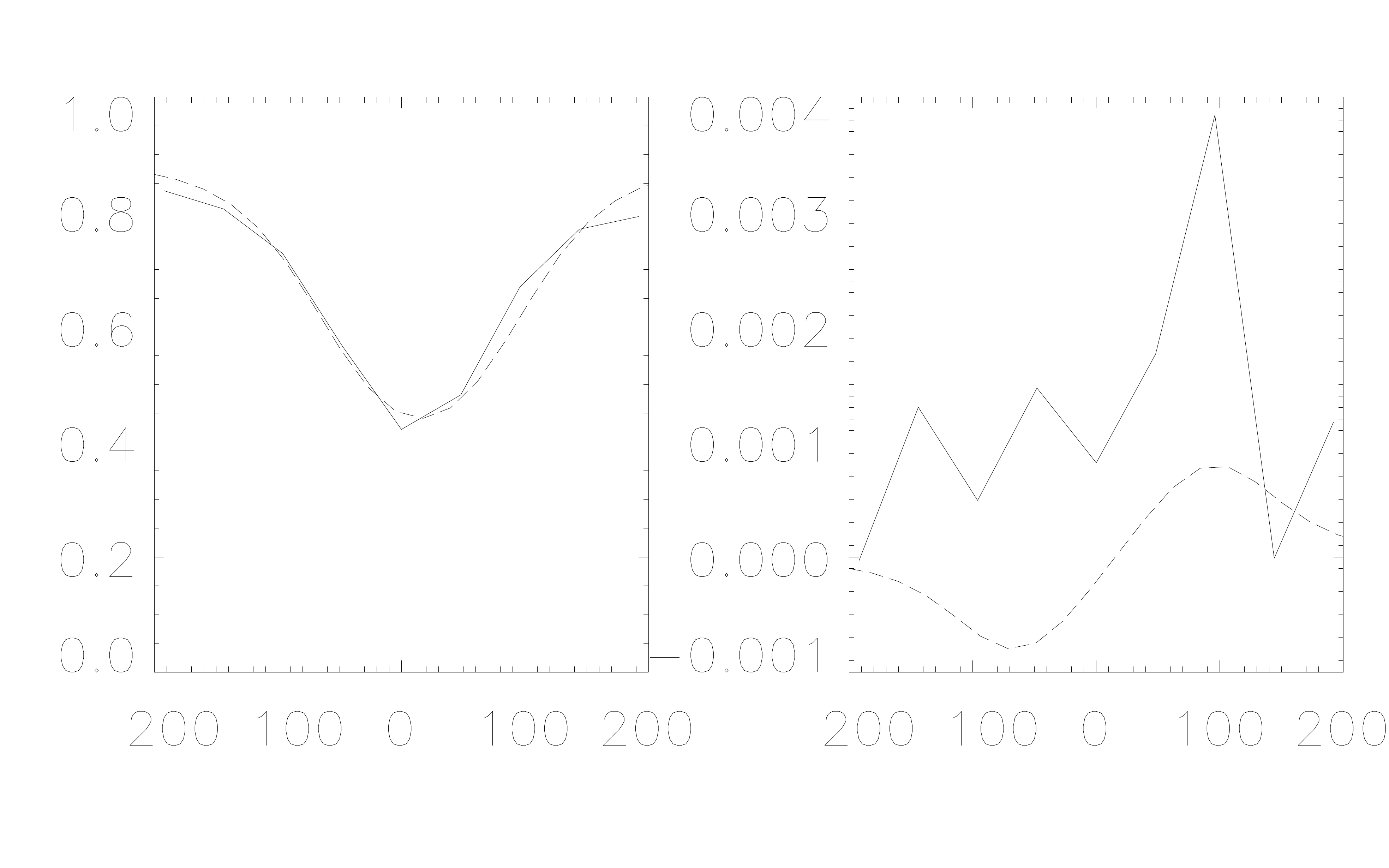}}
\subfigure[ R]{
\includegraphics[bb=45 95 1133 640,clip,width=0.240\linewidth]{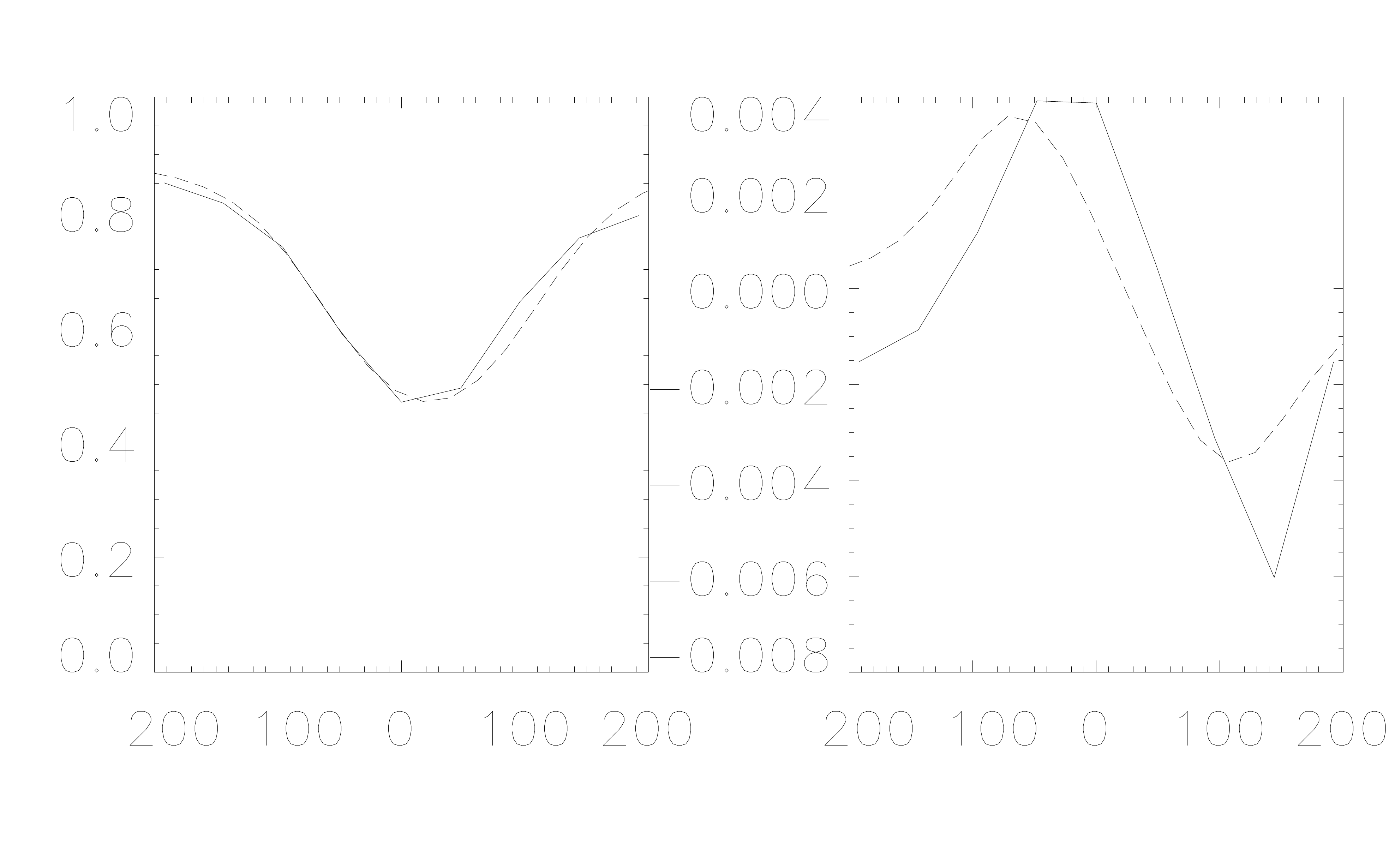}}
\subfigure[ S]{
\includegraphics[bb=45 95 1133 640,clip,width=0.240\linewidth]{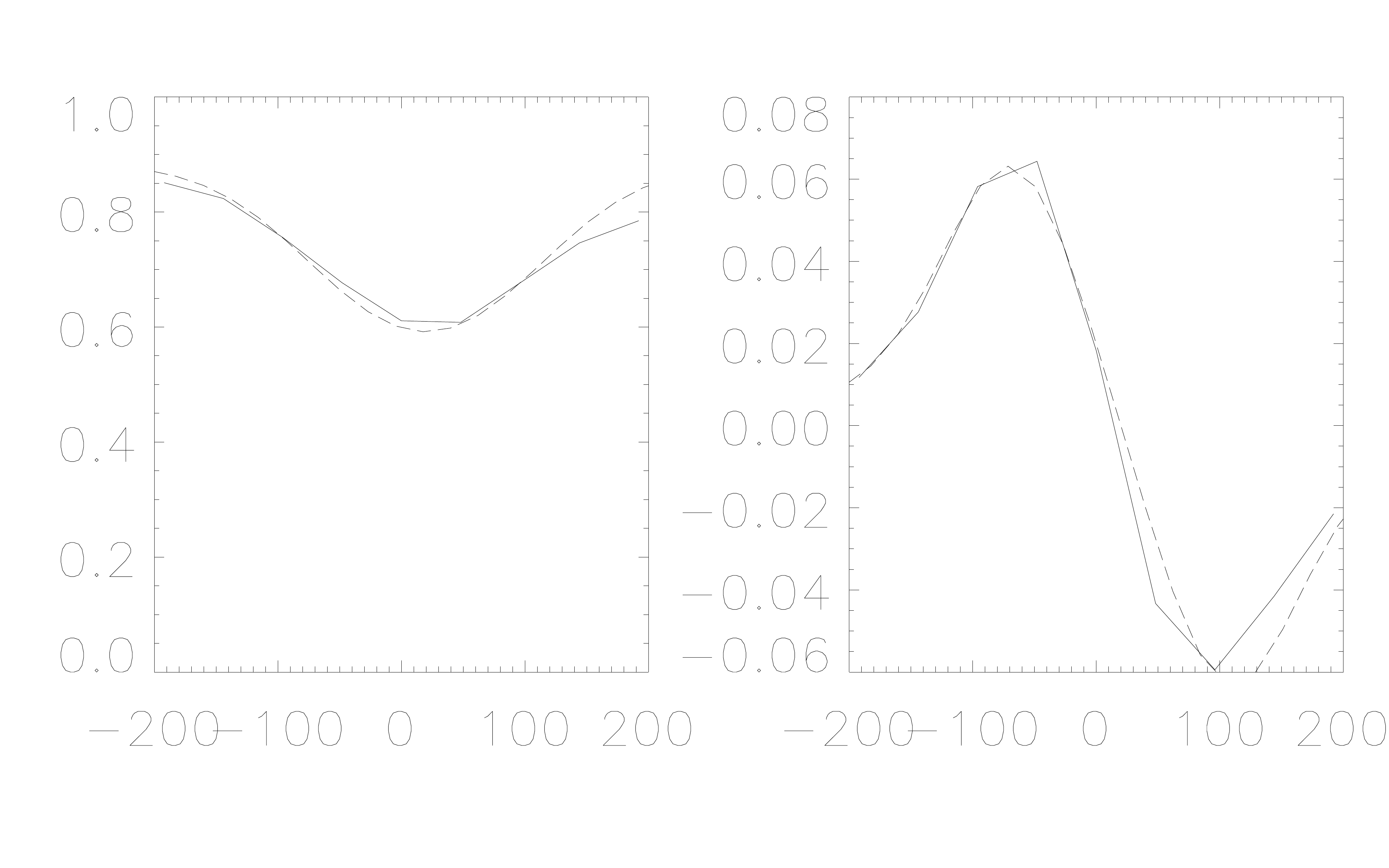}}
\subfigure[ T]{
\includegraphics[bb=45 95 1133 640,clip,width=0.240\linewidth]{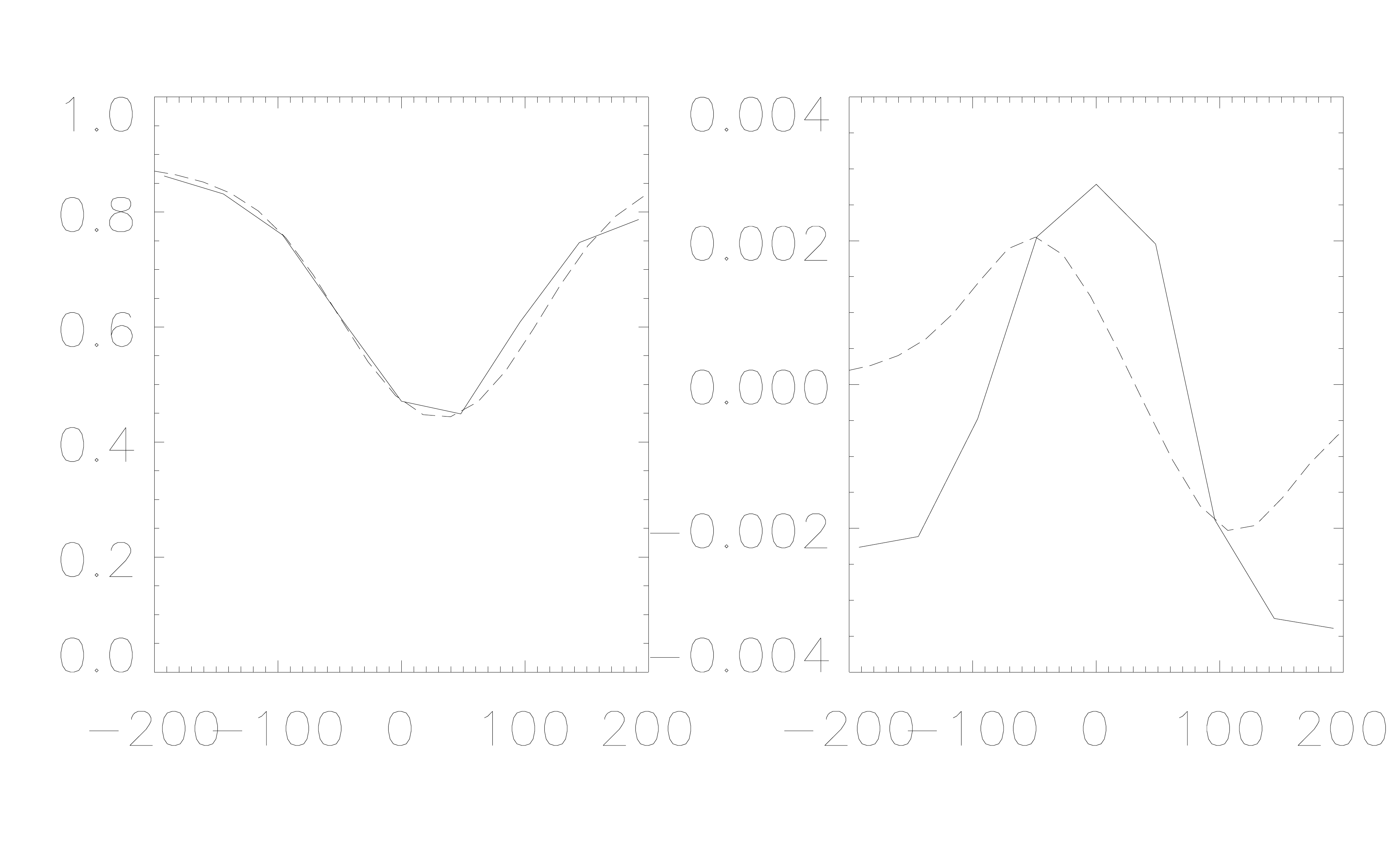}}
\caption{Observed (solid) and fitted (dashed) Stokes $I$ and $V$
  profiles for upflow features (A--T) indicated in
  Fig.~\ref{fig:zoommosaic2}. Both Stokes $I$ and $V$ are normalized to the continuum intensity, the wavelength of which is outside the range plotted. Wavelengths along the x-axis are in units of m{\AA}.}
\label{fig:fitobs2}
\end{figure*}

From the \vlos\ map in Fig.~\ref{fig:zoommosaic2} it is clear that there are several small-scale
upflows often surrounded by ring-shaped downflows. In order to study
these we randomly identified and selected strong upflows using the
\vlos\ map. These upflow features are marked with circles labeled A--T.

Features A--D, N and R are all small scale upflows occurring in
regions where $|\Blos|$ is in the range 10--230~G. The upflows, based
primarily on the Stokes $I$ profiles, for all these cases are fairly
strong, ranging from $-900$~\mps\ to $-1.4$~k\mps. The surrounding
downflows range between 550~\mps\ and 1~k\mps. In cases A, B, D, N and
R we see a weakening in the LOS magnetic field at the location of the
upflow. In C we see the \Blos\ drop off, the reason being its location
at the interface of a strong field and a field-free region. From the
corresponding $\Ic$ image it appears that A, B, D, and R are
probably granular fragments. N and C seem to be located in
intergranular lanes but it is difficult to label these features from
the $\Ic$ image. In the \Imin\ map, N and C appear completely dark
and they are obviously different from the brighter features A, B and D
in this respect. We note, that for some of these features (especially
B, but also N and R), the observed Stokes $V$ profiles appear strongly
redshifted with respect to the fitted profiles. For feature B, this
redshift is approximately 2~k\mps\ such that the Stokes V profile
actually indicates a downflow of about 500~\mps. The very weak Stokes
$V$ profile measured, peaking at less than 0.7\%, indicates the
possibility that we actually measure polarized straylight from the
neighboring magnetic feature.

Features F--H, J--L and S represent small-scale upflows in regions of
where $|\Blos|$ is in the range 60~G to 1.1~kG. For these features we
see upflows ranging from $-500$~\mps\ to $-1.0$~k\mps. The surrounding
downflows range from 200~\mps\ to 1.1~k\mps. In all of these cases,
with the exception of H, we see that the peaking of \vlos\ coincides
with peaking of \Blos. F appears to be part of the ribbon R0 of
Fig.~\ref{fig:zoommosaic} and G a part of flower F0. J is part of a
flower-like feature and L is a part of a ribbon-like feature. K
happens to be a part of O0 feature of Fig.~\ref{fig:zoommosaic}. H
seems to be a fragment of a granule which is adjacent to a ribbon.

Features marked E, I, M, O and P correspond to upflows in regions of
$|\Blos|$ ranging from 5~G to 650~G. The upflows range from
$-530$~\mps\ to $-1.45$~k\mps. The surrounding downflows are in the
range 550~\mps\ to 700~\mps. E, I, O, and M appear to be fragments of
granules. P appears to be an isolated bright point with an upflow at
its center and a strong downflow of 1.4~k\mps\ at its boundary. The
field peaks at the center where the upflow is located.

Finally we have feature like Q and T which are located in very weak
field regions with \Blos\ ranging between 0 and 40 G. The upflows
range between $-550$~\mps\ and $-900$~\mps. The surrounding downflows
are in the range 550~\mps\ to 1.0~k\mps. These are most likely
granular fragments.

From this we have gained additional support for flowers and ribbons
being related to small-scale upflows. Fig.~\ref{fig:zoommosaic2} shows
the observed and fitted stokes profiles for A--T.

\begin{figure*}[!htbp]
\subfigure[A ]{
\includegraphics[bb=11 47 453 323,clip,width=0.24\linewidth]{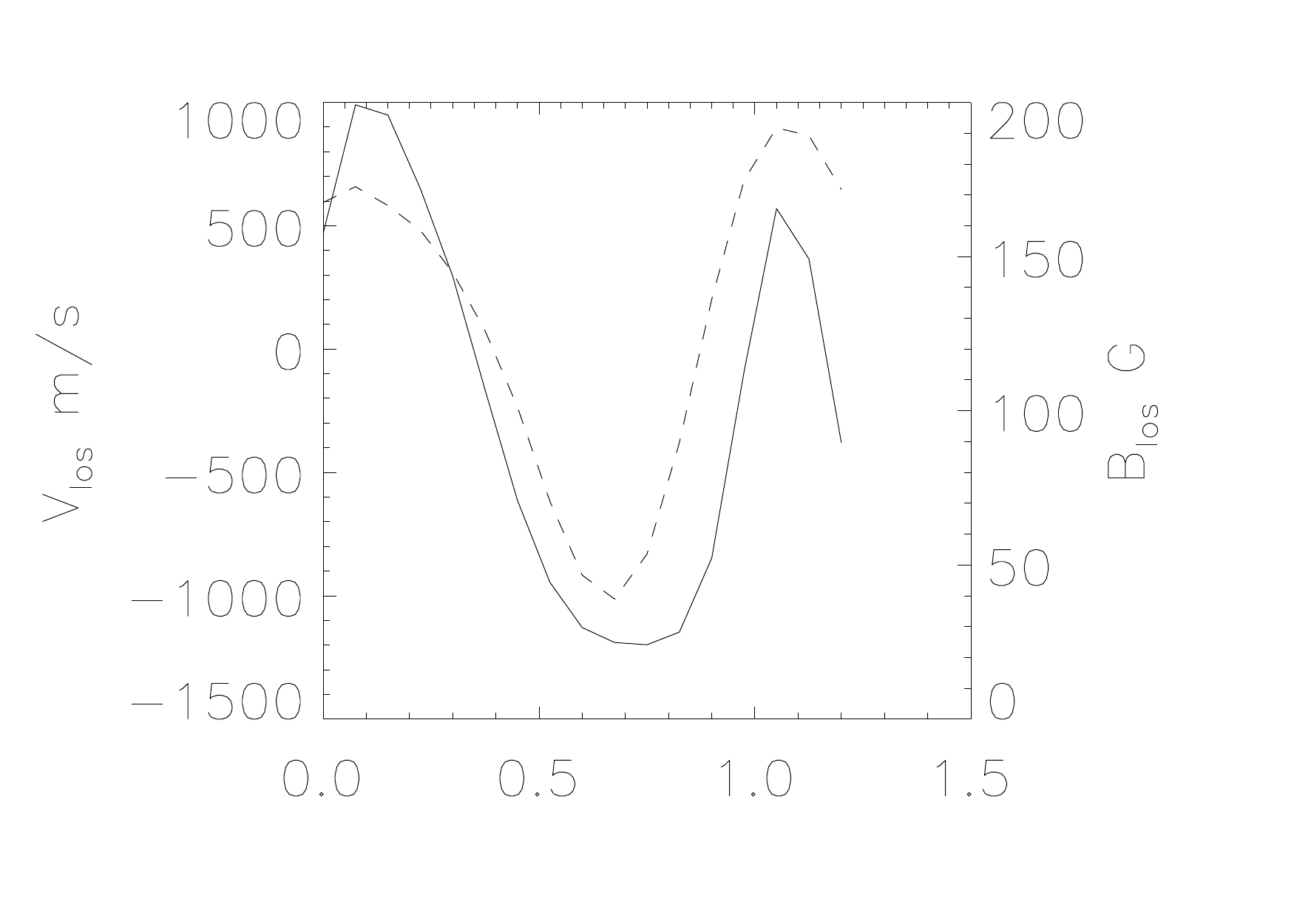}}
\subfigure[B ]{
\includegraphics[bb=11 47 453 323,clip,width=0.24\linewidth]{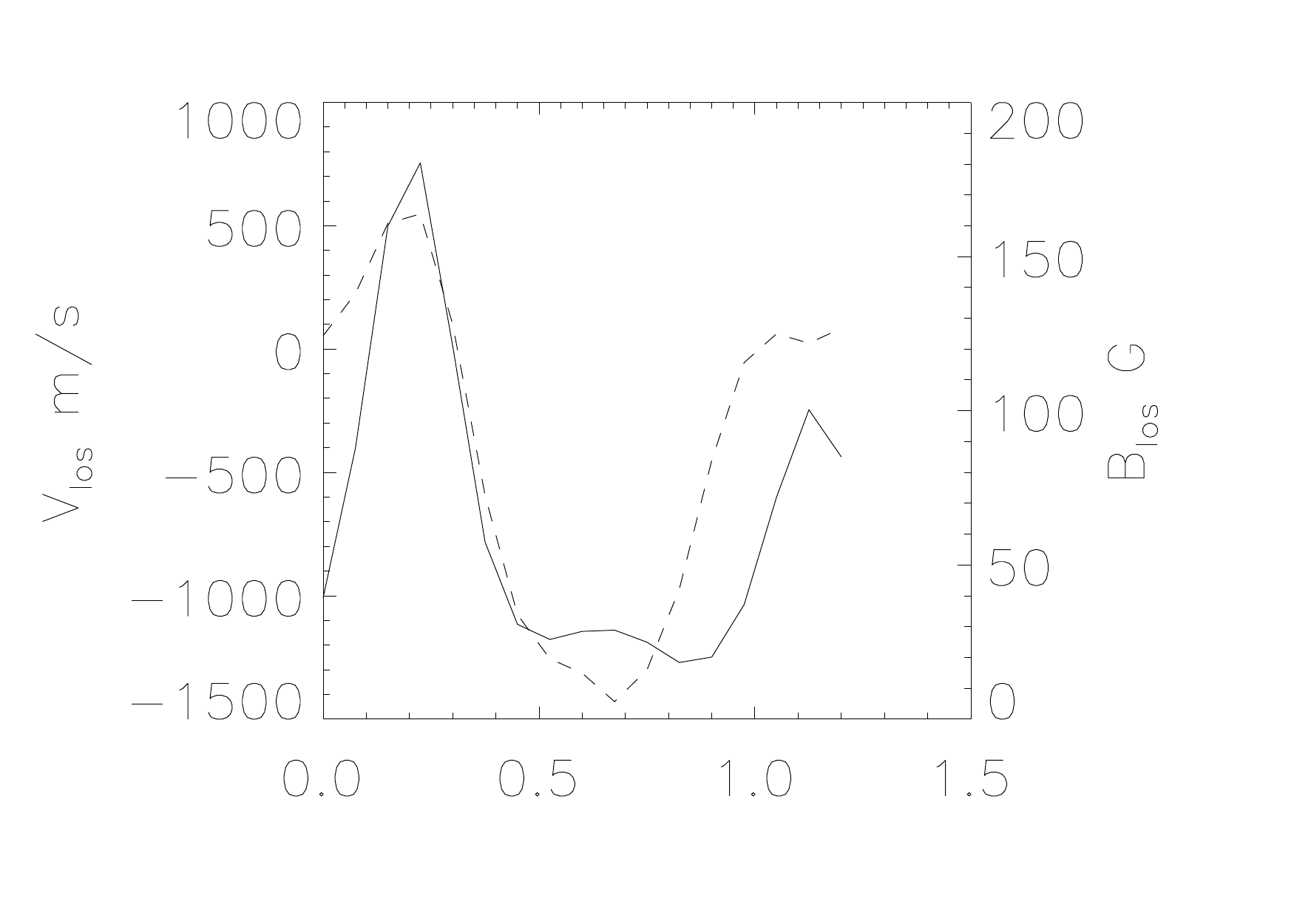}}
\subfigure[C]{
\includegraphics[bb=11 47 453 323,clip,width=0.24\linewidth]{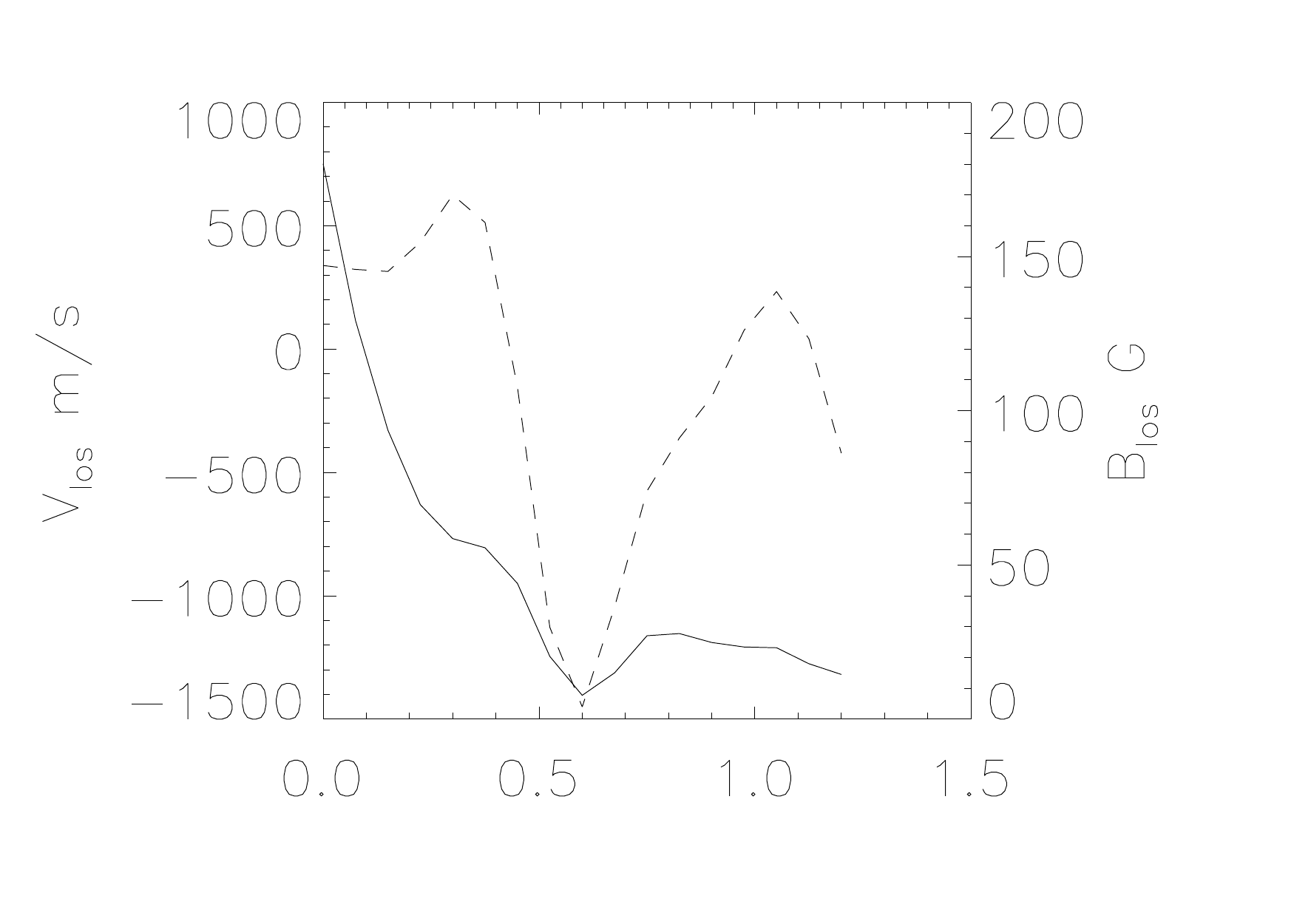}}
\subfigure[D ]{
\includegraphics[bb=11 47 453 323,clip,width=0.24\linewidth]{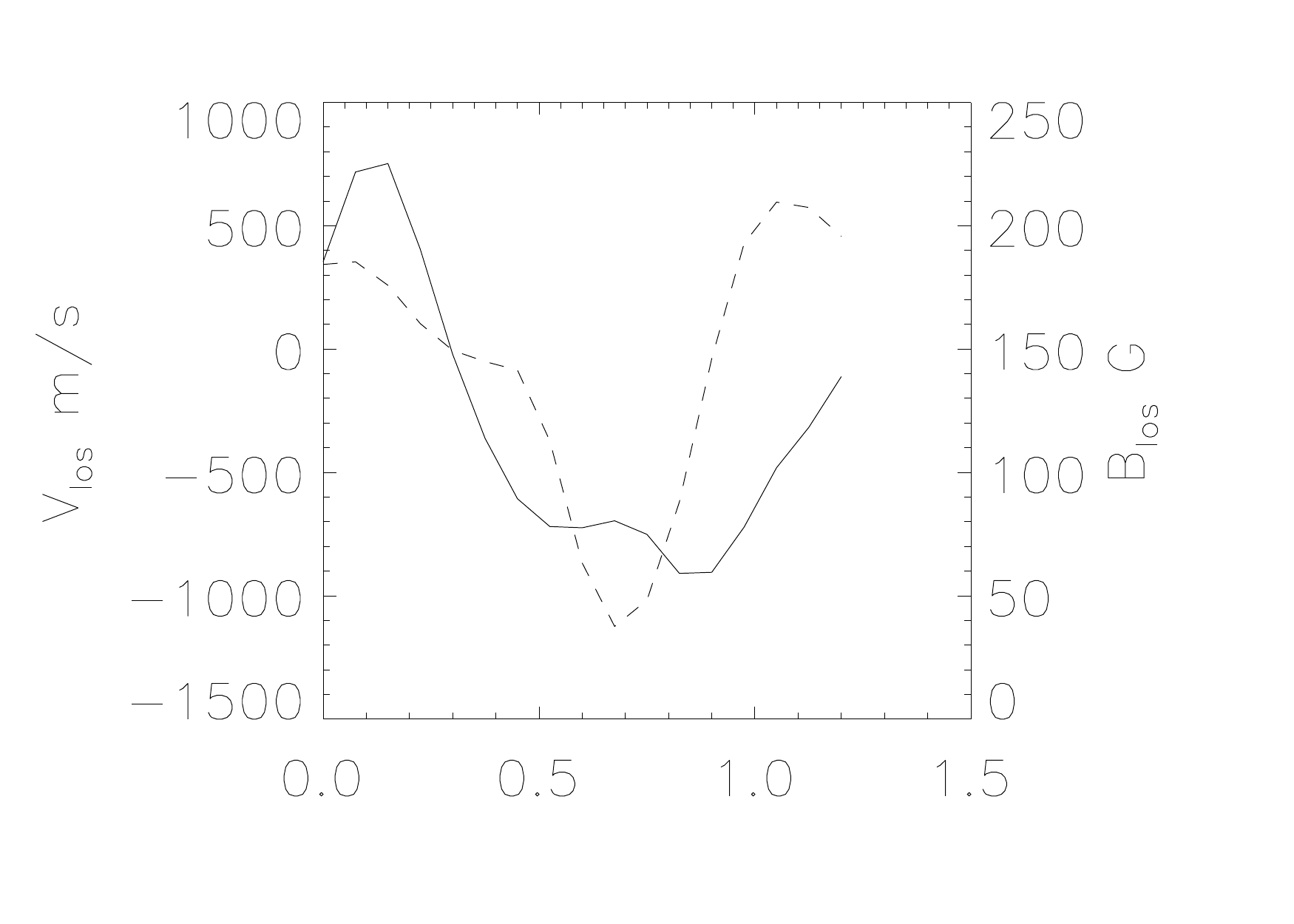}}
\subfigure[E ]{
\includegraphics[bb=11 47 453 323,clip,width=0.24\linewidth]{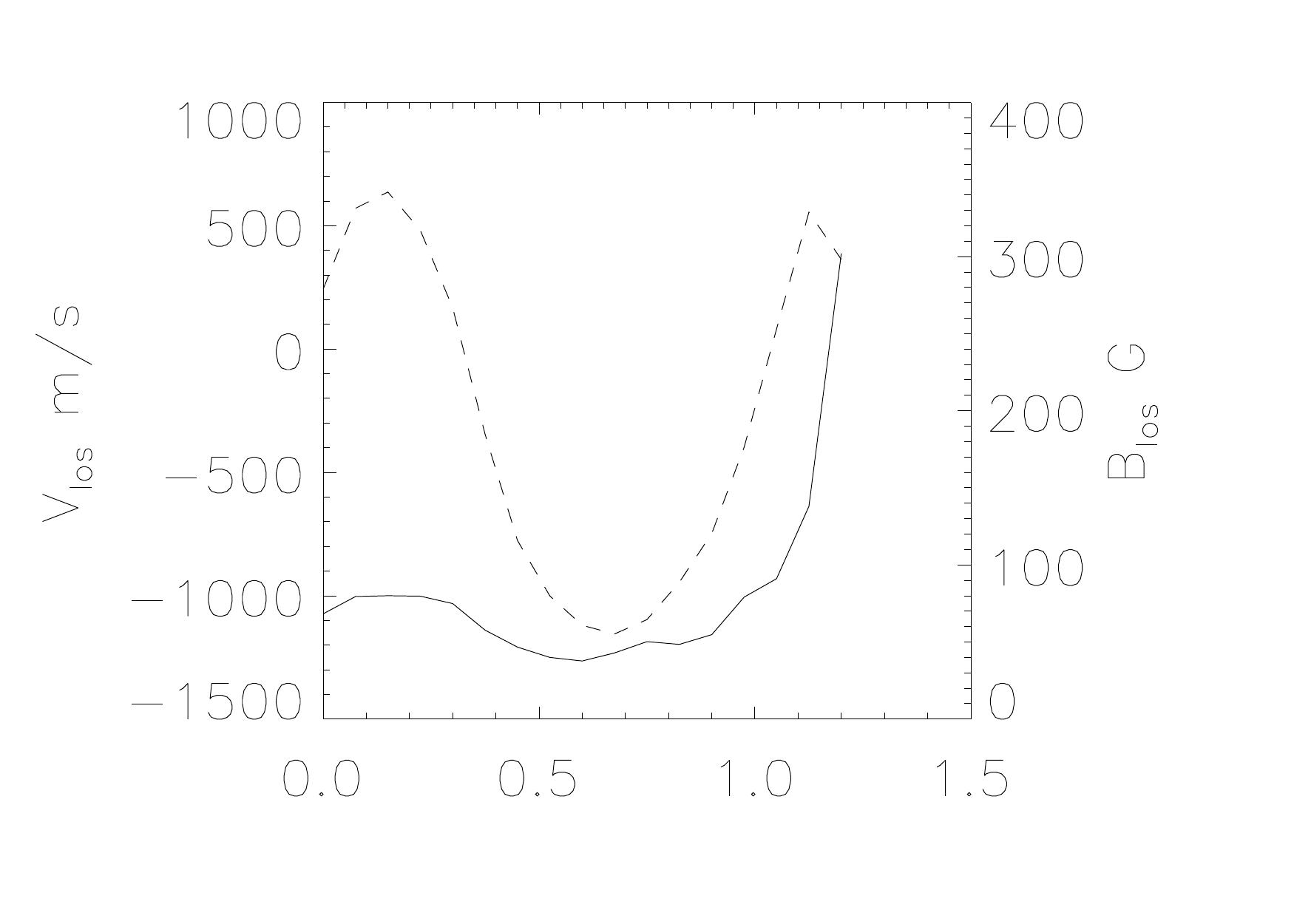}}
\subfigure[ F]{
\includegraphics[bb=11 47 453 323,clip,width=0.24\linewidth]{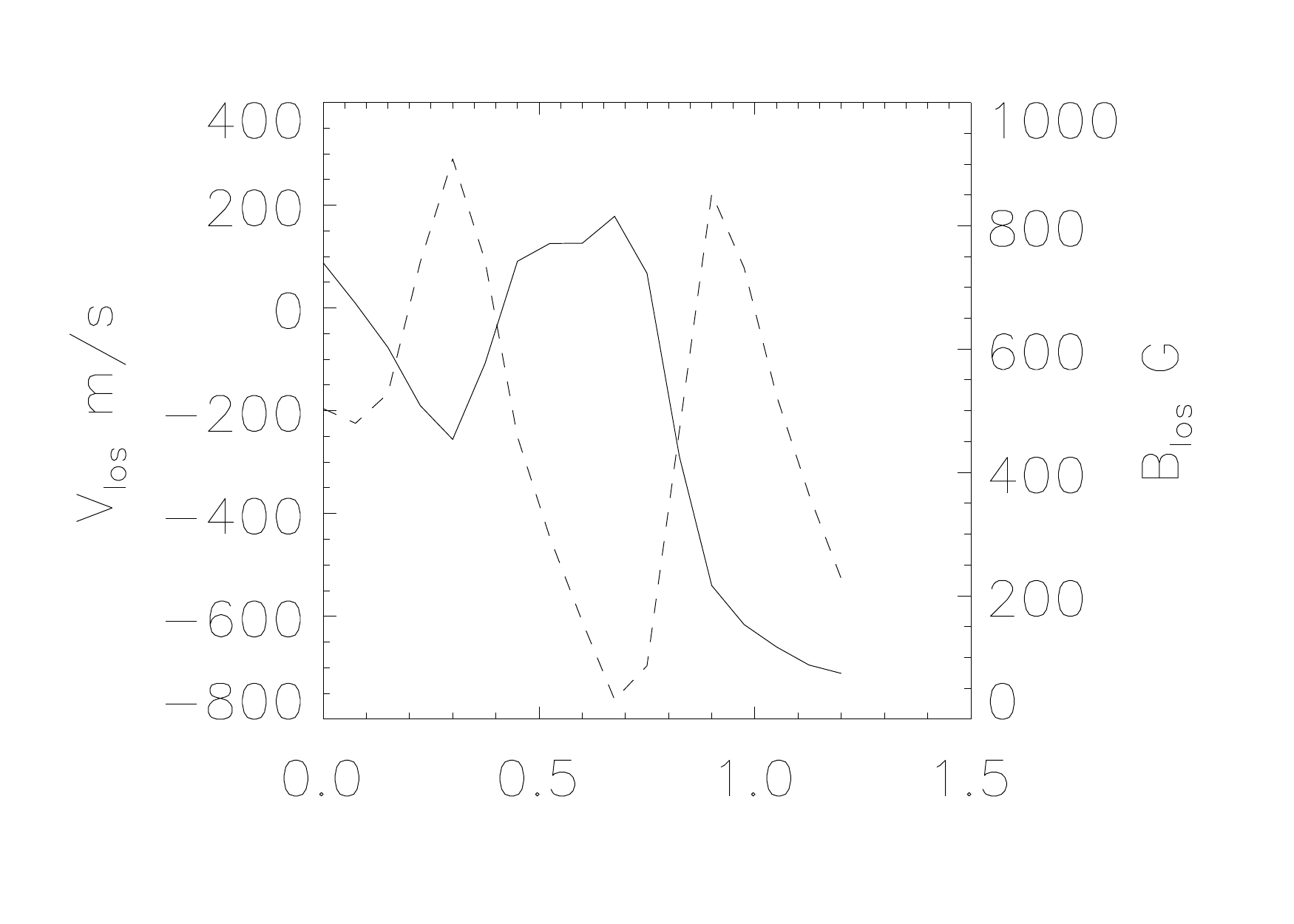}}
\subfigure[G ]{
\includegraphics[bb=11 47 453 323,clip,width=0.24\linewidth]{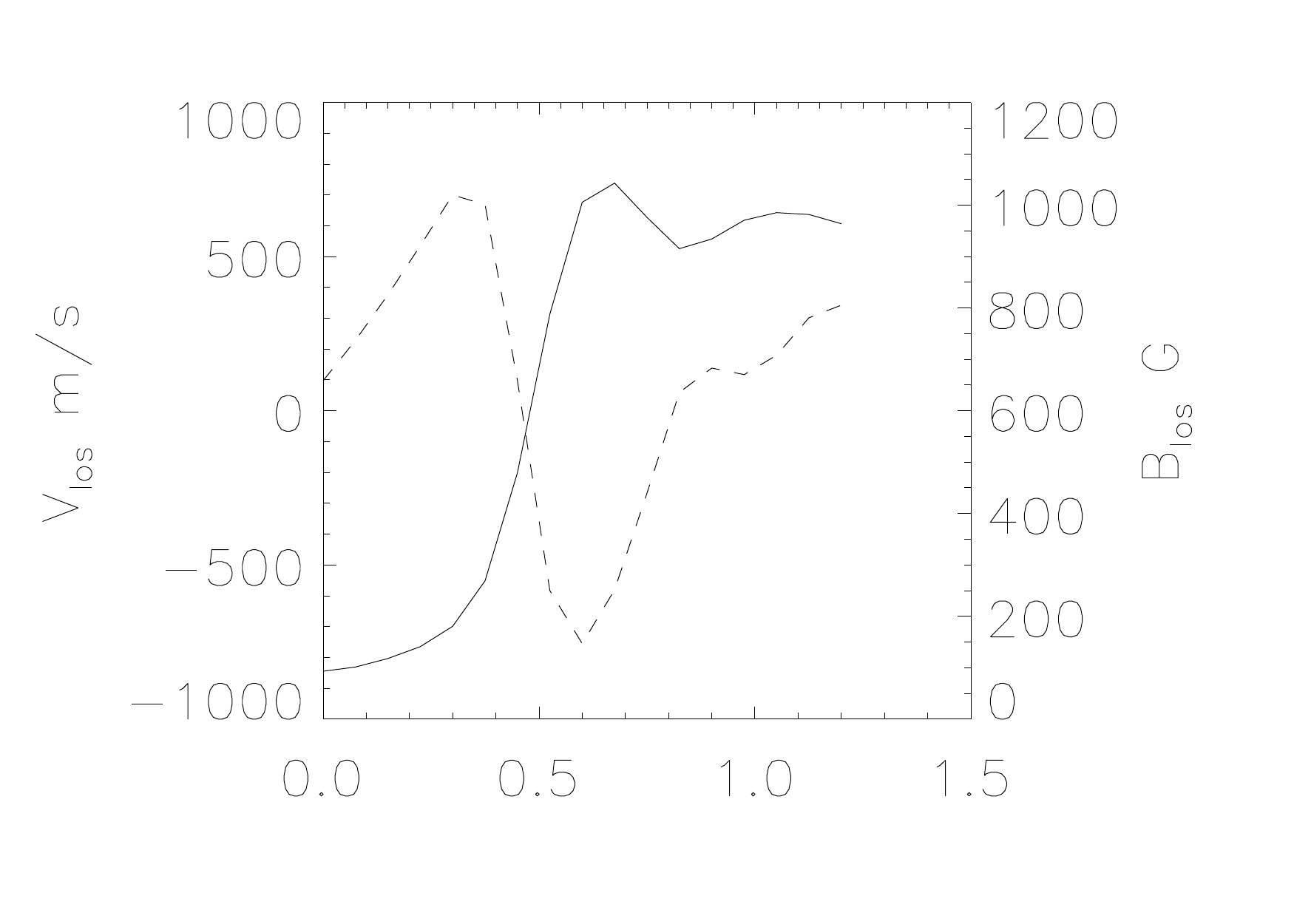}}
\subfigure[ H]{
\includegraphics[bb=11 47 453 323,clip,width=0.24\linewidth]{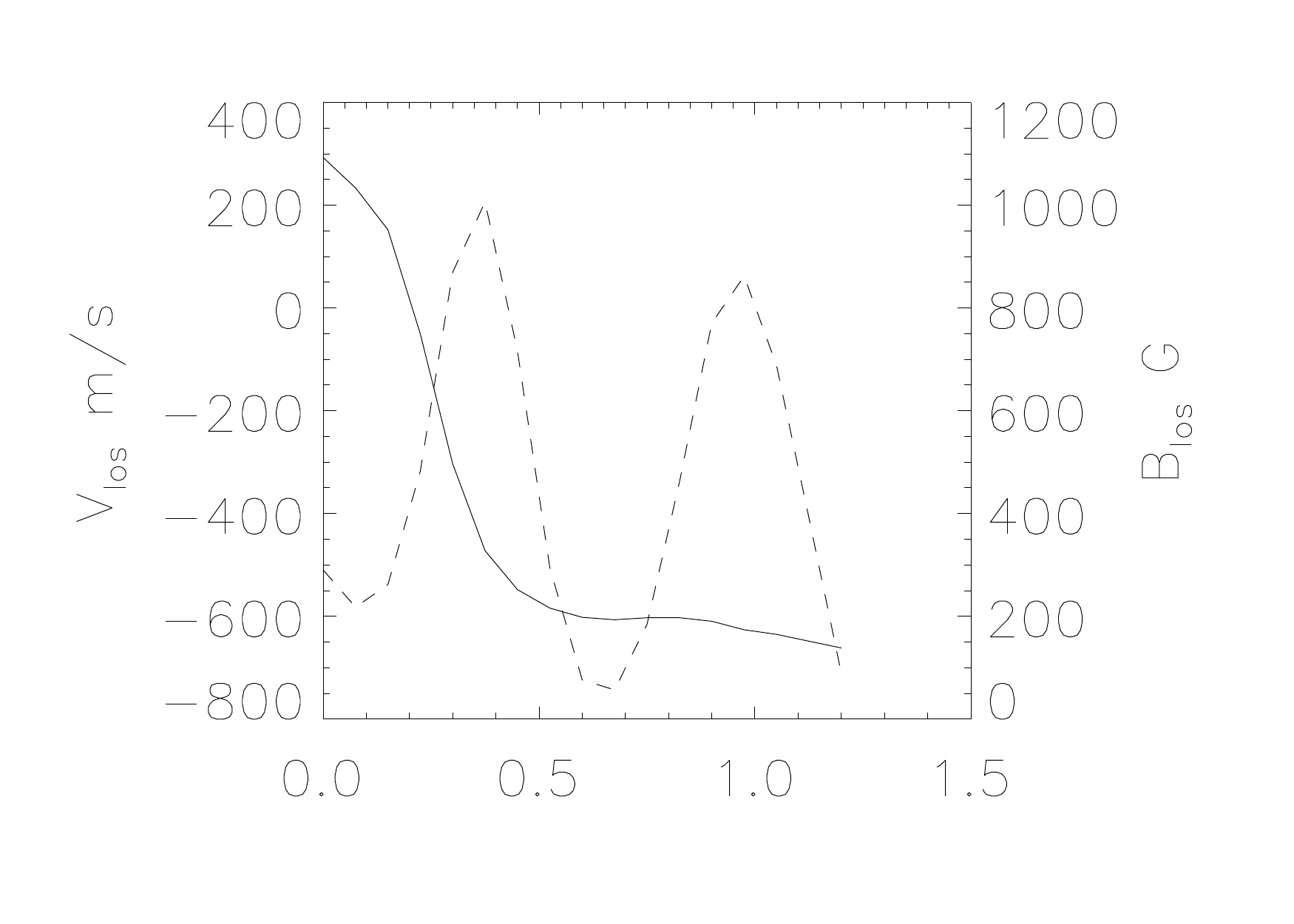}}
\subfigure[ I]{
\includegraphics[bb=11 47 453 323,clip,width=0.24\linewidth]{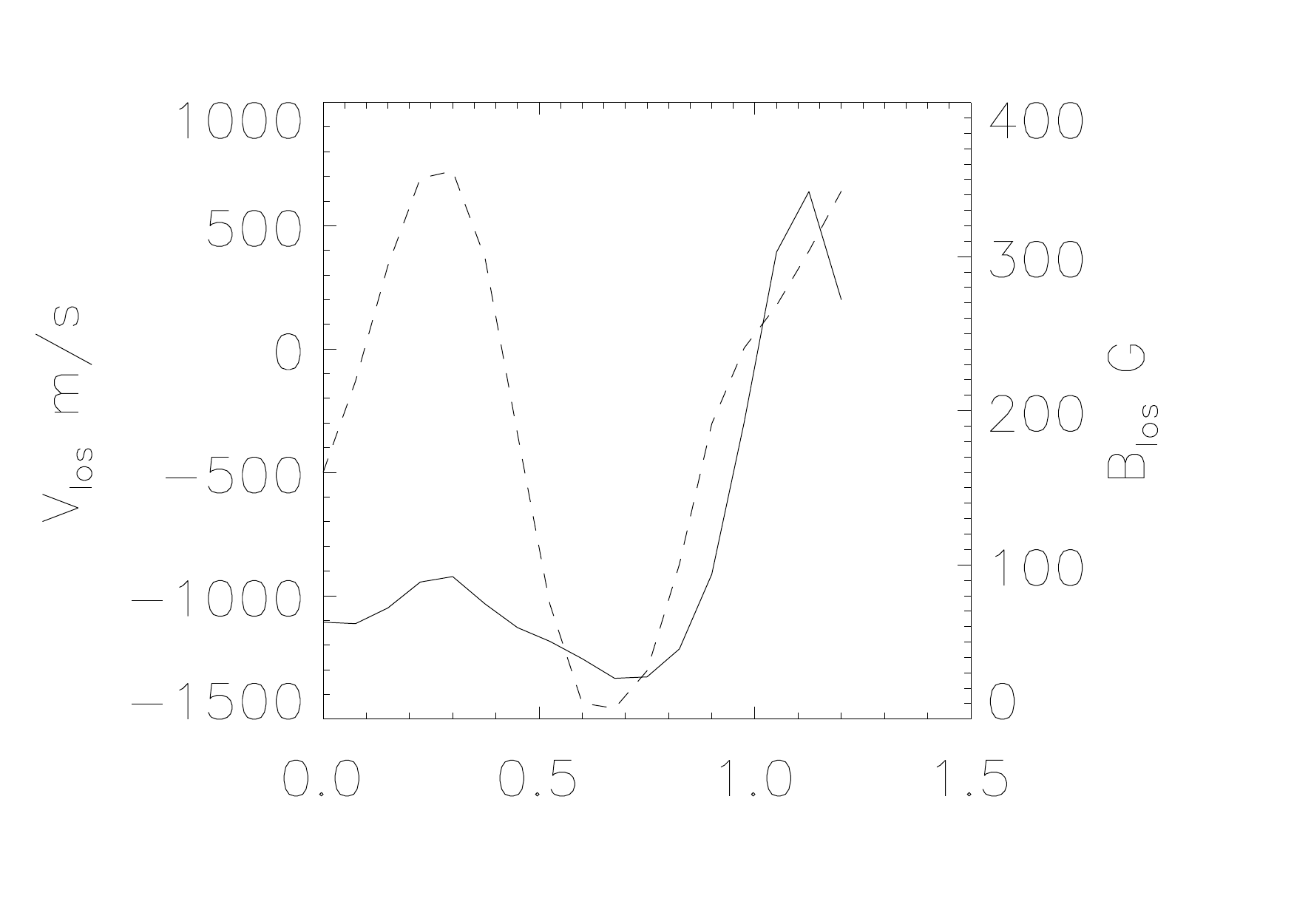}}
\subfigure[ J]{
\includegraphics[bb=11 47 453 323,clip,width=0.24\linewidth]{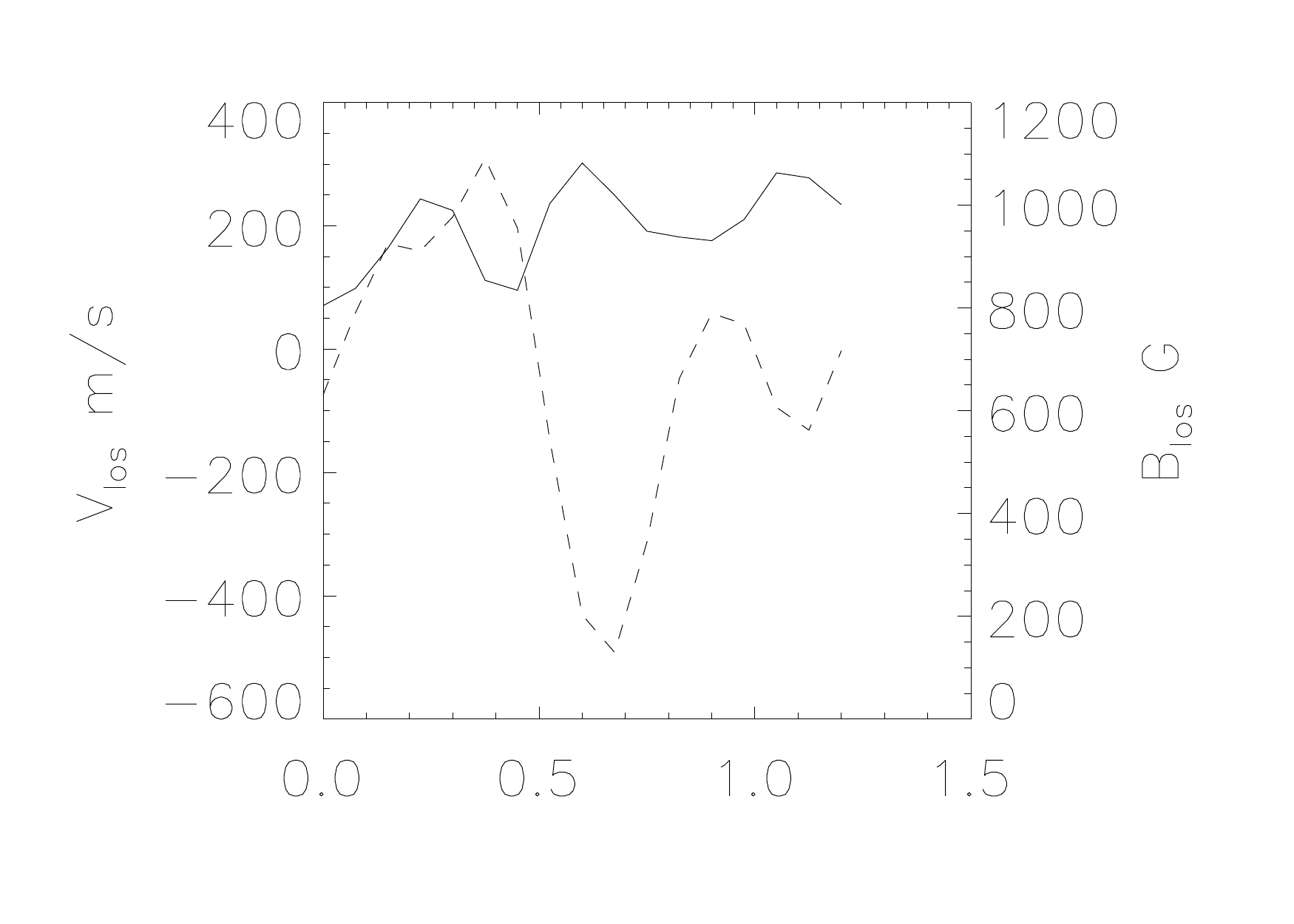}}
\subfigure[ K]{
\includegraphics[bb=11 47 453 323,clip,width=0.24\linewidth]{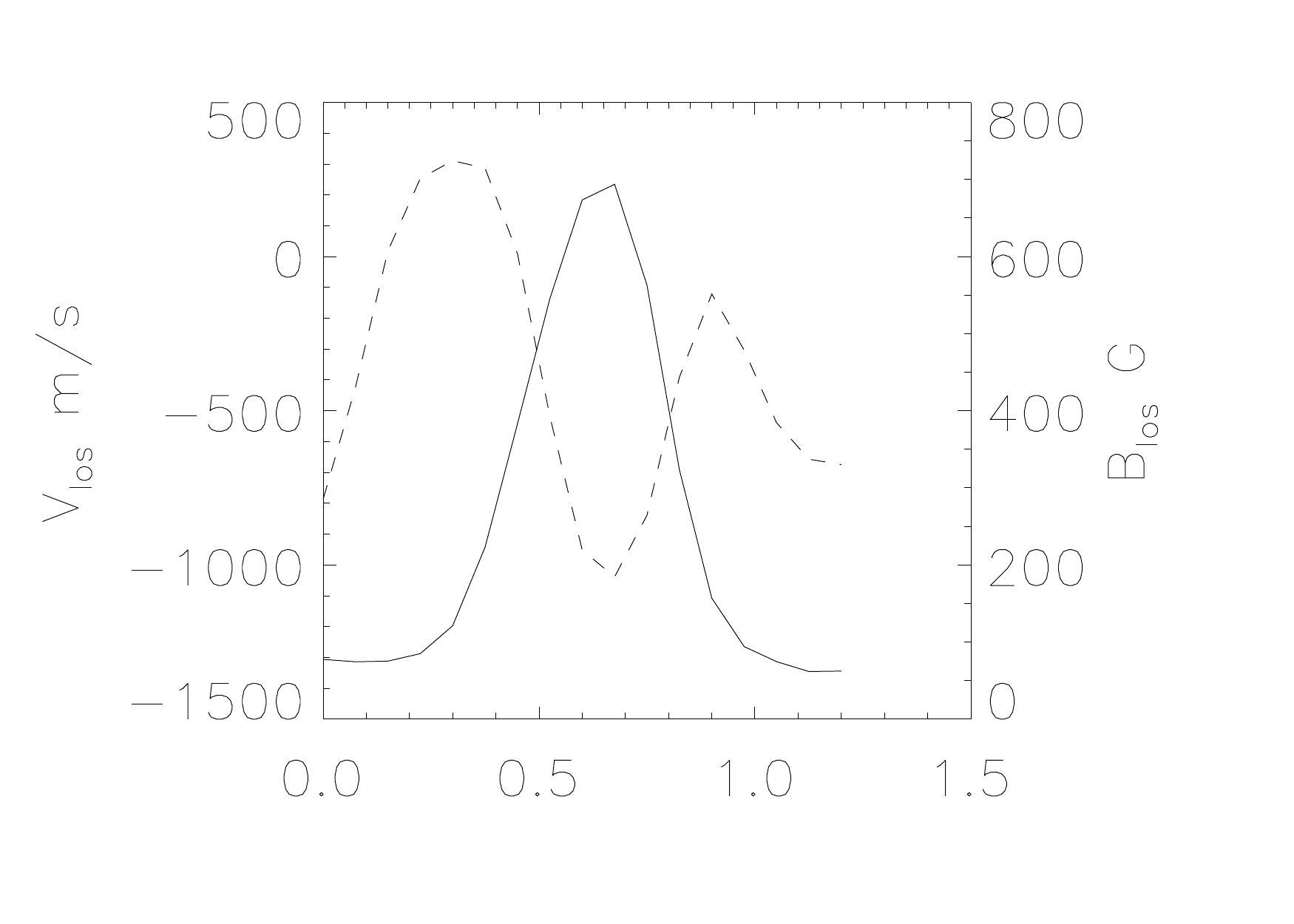}}
\subfigure[L]{
\includegraphics[bb=11 47 453 323,clip,width=0.24\linewidth]{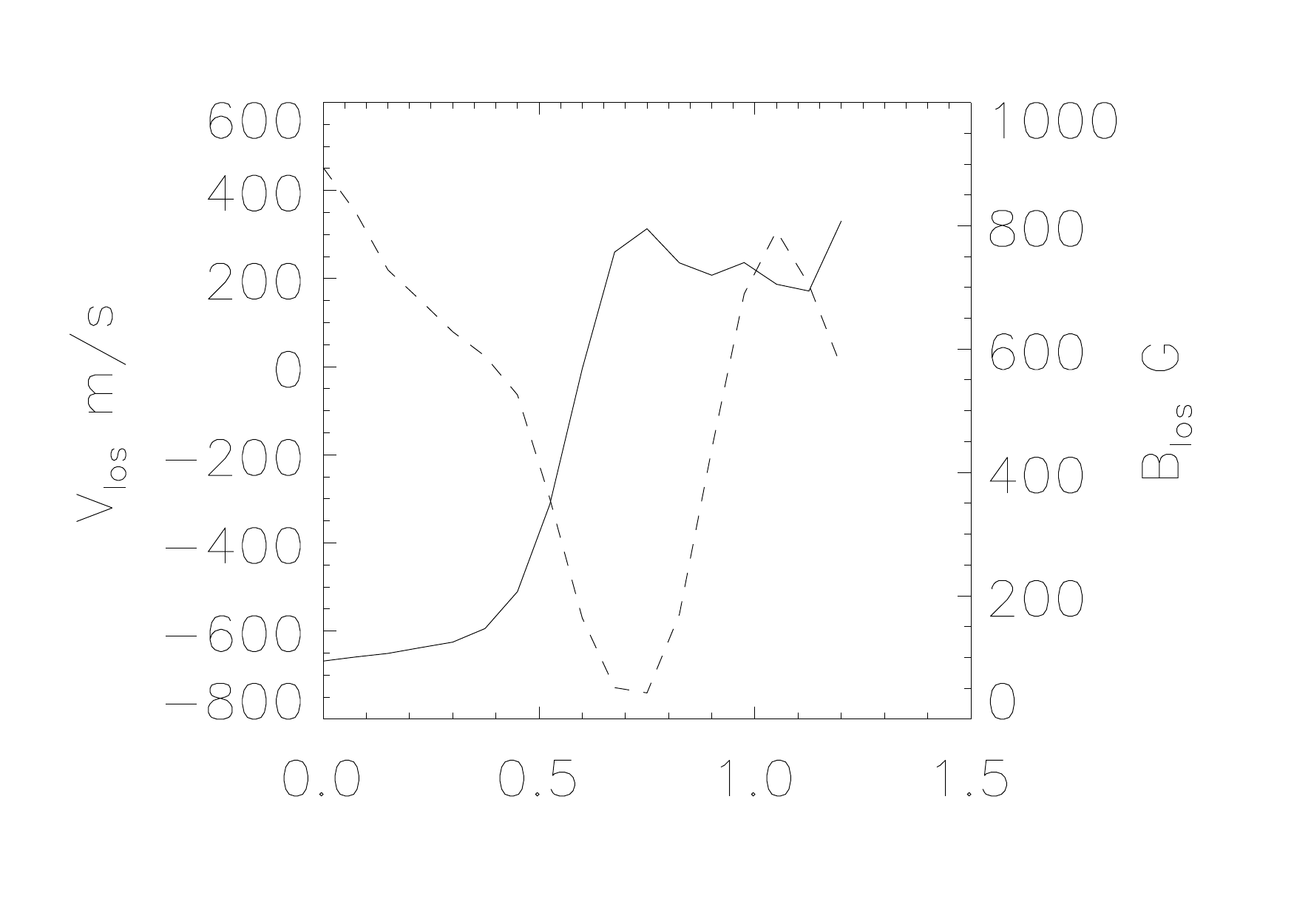}}
\subfigure[ M]{
\includegraphics[bb=11 47 453 323,clip,width=0.24\linewidth]{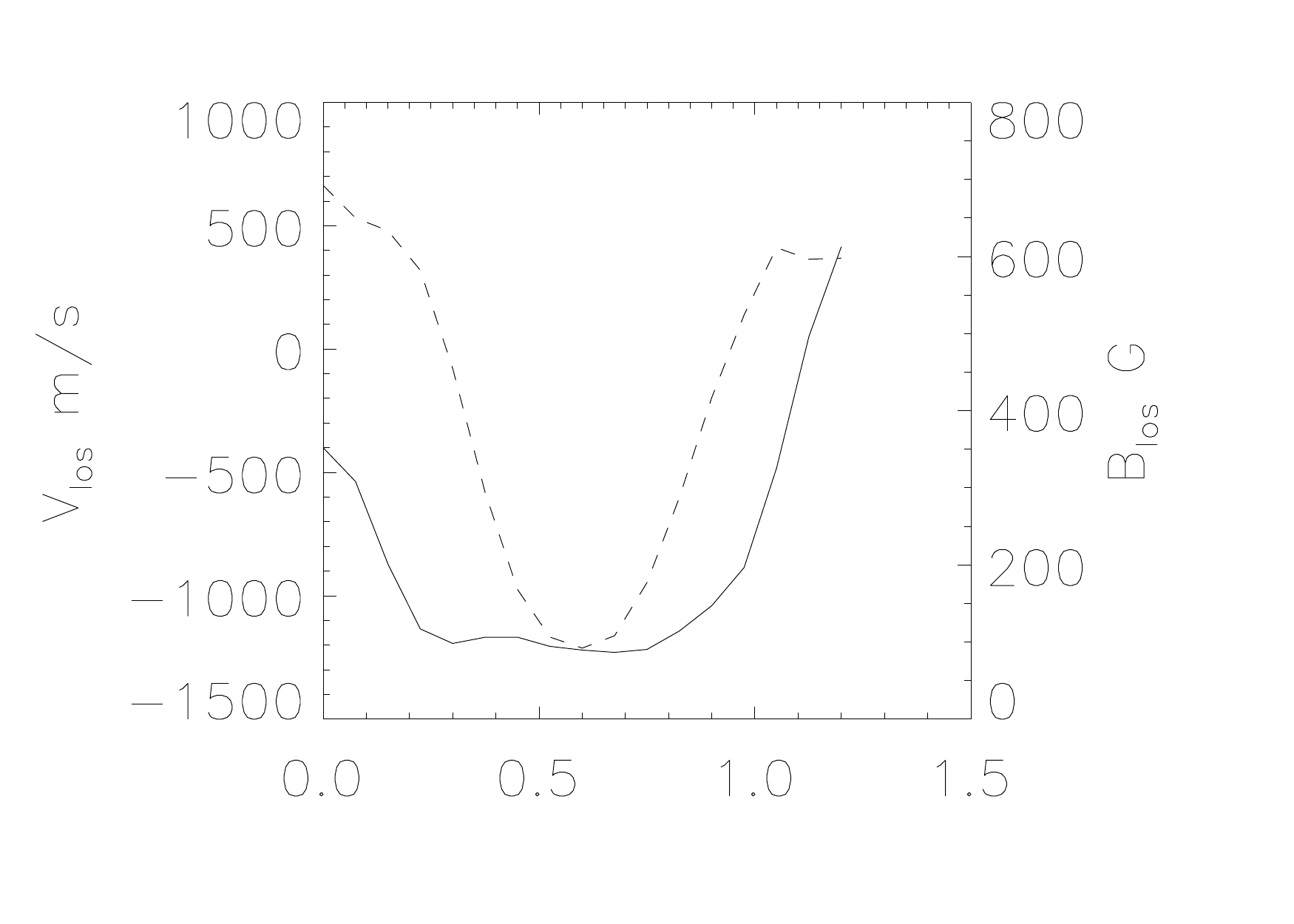}}
\subfigure[ N]{
\includegraphics[bb=11 47 453 323,clip,width=0.24\linewidth]{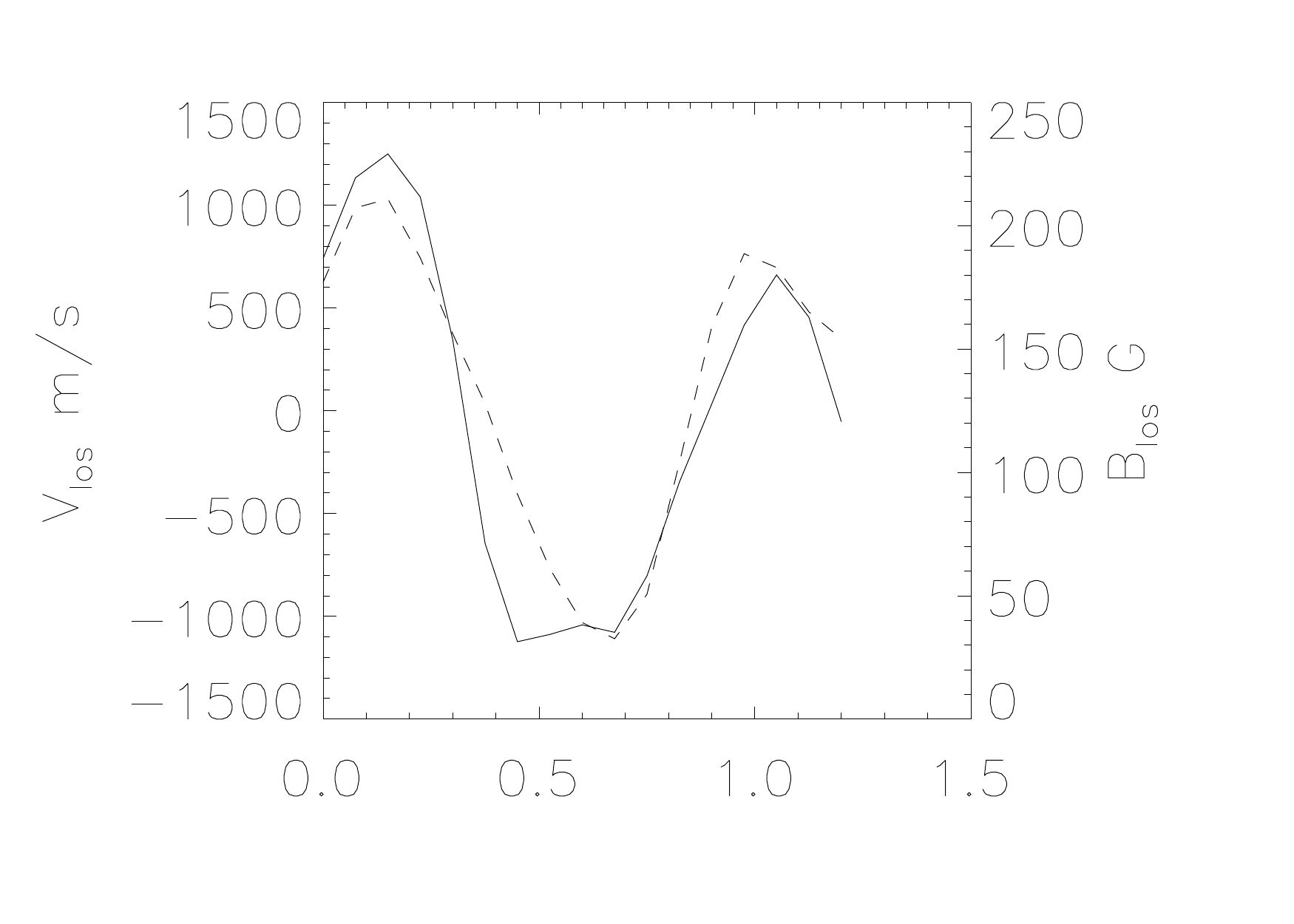}}
\subfigure[ O]{
\includegraphics[bb=11 47 453 323,clip,width=0.24\linewidth]{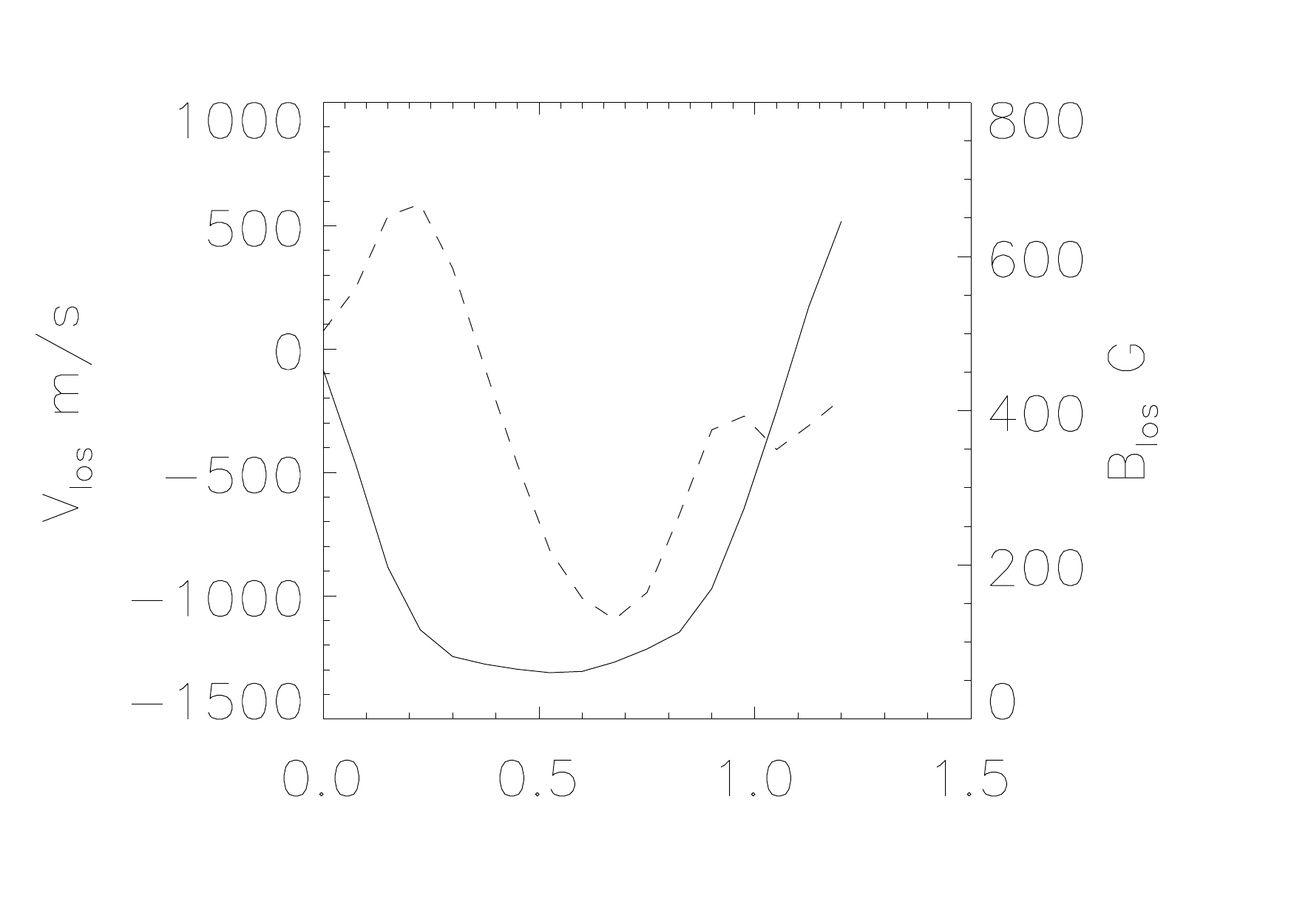}}
\subfigure[ P]{
\includegraphics[bb=11 47 453 323,clip,width=0.24\linewidth]{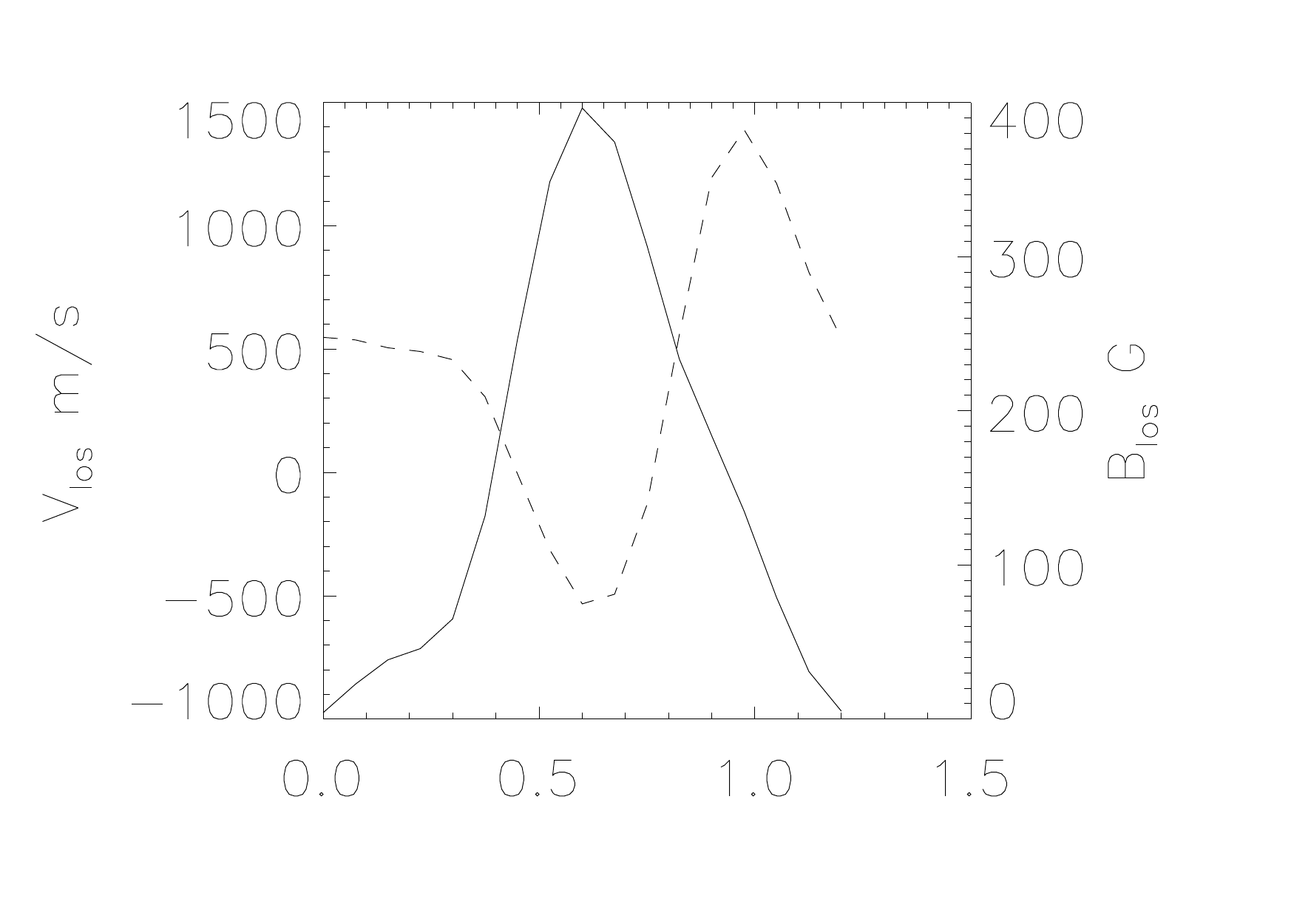}}
\subfigure[ Q]{
\includegraphics[bb=11 47 453 323,clip,width=0.24\linewidth]{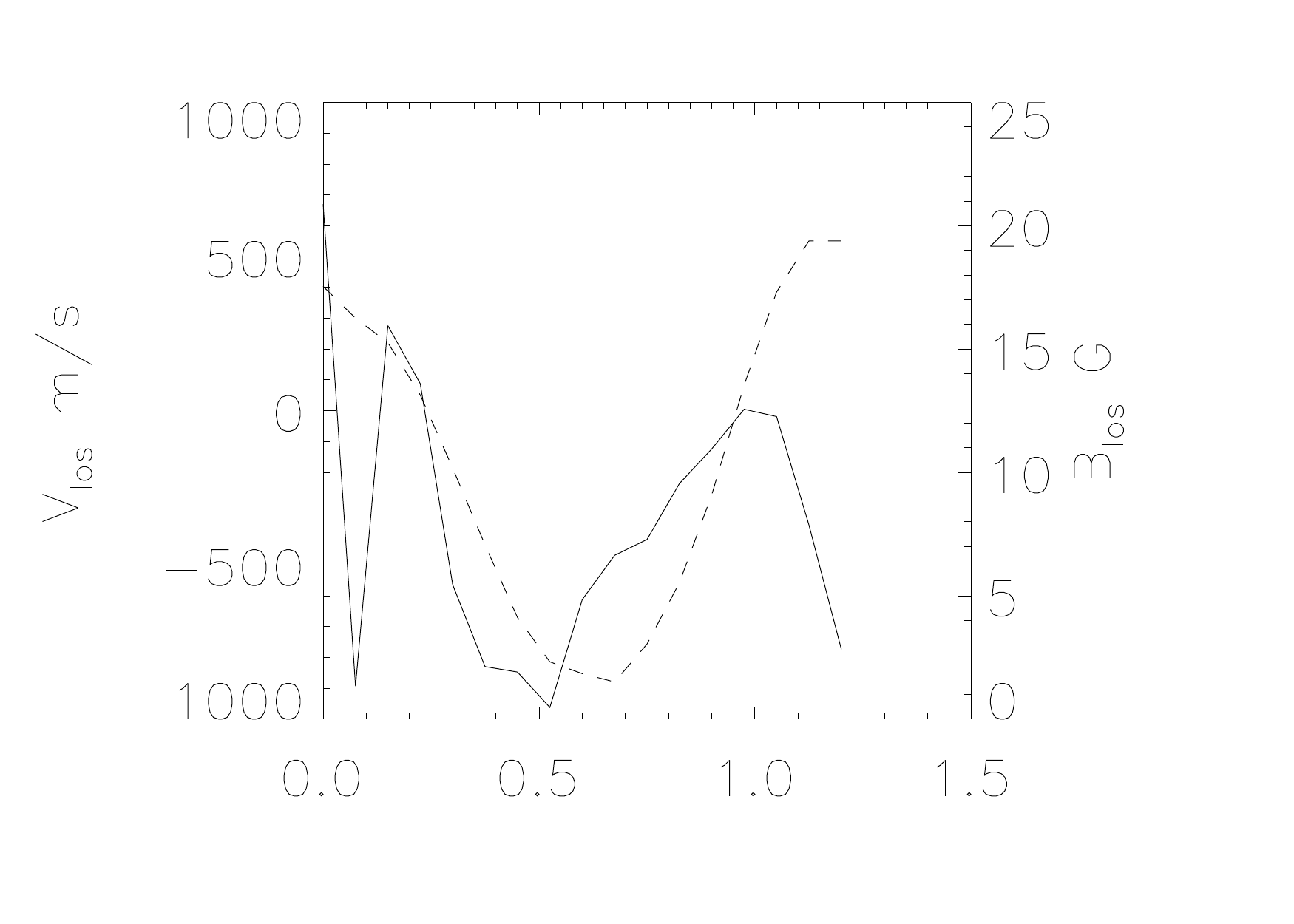}}
\subfigure[ R]{
\includegraphics[bb=11 47 453 323,clip,width=0.24\linewidth]{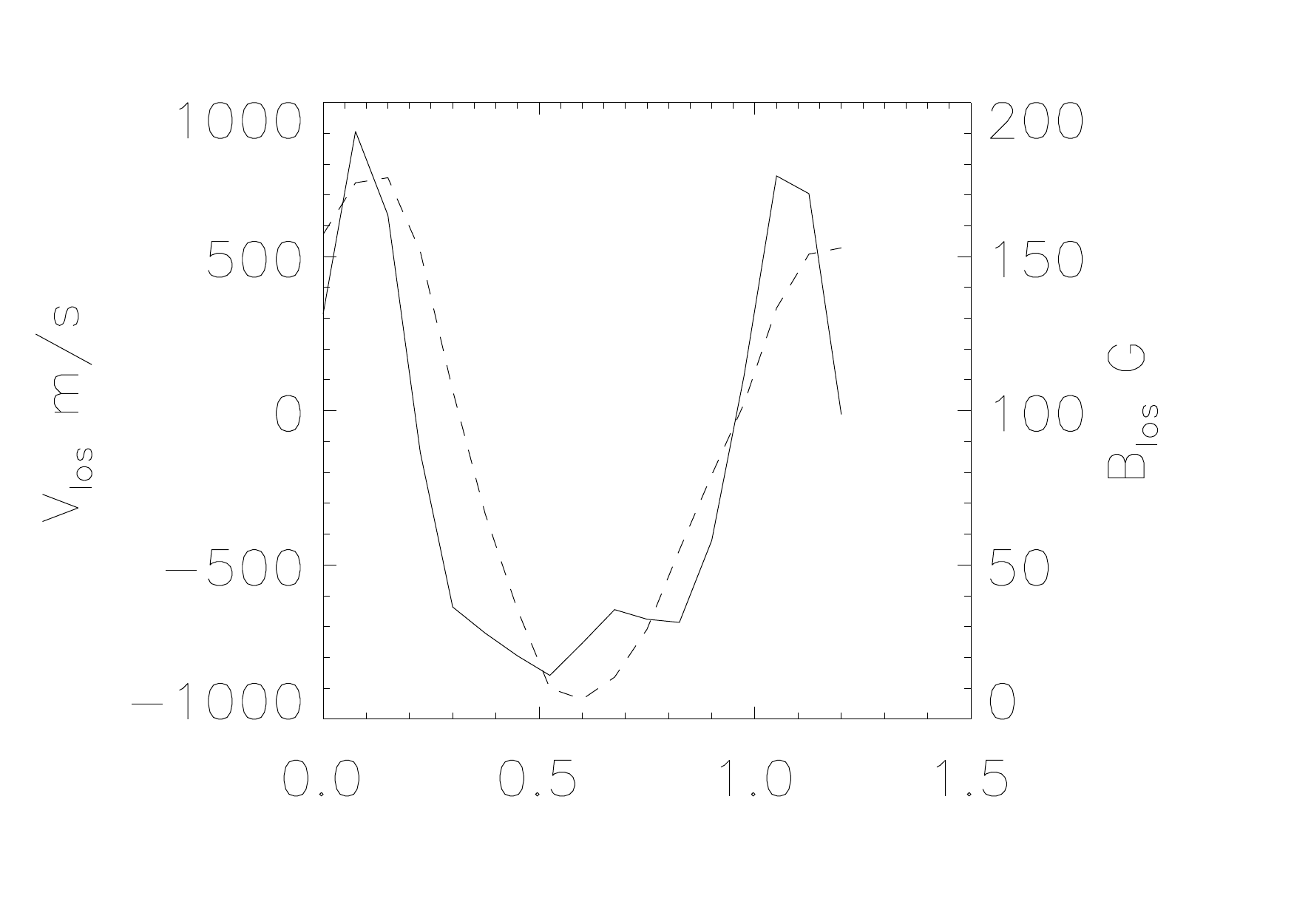}}
\subfigure[ S]{
\includegraphics[bb=11 47 453 323,clip,width=0.24\linewidth]{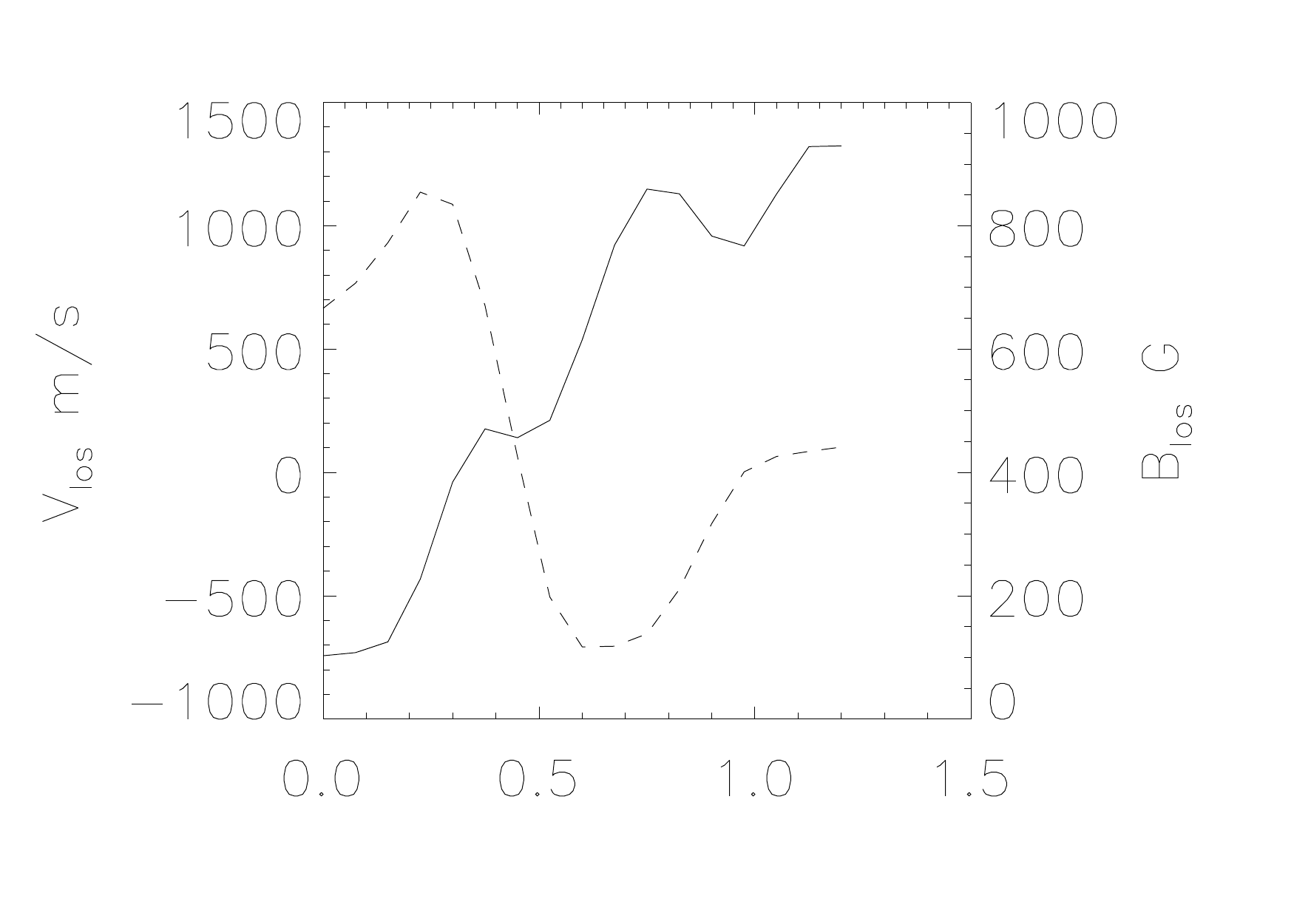}}
\subfigure[ T]{
\includegraphics[bb=11 47 453 323,clip,width=0.24\linewidth]{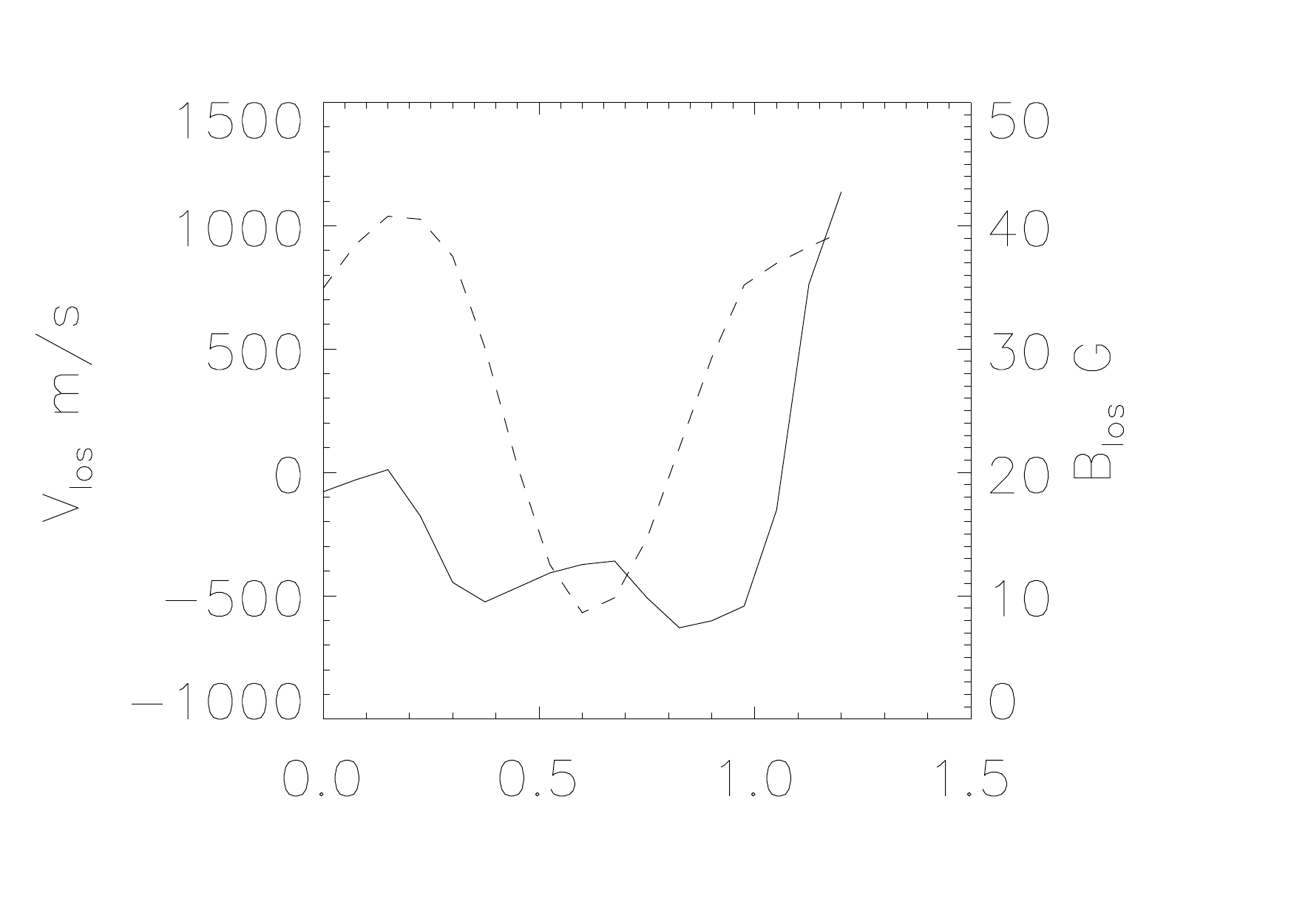}}
\caption{Horizontal cuts of LOS velocity (dashed) and strength of LOS
  magnetic field $|\Blos|$ (solid) for upflow features (A--T)
  indicated in Fig.~\ref{fig:zoommosaic2}. Distances along the x-axis are in units of arcsec.}
\label{fig:vlos-blos2}
\end{figure*}

\subsubsection{Discussion}
\label{sec:discussion}

From these plots we conclude that the various features we have
described have specific signatures in \Blos\ and \vlos. Ribbons and
flowers seem to possess strong upflows while isolated bright points
are associated with strong downflows. Strings, on the other hand seem
to show downflows of intermediate strength. We have shown only a few
examples for each type here but have examined also several other
features in the data. Although some exceptions can be found, the
differences in flow patterns described for these features appear
systematic. We have also found various small-scale distinct features
which do not fit into any of the categories defined previously.

There have been a few recent studies of magnetic upflows and downflows
in magnetic elements at 0\farcs1--0\farcs2 resolution by
\citet{berger04solar} and \citet{rouppe05solar}. Upflows were found
within a few ribbons studied by \citet{rouppe05solar}.
\citet{langangen07velocities} made a spectroscopic study of magnetic
features such as flowers and found weak upflows of about 150~\mps\ in
their centers, similar to what we find here.
\citet{2004ApJ...604..906R} studied magnetic fine structure in a
region close to disk center based on magnetograms in the blue wing of
the \ion{Fe}{i}~630.25~nm and LOS velocities estimated from
filtergrams in the blue and red wings of the weak \ion{C}{i} line at
538~nm and the \ion{Fe}{i} line at 557.6~nm. Without making the same
detailed distinction between different types of magnetic structure as
made here, he found downflows in the range a few hundred \mps\ up to 1~k\mps, located at the edges of small flux
concentrations and chains of bright points. He also found that the
downflows appear significantly narrower and faster at deeper layers
and very weak flows within the flux concentrations at heights
corresponding to the 630.25 and 557.6~nm lines. 
The average velocities within the bright structures were found to be close to zero at higher layers and a few hundred \mps\ in the deeper layers.

The ubiquitous presence of downflows near the boundaries of strong
fields found in the present study appear consistent with the
simulations of \citet{2005A&A...429..335V}, showing stronger such
downflows. In other simulations made at different average field
strengths \citep{2004RvMA...17...69V}, velocity maps were made for
heights corresponding to $\tau_{500}=1$. The morphology of our velocity maps look like a mixture of the simulated velocity maps obtained with average field strengths of $B=400$~G and $B=800$~G.

\section{Conclusion}
\label{sec:conclusion}
In this paper, we analyze spectropolarimetric SST/CRISP data, obtained at a spatial resolution close to 0\farcs15, from a unipolar ephemeral region with small pores close to sun center. We use MOMFBD methods \citep{lofdahl02multi-frame,noort05solar} to restore and align the polarimetric images recorded at different wavelengths and the Milne--Eddington inversion code Helix \citep{2004A&A...414.1109L} to retrieve LOS velocities and the LOS component of the magnetic field. These inversions were made with the magnetic filling factor set to unity, such that LOS magnetic fields obtained and discussed correspond to measured flux densities.

Using this data, we investigate signatures of magneto-convection
associated with strong average magnetic fields. We use a 200~G
threshold for the LOS magnetic field to define such regions of strong field but emphasize that the average field strength within them is approximately 600--800~G, the upper limit taking into account a likely underestimate of the field strength due to our neglect of spatial straylight. This data corresponds to \emph{unipolar strong field over extended contiguous regions} of typically 2\arcsec\ width or more across the smallest dimension. The field strength within a larger 6$\times$6~Mm box, used in MHD simulations relevant to the present work, reaches up to at least 600~G. We discuss statistical relations between measured intensities, LOS velocities and LOS magnetic fields. In addition, we identify specific structures such as pores, bright points, strings, ribbons and flowers \citep{berger04solar} and discuss their LOS velocity and magnetic field signatures. We compare our results to earlier analysis of similar data and to 3D MHD simulations.

Within the strong flux concentrations, we find a small-scale granular velocity pattern resembling that of field-free granules but with scales about 4 times smaller. Based on a zero-point for LOS velocities determined by assuming the pores to be at rest, this small-scale velocity field averages to nearly zero (70~\mps) and mostly corresponds to upflows at the centers of magnetic ``granules'', surrounded by downflows. These LOS velocities correlate, though weakly, with the continuum intensity such that upflows are on the average brighter than downflows. The weak correlation between continuum intensity and LOS velocity found within the strong flux concentrations, compared to what is the case for field-free convection, could be taken as evidence against the interpretation of a convective origin for this velocity field. However, a convective flux below the visible surface is obviously needed to explain the high radiative energy flux in these extended (approximately 2\arcsec across the smallest dimension) magnetic structures. This suggests to us that the main reason for the weak correlation rather is that the nature of small-scale magneto-convection is distinctly different, in particular as regards leaving ``tell-tales'' of convection in the line-forming layers above the photosphere, from that of large-scale field-free convection. Given the uncertainty of an appropriate threshold to define the mask outlying this small-scale velocity pattern and the likely influence of straylight on our estimates of \Blos, we therefore conclude that there is a \emph{transition to small-scale magneto-convection} when the field covers a sufficiently large area and reaches an average strength of 600--800~G. The measured RMS velocity of this magneto-convection is 400--500~\mps\, compared to 700~\mps\ for the nearly field-free surrounding granulation. The measured RMS velocity decreases with increasing field strength for dark as well as bright structures. For $|\Blos|>400$~G, the signed average LOS velocity decreases systematically with increasing $|\Blos|$ such that dark and bright structures on the average show upflows when $|\Blos|>700$~G. The RMS continuum intensity variations within the mask are nearly as strong inside the mask as outside the mask.

The \emph{boundaries} of the magnetic regions, defined in terms of a LOS field of about 200~G, mark the transition between normal-looking granules and small-scale (abnormal) granulation. Here, we find predominantly downflows. In particular, there is a population of \emph{bright downflows} that are over 30 times more frequent than in a field-free area of the same size. At this magnetic boundary, the Stokes $V$ profiles are asymmetric with the blue peak stronger than the red peak, in agreement with synthetic $V$ profiles calculated from simulations with 250~G average field strength \citep{2007A&A...469..731S}. We identify this population with strong downflows occurring at the interfaces between strong magnetic field and field-free granules, seen in simulations also with strong (400--800~G) average field \citet{2004RvMA...17...69V}, reproducing the small-scale magnetic granulation pattern seen in our data. The similarity of the correlation plots shown in Fig.~\ref{fig:vlosplots}, with contributions mostly from extended regions of strong field, and those of \citet{2004ApJ...604..906R}, made from selected bright points and other small-scale flux concentrations, as well as similar plots made from 250~G simulations \citep{2007A&A...469..731S}, suggests that the dynamics close to their \emph{boundaries} is similar for small as well as large flux concentrations.

The weak field returned by Helix close to the boundaries of the flux concentrations can be interpreted as partly due to magnetic canopies expanding over the more field-free adjacent photosphere with normal-looking granules. The gradually weaker field away from the boundaries obtained from the inversions is then not only from polarized straylight but also an effect of the magnetopause gradually shifting to smaller optical depths away from the flux concentration, as discussed by e.g., \citet{1997ApJ...474..810M}.

A detailed comparison with our results and those from simulations as regards the \emph{interior} parts of flux concentrations, where the field is strong over extended regions, has not been possible. This requires the quantities returned by Helix to be compared to simulation data at an optical depth $\tau_{500}=0.01$--0.1, but presently only simulations with strong magnetic field have been analyzed at a height
corresponding to $\tau_{500}=1$.

We also compare our correlation plots between the LOS velocity and LOS magnetic field with measurements made at lower spatial resolution based on HINODE/SOT spectropolarimetric data \citep{2008A&A...481L..29M}. We found good qualitative agreement as to the variation of the average and RMS LOS velocity with field strength, but the RMS velocities measured with the present data are much higher than measured with Hinode. To some extent this difference can be attributed to differences in the methods used to measure LOS velocities -- Morinaga et al. use fits to the line core of the 630.15~nm line, whereas we use the less opaque 630.25~nm line and velocities fitted by Helix to the entire line profiles, representing deeper layers. It is however obvious that the higher spatial resolution of the SST is crucial for resolving these very small-scale magnetic velocity fields.

Using the line minimum intensity map, we identify bright points, strings, ribbons and flowers \citep{berger04solar} and discuss their LOS velocity and magnetic field signatures. We find preferred upflows in ribbons and flowers while isolated bright points and strings mostly show downflows. This is in good agreement with earlier studies, e.g. \citep{rouppe05solar} and \citep{langangen07velocities}. Most of these specific structures identified are located within the 200~G mask used to outline the small-scale magneto-convection pattern seen in the LOS velocity map; they are thus part of that convection.

The data presented clearly demonstrate the importance of high spatial resolution: the convection features shown in the present SST/CRISP data are barely resolved and would be difficult, if not impossible, to resolve with a significantly smaller telescope. A somewhat surprising shortcoming is the lack of simulation data with strong average magnetic fields (400--800~G) and properties extracted from heights relevant to the formation of the 630.25~nm iron line. The simulation results discussed by \citet{2004RvMA...17...69V} are presented only for $\tau_{500}=1$. Such future comparisons obviously are desirable and will need to take into account the limited spatial resolution as well as straylight of the telescope used to record the data. We finally note that the Stokes $Q$ and $U$ profiles, although mostly weak, show up clearly at some locations, allowing additional consistency tests such as investigations of their asymmetries relative to those of the Stokes $V$ profiles \cite[Appendix B]{1997ApJ...474..810M}.

\begin{acknowledgements}
  The authors acknowledge A. Lagg for his efforts to adopt Helix to CRISP data, for his aid in installing Helix and for training one of us (GN) in its use. Mats L{\"o}fdahl is thanked for help with the data reduction and comments on the manuscript. Vasco Henriques and Pit S{\"u}tterlin are acknowledged for assistance during the observations and Dan Kiselman and Pit S{\"u}tterlin for comments on an early version of the manuscript. {\AA}ke Nordlund is thanked for several valuable comments, in particular relating to Sect.~3.2. The Swedish 1-m Solar Telescope is operated on the island of La Palma by the Institute for Solar Physics of the Royal Swedish Academy of Sciences in the Spanish Observatorio del Roque de los Muchachos of the Instituto de Astrof\'isica de Canarias.
\end{acknowledgements}

\end{document}